\documentclass[aps,
twocolumn,
a4paper,showkeys,footinbib,longbibliography]{revtex4-1}
\usepackage[utf8]{inputenc}
\usepackage{filecontents}
\usepackage{natbib}
\usepackage{amsmath,amssymb,bm,amsthm}
\usepackage{subfigure}
\usepackage{graphicx}
\usepackage{dcolumn}
\usepackage{bm}
\usepackage[mathlines]{lineno}
\usepackage{booktabs}
\usepackage{color}
\newcounter{one}
\usepackage{url}

\usepackage{textcase}

\usepackage{colortbl}
\usepackage{tabularx}
\usepackage{verbatim}
\usepackage{multirow}

\usepackage[T1]{fontenc} 
\usepackage{lmodern}
\usepackage{bbm}
\usepackage[utf8]{inputenc}
\usepackage{amsfonts}
\usepackage{array}

\usepackage{bbm}
\usepackage{enumitem}
\usepackage{umoline}
\usepackage[usenames,svgnames]{xcolor}
\usepackage{natbib}
\usepackage[hyperindex,breaklinks]{hyperref}
\hypersetup{
     colorlinks=true,       		
     linkcolor=Navy,          	
     citecolor=Navy,            
     filecolor=Navy,      		
     urlcolor=Navy,           	
    runcolor=cyan,
 }
\usepackage{graphicx}
\usepackage{amsfonts}
\usepackage{amssymb}
\usepackage{amsmath}
\usepackage{bbm}
\usepackage{enumitem}

\newcommand{\bra}[1]{\langle #1 |}
\newcommand{\ket}[1]{| #1 \rangle}

\newcommand{\tr}[0]{ {\rm tr}}

\newcommand{\half}[1]{{ \rm h}}
\newcommand{\Oorderof}{\mathcal{O}}
\newcommand{\orderof}[1]{\Oorderof(#1)} 
 
\newcommand{\tOrder}[0]{\tilde{\mathcal{O}}} 

\newcommand{\for}[0]{\quad \textrm{for} \quad}

\newcommand{\dist}{d}

\newcommand{\co}{{\rm c}}

\newcommand{\const}{{\rm const}}

\newcommand{\diam}{{\rm diam}}

\newcommand{\poly}{{\rm poly}}

\newcommand{\ad}{{\rm ad}}

\def\beq{\begin{equation}}
\def\eeq{\end{equation}}
\def\nbeq{\begin{equation*}}
\def\neeq{\end{equation*}}
\def\<{\langle}
\def\>{\rangle}

\def\tr{{\rm tr}}

\newtheorem{theorem}{Theorem}

\newtheorem{lemma}[theorem]{Lemma}
\newtheorem{corol}[theorem]{Corollary}
 
\newtheorem{definition}{Definition}  
\newtheorem{prop}[theorem]{Proposition} 
\newtheorem{claim}[theorem]{Claim}

\newcommand{\func}{\mathcal{F}}
\newcommand{\gunc}{\mathcal{G}}

\newcommand{\locd}{D_{\rm loc}}

\newcommand{\M}{q}

\newcommand{\br}[1]{\left(#1\right)}

\newcommand{\ignore}[1]{{}}

\usepackage{mathtools}
\def\multiset#1#2{\ensuremath{\left(\kern-.3em\left(\genfrac{}{}{0pt}{}{#1}{#2}\right)\kern-.3em\right)}}

\newcommand{\sd}{d}
\newcommand{\spin}{\varsigma}

\begin{document}

\title{Improved thermal area law and quasi-linear time algorithm\\ for quantum Gibbs states}

\author{Tomotaka Kuwahara$^{1,2}$}
\email{tomotaka.kuwahara@riken.jp}
\author{\'Alvaro M. Alhambra$^{3}$}
\email{alvaro.alhambra@mpq.mpg.de}
\author{Anurag Anshu$^{4}$}
\email{anuraganshu@berkeley.edu}
\affiliation{$^{1}$
Mathematical Science Team, RIKEN Center for Advanced Intelligence Project (AIP),1-4-1 Nihonbashi, Chuo-ku, Tokyo 103-0027, Japan
}
\affiliation{$^{2}$
Interdisciplinary Theoretical \& Mathematical Sciences Program (iTHEMS) RIKEN 2-1, Hirosawa, Wako, Saitama 351-0198, Japan}

\affiliation{$^3$ Max-Planck-Institut fur Quantenoptik, D-85748 Garching, Germany}

\affiliation{$^{4}$ University of California, Berkeley, CA 94720, USA.}

\begin{abstract}


One of the most fundamental problems in quantum many-body physics is the characterization of correlations among thermal states.
Of particular relevance is the thermal area law, which justifies the tensor network approximations to thermal states with a bond dimension growing polynomially with the system size.
In the regime of sufficiently low temperatures, which is crucially important for practical applications, the existing techniques do not yield optimal bounds. Here, we propose a new thermal area law that holds for generic many-body systems on lattices. We improve the temperature dependence from the original $\orderof{\beta}$ to $\orderof{\beta^{2/3}}$ up to a logarithmic factor, thereby suggesting sub-ballistic propagation of entanglement by imaginary time evolution. This qualitatively differs from the real-time evolution which usually induces linear growth of entanglement. We also prove analogous bounds for the R\'enyi entanglement of purification and the entanglement of formation.
Our analysis is based on a polynomial approximation to the exponential function which provides a relationship between the imaginary-time evolution and random walks.
Moreover, for one-dimensional (1D) systems with $n$ spins, we prove that the Gibbs state is well-approximated by a matrix product operator with a sublinear bond dimension for $\beta=o(\log (n))$.
This proof allows us to rigorously establish, for the first time, a quasi-linear time classical algorithm for constructing an MPS representation of 1D quantum Gibbs states at arbitrary temperatures of $\beta = o(\log(n))$. Our new technical ingredient is a block decomposition of the Gibbs state, that bears resemblance to the decomposition of real-time evolution given by Haah et al., \href{https://ieeexplore.ieee.org/document/8555119}{FOCS'18}.
\end{abstract}

\maketitle


\section{Introduction}

\subsection{Background}

One of the most important challenges in quantum many-body physics is to understand their thermal equilibrium properties.
Recently, with the advent of large quantum simulators~\cite{bernien2017probing,zhang2017observation,king2018observation,PhysRevX.8.031022,arute2019quantum}, 
the size and controllability of quantum Gibbs states accessible for experiments have dramatically improved. 
In fact, recent experiments have even succeeded in implementing imaginary time evolution~\cite{motta2020determining}. 
These developments are of considerably interest for quantum computation because quantum Gibbs states play crucial roles in quantum machine learning~\cite{PhysRevX.8.021050,PhysRevA.96.062327,biamonte2017quantum,10.5555/3370185.3370188,PhysRevLett.122.020504,doi:10.1146/annurev-conmatphys-031119-050651,chia2019sampling,anshu2020sample} and quantum algorithms such as semidefinite program solvers (SDP)~\cite{8104077,brando_et_al:LIPIcs:2019:10603,vanApeldoorn2020quantumsdpsolvers}. 
Beyond quantum computation, understanding and characterizing quantum Gibbs states is relevant to many open problems in quantum statistical physics and condensed matter physics.
Thus, understanding \emph{i)} the nature of entanglement structures in quantum Gibbs states and \emph{ii)} their simulability via tensor network methods is of great interest.

It is now widely accepted that the area law plays a crucial role~\cite{RevModPhys.82.277,PhysRevLett.90.227902} in the characterization of low-temperature physics of many-body systems. 
This states that the entanglement entropy between two subsystems is at most as large as the size of their boundaries. A similar notion also applies to finite temperature systems.
Although a rigorous proof of the area law at the zero temperature appears to be a notoriously challenging problem~\cite{Hastings_2007,Aharonov_2011,PhysRevB.85.195145,arad2013area,brandao2013area,abrahamsen2019polynomial,anshu2019entanglement,kuwahara2019area}, an analogous area law at finite temperatures has been proved by Wolf et al.~\cite{PhysRevLett.100.070502} in a simple and elegant manner. The authors proved the following inequality:
 \begin{align}
I(L:R)_{\rho_\beta} \le 2 \beta \|H_{\partial L} \| \propto \beta |\partial L| \label{mutual_information_wolf},
\end{align}
with $\|\cdots\|$ being the operator norm and $\partial L$ being the surface region of $L$,
where $I(L:R)_{\rho_\beta}$ is the mutual information between the subsets $L$ and $R$ (see Eq.~\eqref{mutual_information_def} below) and $H_{\partial L} $ denotes the boundary interaction Hamiltonian. The upper bound~\eqref{mutual_information_wolf} roughly denotes that the correlations between two complementary regions is concentrated around a distance $\orderof{\beta}$  of their boundary.

\begin{figure}
\centering
{
\includegraphics[clip, scale=0.3]{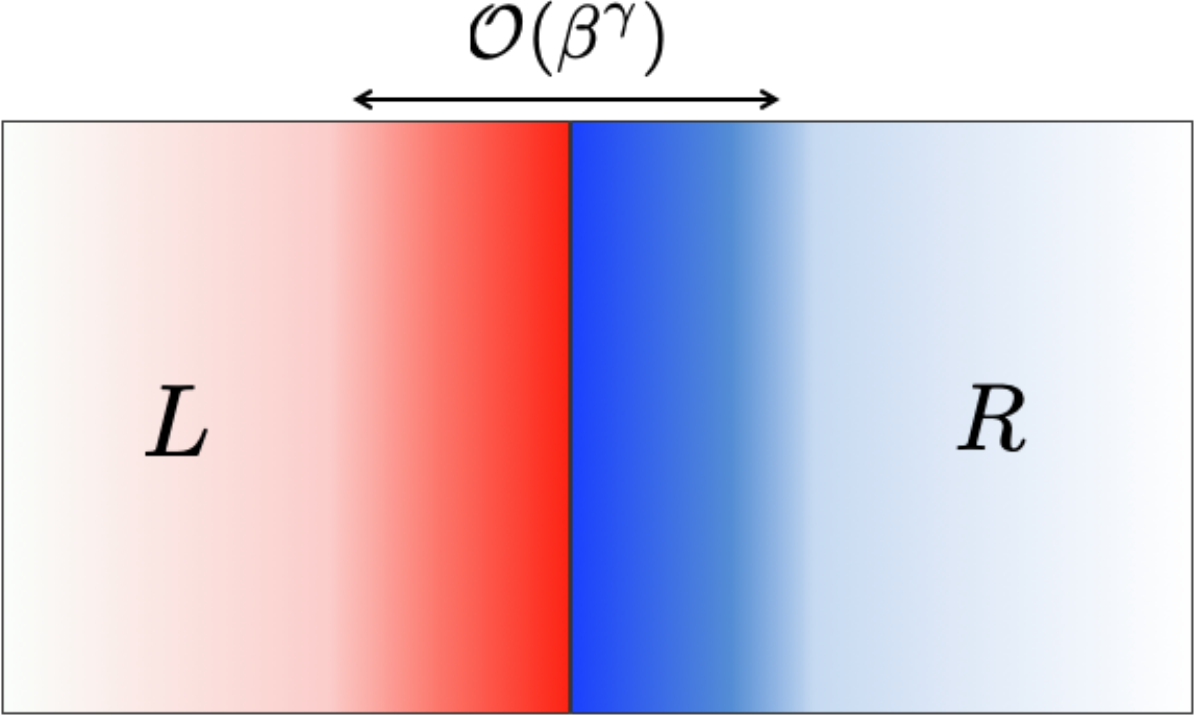}
}
\caption{Schematic depiction of our problem. By decomposing the total system into $L$ and $R$, we consider the mutual information $I(L:R)_{\rho_\beta}$ between $L$ and $R$. 
Then, the thermal area law in Ref.~\cite{PhysRevLett.100.070502} gives $I(L:R)_{\rho_\beta} \lesssim \beta |\partial L|$ ($\gamma=1$ in the above picture).
We aim to establish a new thermal area law in the form of $I(L:R)_{\rho_\beta} \lesssim \beta^\gamma |\partial L|$ with $\gamma<1$. 
In particular, it is a highly non-trivial and fundamental question to identify the best exponent $\gamma_c$ for which the thermal area law holds in generic many-body systems.
Our main result provides the non-trivial upper bound of $\gamma_c \le 2/3$.
}
\label{fig:quasi_linear_alg}
\end{figure}

The thermal area law~\eqref{mutual_information_wolf} is optimal at high temperatures ($\beta \approx \mathcal{O}\left(1\right)$) because the dependence on $|\partial L|$ cannot be improved. One may similarly expect that at low temperatures ($\beta \gg 1$), linear dependence on $\beta$ should be optimal. 
This is suggested by the theory of \emph{belief propagation}~\cite{PhysRevB.76.201102}, which indicates that the non-local quantum effects can be induced in a length scale of $\orderof{\beta}$. However, there are no definite numerical or theoretical examples that achieve the upper bound~\eqref{mutual_information_wolf}.
Indeed, for specific systems~\cite{PhysRevA.78.022103,Bernigau_2015,barthel2017one}, we can get much better area-law bounds than~\eqref{mutual_information_wolf}.
This motivates the possibility of the following improvement of the thermal area law:
\begin{align}
I(L:R)_{\rho_\beta} \lesssim \beta^{\gamma} |\partial L|   \quad (\gamma < 1)\label{mutual_information_improve}.
\end{align}
Any improvement along these lines is intimately associated with new advances in our understanding of the low-temperature physics. For instance, the widely known relation between area laws and tensor networks suggests that the identification of the minimum $\gamma_c$ would also lead to optimal representations of Gibbs states. This would result in faster algorithms for computing local expectation values and evaluating the partition functions.

We now turn our attention to the simulability of the quantum Gibbs state.
There exists a large number of  classical~\cite{PhysRevLett.102.190601,Stoudenmire_2010,PhysRevB.92.125119,PhysRevLett.97.187202,TANG2013557,RevModPhys.73.33,PhysRevB.43.5950,RevModPhys.77.259,PhysRevLett.93.207204,PhysRevLett.93.207205,PhysRevB.78.155117,PhysRevLett.106.127202,PhysRevLett.107.070601,PhysRevB.92.035152,PhysRevB.94.165116,PhysRevX.8.031082,PhysRevB.76.201102,PhysRevA.77.052318,PhysRevB.81.054106} and quantum~\cite{PhysRevLett.103.220502,PhysRevLett.105.170405,temme2011quantum,Yung754,PhysRevLett.116.080503, Kastoryano2016,10.5555/3179483.3179486,Brandao2019,10.1145/3313276.3316366} algorithms to study the properties of the quantum Gibbs states. 
At high temperatures ($\beta=\orderof{1}$), the Gibbs states have numerous analytical properties, such as the exponential decay of bipartite correlations~\cite{Araki1969,Gross1979,Park1995,ueltschi2004cluster,PhysRevX.4.031019,frohlich2015some}, the large deviation principle~\cite{ref:Marco_LD,Netocny2004,kuwahara2019gaussian}, and the approximate quantum Markov property~\cite{kato2016quantum,CMI_clustering}. As a consequence, in this temperature regime, the Gibbs states are proved to be generated by a finite depth quantum circuit~\cite{Brandao2019,CMI_clustering}, and the quantum partition function can be computed in polynomial time~\cite{Mehdi,crosson2020classical,CMI_clustering,mann2020efficient}.

Unfortunately, at lower temperatures, computational complexity theory results severely limit the applicability of the algorithms discussed above. Indeed, computing the partition function of Gibbs states in two and higher dimensions is already known to be NP-hard~\cite{Barahona_1982,Goldberg2015} (see also~\footnote{
More in detail, we mean the following statement. Let $Z$ be a partition function as $Z:=\tr(e^{-\beta H})$ with $H$ the system Hamiltonian. 
Then, there exists a constant $\zeta>0$ such that approximating $\log(Z)$ up to an error $\zeta n$ ($n$: system size) is the NP-hard problem~\cite{Goldberg2015}.
}) except for special cases (e.g., ferromagnetic spin systems~\cite{doi:10.1137/0222066,PhysRevLett.119.100503}). 
This is a serious bottleneck for several practical applications in which the Gibbs states are employed at low temperatures. For example, in the quantum algorithm for semi-definite programming~\cite{8104077}, the quantum Gibbs states with $\beta=\orderof{\log(n)}$ ($n$: system size, $\beta$: the inverse temperature) must be sampled. Similar challenges are faced in the imaginary time evolution, the implementation of which is a central aim of near-term quantum devices~\cite{motta2020determining,PhysRevLett.121.170501,PhysRevB.100.094434,Yuan2019theoryofvariational,McArdle2019,yeter2019practical,love2020cooling,nishi2020implementation}. 
Thus far, below a threshold temperature where the cluster expansion technique does not work~\footnote{Above this threshold temperature, one can ensure that there are no quantum/classical phase transitions.}, little is known about the universal properties of Gibbs states that may hold independent of the system's details.
This provides a strong motivation to identify the optimal thermal area laws.

\subsection{Description of the main results}
For the first main result of the present study, we prove the inequality~\eqref{mutual_information_improve} for $\gamma=2/3$.
On the other hand, we also prove the lower bound of $\gamma_c \geq 1/5$, using the example constructed in \cite{Gottesman_2010} (see Sec.~\ref{append:gammaclowbnd}), which means 
$$1/5\le \gamma_c \le 2/3.$$
There are two remarks: i) the result is applicable only to finite dimensional lattice, while there may be a counterexample in general graph systems \cite{6979009}, and  ii) in high-dimensional cases, the obtained result is slightly weaker, as is given by $I(L:R)_{\rho_\beta} \lesssim \beta^{\gamma} |\partial L| \log (\beta^{\gamma} |\partial L|)$ (see Theorem~\ref{main_thm_area_law2}).
To understand why the result is counter-intuitive at first sight, let us consider the case of real-time evolution $e^{iHt}$. The small-incremental-entangling (SIE) theorem~\cite{PhysRevA.76.052319,PhysRevLett.97.150404,PhysRevLett.111.170501,Marien2016} predicts the linear increase of the entanglement with respect to time, which translates to the fact that the Schmidt rank of the operator $e^{iHt}$ grows as $e^{\orderof{t}}$. This suggests the same linear dependence for the imaginary time evolution operator $e^{-\beta H}$. 
However, the inequality~\eqref{mutual_information_improve} shows that the scaling of the exponent is sub-linear in $\beta$. This means that the entropy growth due to the imaginary time evolution is more diffusive in nature. We explain this difference in Subsection \ref{sec:Physical intuition from the random walk behavior}, which can be traced back to a better polynomial approximation to $e^{-x}$ compared with $e^{-i x}$~\cite{TCS-065}. This polynomial approximation is caused by a random walk interpretation of the Chebyshev basis expansion of $e^{-x}$ (see Sec.~\ref{sec:Physical intuition from the random walk behavior}), which is not available for $e^{-i x}$. This random walk interpretation further suggests that the entropy production in the imaginary time evolution is diffusive. 


The improved area law is not only of fundamental interest but also provides important insights regarding the efficient representation of the quantum Gibbs states.
In previous studies~\cite{PhysRevB.73.085115,PhysRevB.91.045138,PhysRevX.4.031019}, the approximations by matrix product operators/projected entangled pair operators (MPO/PEPO) have been investigated through cluster expansion techniques.
Furthermore, Ref.~\cite{PhysRevB.91.045138} has explicitly given the PEPO/MPO construction scheme with the bond dimensions of 
\begin{align}
D= (n/\epsilon)^{\orderof{\beta}} \quad (\textrm{$\epsilon$: approximation error}). \label{Bond_dim_est}
\end{align}
If we use the cluster expansion technique, this is expected to be the best estimation. 
However, the polynomial-size bond dimension of $n^{\orderof{\beta}}$ may still be a significant overestimation. 
Improvements are strongly motivated by the practical use of tensor network techniques in approximating thermal states~\cite{PhysRevLett.102.190601,PhysRevLett.93.207204,PhysRevX.8.031082}, which appears to be much more successful than is guaranteed by the current analytical bounds. 

Our second main result focuses on classical algorithms for approximating thermal states in one dimension (1D). By applying our new analyses, we establish a sub-linear dependence of the bond dimension of the MPO approximation to the thermal state as 
\begin{align}
D= e^{\tOrder\br{\beta^{2/3}}+\tilde{\mathcal{O}}\left(\sqrt{\beta \log (n/\epsilon)}\right)} \label{Bond_dim_est_2}
\end{align}
with $\epsilon$ the approximation error, where we write $\orderof{n\log(n)}$ as $\tilde{\mathcal{O}}(n)$ by using the notation $\tilde{\mathcal{O}}$. 
The estimated bond dimension is smaller than any power of $(n/\epsilon)$ and is well suited for numerical simulations. 

Finally, we consider the computational complexity of the construction of the MPO, which approximates 1D quantum Gibbs states.
Establishing provably efficient quasi-linear algorithms for physical systems is a central target in the field of Hamiltonian complexity~\cite{0034-4885-75-2-022001,gharibian2015quantum}.
The general difficulty lies in that the existence of an efficient MPO description~\eqref{Bond_dim_est_2}  does not necessarily imply an efficient algorithm to find such a description~\cite{landau2015polynomial,Arad2017}.
So far, the state-of-the-art algorithm \cite{PhysRevB.91.045138} is based on cluster expansion, and MPO construction requires a computation cost which is proportional to $n\times (n/\epsilon)^{\orderof{\beta}}$, where the estimated exponent of $(n/\epsilon)$ is usually impractically large. 
However, most classical heuristic algorithms employed in practice usually require only (quasi-)linear computational time with respect to the system size~\cite{PhysRevLett.102.190601,Stoudenmire_2010,PhysRevB.92.125119,PhysRevLett.97.187202,TANG2013557,RevModPhys.73.33,PhysRevB.43.5950,RevModPhys.77.259,PhysRevLett.93.207204,PhysRevLett.93.207205,PhysRevB.78.155117,PhysRevLett.106.127202,PhysRevLett.107.070601,PhysRevB.92.035152,PhysRevB.94.165116,PhysRevX.8.031082,PhysRevB.76.201102,PhysRevA.77.052318,PhysRevB.81.054106}. 
We, for the first time, give a quasi-linear time algorithm that constructs the approximate MPO, with a runtime of
$$
n \times e^{\tOrder(\beta)+ \tOrder\left( \sqrt{\beta \log (n/\epsilon)}\right)},  
$$
which is quasi-linear in $(n/\epsilon)$ for arbitrary $\beta=o(\log (n))$.

The rest of this paper is organized as follows. 
In Sec.~\ref{sec:set up}, we formulate the precise setting and notations used throughout the paper.
In Sec.~\ref{sec:thermal area law}, we state the main theorems on the area law and the MPO approximation. 
In addition, in Sec.~\ref{sec:Physical intuition from the random walk behavior}, we show the relationship between imaginary time evolution and the random walk, and in Sec.~\ref{append:gammaclowbnd} we show the lower bound on the critical $\gamma_c$.  
In Sec.~\ref{sec:quasi_linear_time_algorithm}, we give the quasi-linear algorithm to compute the MPO approximation of the 1D quantum Gibbs states.
We also provide a brief explanation regarding why the algorithm works well.
In Sec.~\ref{sec:Further discussions}, we discuss several physical implications from our analytical techniques.  
The proofs of the main statements are given in Sec.~\ref{sec:proof_Main statement and the proof} 
Finally, in Sec.~\ref{sec:Conclusion}, we summarize the paper, along with a brief discussion. 
To concentrate on the physics, we have provided the more intricate aspects of the proofs in Appendices.


\section{Setup and notation} \label{sec:set up}

We consider a quantum system with $n$ qudits, each of which has a $\spin$-dimensional Hilbert space. 
We denote the Hilbert space dimension of a subset $S \subseteq \Lambda$, where $\Lambda$ is a lattice, by $\mathcal{D}_S$. For the present discussion,
let us restrict ourselves to the case of 1D lattice; we consider higher dimensional lattices later (see Sec.~\ref{sec:set up_high_D}). We define the Hamiltonian $H$ as follows:
\begin{align}
H = \sum_{i=1}^n h_{i,i+1},\quad  \|h_{i,i+1}\| +\|h_{i-1,i}\|  \le g , \label{def:Hamiltonian}
\end{align}
where $h_{i,i+1}$ contains interactions between $i$ and $i+1$, and $\|\cdots\|$ is the operator norm. 
By taking the energy units appropriately, we set $g=1$. 
Here, we assume two-body interactions of the Hamiltonian, but the generalization to arbitrary $k$-body interactions (i.e., $k$-local Hamiltonian) with $k=\orderof{1}$ is straightforward (see~Appendix~\ref{sec:More detailed setup})
For an arbitrary operator $O$, we define the Schmidt rank ${\rm SR} (O,i)$ as the minimum integer such that
 \begin{align}
 \label{Scgmidt_rank_def}
O = \sum_{m=1}^{{\rm SR} (O,i)} O_{\le i,m} \otimes O_{>i,m},
\end{align}
where $\{O_{\le i,m}\}$ and $\{O_{>i,m}\}$ are operators acting on subsets $\{j\}_{j\le i}$ and $\{j\}_{j\ge i+1}$, respectively.
Note that the Schmidt rank ${\rm SR} (h_{i,i+1},i)$ is always smaller than the local Hilbert dimension $\spin$ (i.e., ${\rm SR} (h_{i,i+1},i) \le \spin$).

Throughout the paper, we have focused on the Gibbs state $\rho_\beta$ with an inverse temperature $\beta$:
 \begin{align}
\rho_\beta =\frac{e^{-\beta H}}{\tr(e^{-\beta H})}. \notag 
\end{align}
To extend the concept of the entanglement area law from the ground states to finite temperatures, we often utilize the mutual information $I(L:R)_{\rho}$ as in  Ref.~\cite{PhysRevLett.100.070502}. 
The mutual information $I(L:R)_{\rho}$ reduces to entanglement entropy when $\rho$ is given by a quantum pure state.
For an arbitrary decomposition of the total system $\Lambda$ into
$
\Lambda = L \cup R,
$
it is defined as 
 \begin{align}
I(L:R)_{\rho_\beta} := S(\rho^L_\beta)+S(\rho^R_\beta) -S(\rho_\beta) ,
\label{mutual_information_def}
\end{align}
where $S(\cdots)$ is the von Neumann entropy, i.e., $S(\rho):=-\tr [\rho \log (\rho)]$, and $\rho^L_\beta$ $(\rho^R_\beta)$ is the reduced density matrix in subsets $L$ ($R$).
We define the subsets $L$ and $R$ as $L=\{1,2,\ldots,i_0\}$ and $R=\{i_0+1,i_0+2,\ldots,n\}$, respectively. 
Then,  the boundary Hamiltonian $H_{\partial L}$ is given by $h_{i_0,i_0+1}$, which gives the previously known thermal area law~\eqref{mutual_information_wolf} of
\begin{align}
I(L:R)_{\rho_\beta} \le 2\beta\|h_{i_0,i_0+1}\|  \label{mutual_information_def_trivial_bound}.
\end{align}

For a more detailed characterization of the structure of the quantum Gibbs state, we focus on the matrix-product-operator (MPO) representation.
We aim to approximate the Gibbs state $\rho_\beta$ by the following operator:
\begin{align}
M_D = \sum_{\substack{s_1,s_2,\ldots,s_n =1 \\ s_1',s_2',\ldots,s_n' =1}}^\spin& \tr \left(A_{1}^{[s_1,s_1']}A_{2}^{[s_2,s_2']} \cdots A_{n}^{[s_n,s_n']}\right)  \notag \\
&\ket{s_1,s_2,\ldots,s_n}\bra{s_1',s_2',\ldots,s_n'} , 
\label{Matrix_Product_Op_Def}
\end{align}
where each of the matrices $\{A_i^{[s_i,s_i']}\}_{i,s_i,s_i'}$ is described by the $D\times D$ matrix.
We refer to the matrix size $D$ as the bond dimension.
By choosing $D$ to be sufficiently large as $D=e^{\orderof{n}}$, we can describe arbitrary operators in the form of MPO; however, only a relatively small bond dimension is often required in practical applications (e.g., $D=o(n)$).  
To relate this to the mutual information, notice that $I(L:R)$ can in general be bounded by the bond dimension of the purification. For the subclass of MPDOs with local purifications \cite{PhysRevLett.100.070502}, this cannot exceed $2\log D$ for an arbitrary decomposition $\Lambda=L\sqcup R$, although no upper bound exists for general MPOs \cite{de2013purifications,de2016fundamental}. To circumvent this difficulty, we directly give a bound on the bond dimension of a purification that scales as Eq. \eqref{Bond_dim_est_2} (see Sec.~\ref{proof_thm:Renyi entanglement of purification}).
The primary problem is to estimate how large the bond dimension needs to be to achieve a certain precision error.

In quantitatively estimating the approximation error, we utilize the Schatten $p$ norm, which is defined for arbitrary operator $O$ as follows:
\begin{align}
\|O\|_p := \left[\tr (O^\dagger O)^{p/2} \right]^{1/p}. \label{def:Schatten p norm}
\end{align}
Note that $\|O\|_1$ corresponds to the trace norm and $\|O\|_\infty$ corresponds to the standard operator norm, which we denote by $\|O\|$ for simplicity.
When $O$ is a density operator, it is a common practice to consider the trace norm (i.e., $p=1$) for the approximation error. 
However, for estimating approximation errors in the present context, calculations in terms of general Schatten $p$ norm are crucially important.  
For example, let us consider the situation where we have obtained a good approximation for $O$ by $\tilde{O}$ and are interested in approximating $O^s$ by $\tilde{O}^s$. 
In order to achieve $\|O^s-\tilde{O}^s\|_1 \ll 1$, we need to prove $\|O-\tilde{O}\|_s \ll 1$; evidently, approximation solely in terms of the trace norm is not sufficient. 
This point is clarified in Lemmas~\ref{lem:Approximation of square operators_0} and \ref{lem:Approximation of qth power of operators_0} in Appendix~\ref{appdx_Basic analytical tools}, which are based on the analyses in Ref.~\cite{PhysRevB.91.045138}. 
This kind of the technique is crucial in developing a quasi-linear time algorithm for the quantum Gibbs states (Sec.~\ref{sec:quasi_linear_time_algorithm}).

The state-of-the-art results~\cite{PhysRevB.91.045138} ensure the existence of $M_D$ such that $\|\rho_\beta- M_D\|_1 \le \epsilon$ with the bond dimension as in Eq.~\eqref{Bond_dim_est}.
The bond dimension $D$ is roughly related to the mutual information $I(L:R)_{\rho_\beta}$ as $I(L:R)_{\rho_\beta} \lesssim \log(D)$, and thus, the estimation~\eqref{Bond_dim_est} implies the area-law bound of~\eqref{mutual_information_def_trivial_bound}.

\subsection{High-dimensional setup} \label{sec:set up_high_D}

In extending to the high-dimensional systems, 
we consider a quantum system on a $\sd$-dimensional rectangular lattice with $\sd$ the spatial dimension (we note that our analysis can also be applied to other lattices).
For simplicity of notations, we consider nearest-neighbor interactions as follows:
\begin{align}
H = \sum_{\langle i,j \rangle} h_{i,j},\quad  \max_{i\in \Lambda}\sum_{j} \|h_{i,j}\| \le g, \label{def:Hamiltonian_gene}
\end{align}
where $\langle i,j \rangle$ denotes the pairs of adjacent qudits, and $\|\cdots\|$ is the operator norm. 
By taking the energy unit appropriately, we set $g=1$. 

For convenience, we consider a vertical cut of the total system (see Fig.~\ref{fig:quasi_linear_alg} for example);
however, the same argument can be applied to a rectangular cut.  
For any partition $\Lambda=L\sqcup R$, we define an upper bound with the size of the surface region as $|\partial \Lambda|$, which is as large as $\orderof{n^{1-1/\sd}}$.
Note that $|\partial \Lambda|=1$ in the 1D lattice.

\section{Improved thermal area law} \label{sec:thermal area law}

We first show our main result in the thermal area law. 
The following theorem holds for arbitrary lattice dimensions:
\begin{theorem} \label{main_thm_area_law2}
For an arbitrary cut $\Lambda=L\cup R$, the mutual information $I(L:R)_{\rho_\beta}$ is upper-bounded by
 \begin{align}
I(L:R)_{\rho_\beta} \le  C \beta^{2/3} |\partial \Lambda| \left( \log^{2/3} (|\partial \Lambda|) + \log(\beta) \right), 
\label{mutual_information_def_improved_bound}
\end{align}
where $C$ is a constant of $\orderof{1}$. In particular, for one-dimensional systems ($|\partial \Lambda|=1$), we have 
 \begin{align}
I(L:R)_{\rho_\beta} \le  C \beta^{2/3}\log(\beta)= \tOrder(\beta^{2/3}) . 
\label{mutual_information_def_improved_bound_1D}
\end{align}
\end{theorem}
\noindent 
We show the proof in Sec.~\ref{proof_thm:Renyi entanglement of purification}. 
The above result has a logarithmic correction of $\log( |\partial \Lambda|)$ to the area law in high dimensions.
Even then, for $\beta \gtrsim\log^2( |\partial \Lambda|)$, our result provides a qualitatively better upper bound than the previous one~\eqref{mutual_information_wolf}.
We expect that this correction should be removed using refined analyses on the Schmidt rank for polynomials of the Hamiltonian.

Moreover, for the MPO representation of the 1D quantum Gibbs states, we obtain the following theorem
\begin{theorem} \label{main_thm_MPO_approximation}
For arbitrary 1D quantum Gibbs state $\rho_\beta$, there exists a MPO $M_D$ as in Eq.~\eqref{Matrix_Product_Op_Def} such that for the Schatten-$p$ norm with $p=1$ and $p=2$ 
\begin{align}
\| \rho_\beta -M_D \|_p \le \epsilon \|\rho_\beta \|_p \label{MPO_approx}
\end{align}
with 
 \begin{align}
D \le \exp[\tilde{q}^\ast_\epsilon \log (\tilde{q}^\ast_\epsilon)], 
\label{MPO_approx_bond}
\end{align}
where $\tilde{q}^\ast_\epsilon :=C_0' \max \left(\beta^{2/3}, [\beta\log(\beta n/\epsilon)]^{1/2} \right)$.
\end{theorem}
\noindent
This proof is shown in Sec.~\ref{sec_thm:MPO_approx}. We believe that the above MPO could also be used to construct a quantum circuit with depth polynomial in the stated bond dimensions. 
As far as we know, explicit constructions of quantum circuits for generic MPOs (or MPSs) have been an open problem except for special cases~\cite{PhysRevLett.95.110503,PhysRevResearch.1.023025,PhysRevA.101.032310}.

We remark on the generalization to high-dimensional cases. 
As for the MPO representation, we can improve the $\beta$-dependence of the bond dimension in high dimensions. 
However, the MPO representation for high-dimensional systems is not useful since the bond dimension is inherently sub-exponentially large with respect to the system size.
For an arbitrary bi-partition $\Lambda=L\sqcup R$, the bond dimension scales as $e^{\orderof{\beta^{2/3}}|\partial \Lambda|}=
e^{\orderof{\beta^{2/3}} n^{1-1/d}}$.
In order to obtain a meaningful representation for high-dimensional Gibbs state, we need to consider the projected entangled pair operators (PEPO)~\cite{PhysRevB.73.085115,PhysRevB.91.045138,PhysRevX.4.031019}.
We expect that the bond dimension of the PEPO might be also sublinear as~\eqref{MPO_approx_bond} in order to achieve a good approximation~\eqref{MPO_approx}.
So far, this remains open and one of the most important future directions (see also Sec.~\ref{sec:Conclusion}).

\begin{figure}[tt]
\centering
\subfigure[Real time evolution]
{\includegraphics[clip, scale=0.31]{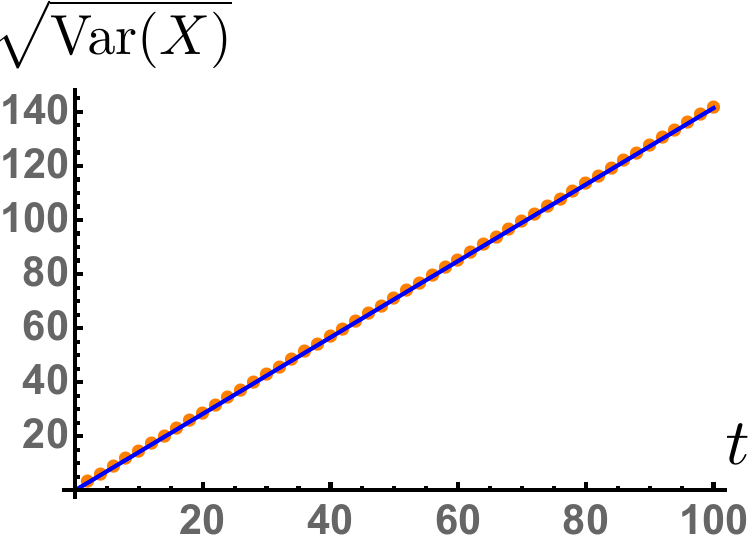}}
\subfigure[Imaginary time evolution]
{\includegraphics[clip, scale=0.31]{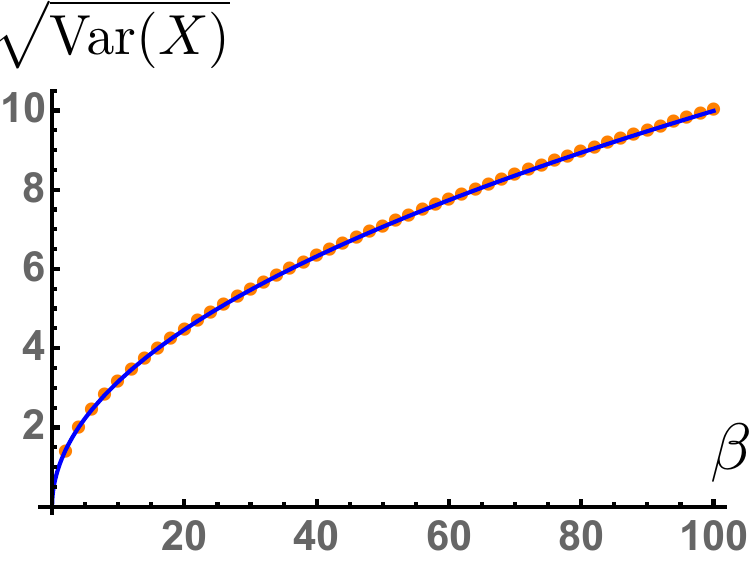}}
\caption{
Comparison between real-time evolution and imaginary time evolution in the tight-binding model~\eqref{tight_binding} with $R=500$.
In (a) and (b), we plot the fluctuation of the position~$\sqrt{{\rm Var}(X)}$ after the real and imaginary time evolutions, respectively.
Here, the state at the initial time is given by $\ket{0}$. The fitting functions for (a) and (b) are given by $\sqrt{2} t$ and $0.998769\beta^{1/2}$, respectively.
This clearly indicates that the real-time evolution induces a ballistic propagation, whereas the imaginary-time evolution induces a diffusive propagation.
}
\label{fig:im_time_re_time}
\end{figure}

Now, we discuss the key principles that allow us to improve the original thermal area law (See Appendix~\ref{sec:proof_Main proposition and the proof} for the details). Our analysis utilizes various recent techniques employed in the proofs of the area law for ground states~\cite{arad2013area,anshu2019entanglement,kuwahara2019area}. Inspired by these studies, we construct an approximation of the quantum Gibbs state using an appropriate polynomial of low degree \cite{TCS-065} and then perform a Schmidt rank analysis adapted from \cite{arad2013area}. As mentioned in the introduction, the main insight is that the polynomial used by us satisfies the random walk property, which we explain below.

\subsection{Physical intuition from the random walk behavior}\label{sec:Physical intuition from the random walk behavior}

Before the main discussion, let us consider an illustrative example of the random walk behavior in imaginary-time evolution.  
We here consider a one-particle tight-binding model as 
 \begin{align}
H= \sum_{x=-R}^R (\ket{x}\bra{x+1} + \ket{x+1}\bra{x} -2 \ket{x}\bra{x}) , \label{tight_binding}
\end{align}
where $\ket{x}$ is the state of the particle on site $x$. 
Then, the real-time Schr\"odinger equation gives the ballistic propagation of the particle. 
We consider a time-evolved quantum state $\ket{0(t)}=e^{-iHt}\ket{0}$, where the initial state $\ket{0}$ is the localized state on $x=0$. 
In Fig.~\ref{fig:im_time_re_time} (a), we show the fluctuation of the position, which is given by the square root of the variance ${\rm Var}(X):= \bra{0(t)} X^2 \ket{0(t)} - (\bra{0(t)} X \ket{0(t)})^2$, where $X=\sum_{x=-R}^R x \ket{x}\bra{x}$. 
In contrast, the imaginary-time Schr\"odinger equation is formally equivalent to the random walk differential equation. 
Hence, the fluctuation for the state $\ket{0(-i\beta)}=e^{-\beta H}\ket{0}$ grows diffusively with time $t$ [see Fig.~\ref{fig:im_time_re_time} (b)].
This indicates that the imaginary time evolution may generally induce a diffusive propagation of information in quantum many-body systems. 
In the following sections, we mathematically justify this intuition.  

\begin{figure}[tt]
\centering
\includegraphics[clip, scale=0.44]{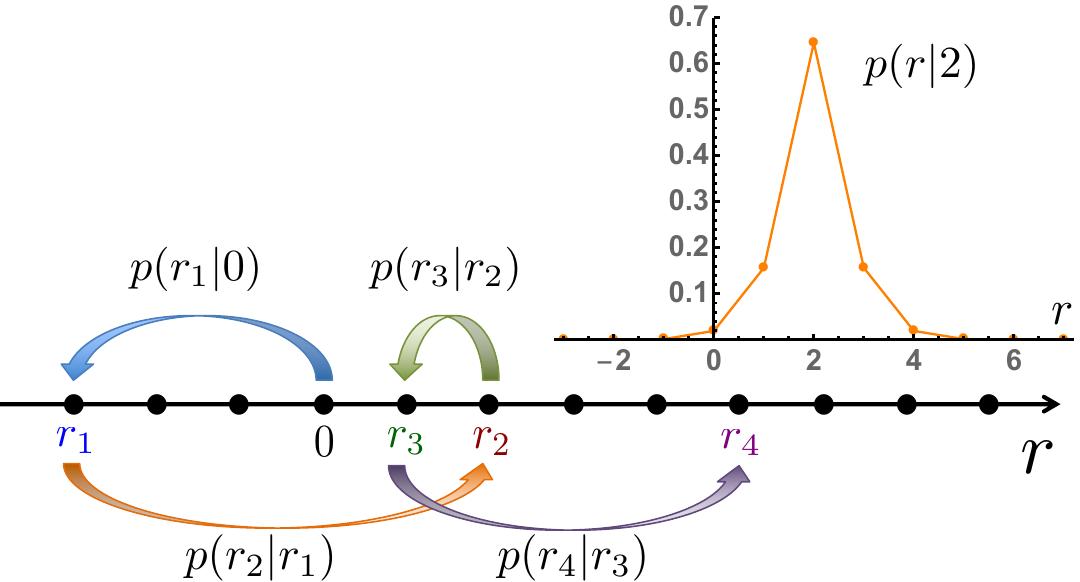}
\caption{Schematic picture of the random walk. The exponential function $e^{-b(1+y)/2}$ is given by the expectation of $T_{r_b}(x)$ with the probability $P(r_b)$, as in Eq.~\eqref{random_walk_exponential}.
The probability $P(r_b)$ is generated from the $b$-step random walk. In each step, the probability from $r$ to $r'$ is given by $p(r'|r)$, which is a symmetric function around $r$. In the picture, we give the numerical plot of $p(r|2)$, where the shape of $p(r|r')$ does not depend on $r'$. 
}
\label{fig:Random_walk}
\end{figure}

Suppose $x$ is fixed to be in a range $[0, b]$. As shown in \cite{TCS-065}, $e^{-x}$ can be approximated by a polynomial of degree $\orderof{b^{1/2}}$, for a constant error. This is the consequence of a random walk that is concentrated around degree $\orderof{b^{1/2}}$ after $b$ steps. Let us introduce $y\in [-1,1]$ such that $x=b(1+y)/2$ and $e^{-x} = \br{e^{-\frac{1}{2}(1+y)}}^b$. Below, we will show that the exponential function $e^{-b(1+y)/2}$ ($b \in \mathbb{N}$) can be expanded in terms of the Chebyshev polynomials as (see also Fig.~\ref{fig:Random_walk}):
\begin{widetext}
\begin{align}
 \br{e^{-\frac{1}{2}(1+y)}}^b &=\sum_{r_b=-\infty}^\infty P(r_b) T_{r_b} (y) \notag \\
 &= \sum_{r_b=-\infty}^\infty \left( p(r_b | r_{b-1}) \sum_{r_{b-1}=-\infty}^\infty p(r_{b-1} | r_{b-2}) \cdots \sum_{r_{2}=-\infty}^\infty p(r_{2} | r_{1}) \sum_{r_{1}=-\infty}^\infty p(r_{1} | r_{0})\right) T_{r_b} (y)
 \label{random_walk_exponential}
\end{align}
with $r_0=0$, 
where $T_{r}(x)$ is the Chebyshev polynomial and $p(r | r')$ is a random walk probability from $r_{b-1}$ to $r_b$ which is defined below.
\end{widetext}

%

For its application to $e^{-\beta H}$, we choose $b=\beta\|H\|$. Because the Schmidt rank and polynomial degree are closely related \cite{arad2013area}, we get a diffusive interpretation of the Schmidt rank of $e^{-\beta H}$. 
We thus infer a sublinear $\beta$-dependence of the mutual information, namely $\gamma_c< 1$. 
There are two main issues while achieving this value. First, the above polynomial gives an approximation to $e^{-\beta H}$ only in the operator norm, whereas we are searching for an approximation in a family of norms. Second, even for a constant error approximation in the operator norm, degree $\sqrt{b}=\sqrt{\beta \|H\|}$ scales with the system size. We solve both the problems using the quantum belief propagation in 1D and a refined version of Suzuki-Trotter decomposition in higher dimensions, which allows us to reduce the problem to a local Hamiltonian $H_S$, where $S$ is a much smaller region. The loss incurred because of the belief propagation and the conversion from the operator norm to other norms leads to our main result of $\gamma_c\leq 2/3$.

\subsubsection{Derivation of Eq.~\eqref{random_walk_exponential}}

As a first step, we expand 
\begin{align}
e^{-\frac{1}{2}(1+y)}=  \sum_{j=0}^{\infty}\frac{e^{-1/2}}{2^j j!}\cdot (-y)^j,
 \label{first_expansion_Cheby}
\end{align}
 which is an expectation of $(-y)^j$ according to the distribution $q(j):=\frac{e^{-1/2}}{2^j j!}$. Next, we introduce Chebyshev polynomials $T_r(y)$ (for an integer $r$) and utilize the observation from \cite{TCS-065} that for $j>0$ and integer $k$, 
\begin{align}
&(-y)^jT_r(-y)=\sum_{r'=-\infty}^\infty B_j(r'|r)T_{r'}(-y), \notag \\
&B_j(r'|r) = 2^{-j}\binom{j}{(j+r'-r)/2},
 \label{second_expansion_Cheby}
\end{align}
where we set $\binom{j}{s+1/2}=0$ ($s\in \mathbb{N}$) and $\binom{j}{s}=0$ for $s<0$ and $s>j$.
Here, $B_j(r'|r)$ is the binomial distribution which is centered at $r$ with a variance of $\sqrt{j}$ (see also the footnote~\footnote{
We start from the basic formula for the Chebyshev polynomials: $xT_r(x)= [T_{r+1}(x)+T_{r-1}(x)]/2$. 
Let $Y$ be a random variable taking values $1$ or $-1$ with the probability $1/2$. 
We then obtain $xT_r(x) =\mathbb{E}_{Y_1} [T_{r+Y_1}(x)]= [T_{r+1}(x)+T_{r-1}(x)]/2$. 
In the same way, we can obtain  $x^2T_r(x) =\mathbb{E}_{Y_1} [xT_{r+Y_1}(x)]=\mathbb{E}_{Y_1,Y_2} [T_{r+Y_1+Y_2}(x)]$.
By repeating the process, we obtain $x^jT_r(x) =\mathbb{E}_{Y_1,\ldots Y_j} [T_{r+D_j}(x)]$ with $D_j=Y_1+Y_2+\cdots +Y_j$. 
The probability distribution of $D_j$ obeys the binomial distribution as $2^{-j} \binom{j}{(j+D_j)/2}$, which gives Eq.~\eqref{second_expansion_Cheby}.}). 
Now, we have all the tools to set-up the random walk over integers. 
By combining Eqs.~\eqref{first_expansion_Cheby} and \eqref{second_expansion_Cheby}, 
we start with the first random walk step of
\begin{align}
e^{-\frac{1}{2}(1+y)}= \sum_{r_1=-\infty}^{\infty} p(r_1|0) T_{r_1}(y),
\end{align}
where the symmetric distribution $p(r_1|0)$ (with mean $0$ and variance $\orderof{1}$) is defined using $p(r_1|0):=\sum_{j=0}^\infty q(j)B_j(r_1|0)$. The subsequent steps are obtained by writing 
\begin{align*}
T_{r_1} (y)e^{-\frac{1}{2}(1+y)} &=  \sum_{j=0}^{\infty}\frac{e^{-1/2}}{2^j j!}\cdot T_{r_1}(y)(-y)^j\\
                            &= \sum_{r_2=-\infty}^{\infty} p(r_2|r_1) T_{r_2}(y)
\end{align*}
with $p(r_2|r_1):=\sum_{j=0}^\infty q(j)B_j(r_2|r_1)$.
One can show that the function $p(r_2|r_1)$ is symmetric around its mean $r_1$ and has a variance of $0.5$ (see Fig.~\ref{fig:Random_walk} for the shape of $p(r|2)$). 
By repeating the process, we can arrive at the equation~\eqref{random_walk_exponential}.
Thus, $\br{e^{-\frac{1}{2}(1+y)}}^b$ is an expectation over $T_{r}(y)$, according to a distribution obtained by performing $b$ steps of a symmetric random walk with constant variance. It is now clear that the degree is strongly concentrated around $\orderof{b^{1/2}}$. 
This random walk behavior is not available for $e^{ix}$ because the distribution $p(r_2|r_1)$ is not given by a real number. 
It leads to $\orderof{b}$ approximate degree for real-time evolution.


\subsection{Lower bound on the critical $\pmb{\gamma_c}$}
\label{append:gammaclowbnd}

We here show that the exponent $\gamma$ in \eqref{mutual_information_improve} is at least larger than $1/5$.
According to Ref.~\cite{Gottesman_2010}, there exists a frustration-free local Hamiltonian system with $n$ qudits ($\spin=3$) such that the half-chain entanglement entropy is linear in system size $n$, and the spectral gap $\Delta$ is given as 
\begin{align}
\Delta = \frac{c_\Delta }{n^4\log n},
\end{align}
where $c_\Delta$ is a constant of $\Omega(1)$.
For this Hamiltonian, let us consider a quantum Gibbs state at the inverse-temperature of $\beta=2c_\Delta^{-1} \log(\spin) n^{5}\log n$.
Then, the total weight of the excited state is at most as large as  $\spin^n e^{-\beta \Delta} = e^{-n\log(\spin)}$.
Therefore, this Gibbs state is exponentially close to the ground state.
Using the Fannes inequality~\cite{fannes1973}, the half-chain mutual information in the Gibbs state is 
\begin{align}
I (L:R)_{\rho_\beta} = \Omega(n)  = \frac{\Omega(1)}{\log^{1/5}\beta} \beta^{1/5} ,
\end{align}
which implies that $I (L:R)_{\rho_\beta}$ should be at least larger than $\beta^{1/5}$.


\section{Quasi-linear time algorithm for 1D Gibbs state} \label{sec:quasi_linear_time_algorithm}

\subsection{Main statement}

\begin{figure*}
\centering
{\includegraphics[clip, scale=0.32]{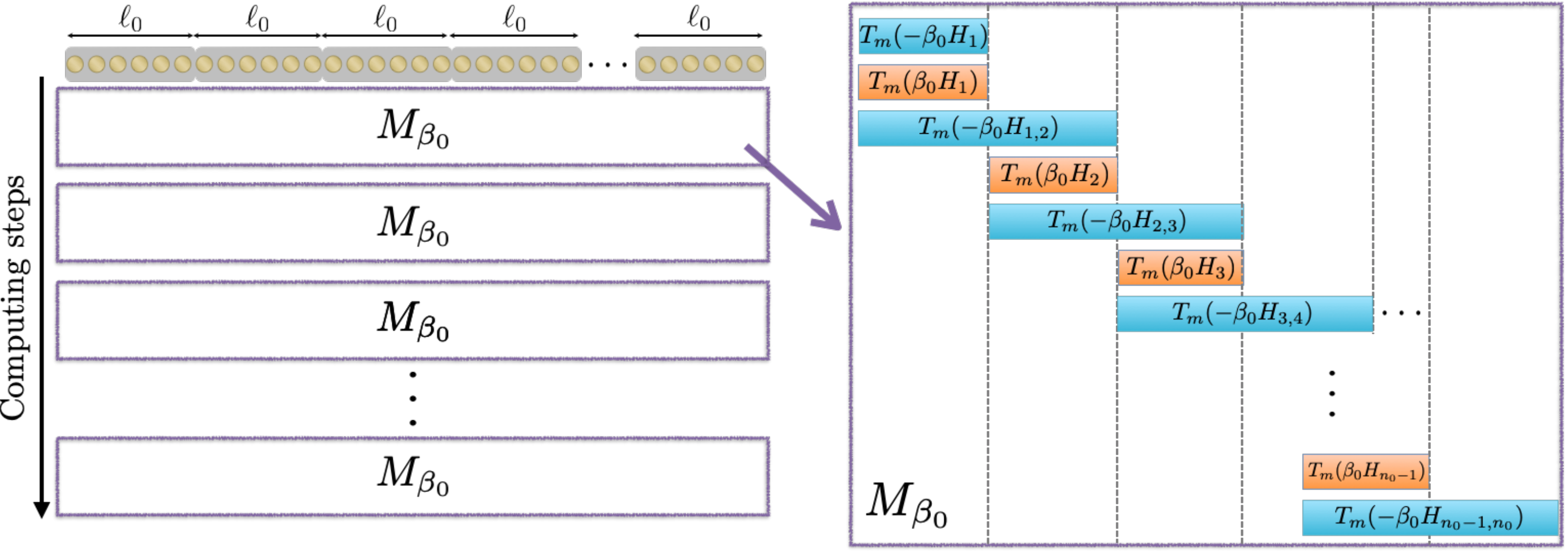}}
\caption{Our algorithm proceeds by iterated approximations of $e^{-\beta_0 H}$, performed $\beta/\beta_0$ times. In each step, we approximate the Gibbs operator $e^{-\beta_0 H}$ by the operator $M_{\beta_0}$. For this, we establish a decomposition of $e^{-\beta_0 H}$ as a product of operators shown on the right-hand side. This uses an imaginary-time version of the Lieb-Robinson bound and the Taylor truncation of the exponential function.}
\label{fig:quasi_linear_alg1}
\end{figure*}

Here, we show that the classical algorithm generating an MPO approximation of the Gibbs state $\rho_\beta$ is possible with a run time of $\orderof{n^{1+o(1)}}$ as long as $\beta = o(\log(n))$ 
We prove the following theorem:
\begin{theorem} \label{thm:quasi-linear_algorithm}
For arbitrary $\beta$, we can efficiently compute a matrix product operator $M_\beta$ which approximates $e^{-\beta H}$ in the sense that
\begin{align}\label{eq:approx-MPO}
\| M_\beta  - e^{-\beta H} \|_p  \le  \epsilon \| e^{-\beta H}\|_p \quad (\epsilon\le1), 
\end{align}
where the bond dimension of $M_\beta$ is given by $\exp(Q_\epsilon)$. 
Also, the computational time to calculate $M_\beta$ is $n\beta \exp(Q_\epsilon)$ with  
\begin{align}
\label{Def_Q_epsilon}
Q_\epsilon: = C \max \br { \beta , \sqrt{\beta \log (n/\epsilon)}} \log [ \beta \log (n/\epsilon)] ,
\end{align}
where $C$ is an $\orderof{1}$ constant. 
When $\beta \lesssim \log (n/\epsilon)$ and $\epsilon=1/\poly(n)$, the time complexity is given by
\begin{align}
\label{eq:time-complexity}
n \exp \left[\tOrder \br{ \sqrt{\beta \log (n)}} \right].
\end{align}
\end{theorem}

We  compare the bond dimensions of $M_{\beta_0}^{(\beta/\beta_0)}$ with that of the theoretical bound in~\eqref{MPO_approx_bond}.
For $\beta \lesssim \log(n)$, the both estimations are in the form of $e^{\tOrder \br{ \sqrt{\beta \log (n)}}}$, whereas for $\beta \gg \log(n)$, the estimation~\eqref{MPO_approx_bond} gives a slightly better bound. 

\subsection{MPO for ground space}

We also discuss the consequences regarding the calculation of quantum ground states.
Let us assume the following condition for the density of states in an energy shell $(E-1,E]$ for the low-energy regime~\cite{PhysRevB.76.035114,PhysRevA.80.052104,PhysRevLett.124.200604}:
\begin{align}
{\cal N}_{E, 1} & \le n^{c E} \label{density_of_state_DE}  
\end{align}
with $c$ a constant of $\orderof{1}$,  
where ${\cal N}_{E,1}$ is the number of eigenstates within the energy shell of $(E-1,E]$. 
This condition is typically observed for quantum Hamiltonians which have a spectral gap between the ground state and the first excited state~\cite{PhysRevB.76.035114}. 
Under this assumption, the quantum Gibbs state is approximated by the ground state up to an error of $1/\poly(n)$ for $\beta = \orderof{\log(n)}$; i.e., $\|\rho_\beta - \rho_\infty\|_1 = 1/\poly(n)$.
Then, the computation of the quantum Gibbs state for $\beta=\orderof{\log(n)}$ is closely related to the computation of ground states. 

By applying $\beta=\orderof{\log(n)}$ to \eqref{eq:time-complexity}, we obtain the time complexity of an almost polynomial form, as $n^{\orderof{\log\log(n)}}$.
This result rigorously justifies the empirical success of the imaginary TEBD methods in the computation of the ground states~\cite{PhysRevB.78.155117, PhysRevLett.107.070601,PhysRevB.94.165116,PhysRevX.8.031082}.
Our estimation, however, is still slightly worse than the polynomial form (i.e., $n^{\orderof{1}}$). 
In the case of the gapped ground states, the existing algorithms~\cite{landau2015polynomial,Arad2017} have already achieved polynomial computational costs without the assumption~\eqref{density_of_state_DE}.
Any small improvement of \eqref{eq:time-complexity} will allow us to obtain a quasi-linear time algorithm for the computation of the ground states under the assumption of \eqref{density_of_state_DE}.

\subsection{Details of the algorithm and proof of Theorem~\ref{thm:quasi-linear_algorithm}}

 The algorithm proceeds as follows. Suppose we are at a high temperature $\beta_0\le 1/16$. First, the 1D Hamiltonian is split into blocks of length $l_0 = \orderof{\log (n/\epsilon)} $, as
\begin{align}
H = \sum_{j=1}^{n_0} H_j ,\quad H_j=\sum_{s=(j-1) \ell_0+1}^{j \ell_0} h_{j,j+1} ,
\end{align}
where $n_0$ is the number of blocks.
We then write $e^{-\beta H}$ as follows:
\begin{align}
e^{-\beta H} = \prod_{j=1}^{n_0} e^{\beta_0 H_{1:j-1}} e^{-\beta_0 H_{1:j}}=:  \prod_{j=1}^{n_0} \Phi_j ,
\end{align}
where $H_{1:j}=\sum_{s\le j} H_s$ and $H_{1:0} = \hat{0}$.    
Here, the operator $\Phi_j$ is the non-local operator on the qudits $\{1,2,\ldots, j\ell_0 \}$.
We first approximate $\Phi_j$ by the following operator on the local region:
\begin{equation}
\tilde{\Phi}_j=e^{\beta_0 H_{j-1}}e^{-\beta_0 (H_{j-1}+H_j)}.
\end{equation}
The second approximation is the low-degree polynomial expression of $\tilde{\Phi}_j$:
\begin{equation}
\label{Taylor_exp_Phi}
\tilde\Phi^{(m)}_j=T_m(\beta_0 H_{j-1}) T_m(-\beta_0 (H_{j-1}+H_j)),
\end{equation}
where $T_m (x) =\sum_{s=0}^m x^m/m!$ is the truncated Taylor expansion of order $m=\orderof{\log (n/\epsilon)}$.

Using the above notation, we can approximate the high-temperature Gibbs state by
\begin{align}
\label{def_M_beta_0}
    M_{\beta_0}:= \prod_{j=1}^{n_0} \tilde\Phi^{(m)}_j. 
\end{align}
We have illustrated this construction in Fig.~\ref{fig:quasi_linear_alg1}.  
We notice that our construction resembles the decomposition of the real-time evolution developed in \cite{HaahHKL18}. 
Crucially, this approximation is justified using an imaginary time version of the Lieb-Robinson bound (see Appendix~\ref{sec:app_quasi_poly_alg_prop} for the proof):
\begin{prop} \label{Prop:small_beta}
For $\beta \le 1/16$,  Eq.~\eqref{def_M_beta_0} gives the approximation of the Gibbs state up to an error of   
\begin{align}
\label{ineq_M_beta_high_temp_0}
\| M_\beta e^{\beta H} -1 \| \le  \epsilon  ,
\end{align}
where $M_\beta$ has the bond dimension of $e^{\tOrder \left(\sqrt{\log(n/\epsilon)}\right) }$.
The sufficient computational time for the construction is given by
\begin{align}
n \exp\left[\tOrder \left(\sqrt{\log(n/\epsilon)}\right) \right]. \label{comp_time_high_temp}
\end{align}
We notice that the inequality~\eqref{ineq_M_beta_high_temp_0} immediately reduces to
\begin{align}
\label{ineq_M_beta_high_temp}
\| M_\beta - e^{-\beta H} \|_p \le  \epsilon \|e^{-\beta H}\|_p 
\end{align}
for an arbitrary positive $p$
\end{prop}

\noindent 
The computational time~\eqref{comp_time_high_temp} is qualitatively explained as follows.
The operator $M_{\beta_0}$ is a product of degree-$m$ polynomials $T_m(x)$. From Ref.~\cite{arad2013area}, the Schmidt rank of each of $\{\Phi^{(m)}_j\}_{j=1}^{n_0}$ in Eq.~\eqref{def_M_beta_0} is upper-bounded by $m^{\orderof{\sqrt{m}}} \sim \log (n)^{\sqrt{\log n}}$ along every cut.
Because $\{\Phi^{(m)}_j\}_{j=1}^{n_0}$ are locally defined, for every cut, constant number of operators in $\{\Phi^{(m)}_j\}_{j=1}^{n_0}$ contribute to the Schmidt rank. Therefore, the computational time to construct $M_{\beta_0}$ is at most $n e^{\tOrder\br{\sqrt{\log(n/\epsilon)} }}$.
 
 To extend this to arbitrary $\beta$, we utilize the following upper bound (see Lemma~\ref{lem:Approximation of qth power of operators_0} in Appendix~\ref{appdx_Basic analytical tools}), which slightly extends the analyses in Ref.~\cite{PhysRevB.91.045138}:
 \begin{align}
\|e^{-2q\beta_0 H} - (M_{\beta_0}^\dagger M_{\beta_0})^q \|_p \le 3\epsilon_0 q e^{3\epsilon_0 q}\|e^{-\beta H}\|_p 
\end{align}
for arbitrary positive integers $q$ and $p$, 
where $M_{\beta_0}$ satisfies the inequality~\eqref{ineq_M_beta_high_temp} with $\epsilon=\epsilon_0$ for arbitrary $p\in \mathbb{N}$.
We get $e^{-\beta H} = \br{e^{-\beta_0 H}}^{(\beta/\beta_0)}$ and then multiply the above MPO construction $\beta/\beta_0$ times, where $\beta_0$ is appropriately chosen so that $\beta/\beta_0$ becomes an even integer (i.e., $q= \beta/(2\beta_0)$). 
 To make $3\epsilon_0 q e^{3\epsilon_0 q} \le \epsilon$ $(\le 1)$, we need to choose $\epsilon_0=\epsilon/(6q)= \epsilon\beta _0 /(3\beta)$.

 By extending the Schmidt rank estimation in Ref.~\cite{arad2013area}, 
 we can ensure that the Schmidt rank of $M_{\beta_0}^{(\beta/\beta_0)}$ is at most as large as $\exp(Q_\epsilon)$.
In more detail, we can prove the following lemma (see Appendix~\ref{sec:app_quasi_poly_alg_lemma} for the proof):
 \begin{lemma}\label{lem:Schmidt_rank_M_beta_q}
Let $M_\beta$ be an approximate operator that has been defined in Eq.~\eqref{def_M_beta_0}. 
Then, for arbitrary $q\in \mathbb{N}$, the Schmidt rank of the power of $M_\beta$ is upper-bounded by 
\begin{align}
\label{upp_SR_M_beta_q}
\log[ {\rm SR} (M_\beta^q)] \le C' \max (q, \sqrt{mq}) \log(mq)
\end{align}
for an arbitrary cut, 
where $C'$ is an $\orderof{1}$ constant.
\end{lemma}
\noindent
Because $m$ has been chosen as $m=\orderof{\log(n/\epsilon)}$, this upper bound is proportional to $Q_\epsilon$ for $q=\orderof{\beta}$.
Therefore, the quantum Gibbs state $e^{-\beta H}$ is well approximated by the MPO, with its bond dimensions of $\exp(Q_\epsilon)$.  

We have already prepared the MPO form of $M_{\beta_0}$ in Proposition~\ref{Prop:small_beta}. 
Using the standard results regarding the canonical form of MPOs \cite{PhysRevLett.93.040502,SCHOLLWOCK201196}, 
we can efficiently calculate $M_{\beta_0}^q$ ($q\lesssim \beta$) from $M_{\beta_0}^{q-1}$ in a computational time of at most $\poly(\exp(Q_\epsilon))$.
We notice that in each of the steps, we can compress the MPO without any truncation error so that the bond dimension of $M_{\beta_0}^q$ is smaller than the bound in \eqref{upp_SR_M_beta_q}. 
By recursively constructing $M_{\beta_0}^q$, the computation of $M_{\beta_0}^{(\beta/\beta_0)}$ requires the time steps as many as \eqref{eq:time-complexity}.
We thus prove Theorem~\ref{thm:quasi-linear_algorithm}. $\square$

Finally, let us compare our method with the imaginary-time-evolving block decimation (TEBD) methods~\cite{PhysRevLett.93.207204,PhysRevX.8.031082,PhysRevLett.102.190601}, which proceed by truncation of the Schmidt rank at each imaginary-time Trotter step. 
A major limitation of these studies is the lack of rigorous justification of the Schmidt rank truncation, as explained below.

In the TEBD algorithms~\cite{PhysRevX.8.031082}, we start with a matrix product operator $M_1$ which gives the approximation of $e^{-\beta_1 H}$ for a certain $\beta_1$. 
We then connect two MPOs as $M_1^\dagger M_1$, which is expected to approximate $e^{-2\beta_1 H}$. 
To ensure the precision of approximation, we use Lemma~\ref{lem:Approximation of square operators_0} for the MPO $M_1$, which necessitates the approximation in terms of general Schatten $p$ norm. 
Now, the main technical difficulty comes from the Schmidt rank truncation of $M_1^\dagger M_1$, which gives MPO $M_2$ in the next step.  
After the Schmidt rank truncation, we connect $M_2^\dagger M_2$ to approximate the Gibbs state $e^{-4\beta_1 H}$. 
However, to ensure the good approximation from Lemma~\ref{lem:Approximation of square operators_0}, 
we have to truncate the Schmidt rank of $M_1^\dagger M_1$ so that $M_2$ is close to $M_1^\dagger M_1$ in terms of the general Schatten $p$ norm. 
The Schmidt rank truncation based on the singular value decomposition only ensures the approximation in terms of Schatten $2$-norm, as in Ref.~\cite[Lemma~1]{VC06mps}.
So far, we have no mathematical tools to perform Schmidt rank truncation, which guarantees approximation in terms of the general Schatten $p$-norm. In summary, even though the quantum Gibbs state can be approximated by an MPO with a small bond dimension, it is highly nontrivial to show whether the truncation of the Schmidt rank retains the good approximation.

We can circumvent this problem by constructing $e^{-\beta_0 H}$ as a product of local polynomial approximations, which covers the whole chain. 
Thus, the operator $M_{\beta_0}$ (or $M_{\beta_0}^{(\beta/\beta_0)}$) has \textit{a finitely bounded Schmidt rank}, and 
we do not need to approximate it further using the Schmidt rank truncation. 

\section{Further discussions} \label{sec:Further discussions}

\subsection{R\'enyi entanglement of purification} \label{sec:Renyi entanglement of purification}

To characterize the bipartite correlations beyond mutual information, we also consider the 
R\'enyi entanglement of purification $E_{p,\alpha}$ \cite{TerhalHLD02}, defined as follows:
\begin{definition} \label{def:Renyi entanglement of purification}
Let $\Lambda'$ be a copy of the total system with the Hilbert space $\mathcal{H}'$.
For an arbitrary quantum state $\sigma$, we define $E_{p,\alpha}(\sigma)$ for the partition $\Lambda=L\cup R$ as 
 \begin{align}
&E_{p,\alpha}(\sigma) := \inf_{\ket{\phi} \in \mathcal{H}\otimes \mathcal{H}'} E_\alpha(\phi),  \notag \\
&E_\alpha(\phi) = S_\alpha (\sigma_{L,L'}),
\label{eq:def:Renyi entanglement of purification}
\end{align}
where $\ket{\phi}$ is the purification of $\sigma$ (i.e., $\tr_{\Lambda'}(\ket{\phi}\bra{\phi})=\sigma$), 
$S_\alpha(\cdot)$ is the R\'enyi entropy and $\sigma_{L,L'}:=\tr_{R,R'} (\ket{\phi} \bra{\phi})$, 
namely $S_\alpha(\sigma_{L,L'})=\frac{1}{1-\alpha} \log [\tr (\sigma^\alpha_{L,L'} ) ]$
\end{definition}
\noindent
A bound on the R\'enyi entanglement of purification imposes a stronger restriction to the structure of the quantum state than the mutual information in Eq.~\eqref{mutual_information_def}. For instance, \cite{jarkovsky2020efficient} showed that an upper bound on the entanglement of purification of a 1D system guarantees an efficient approximation by MPOs.

The mutual information $I(L:R)_{\rho_\beta} $ is related to the 
R\'enyi entanglement of purification with $\alpha=1$ (see Ref.~\cite{PhysRevA.91.042323}):
 \begin{align}
I(L:R)_\sigma \le 2 E_{p,1}(\sigma). \label{mutual_info_renyi_purifi}
\end{align}

We also present an upper bound on this quantity as follows (see Sec.~\ref{proof_thm:Renyi entanglement of purification} for the proof): 
\begin{theorem} \label{thm:Renyi entanglement of purification}
For arbitrary non-zero $0<\alpha\le 1$, the R\'enyi entanglement of purification $E_{p,\alpha}$ is upper-bounded as follows:
\begin{align}
E_{p,\alpha}(\rho) \le \tilde{C}_0 \max \left[\beta^{2/3}  \log (\beta)  ,  \frac{ (1-\alpha)\beta}{\alpha} \log\br{\frac{\beta}{\alpha}}   \right].
\label{Ineq:Renyi entanglement of purification}
\end{align}
\end{theorem}
\noindent
The above upper bound implies that for $\alpha<1$, the entanglement scaling may be linear to $\beta$ instead of $\beta^{2/3}$.   
This can be explained as follows. 
To calculate the R\'enyi entanglement of purification, we need to obtain the MPO which has an approximation error $\epsilon$ such that 
$D_\epsilon  \epsilon^\alpha \lesssim 1$, where $D_\epsilon$ is the bond dimension to achieve error $\epsilon$.
From the MPO with this property, we have $E_{p,\alpha}(\rho) \lesssim \log(D_\epsilon)$. 
Because of $D_\epsilon\lesssim e^{\sqrt{\beta \log(1/\epsilon)}}$, the condition $D_\epsilon  \epsilon^\alpha \lesssim 1$ reduces to $\sqrt{\beta \log(1/\epsilon)} - \alpha \log(1/\epsilon) \lesssim 0$ or $\log(1/\epsilon) \gtrsim \beta/\alpha^2$, which gives $\log(D_\epsilon) \approx \beta/\alpha$.  

Let us compare our result to those of previous studies~\cite{PhysRevB.73.085115, PhysRevX.4.031019, PhysRevB.91.045138}. The bond dimension scales as $D=(1/\epsilon)^{\orderof{\beta}}$. By using this estimation, there exists a critical $\alpha_c$ ($=1-\orderof{\beta^{-1}}$) that violates the finite upper bound of $E_{p,\alpha}(\rho)$ for $\alpha < \alpha_c$.

\subsection{Convex combination of matrix product states}

Ref.~\cite{PhysRevB.98.235154} showed that the thermal state can be expressed as a convex combination of MPS with bond dimension scaling doubly exponentially with $\beta$. We can prove the following corollary, which substantially improves their main result (see Sec.~\ref{sec_Proof of Corollary_corol_convex_comb} for the proof):
\begin{corol}\label{corol_convex_comb}
The quantum Gibbs state $\rho_\beta$ is given by a convex combination of the matrix product states in the following sense:
\begin{align} 
\left \| \rho_\beta - \sum_{i=1}^{\mathcal{D}_\Lambda} p_i \ket{M_i} \bra{M_i} \right \|_1 \le \epsilon ,
\label{convex_combination_MPS}
\end{align}
where $\{\ket{M_i}\}$ are matrix product states with the bond dimension of
 \begin{align}
&D = \exp[\tOrder( \tilde{q}^\ast_\epsilon)], 
\label{bond_D_M_i_0}
\end{align}
where $\tilde{q}^\ast_\epsilon$ has been defined in Theorem~\ref{main_thm_MPO_approximation}.
\end{corol}
This result can be used to justify the METTS algorithm~\cite{PhysRevLett.102.190601} and the algorithm of \cite{leviatan2017quantum} (see also Ref.~\cite{PhysRevB.98.235154} for more detailed motivations to study the convex combinations of MPS). 
Using our bounds, we provide further analytical evidence regarding why these work in practice.

A related quantity in the study of mixed-state entanglement is the \emph{entanglement of formation}. It captures the `average bond dimension' in the convex combination shown in Equation \eqref{convex_combination_MPS} and can be defined as follows.

\begin{definition} \label{def:Renyi entanglement of formation}
Let $\Lambda'$ be a copy of the total system with the Hilbert space $\mathcal{H}'$.
For arbitrary quantum state $\sigma$, we define $E_{f,\alpha}(\sigma)$ for the partition $\Lambda=L\cup R$ as 
 \begin{align}
&E_{f,\alpha}(\sigma) := \inf \sum_i p_i S_\alpha (\sigma^{(i)}_{L}),
\label{eq:def:Renyi entanglement of formation}
\end{align}
where again $S_\alpha(\cdot)$ is the R\'enyi entropy, and the minimization is over all pure-state decompositions $\sigma=\sum_i p_i \sigma^{(i)}$ .
\end{definition}

Entanglement of formation is upper bounded by the entanglement of purification \cite{TerhalHLD02} (see also the footnote~\footnote{At the inequality (9) in Ref.~\cite{TerhalHLD02}, the upper bound of $E_{f,\alpha}(\sigma) \le E_{p,\alpha}(\sigma)$ is obtained for the case of $\alpha=1$. However, the proof therein can be easily extended to generic $\alpha$.}):
\begin{equation}
E_{f,\alpha}(\sigma) \le E_{p,\alpha}(\sigma),
\end{equation}
where the equality holds for pure states. 
When $\sigma$ is given by the quantum Gibbs state $\rho_\beta$, an upper bound follows from Theorem \ref{thm:Renyi entanglement of purification},
\begin{equation}
E_{f,\alpha}(\rho_\beta) \le  \tilde{C}_0 \max \left[\beta^{2/3}  \log (\beta)  ,  \frac{ (1-\alpha)\beta}{\alpha} \log\br{\frac{\beta}{\alpha}}   \right] .
\notag 
\end{equation}

\subsection{Real time evolution} \label{sec:Improved entanglement rate in real-time evolution}

Our analyses can be partially applied to real-time evolution.
In this case, we approximate the unitary time evolution $e^{-iHt}$ instead of the quantum Gibbs state $e^{-\beta H}$. 
The most essential difference is that the random-walk-like behavior [i.e., Eq.~\eqref{random_walk_exponential}] cannot be justified.
Mathematically, the polynomial approximation based on Eq.~\eqref{random_walk_exponential} (see also~Lemma~\ref{sup_lem:low_deg_poly}) is only applicable to imaginary time evolution.
Hence, the MPO approximation of $e^{-iHt}$ requires the bond dimensions of $e^{\orderof{t}}$ instead of $e^{\orderof{t^{2/3}}}$. 
This is expected and consistent with the numerical calculations and the theoretical upper bound~\cite{PhysRevA.76.052319,PhysRevLett.97.150404,PhysRevLett.111.170501,Marien2016}.

Still, our results on the quasi-linear time algorithm can be also applied to real-time evolution, where we utilize only the Taylor expansion~\eqref{Taylor_exp_Phi}. 
Let us approximate the unitary time evolution $e^{-iHt}$ by using a MPO $M_t$. 
For an arbitrary quantum state $\ket{\psi}$, we obtain 
\begin{align}
\| ( M_t  - e^{-i Ht})\ket{\psi} \|  \le  \| M_t  - e^{-i Ht}\|_\infty. \notag 
\end{align}
Recall that the Schatten norm with $p=\infty$ is equivalent to the operator norm. Hence, by applying
Theorem~\ref{thm:quasi-linear_algorithm} to the case of $\beta=it$ and $p=\infty$, we can obtain the following corollary:
\begin{corol} \label{corol:quasi-linear_algorithm_real}
For arbitrary $t$, we can efficiently compute a matrix product operator $M_t$ that approximates $e^{-i Ht}$ in the sense that
\begin{align}\label{eq:approx-MPO1}
\| M_t  - e^{-i Ht} \|  \le  1/\poly(n) , 
\end{align}
where the bond dimension of $M_t$ is given by $\exp \left[\tOrder(|t|)+ \tOrder \br{ \sqrt{|t| \log (n)}} \right]$.  
The computational time to calculate $M_t$ is given by  
\begin{align}
\label{comp_real_time}
n \exp \left[\tOrder(|t|)+ \tOrder \br{ \sqrt{|t| \log (n)}} \right].
\end{align}
\end{corol}
For $|t|\lesssim \log(n)$, our result gives a quasi-linear computational cost for $n$; thus, it is better than the previous computational cost 
$e^{\orderof{t} + 
\orderof{\log (n/\epsilon )}}$, which is derived from the Lieb-Robinson bound~\cite{PhysRevLett.97.157202,PhysRevB.77.144302}.
However, for $|t|\gtrsim \log(n)$, the computational cost~\eqref{comp_real_time} grows exponentially with $t$, and has the same limitation as the previous methods.

\subsection{Entanglement rate by imaginary time evolution} \label{sec:imaginary_time_evo}

The quantum Gibbs state is regarded as an imaginary time evolution of the uniformly mixed state, namely $\rho_\beta \propto e^{-\beta/2} \rho_{\beta=0} e^{-\beta/2}$.  
Thus, the entropy-production rate of the imaginary time evolution is sublinear with respect to $\beta$. 
Can we extend it to general quantum states instead of the uniformly mixed state?   
Clearly, when we consider the arbitrary quantum state $\ket{\psi}$, the answer is no; that is, the entanglement generation by $e^{-\beta H}$ for a given cut (e.g., $\Lambda=L\sqcup R$) is usually unbounded.  Even if there are no interactions between $L$ and $R$ or $e^{-\beta H}=e^{-\beta H_L} \otimes e^{-\beta H_R}$, the entanglement rate can be non-zero if an initial state is arbitrarily chosen. For example, let us consider the initial state $\ket{\psi}$ as 
\begin{align}
\ket{\psi} =C \sum_{i} e^{\beta ( E_{L,i} + E_{R,i})} \ket{E_{L,i}} \otimes \ket{E_{R,i}}, \notag 
\end{align}
where $C$ is a normalization constant and $\ket{E_{L,i}}$ ($\ket{E_{R,i}}$) is the eigenstate of $H_L$ ($H_R$) with corresponding eigenvalue $E_{L,i}$ ($E_{R,i}$). 
Then, we have 
\begin{align} 
\frac{e^{-\beta H_L} \otimes e^{-\beta H_R} \ket{\psi}}{\|e^{-\beta H_L} \otimes e^{-\beta H_R} \ket{\psi} \|} \propto \sum_{i} \ket{E_{L,i}} \otimes \ket{E_{R,i}},
\notag 
\end{align}
which is the maximally entangled state. 
In the above case, the entanglement entropy is significantly increased by the Hamiltonian with no boundary-boundary interactions.

In order to obtain a non-trivial result, we here consider the imaginary time evolution for a product state $\ket{{\rm P}_{L, R}}$ as
\begin{align}
\label{imaginary_time_product_state}
\ket{{\rm P}_{L, R} (\beta)}:= \frac{e^{-\beta H} \ket{{\rm P}_{L, R}}}{\|e^{-\beta H} \ket{{\rm P}_{L, R}}\|}.
\end{align}
This setup is feasible in experimental settings~\cite{McArdle2019,motta2020determining}. 
When we consider the real-time evolution (i.e., $\beta=it$), the SIE theorem~\cite{PhysRevLett.111.170501} gives the upper bound for the entanglement rate as $\orderof{t}$.
In contrast, no theoretical studies have given an upper bound of the entanglement generation by the imaginary time evolution. 
It is an intriguing open problem whether or not the entanglement rate is finitely bounded for large $\beta$.

Using our current analyses, we can partially answer this question.  
To approximate $\ket{{\rm P}_{L, R}(\beta)}$, we use an operator $O_D$ that satisfies ${\rm SR}(O_D) =D$ for the cut of $\Lambda=L\sqcup R$ and approximates $\ket{{\rm P}_{L, R}(\beta)}$ as   
$\ket{{\rm P}_{L, R}(\beta)} \approx O_D \ket{{\rm P}_{L, R}}$. 
We aim to estimate the approximation error of $\ket{{\rm P}_{L, R}(\beta)}$ depending on the Schmidt rank $D$.  
Let us set the ground-state energy of $H$ equal to zero.
Then, from the inequality~\eqref{ineq_hat_rho_p} in Proposition~\ref{main_thm_area_law} with $p=\infty$, there exists $O_D$ such that  
\begin{align}
&\| \ket{{\rm P}_{L, R}(\beta)}-O_D \ket{{\rm P}_{L, R}}\| \le \frac{\epsilon}{\|e^{-\beta H} \ket{{\rm P}_{L, R}}\|},\notag  \\
&D=e^{\tOrder\br{\beta^{2/3}} + \tOrder\br{ \sqrt{\beta\log(\beta/\epsilon)}}}.
\end{align}
If $\|e^{-\beta H} \ket{{\rm P}_{L, R}}\| =\orderof{1}$, the entanglement entropy of ${\rm P}_{L, R}$ satisfies the same inequality as \eqref{mutual_information_def_improved_bound} and scales as $\beta^{2/3}$. 
However, in general, the quantity $\|e^{-\beta H} \ket{{\rm P}_{L, R}}\|$ is exponentially small for $n$, and hence, the value of $\epsilon$ should be as small as $e^{\orderof{n}}$, 
which gives the entanglement scaling as $\sqrt{n\beta}$.
This is still non-trivial but is rather worse than the expected scaling of $\beta^{2/3}$.

To improve the bound, a refined approximation error is required, which is given by the following form of 
\begin{align}
\label{O_D_e^beta_H_1}
\|O_D e^{\beta H} -1 \| \le \epsilon  
\end{align}
instead of the approximation $\|e^{-\beta H} -O_D \|_p \le \epsilon \|e^{-\beta H}\|_p$ for Schatten-$p$ norm. 
The approximation of the form of \eqref{O_D_e^beta_H_1} can be derived for sufficiently high temperatures (see Proposition~\ref{Prop:small_beta}). If we can extend the Proposition~\ref{main_thm_area_law} in Sec.~\ref{sec:proof_Main statement and the proof} to the form~\eqref{O_D_e^beta_H_1}, 
we will be able to prove that the entanglement rate by the imaginary time evolution~\eqref{imaginary_time_product_state} is upper-bounded by $\tOrder\br{\beta^{2/3}}$.

Finally, we mention that in various cases the entanglement rate can be smaller than $\tOrder\br{\beta^{2/3}}$. 
In particular, when the ground state is non-critical (or gapped), the imaginary time evolution for $e^{-\beta H} \ket{{\rm P}_{L, R}}$ is expected to rapidly converge to the ground state~\cite{PhysRevLett.93.207204,PhysRevLett.93.207205,PhysRevB.78.155117,PhysRevLett.106.127202,PhysRevLett.107.070601,PhysRevB.92.035152,PhysRevB.94.165116,PhysRevX.8.031082}. 
Indeed, in the case where the Hamiltonian is gapped and defined on a spin chain, there exists a product state which has an $\orderof{1}$ overlap with the ground state~\cite{PhysRevB.85.195145,arad2013area,kuwahara2019area}.
If we choose it as the initial state $\ket{{\rm P}_{L, R}}$, the entanglement entropy for  $e^{-\beta H} \ket{{\rm P}_{L, R}}/\|e^{-\beta H} \ket{{\rm P}_{L, R}}\|$ approaches a constant value (i.e., the entanglement entropy of the ground state) exponentially fast with $\beta$.
Thus, the entanglement rate should be much smaller than $\orderof{\beta^{2/3}}$.
It is an intriguing question to investigate which class of quantum many-body systems shows a non-trivial entanglement rate for the imaginary time evolution.

%



\section{Proofs of the main theorems} \label{sec:proof_Main statement and the proof}

Here we prove Theorem~\ref{main_thm_area_law2} (Theorem~\ref{thm:Renyi entanglement of purification}) and Theorem~\ref{main_thm_MPO_approximation} regarding the thermal area law. 
For simplicity, we focus on one-dimensional systems; however, the essence of the proof is the same in high-dimensional cases (see Appendix~\ref{sup_sec_Improved thermal area law in high dimensions}). 
In Sec.~\ref{sec_Proof of Corollary_corol_convex_comb}, we also prove Corollary \ref{corol_convex_comb}, which is based on Theorem~\ref{main_thm_MPO_approximation}

Both Theorems \ref{main_thm_area_law2} and \ref{main_thm_MPO_approximation} are based on the following basic approach. 
We aim to approximate the Gibbs state $\rho_\beta$ by another operator $\hat{\rho}_\beta$ which has a smaller Schmidt rank for a given cut $\Lambda = L \cup R$.
This is formalized in the following proposition, which plays a central role in deriving our main results:
\begin{prop} \label{main_thm_area_law}
Let $\epsilon$ be an arbitrary error such that $\epsilon \le e$.
Then, there exists an operator $\hat{\rho}_\beta$ which approximates $\rho_\beta$ as follows: 
\begin{align}
\| \rho_\beta -\hat{\rho}_\beta \|_p \le \epsilon \|\rho_\beta \|_p \label{ineq_hat_rho_p}
\end{align}
for arbitrary $p \in \mathbb{N}$,  and
 \begin{align}
{\rm SR}(\hat{\rho}_\beta,i_0) \le \exp[q^\ast_\epsilon \log (q^\ast_\epsilon)]
\label{ineq_on_schmidt_rank_thm}
\end{align}
with
\begin{align}
\label{def_q_ast}
q^\ast_\epsilon =C_0 \max \left(\beta^{2/3}, [\beta\log(\beta/\epsilon)]^{1/2} \right),
\end{align}
where $C_0$ is a constant of $\orderof{1}$.
\end{prop}
\noindent
The proof is shown in Appendix~\ref{sec:proof_Main proposition and the proof}. For sufficiently small $\epsilon$, this estimation gives a sublinear dependence of the Schmidt rank with respect to $(1/\epsilon)$. 
For example, for $\epsilon=1/\poly(n)$, we have ${\rm SR}(\hat{\rho}_\beta,i_0) \le n^{\log^{-1/2} (n)}$, which is slower than any power of $n$.
If the Schmidt rank of the approximating operator $\hat{\rho}_\beta$ exceeds $e^{\tOrder(\beta^{2/3})}$, the error $\epsilon$ decays super-polynomially as a function of the Schmidt rank.

\subsection{Proof of Theorems~\ref{main_thm_area_law2} and \ref{thm:Renyi entanglement of purification}}
\label{proof_thm:Renyi entanglement of purification}

Theorems~\ref{main_thm_area_law2} and \ref{thm:Renyi entanglement of purification} give upper bounds on the mutual information and the R\'enyi entanglement of purification, respectively.
Due to the inequality~\eqref{mutual_info_renyi_purifi},
Theorem~\ref{thm:Renyi entanglement of purification} includes the Theorem~\ref{main_thm_area_law2}, once we consider $E_{p,1}(\rho)$.
Hence, we only need to prove Theorem~\ref{thm:Renyi entanglement of purification}. 

We start from the purification in the form of 
\begin{align}
\ket{\psi} =Z^{-1/2} (e^{-\beta H/2}\otimes \hat{1}) \sum_{j=1}^{\mathcal{D}_\Lambda} \ket{j}_{\Lambda}\otimes \ket{j}_{\Lambda'},
\label{purification_gibbs}
\end{align}
where $\{\ket{j}\}_{j=1}^{\mathcal{D}_\Lambda}$ is an arbitrary orthonormal basis, and we denote the partition function $\tr(e^{-\beta H})$ by $Z$.
Note that from the above definition $\tr_{\Lambda'} (\ket{\psi} \bra{\psi}) = e^{-\beta H}/Z$.
Then, from the definition of the R\'enyi entanglement of purification~\eqref{eq:def:Renyi entanglement of purification}, we have 
\begin{align}
E_{p,\alpha}(\rho) \le  E_\alpha(\psi) .
\end{align}
Next, we estimate an upper bound on $E_\alpha(\psi)$.

From Proposition~\ref{main_thm_area_law}, we can find an approximation $\hat{\rho}_{\beta/4}$ of $e^{-\beta H/4}$ such that 
\begin{align}
\| e^{-\beta H/4}- \hat{\rho}_{\beta/4} \|_p \le \epsilon  \| e^{-\beta H/4}\|_p,  \label{hat_rho_approx_p}
\end{align}
for all $p$, 
where the Schmidt rank of $\hat{\rho}_{\beta/4}$ is upper-bounded by \eqref{ineq_on_schmidt_rank_thm}.
 Define $\ket{\tilde{\psi}}$ as 
\begin{align}
\ket{\tilde{\psi}} :=\tilde{Z}^{-1/2}\hat{\rho}_{\beta/4}^\dagger \hat{\rho}_{\beta/4}\otimes \hat{1} \sum_{j=1}^{\mathcal{D}_\Lambda} \ket{j}\otimes \ket{j},
\label{ket_tilde_psi_def}
\end{align}
where  we define $\tilde{Z}:= \tr \bigl( \hat{\rho}_{\beta/4}^\dagger \hat{\rho}_{\beta/4}\hat{\rho}_{\beta/4} \hat{\rho}_{\beta/4}^\dagger \bigr)$.
Using the inequality \eqref{ineq_Approximation of square operators_0} with $p=2$, $O=e^{-\beta H/4}$ and $\tilde{O}=\hat{\rho}_{\beta/4}$, we first obtain for $\tilde{Z}^{1/2}=\| \hat{\rho}_{\beta/4}^\dagger \hat{\rho}_{\beta/4} \|_2 $
\begin{align}
\tilde{Z}^{1/2}&\le \| \hat{\rho}_{\beta/4}^\dagger \hat{\rho}_{\beta/4} - e^{-\beta H/2} \|_2 +  \| e^{-\beta H/2} \|_2  \notag \\
&\le (3\epsilon +1)  \| e^{-\beta H/2} \|_2 = (3\epsilon +1)  Z^{1/2}, 
\label{tilde_Z_1/2_Z_1/2}
\end{align}
where we use the triangle inequality in the first inequality.
We then obtain the fidelity between $\ket{\psi}$ and $\ket{\tilde{\psi}} $ as follows:
\begin{align}
\langle \tilde{\psi}\ket{\psi} 
&=Z^{-1/2}\tilde{Z}^{-1/2}  \|\hat{\rho}_{\beta/4} e^{-\beta H/4}\|_2^2 \notag \\
&\ge \frac{Z^{-1}}{3\epsilon +1}\|\hat{\rho}_{\beta/4} e^{-\beta H/4}\|_2^2,
\label{Ineq:langle tilde_psi_ket_psi_ =Z^-1/2_tilde_Z_^-1/2}
\end{align}
where we apply the inequality~\eqref{tilde_Z_1/2_Z_1/2} to $\tilde{Z}$ in the last inequality.
From the triangle inequality, we obtain the upper bound of $\|\hat{\rho}_{\beta/4} e^{-\beta H/4}\|_2$ in the following form:
\begin{align} 
&\|\hat{\rho}_{\beta/4} e^{-\beta H/4}\|_2  \notag \\
&\ge  \| e^{-\beta H/2}\|_2 -\| (\hat{\rho}_{\beta/4} -e^{-\beta H/4}) e^{-\beta H/4}\|_2  \notag \\ 
&\ge  \| e^{-\beta H}\|_1^{1/2}  -\| \hat{\rho}_{\beta/4} -e^{-\beta H/4}\|_4 \cdot \| e^{-\beta H/4}\|_4 \notag \\
&\ge \| e^{-\beta H}\|_1^{1/2} -  \epsilon \|e^{-\beta H/4}\|_4^2  =(1-\epsilon ) Z^{1/2} ,
\label{Ineq:|hat_rho_beta/4_e^-_beta H/4_|_2}
\end{align}
where we use $\| e^{-\beta H/2}\|_2=\| e^{-\beta H}\|_1^{1/2}$ and the H\"older inequality in the second inequality, in the third inequality 
we use the inequality~\eqref{hat_rho_approx_p}, and the last equation is derived from $\| e^{-\beta H/4}\|_4=\| e^{-\beta H}\|_1^{1/4}$.
By applying inequality~\eqref{Ineq:|hat_rho_beta/4_e^-_beta H/4_|_2} to \eqref{Ineq:langle tilde_psi_ket_psi_ =Z^-1/2_tilde_Z_^-1/2}, we obtain the inequality of 
\begin{align}
\langle \tilde{\psi}\ket{\psi} \ge \frac{(1-\epsilon )^2}{3\epsilon +1} \ge 1-5\epsilon,
\end{align}
which implies 
\begin{align}
\| \ket{\psi} - \ket{\tilde{\psi}} \|^2  \le 2- 2 \langle \tilde{\psi}\ket{\psi} \le 10\epsilon, \label{norm_diff_psi_tilde_psi}
\end{align}
In the following, using the above upper bound, we estimate the upper bound of R\'enyi entanglement entropy for arbitrary $\alpha>0$.
We consider the cases of $\alpha=1$ and $\alpha< 1$ separately.

\subsubsection{Case of $\alpha=1$}

We first consider the case of $\alpha= 1$.
We define $\ket{\tilde{\psi}_s}$ as an approximation of $\ket{\psi}$ which satisfies 
\begin{align}
\| \ket{\psi} - \ket{\tilde{\psi}_s} \|^2  \le 1/s^2,
\label{cond_tilde_psi_s_1}
\end{align}
where we use Eq.~\eqref{ket_tilde_psi_def} for the representation of $\ket{\tilde{\psi}_s}$. 
From Theorem~1, the Schmidt rank of $\ket{\tilde{\psi}_s}$, say $D_s$, is upper-bounded from above by
\begin{align}
D_s \le e^{q_s \log (q_s)} \label{D_s_upp_bound_1}
\end{align}
with
$
q_s =\tilde{C} \max \left(\beta^{2/3}, [\beta\log(s)]^{1/2} \right).
$
We define $\bar{s}$ as an integer such that
\begin{align}
q_s \begin{cases}
= \tilde{C}\beta^{2/3} &\for s \le \bar{s},    \notag \\
= \tilde{C} [\beta\log(s)]^{1/2} &\for s > \bar{s} ,
\end{cases}
\end{align}
where $\bar{s}$ is in the order of $\exp[\orderof{\beta^{1/3}}]$.

Let us denote the Schmidt decomposition of $\ket{\psi}$ in Eq.~\eqref{purification_gibbs} as follows:
 \begin{align}
\ket{\psi} = \sum_{m=1}^{D_\psi} \mu_m  \ket{\psi_{L,L',m}} \otimes \ket{\psi_{R,R',m}}, \label{Schmmidt_psi_ket}
\end{align}
where $\ket{\psi_{L,L',m}}$ and $\ket{\psi_{R,R',m}}$ are defined on the Hilbert space of $L\sqcup L'$ and $R\sqcup R'$, respectively.
From the above representation, we obtain the R\'enyi entropy with $\alpha=1$ as 
 \begin{align}
S_1(\ket{\psi}) =-\sum_{m=1}^\infty \mu_m^{2} \log (\mu_m^{2})  ,
\end{align}
which is equal to the standard entanglement entropy.

To estimate $S_1(\ket{\psi})$, we utilize the Eckart-Young theorem.
By applying the inequality~\eqref{sup_thm:The Eckart-Young theorem} to $\ket{\psi}$ and $\ket{\tilde{\psi}_s}$,  
we obtain the following inequality:
 \begin{align}
\sum_{m>D_s} \mu_m^2 \le \| \ket{\psi} -\ket{\tilde{\psi}_s}\|^2 \le 1/s^2  , \label{m>D_s_sum_upp}
\end{align}
where, in the second inequality, we use the condition~\eqref{cond_tilde_psi_s_1}.
To upper-bound the R\'enyi entropy, we first define 
 \begin{align}
\Gamma_s^2 := \sum_{m=D_{s}+1}^{D_{s+1} } \mu_{m}^2, \label{Def:gamma_s^2}
\end{align}
where we define $D_{0}=0$. 
We then obtain 
 \begin{align}
&S_1(\ket{\psi})  \notag \\
&= -\sum_{m=1}^{D_{\bar{s}} }\mu_{m}^2 \log (\mu_{m}^2 ) -\sum_{s=\bar{s}}^\infty  \sum_{m=D_{s}+1}^{D_{s+1} }\mu_{m}^2 \log (\mu_{m}^2 )\notag \\
&\le \log(D_{\bar{s}})  - \sum_{s=\bar{s}}^\infty\sum_{m=D_{s}+1}^{D_{s+1} } \Gamma_s^2 \log  \frac{\Gamma_s^2}{D_{s+1}-D_{s}},
\label{upp_bound_entropy_Gamma_p}
\end{align}
where we use the fact that the uniform distribution maximizes $\sum_{m=D_{s}+1}^{D_{s+1} } \mu_{m}^2 \log (\mu_{m}^2 )$, i.e., $\mu^2_{D_s+1}=\mu^2_{D_s+2}=\cdots =\mu^2_{D_{s+1}}=\Gamma_s^2/(D_{s+1}-D_{s})$. 
Because of the inequalities~\eqref{D_s_upp_bound_1} and \eqref{m>D_s_sum_upp}, we have $\Gamma_s^2\le 1/s^2$
  \begin{align}
S_1(\ket{\psi}) 
\le& \tilde{C}\beta^{2/3} \log(\tilde{C}\beta^{2/3}) + \sum_{s=\bar{s}}^\infty (1/s)^2 \log (3s^2)   \notag \\ 
&+ \sum_{s=\bar{s}}^\infty \frac{\tilde{C} [\beta\log(s)]^{1/2}  \log (\tilde{C} [\beta\log(s)]^{1/2} )}{s^2} ,\notag 
\end{align}
where we apply  the inequality $-x \log x \le -x \log (x/3) \le -y \log (y/3)$ for $0< x \le y \le 1$ to $-\Gamma_s^2 \log (\Gamma_s^2)$.
Using $\bar{s}=\exp[\orderof{\beta^{1/3}}]$, the second and the third terms become less dominant in comparison with the first term when $\beta$ is large.
We thus obtain the main inequality~\eqref{Ineq:Renyi entanglement of purification} in the theorem for $\alpha=1$.

\subsubsection{Case of $\alpha<1$}

We follow the same analyses as in the case of $\alpha=1$.
In this case, we define $\ket{\tilde{\psi}_s'}$ as an approximation of $\ket{\psi}$ which satisfies 
\begin{align}
\| \ket{\psi} - \ket{\tilde{\psi}_s'} \|^2  \le s^{-2/\alpha},
\label{diffenrence_generic_alpha}
\end{align}
where the Schmidt rank of $\ket{\tilde{\psi}_s'}$, say $D_s$, is upper-bounded from above by
\begin{align}
D'_s \le e^{q_s' \log (q_s')} \label{D_s_upp_bound}
\end{align}
with
$
q_s' =\tilde{C} \max \left(\beta^{2/3}, [\alpha^{-1}\beta\log(s)]^{1/2} \right).
$
We define $\bar{s}'$ as an integer such that
\begin{align}
q_s' \begin{cases}
= \tilde{C}\beta^{2/3} &\for s \le \bar{s}',    \notag \\
= \tilde{C} [\alpha^{-1} \beta\log(s)]^{1/2} &\for s > \bar{s}' ,
\end{cases}
\end{align}
where we have $\bar{s}'=\exp[\orderof{\alpha \beta^{1/3}}]$.

Using the Schmidt decomposition as in Eq.~\eqref{Schmmidt_psi_ket}, the $\alpha$-R\'enyi entropy is given by
 \begin{align}
S_{\alpha}(\ket{\psi}) 
&= \frac{1}{1-\alpha} \log \left(\sum_{s=0}^\infty \sum_{m=D'_{s}+1}^{D'_{s+1} } \mu_{m}^{2\alpha}  \right) . 
\end{align}
For $\alpha<1$, we obtain the upper bound of
 \begin{align}
\sum_{m=D'_{s}+1}^{D'_{s+1} } \mu_{m}^{2\alpha} 
\le  (D'_{s+1}-D'_s) \left( \frac{\Gamma_s^{'2}}{D'_{s+1}-D'_s}\right)^{\alpha} \le  \frac{D_{s+1}^{'1-\alpha}}{s^2},  \notag 
\end{align}
where we adopt the similar notation~\eqref{Def:gamma_s^2} for $\Gamma'_s$, and to derive $\Gamma_s^{'2} \le s^{-2/\alpha}$, 
we use the condition~\eqref{diffenrence_generic_alpha} and the Eckart-Young theorem as in \eqref{m>D_s_sum_upp}.
Therefore, we have the following upper bound for the summation  
 \begin{align}
&\sum_{s=0}^\infty  \sum_{D_{s}<m \le D_{s+1} }\mu_{m}^{2\alpha}
\le D_{\bar{s}}^{'1-\alpha} +  \sum_{s\ge \bar{s}}  \frac{D_{s+1}^{'1-\alpha}}{s^2} \notag \\
& \le e^{(1-\alpha)\tilde{C}\beta^{2/3}  \log (\tilde{C}\beta^{2/3} )}  \notag \\
&+  \sum_{s\ge \bar{s}} \frac{\exp\left((1-\alpha)\tilde{c}_\alpha\log^{1/2}(s)  \log [\tilde{c}_\alpha\log^{1/2}(s) ] \right)}{s^2}, \notag 
\end{align}
where we define $\tilde{c}_\alpha:= \tilde{C} \sqrt{\beta/\alpha}$.
For the estimation of the summation for $\sum_{s\ge \bar{s}}$, we also use the inequality of 
 \begin{align}
&\int_1^\infty  \frac{\exp\left((1-\alpha) \tilde{c}_\alpha\log^{1/2}(x)  \log [\tilde{c}_\alpha\log^{1/2}(x) ] \right)}{x^2}dx   \notag \\
=&\int_0^\infty  2 t e^{-t^2 + (1-\alpha) \tilde{c}_\alpha t \log (\tilde{c}_\alpha t) } dt  \le e^{\tilde{C}_1 (1-\alpha)^2\tilde{c}_\alpha^2 \log^2 (\tilde{c}_\alpha) } , \notag 
\end{align}
where $\tilde{C}_1$ is a constant of $\orderof{1}$.
By combining the above inequalities together, we obtain 
 \begin{align}
&\frac{1}{1-\alpha}\log \left( \sum_{s=0}^\infty  \sum_{D_{s}<m \le D_{s+1} }\mu_{m}^{2\alpha}  \right) \notag \\
&\le \tilde{C}_0 \max \left(\beta^{2/3}  \log (\beta)  ,  (1-\alpha) (\beta/\alpha) \log(\beta/\alpha)   \right). \notag
\end{align}
This gives the main inequality~\eqref{Ineq:Renyi entanglement of purification} in the theorem for $\alpha<1$.
This completes the proof. $\square$

\subsection{Proof of Theorem~\ref{main_thm_MPO_approximation}} \label{sec_thm:MPO_approx}

Here, we prove Theorem~\ref{main_thm_MPO_approximation}, which gives the MPO approximation of quantum Gibbs state.

We first prove the case of $p=2$.
Let $\hat{\rho}_ {\beta}$ be an approximation of $\rho_{\beta}$ such that for a given cut $\Lambda=L\sqcup R$.
We define the Schmidt rank of $\hat{\rho}_{\beta}$ as $D_{\epsilon_0}$, which satisfies the inequality~\eqref{ineq_on_schmidt_rank_thm} with the approximation error $\epsilon_0$, namely
\begin{align}
\| \rho_{\beta} -\hat{\rho}_\beta \|_p \le \epsilon_0 \|\rho_\beta \|_p 
\label{Young_Eckardt_base}
\end{align}
with 
\begin{align}
{\rm SR} (\hat{\rho}_\beta , i_0) := D_{\epsilon_0} \le \exp[q^\ast_\epsilon \log (q^\ast_\epsilon)],
\end{align}
where $q^\ast_\epsilon$ has been defined in Eq.~\eqref{def_q_ast}.
For the  cut, we define the Schmidt decomposition of $\rho_\beta$ as follows:
\begin{align}
\rho_{\beta} =\sum_{m} \mu_m \Phi_{L,m} \otimes \Phi_{R,m} \quad (\mu_m >0),
\end{align}
where $\{\Phi_{L,m}\}$ ($\{\Phi_{R,m}\}$) are orthonormal operator bases which satisfy
\begin{align}
\| \Phi_{L,m}\|_2 = 1 , \quad \tr (\Phi_{L,m}\Phi_{L,m'})=0 
\end{align}
for $m\neq m'$.
Note that from the above definition, we have
\begin{align}
\| \rho_{\beta}\|_2^2=  \sum_{m} \mu_m^2 .
\end{align}

By applying the Eckart-Young theorem~\eqref{sup_thm:The Eckart-Young theorem_op} to $\rho_\beta$ and $\hat{\rho}_\beta$, we obtain
\begin{align}
\sum_{m>D_{\epsilon_0}} \mu_m^2  \le \|\rho_{\beta} - \hat{\rho}_{\beta}\|_2^2 \le \epsilon_0^2\|\rho_{\beta} \|_2^2,
\label{Young_Eckart_basic_MPO}
\end{align}
where we use the inequality~\eqref{Young_Eckardt_base} with $p=2$.
Then, from Lemma~1 in Ref.~\cite{VC06mps}, there exists an MPO $M_{D_\epsilon}$ such that
\begin{align}
\|\rho_\beta -M_{D_{\epsilon_0}} \|_2^2 \le 2\epsilon_0^2 n\|\rho_\beta \|_2^2 .
\end{align}
Therefore, by choosing $\epsilon_0=[\epsilon/(2n)]^{1/2}$, we obtain the desired approximation error~\eqref{MPO_approx}, and the bond dimension $D_{\epsilon_0}$ satisfies the inequality~\eqref{MPO_approx_bond}.

Second, we prove the case of $p=1$.
For this, we consider the purification of the quantum Gibbs state $\rho_{\beta/2}$ as in Eq.~\eqref{purification_gibbs}, which is denoted by $\ket{\psi}$:
\begin{align}
\ket{\psi} &=Z^{-1/2} (e^{-\beta H/2}\otimes \hat{1}) \sum_{j=1}^{\mathcal{D}_\Lambda} \ket{j}_{\Lambda}\otimes \ket{j}_{\Lambda'} \notag \\
&= \sum_{m=1}^{D_\psi} \nu_m  \ket{\psi_{L,L',m}} \otimes \ket{\psi_{R,R',m}},
\end{align}
Where, in the second equation, we use an expression of the Schmidt decomposition similar to that of Eq.~\eqref{Schmmidt_psi_ket}.
If we can obtain a matrix product state (MPS) $\ket{M_{\tilde{D}_\epsilon}}$ such that
\begin{align}
\| \ket{\psi} - \ket{M_{\tilde{D}_\epsilon}} \| \le \epsilon, \label{MPS_approx_psi}
\end{align}
we obtain 
\begin{align}
\left \| \tr_{\Lambda'} \left(\ket{\psi}\bra{\psi} - \ket{M_{\tilde{D}_\epsilon}} \bra{M_{\tilde{D}_\epsilon}} \right) \right\|_1 = 
\left \| \rho_\beta - M_{\tilde{D}^2_\epsilon} \right\|_1 \le \epsilon.
\end{align}
where we define $M_{\tilde{D}^2_\epsilon}:=\tr_{\Lambda'}(\ket{M_{\tilde{D}_\epsilon}}\bra{M_{\tilde{D}_\epsilon}} )$. Note that $\ket{M_{\tilde{D}_\epsilon}} \bra{M_{\tilde{D}_\epsilon}}$ is given by a MPO with the bond dimension of $\tilde{D}_\epsilon^2$.

Our task is now to find an MPS $\ket{M_{\tilde{D}_\epsilon}}$ which satisfies \eqref{MPS_approx_psi}.
For this purpose, we consider the purification of $\hat{\rho}_{\beta/4}^\dagger \hat{\rho}_{\beta/4}$ as in Eq.~\eqref{ket_tilde_psi_def}, which we denote by $\ket{\tilde{\psi}}$.
Here, $\hat{\rho}_{\beta/4}$ gives the approximation of $\rho_{\beta/4}$ as 
\begin{align}
\| \rho_{\beta/4} -\hat{\rho}_{\beta/4} \|_2 \le \epsilon_1 \|\rho_{\beta/4} \|_2 
\end{align}
with ${\rm SR} (\hat{\rho}_{\beta/4},i_0)=D_{\epsilon_1}\le \exp[q^\ast_{\epsilon_1} \log (q^\ast_{\epsilon_1})]$ for a given cut $\Lambda=L\sqcup R$.
The Schmidt rank of $\ket{\tilde{\psi}}$ along the cut is upper-bounded by $D_{\epsilon_1}^2$. 
In contrast, from the inequality~\eqref{norm_diff_psi_tilde_psi}, we obtain 
\begin{align}
\| \ket{\psi} - \ket{\tilde{\psi}} \|^2 \le 10\epsilon_1,
\end{align}
and hence, the Eckart-Young theorem gives the same inequality as \eqref{Young_Eckart_basic_MPO}:
\begin{align}
\sum_{m>D_{\epsilon_1}^2} \nu_m^2  \le \| \ket{\psi} - \ket{\tilde{\psi}} \|^2 \le 10\epsilon_1.
\label{Young_Eckart_basic_MPO1}
\end{align}
Thus, from Lemma~1 in Ref.~\cite{VC06mps}, there exists an MPS $\ket{M_{D_{\epsilon_1}^2}}$ such that
\begin{align}
\| \ket{\psi} -\ket{M_{D_{\epsilon_1}^2}} \| \le \sqrt{20 n\epsilon_1}.
\end{align}
To obtain the approximation error $\epsilon$, we need to choose $\epsilon_1=\epsilon/(20n)$.
Therefore, if we choose $D=D_{\epsilon^2/(400n^2)}$, there exists an MPO $M_D$ that satisfies the inequality~\eqref{MPO_approx} with $p=1$. 
Note that the bond dimension $D_{\epsilon^2/(400n^2)}$ satisfies the inequality~\eqref{MPO_approx_bond} by choosing $C_0'$ appropriately. 
This completes the proof of Theorem~\ref{main_thm_MPO_approximation}. $\square$

\subsection{Proof of Corollary~\ref{corol_convex_comb}} \label{sec_Proof of Corollary_corol_convex_comb}

Here, we prove that the quantum Gibbs state is well approximated by a convex combination of matrix product states as in \eqref{convex_combination_MPS}:
\begin{align} 
\rho_\beta \approx  \sum_{i=1}^{\mathcal{D}_\Lambda} p_i \ket{M_i} \bra{M_i}  .
\label{convex_combination_MPS_sup}
\end{align}
We then show that the approximation error $\epsilon$ is achieved by taking the bond dimension as in Eq.~\eqref{bond_D_M_i_0}.  

The proof is based on Theorem~\ref{main_thm_MPO_approximation}. 
We first consider the MPO approximation of $e^{-\beta H/2}$ as follows:
\begin{align}
\|e^{-\beta H/2} - M_{\beta/2} \|_2 \le \frac{\epsilon}{6} \|e^{-\beta H/2}\|_2,
\end{align}
where the bond dimension of $M_{\beta/2}$ is given by Eq.~\eqref{MPO_approx_bond} [or Eq.~\eqref{bond_D_M_i_0}].
By using Lemma~\ref{lem:Approximation of square operators_0} with $p=1$, we get 
\begin{align}
\label{inequ_e_-beta_H_M_beta_P_i_01}
\left \|e^{-\beta H} - M_{\beta/2} M_{\beta/2}^\dagger \right \|_1 \le  \frac{\epsilon}{2} \|e^{-\beta H}\|_1.
\end{align}
By inserting $\hat{1}=\sum_{i=1}^{\mathcal{D}_\Lambda} \ket{P_i} \bra{P_i}$ with $\{P_i\}_{i=1}^{D_\Lambda}$ the product-state basis, we obtain 
\begin{align}
\left\| \frac{e^{-\beta H}}{\tr(e^{-\beta H})}- \sum_{i=1}^{\mathcal{D}_\Lambda}\frac{ M_{\beta/2}\ket{P_i} \bra{P_i} M_{\beta/2}^\dagger }{\tr(e^{-\beta H})}\right \|_1 \le  \frac{\epsilon}{2},
\label{inequ_e_-beta_H_M_beta_P_i}
\end{align}
where we use $\|e^{-\beta H}\|_1= \tr(e^{-\beta H})$.

We now define 
\begin{align}
&\ket{M_i} := \frac{M_{\beta/2}\ket{P_i}}{\|M_{\beta/2}\ket{P_i}\|}  ,\quad  p_i := \frac{\|M_{\beta/2}\ket{P_i}\|^2 }{\| M_{\beta/2} \|^2_2}, \notag \\
&\sigma_\beta:=  \sum_{i=1}^{\mathcal{D}_\Lambda} p_i \ket{M_i} \bra{M_i},
\end{align}
where $\sigma_\beta$ is the normalized quantum state and satisfies $\|\sigma_\beta\|_1=1$ because of $\sum_i \|M_{\beta/2}\ket{P_i}\|^2 = \tr(M_{\beta/2} M_{\beta/2}^\dagger)=\| M_{\beta/2} \|^2_2$.
The MPO $M_{\beta/2}$ has the bond dimension of \eqref{bond_D_M_i_0}, and hence, the quantum state $M_{\beta/2}\ket{P_i}$ is also given by a matrix product state with \eqref{bond_D_M_i_0}.
We obtain the norm difference between $\rho_\beta$ and $\sigma_\beta$ as 
\begin{align}
&\left \|\rho_\beta - \sigma_\beta \right \|_1  \notag \\
&\le  \left \|\rho_\beta - \frac{\| M_{\beta/2} \|^2_2}{\tr(e^{-\beta H})}\sigma_\beta \right \|_1 + 
\left \|\sigma_\beta - \frac{\| M_{\beta/2} \|^2_2}{\tr(e^{-\beta H})}\sigma_\beta \right \|_1  \notag \\
&\le \frac{\epsilon}{2} + \left |1 - \frac{\| M_{\beta/2} \|^2_2}{\tr(e^{-\beta H})} \right| \cdot \|\sigma_\beta \|_1 \le \epsilon,  
\end{align}
where we use \eqref{inequ_e_-beta_H_M_beta_P_i} for the first term, and for the second term we use $\|\sigma_\beta\|_1=1$ and 
\begin{align}
\left |\tr(e^{-\beta H}) - \| M_{\beta/2} \|^2_2 \right| &= \left |\tr \br{e^{-\beta H} - M_{\beta/2} M_{\beta/2}^\dagger}\right| \notag \\
& \le  \left \|e^{-\beta H} - M_{\beta/2} M_{\beta/2}^\dagger \right \|_1 \notag \\
&\le  \frac{\epsilon}{2} \tr (e^{-\beta H}).
\end{align}
We thus prove the inequality~\eqref{convex_combination_MPS}.
This completes the proof. $\square$

\section{Conclusion} \label{sec:Conclusion}
We have shown two main results in this work.
The first one is the improved thermal area law that gives a scaling of $\tOrder\br{\beta^{2/3}}$ over all lattices (Theorem~\ref{main_thm_area_law2}).
This scaling behavior is qualitatively explained by the fact that the imaginary time evolution is intrinsically related to the random walk as in Eq.~\eqref{random_walk_exponential}.
In the 1D case, we also give an MPO representation of the quantum Gibbs state with a sublinear bond dimension with respect to the system size $n$ (Theorem~\ref{main_thm_MPO_approximation}). 
The second one is a quasi-linear time algorithm for preparing an MPO approximation to the 1D thermal state (Theorem~\ref{thm:quasi-linear_algorithm}), which improves upon all the prior rigorous constructions.
It also justifies the quasi-linear runtime of several heuristic algorithms inspired by the MPO-based techniques.
Moreover, our algorithm can be applied to the computation of the ground state under the low-energy-density assumption of \eqref {density_of_state_DE}.
Our first technical insight is the use of polynomial approximations of the exponential function, which are based on Taylor truncation and Chebyshev expansion~\eqref{random_walk_exponential}. The second technical contribution is a Trotter-Suzuki type decomposition of the Gibbs state (see Fig.~\ref{fig:quasi_linear_alg1}).
It would be interesting to see the possibility to further develop our approximation by using the results in Ref.~\cite{childs2019theory}.

We leave the following questions to be considered in future work.
\begin{itemize}
    \item \textbf{High-dimensional PEPO representation with sublinear bond dimension:} 
Our analytical approach has improved the bond dimension of the MPO for 1D quantum Gibbs states. 
Here, the point is to utilize the estimation in Ref.~\cite{arad2013area} to efficiently encode the polynomial of the Hamiltonian to the MPO representation. 
We expect that the same improvement should be possible in the PEPO approximation for the high-dimensional Gibbs state. 
Even though the PEPO representation of the quantum Gibbs state does not imply an efficient simulation by itself~\cite{Barahona_1982,Goldberg2015}, it is of great importance in the implementation of numerical algorithms employing the PEPO ansatz.
The key question is how to encode the polynomial of the Hamiltonian to a PEPO representation with a non-trivial bond dimension. 
Such a representation will also be useful in the context of area laws for ground states in higher dimensions.

    \item \textbf{Improving the runtime of the algorithm:} 
Our algorithm presented in Theorem~\ref{thm:quasi-linear_algorithm} has a runtime of 
 $n e^{\tOrder\br{\beta} + \tOrder\br{\sqrt{ \beta \log(n)}}}$. We expect that this could be improved to $n e^{\tOrder\br{\beta^{2/3}} + \tOrder\br{\sqrt{ \beta \log(n)}}}$ because this matches the bond dimension of the MPO constructed in Theorem~\ref{main_thm_MPO_approximation}. 
Another challenge is to improve the runtime to the subexponential form with respect to $\log(n)$ for $\beta=\orderof{\log(n)}$. 
This improvement would lead to quasi-linear time algorithms for ground states under the assumption~\eqref {density_of_state_DE}.
The main difficulty lies in constructing a better polynomial approximation to the quantum Gibbs state than $M_{\beta_0}^{(\beta/\beta_0)}$ in Eq.~\eqref{def_M_beta_0}
    
    \item \textbf{Stronger norm inequality for imaginary time evolution:} 
    As discussed in Sec.~\ref{sec:imaginary_time_evo}, we observed that an approximation of the form $\|O_D e^{\beta H} - 1\| \leq \epsilon$ instead of the current one $\|e^{-\beta H} - O_D\|_p \leq \epsilon \|e^{-\beta H} \|_p$ would lead to an imaginary-time version of the SIE theorem.
    
    \item \textbf{Circuit complexity of preparing 1D quantum Gibbs state:} 
As discussed after Theorem~\ref{main_thm_MPO_approximation}, we believe that our MPO approximation could be used to construct a quantum circuit for preparing the quantum Gibbs state. 
So far, the best estimation requires $n^{\orderof{\beta}}$ to prepare the 1D quantum Gibbs states on the quantum computer~\cite{PhysRevLett.105.170405}.
The quantum preparation of the quantum Gibbs state is expected to be easier than the MPO construction on the classical computer. 
Hence, we conjecture that the sufficient number of the elementary quantum gates should be also quasi-linear as in \eqref{eq:time-complexity}.

For instance, the adiabatic algorithm presented in~\cite{PhysRevLett.116.080503} could be used in this context, by establishing the injectivity of the MPO in \eqref{MPO_approx}. 
As another route, we may be able to employ the techniques in~\cite[Appendix~B]{vanApeldoorn2020quantumsdpsolvers}, which implemented the smooth-function of a Hamiltonian (see also~\cite[Sec. 5.3]{10.1145/3313276.3316366} for further discussions). By using this method, which relies on polynomial approximations to $e^{-\beta H}$, the polynomial presented in Theorem~\ref{thm:quasi-linear_algorithm} could be efficiently implemented on a quantum computer.

    \item \textbf{Improving the thermal area law to $\bm{\beta^{1/2}|\partial L|}$:} 
In this work, we identified the critical $\gamma_c$ satisfying \eqref{mutual_information_improve} as $1/5\leq \gamma_c \leq 2/3$. 
From the random walk behavior in Sec.~\ref{sec:Physical intuition from the random walk behavior}, we expect that $\gamma_c$ may be equal to $1/2$ or even smaller, which would suggest the diffusive propagation of information by the imaginary-time evolution.  
For the characterization of entanglement structures of quantum many-body systems at finite temperatures, 
identification of the optimal $\gamma$ is one of the most fundamental future problems. 
\end{itemize}

\begin{acknowledgments}
A.A. would like to thank David Gosset for introducing the excellent survey \cite{TCS-065} on polynomial approximations. 
The authors also would like to thank John Preskill for the suggestion that an improvement to the thermal area law on graph networks might not hold.
The work of T.K. is supported by the RIKEN Center for AIP and JSPS KAKENHI Grant No. 18K13475. Part of the work was done when T.K. was visiting the Perimeter Institute.
TK gives thanks to God for his wisdom.
The work was done when A.A. was affiliated with the Institute for Quantum Computing and the Department of Combinatorics $\&$ Optimization, University of Waterloo  and A.M.A., A.A were with the Perimeter Institute for Theoretical Physics. A.A. was supported by the Canadian Institute for Advanced Research, through funding provided to the Institute for Quantum Computing by the Government of Canada and the Province of Ontario. This research was supported in part by the Perimeter Institute for Theoretical Physics. Research at the Perimeter Institute is supported in part by the Government of Canada through the Department of Innovation, Science and Economic Development Canada and by the Province of Ontario through the Ministry of Colleges and Universities.
\end{acknowledgments}

%
%
%

\bibliography{Area_thermal}

\providecommand{\noopsort}[1]{}\providecommand{\singleletter}[1]{#1}%
\begin{thebibliography}{128}%
\makeatletter
\providecommand \@ifxundefined [1]{%
 \@ifx{#1\undefined}
}%
\providecommand \@ifnum [1]{%
 \ifnum #1\expandafter \@firstoftwo
 \else \expandafter \@secondoftwo
 \fi
}%
\providecommand \@ifx [1]{%
 \ifx #1\expandafter \@firstoftwo
 \else \expandafter \@secondoftwo
 \fi
}%
\providecommand \natexlab [1]{#1}%
\providecommand \enquote  [1]{``#1''}%
\providecommand \bibnamefont  [1]{#1}%
\providecommand \bibfnamefont [1]{#1}%
\providecommand \citenamefont [1]{#1}%
\providecommand \href@noop [0]{\@secondoftwo}%
\providecommand \href [0]{\begingroup \@sanitize@url \@href}%
\providecommand \@href[1]{\@@startlink{#1}\@@href}%
\providecommand \@@href[1]{\endgroup#1\@@endlink}%
\providecommand \@sanitize@url [0]{\catcode `\\12\catcode `\$12\catcode
  `\&12\catcode `\#12\catcode `\^12\catcode `\_12\catcode `\%12\relax}%
\providecommand \@@startlink[1]{}%
\providecommand \@@endlink[0]{}%
\providecommand \url  [0]{\begingroup\@sanitize@url \@url }%
\providecommand \@url [1]{\endgroup\@href {#1}{\urlprefix }}%
\providecommand \urlprefix  [0]{URL }%
\providecommand \Eprint [0]{\href }%
\providecommand \doibase [0]{http://dx.doi.org/}%
\providecommand \selectlanguage [0]{\@gobble}%
\providecommand \bibinfo  [0]{\@secondoftwo}%
\providecommand \bibfield  [0]{\@secondoftwo}%
\providecommand \translation [1]{[#1]}%
\providecommand \BibitemOpen [0]{}%
\providecommand \bibitemStop [0]{}%
\providecommand \bibitemNoStop [0]{.\EOS\space}%
\providecommand \EOS [0]{\spacefactor3000\relax}%
\providecommand \BibitemShut  [1]{\csname bibitem#1\endcsname}%
\let\auto@bib@innerbib\@empty
\bibitem [{\citenamefont {Bernien}\ \emph {et~al.}(2017)\citenamefont
  {Bernien}, \citenamefont {Schwartz}, \citenamefont {Keesling}, \citenamefont
  {Levine}, \citenamefont {Omran}, \citenamefont {Pichler}, \citenamefont
  {Choi}, \citenamefont {Zibrov}, \citenamefont {Endres}, \citenamefont
  {Greiner} \emph {et~al.}}]{bernien2017probing}%
  \BibitemOpen
  \bibfield  {author} {\bibinfo {author} {\bibfnamefont {Hannes}\ \bibnamefont
  {Bernien}}, \bibinfo {author} {\bibfnamefont {Sylvain}\ \bibnamefont
  {Schwartz}}, \bibinfo {author} {\bibfnamefont {Alexander}\ \bibnamefont
  {Keesling}}, \bibinfo {author} {\bibfnamefont {Harry}\ \bibnamefont
  {Levine}}, \bibinfo {author} {\bibfnamefont {Ahmed}\ \bibnamefont {Omran}},
  \bibinfo {author} {\bibfnamefont {Hannes}\ \bibnamefont {Pichler}}, \bibinfo
  {author} {\bibfnamefont {Soonwon}\ \bibnamefont {Choi}}, \bibinfo {author}
  {\bibfnamefont {Alexander~S}\ \bibnamefont {Zibrov}}, \bibinfo {author}
  {\bibfnamefont {Manuel}\ \bibnamefont {Endres}}, \bibinfo {author}
  {\bibfnamefont {Markus}\ \bibnamefont {Greiner}},  \emph {et~al.},\
  }\bibfield  {title} {\enquote {\bibinfo {title} {{\it Probing many-body
  dynamics on a 51-atom quantum simulator}},}\ }\href
  {https://doi.org/10.1038/nature24622} {\bibfield  {journal} {\bibinfo
  {journal} {Nature}\ }\textbf {\bibinfo {volume} {551}},\ \bibinfo {pages}
  {579} (\bibinfo {year} {2017})},\ \bibinfo {note} {article}\BibitemShut
  {NoStop}%
\bibitem [{\citenamefont {Zhang}\ \emph {et~al.}(2017)\citenamefont {Zhang},
  \citenamefont {Pagano}, \citenamefont {Hess}, \citenamefont {Kyprianidis},
  \citenamefont {Becker}, \citenamefont {Kaplan}, \citenamefont {Gorshkov},
  \citenamefont {Gong},\ and\ \citenamefont {Monroe}}]{zhang2017observation}%
  \BibitemOpen
  \bibfield  {author} {\bibinfo {author} {\bibfnamefont {Jiehang}\ \bibnamefont
  {Zhang}}, \bibinfo {author} {\bibfnamefont {Guido}\ \bibnamefont {Pagano}},
  \bibinfo {author} {\bibfnamefont {Paul~W}\ \bibnamefont {Hess}}, \bibinfo
  {author} {\bibfnamefont {Antonis}\ \bibnamefont {Kyprianidis}}, \bibinfo
  {author} {\bibfnamefont {Patrick}\ \bibnamefont {Becker}}, \bibinfo {author}
  {\bibfnamefont {Harvey}\ \bibnamefont {Kaplan}}, \bibinfo {author}
  {\bibfnamefont {Alexey~V}\ \bibnamefont {Gorshkov}}, \bibinfo {author}
  {\bibfnamefont {Z-X}\ \bibnamefont {Gong}}, \ and\ \bibinfo {author}
  {\bibfnamefont {Christopher}\ \bibnamefont {Monroe}},\ }\bibfield  {title}
  {\enquote {\bibinfo {title} {{\it Observation of a many-body dynamical phase
  transition with a 53-qubit quantum simulator}},}\ }\href
  {https://doi.org/10.1038/nature24654} {\bibfield  {journal} {\bibinfo
  {journal} {Nature}\ }\textbf {\bibinfo {volume} {551}},\ \bibinfo {pages}
  {601} (\bibinfo {year} {2017})}\BibitemShut {NoStop}%
\bibitem [{\citenamefont {King}\ \emph {et~al.}(2018)\citenamefont {King},
  \citenamefont {Carrasquilla}, \citenamefont {Raymond}, \citenamefont
  {Ozfidan}, \citenamefont {Andriyash}, \citenamefont {Berkley}, \citenamefont
  {Reis}, \citenamefont {Lanting}, \citenamefont {Harris}, \citenamefont
  {Altomare} \emph {et~al.}}]{king2018observation}%
  \BibitemOpen
  \bibfield  {author} {\bibinfo {author} {\bibfnamefont {Andrew~D}\
  \bibnamefont {King}}, \bibinfo {author} {\bibfnamefont {Juan}\ \bibnamefont
  {Carrasquilla}}, \bibinfo {author} {\bibfnamefont {Jack}\ \bibnamefont
  {Raymond}}, \bibinfo {author} {\bibfnamefont {Isil}\ \bibnamefont {Ozfidan}},
  \bibinfo {author} {\bibfnamefont {Evgeny}\ \bibnamefont {Andriyash}},
  \bibinfo {author} {\bibfnamefont {Andrew}\ \bibnamefont {Berkley}}, \bibinfo
  {author} {\bibfnamefont {Mauricio}\ \bibnamefont {Reis}}, \bibinfo {author}
  {\bibfnamefont {Trevor}\ \bibnamefont {Lanting}}, \bibinfo {author}
  {\bibfnamefont {Richard}\ \bibnamefont {Harris}}, \bibinfo {author}
  {\bibfnamefont {Fabio}\ \bibnamefont {Altomare}},  \emph {et~al.},\
  }\bibfield  {title} {\enquote {\bibinfo {title} {{\it Observation of
  topological phenomena in a programmable lattice of 1,800 qubits}},}\ }\href
  {https://doi.org/10.1038/s41586-018-0410-x} {\bibfield  {journal} {\bibinfo
  {journal} {Nature}\ }\textbf {\bibinfo {volume} {560}},\ \bibinfo {pages}
  {456--460} (\bibinfo {year} {2018})}\BibitemShut {NoStop}%
\bibitem [{\citenamefont {Hempel}\ \emph {et~al.}(2018)\citenamefont {Hempel},
  \citenamefont {Maier}, \citenamefont {Romero}, \citenamefont {McClean},
  \citenamefont {Monz}, \citenamefont {Shen}, \citenamefont {Jurcevic},
  \citenamefont {Lanyon}, \citenamefont {Love}, \citenamefont {Babbush},
  \citenamefont {Aspuru-Guzik}, \citenamefont {Blatt},\ and\ \citenamefont
  {Roos}}]{PhysRevX.8.031022}%
  \BibitemOpen
  \bibfield  {author} {\bibinfo {author} {\bibfnamefont {Cornelius}\
  \bibnamefont {Hempel}}, \bibinfo {author} {\bibfnamefont {Christine}\
  \bibnamefont {Maier}}, \bibinfo {author} {\bibfnamefont {Jonathan}\
  \bibnamefont {Romero}}, \bibinfo {author} {\bibfnamefont {Jarrod}\
  \bibnamefont {McClean}}, \bibinfo {author} {\bibfnamefont {Thomas}\
  \bibnamefont {Monz}}, \bibinfo {author} {\bibfnamefont {Heng}\ \bibnamefont
  {Shen}}, \bibinfo {author} {\bibfnamefont {Petar}\ \bibnamefont {Jurcevic}},
  \bibinfo {author} {\bibfnamefont {Ben~P.}\ \bibnamefont {Lanyon}}, \bibinfo
  {author} {\bibfnamefont {Peter}\ \bibnamefont {Love}}, \bibinfo {author}
  {\bibfnamefont {Ryan}\ \bibnamefont {Babbush}}, \bibinfo {author}
  {\bibfnamefont {Al\'an}\ \bibnamefont {Aspuru-Guzik}}, \bibinfo {author}
  {\bibfnamefont {Rainer}\ \bibnamefont {Blatt}}, \ and\ \bibinfo {author}
  {\bibfnamefont {Christian~F.}\ \bibnamefont {Roos}},\ }\bibfield  {title}
  {\enquote {\bibinfo {title} {{\it Quantum Chemistry Calculations on a
  Trapped-Ion Quantum Simulator}},}\ }\href {\doibase
  10.1103/PhysRevX.8.031022} {\bibfield  {journal} {\bibinfo  {journal} {Phys.
  Rev. X}\ }\textbf {\bibinfo {volume} {8}},\ \bibinfo {pages} {031022}
  (\bibinfo {year} {2018})}\BibitemShut {NoStop}%
\bibitem [{\citenamefont {Arute}\ \emph {et~al.}(2019)\citenamefont {Arute},
  \citenamefont {Arya}, \citenamefont {Babbush}, \citenamefont {Bacon},
  \citenamefont {Bardin}, \citenamefont {Barends}, \citenamefont {Biswas},
  \citenamefont {Boixo}, \citenamefont {Brandao}, \citenamefont {Buell} \emph
  {et~al.}}]{arute2019quantum}%
  \BibitemOpen
  \bibfield  {author} {\bibinfo {author} {\bibfnamefont {Frank}\ \bibnamefont
  {Arute}}, \bibinfo {author} {\bibfnamefont {Kunal}\ \bibnamefont {Arya}},
  \bibinfo {author} {\bibfnamefont {Ryan}\ \bibnamefont {Babbush}}, \bibinfo
  {author} {\bibfnamefont {Dave}\ \bibnamefont {Bacon}}, \bibinfo {author}
  {\bibfnamefont {Joseph~C}\ \bibnamefont {Bardin}}, \bibinfo {author}
  {\bibfnamefont {Rami}\ \bibnamefont {Barends}}, \bibinfo {author}
  {\bibfnamefont {Rupak}\ \bibnamefont {Biswas}}, \bibinfo {author}
  {\bibfnamefont {Sergio}\ \bibnamefont {Boixo}}, \bibinfo {author}
  {\bibfnamefont {Fernando~GSL}\ \bibnamefont {Brandao}}, \bibinfo {author}
  {\bibfnamefont {David~A}\ \bibnamefont {Buell}},  \emph {et~al.},\ }\bibfield
   {title} {\enquote {\bibinfo {title} {{\it Quantum supremacy using a
  programmable superconducting processor}},}\ }\href
  {https://doi.org/10.1038/s41586-019-1666-5} {\bibfield  {journal} {\bibinfo
  {journal} {Nature}\ }\textbf {\bibinfo {volume} {574}},\ \bibinfo {pages}
  {505--510} (\bibinfo {year} {2019})}\BibitemShut {NoStop}%
\bibitem [{\citenamefont {Motta}\ \emph {et~al.}(2020)\citenamefont {Motta},
  \citenamefont {Sun}, \citenamefont {Tan}, \citenamefont {O’Rourke},
  \citenamefont {Ye}, \citenamefont {Minnich}, \citenamefont {Brand{\~a}o},\
  and\ \citenamefont {Chan}}]{motta2020determining}%
  \BibitemOpen
  \bibfield  {author} {\bibinfo {author} {\bibfnamefont {Mario}\ \bibnamefont
  {Motta}}, \bibinfo {author} {\bibfnamefont {Chong}\ \bibnamefont {Sun}},
  \bibinfo {author} {\bibfnamefont {Adrian~TK}\ \bibnamefont {Tan}}, \bibinfo
  {author} {\bibfnamefont {Matthew~J}\ \bibnamefont {O’Rourke}}, \bibinfo
  {author} {\bibfnamefont {Erika}\ \bibnamefont {Ye}}, \bibinfo {author}
  {\bibfnamefont {Austin~J}\ \bibnamefont {Minnich}}, \bibinfo {author}
  {\bibfnamefont {Fernando~GSL}\ \bibnamefont {Brand{\~a}o}}, \ and\ \bibinfo
  {author} {\bibfnamefont {Garnet Kin-Lic}\ \bibnamefont {Chan}},\ }\bibfield
  {title} {\enquote {\bibinfo {title} {{\it Determining eigenstates and thermal
  states on a quantum computer using quantum imaginary time evolution}},}\
  }\href {https://doi.org/10.1038/s41567-019-0704-4} {\bibfield  {journal}
  {\bibinfo  {journal} {Nature Physics}\ }\textbf {\bibinfo {volume} {16}},\
  \bibinfo {pages} {205--210} (\bibinfo {year} {2020})}\BibitemShut {NoStop}%
\bibitem [{\citenamefont {Amin}\ \emph {et~al.}(2018)\citenamefont {Amin},
  \citenamefont {Andriyash}, \citenamefont {Rolfe}, \citenamefont
  {Kulchytskyy},\ and\ \citenamefont {Melko}}]{PhysRevX.8.021050}%
  \BibitemOpen
  \bibfield  {author} {\bibinfo {author} {\bibfnamefont {Mohammad~H.}\
  \bibnamefont {Amin}}, \bibinfo {author} {\bibfnamefont {Evgeny}\ \bibnamefont
  {Andriyash}}, \bibinfo {author} {\bibfnamefont {Jason}\ \bibnamefont
  {Rolfe}}, \bibinfo {author} {\bibfnamefont {Bohdan}\ \bibnamefont
  {Kulchytskyy}}, \ and\ \bibinfo {author} {\bibfnamefont {Roger}\ \bibnamefont
  {Melko}},\ }\bibfield  {title} {\enquote {\bibinfo {title} {{\it Quantum
  Boltzmann Machine}},}\ }\href {\doibase 10.1103/PhysRevX.8.021050} {\bibfield
   {journal} {\bibinfo  {journal} {Phys. Rev. X}\ }\textbf {\bibinfo {volume}
  {8}},\ \bibinfo {pages} {021050} (\bibinfo {year} {2018})}\BibitemShut
  {NoStop}%
\bibitem [{\citenamefont {Kieferov\'a}\ and\ \citenamefont
  {Wiebe}(2017)}]{PhysRevA.96.062327}%
  \BibitemOpen
  \bibfield  {author} {\bibinfo {author} {\bibfnamefont {M\'aria}\ \bibnamefont
  {Kieferov\'a}}\ and\ \bibinfo {author} {\bibfnamefont {Nathan}\ \bibnamefont
  {Wiebe}},\ }\bibfield  {title} {\enquote {\bibinfo {title} {{\it Tomography
  and generative training with quantum Boltzmann machines}},}\ }\href {\doibase
  10.1103/PhysRevA.96.062327} {\bibfield  {journal} {\bibinfo  {journal} {Phys.
  Rev. A}\ }\textbf {\bibinfo {volume} {96}},\ \bibinfo {pages} {062327}
  (\bibinfo {year} {2017})}\BibitemShut {NoStop}%
\bibitem [{\citenamefont {Biamonte}\ \emph {et~al.}(2017)\citenamefont
  {Biamonte}, \citenamefont {Wittek}, \citenamefont {Pancotti}, \citenamefont
  {Rebentrost}, \citenamefont {Wiebe},\ and\ \citenamefont
  {Lloyd}}]{biamonte2017quantum}%
  \BibitemOpen
  \bibfield  {author} {\bibinfo {author} {\bibfnamefont {Jacob}\ \bibnamefont
  {Biamonte}}, \bibinfo {author} {\bibfnamefont {Peter}\ \bibnamefont
  {Wittek}}, \bibinfo {author} {\bibfnamefont {Nicola}\ \bibnamefont
  {Pancotti}}, \bibinfo {author} {\bibfnamefont {Patrick}\ \bibnamefont
  {Rebentrost}}, \bibinfo {author} {\bibfnamefont {Nathan}\ \bibnamefont
  {Wiebe}}, \ and\ \bibinfo {author} {\bibfnamefont {Seth}\ \bibnamefont
  {Lloyd}},\ }\bibfield  {title} {\enquote {\bibinfo {title} {{\it Quantum
  machine learning}},}\ }\href {https://doi.org/10.1038/nature23474} {\bibfield
   {journal} {\bibinfo  {journal} {Nature}\ }\textbf {\bibinfo {volume}
  {549}},\ \bibinfo {pages} {195} (\bibinfo {year} {2017})}\BibitemShut
  {NoStop}%
\bibitem [{\citenamefont {Crawford}\ \emph {et~al.}(2018)\citenamefont
  {Crawford}, \citenamefont {Levit}, \citenamefont {Ghadermarzy}, \citenamefont
  {Oberoi},\ and\ \citenamefont {Ronagh}}]{10.5555/3370185.3370188}%
  \BibitemOpen
  \bibfield  {author} {\bibinfo {author} {\bibfnamefont {Daniel}\ \bibnamefont
  {Crawford}}, \bibinfo {author} {\bibfnamefont {Anna}\ \bibnamefont {Levit}},
  \bibinfo {author} {\bibfnamefont {Navid}\ \bibnamefont {Ghadermarzy}},
  \bibinfo {author} {\bibfnamefont {Jaspreet~S.}\ \bibnamefont {Oberoi}}, \
  and\ \bibinfo {author} {\bibfnamefont {Pooya}\ \bibnamefont {Ronagh}},\
  }\bibfield  {title} {\enquote {\bibinfo {title} {{\it Reinforcement Learning
  Using Quantum Boltzmann Machines}},}\ }\href@noop {} {\bibfield  {journal}
  {\bibinfo  {journal} {Quantum Info. Comput.}\ }\textbf {\bibinfo {volume}
  {18}},\ \bibinfo {pages} {51–74} (\bibinfo {year} {2018})}\BibitemShut
  {NoStop}%
\bibitem [{\citenamefont {Bairey}\ \emph {et~al.}(2019)\citenamefont {Bairey},
  \citenamefont {Arad},\ and\ \citenamefont
  {Lindner}}]{PhysRevLett.122.020504}%
  \BibitemOpen
  \bibfield  {author} {\bibinfo {author} {\bibfnamefont {Eyal}\ \bibnamefont
  {Bairey}}, \bibinfo {author} {\bibfnamefont {Itai}\ \bibnamefont {Arad}}, \
  and\ \bibinfo {author} {\bibfnamefont {Netanel~H.}\ \bibnamefont {Lindner}},\
  }\bibfield  {title} {\enquote {\bibinfo {title} {{\it Learning a Local
  Hamiltonian from Local Measurements}},}\ }\href {\doibase
  10.1103/PhysRevLett.122.020504} {\bibfield  {journal} {\bibinfo  {journal}
  {Phys. Rev. Lett.}\ }\textbf {\bibinfo {volume} {122}},\ \bibinfo {pages}
  {020504} (\bibinfo {year} {2019})}\BibitemShut {NoStop}%
\bibitem [{\citenamefont {Torlai}\ and\ \citenamefont
  {Melko}(2020)}]{doi:10.1146/annurev-conmatphys-031119-050651}%
  \BibitemOpen
  \bibfield  {author} {\bibinfo {author} {\bibfnamefont {Giacomo}\ \bibnamefont
  {Torlai}}\ and\ \bibinfo {author} {\bibfnamefont {Roger~G.}\ \bibnamefont
  {Melko}},\ }\bibfield  {title} {\enquote {\bibinfo {title} {{\it
  Machine-Learning Quantum States in the NISQ Era}},}\ }\href {\doibase
  10.1146/annurev-conmatphys-031119-050651} {\bibfield  {journal} {\bibinfo
  {journal} {Annual Review of Condensed Matter Physics}\ }\textbf {\bibinfo
  {volume} {11}},\ \bibinfo {pages} {325--344} (\bibinfo {year} {2020})},\
  \Eprint
  {http://arxiv.org/abs/https://doi.org/10.1146/annurev-conmatphys-031119-050651}
  {https://doi.org/10.1146/annurev-conmatphys-031119-050651} \BibitemShut
  {NoStop}%
\bibitem [{\citenamefont {Chia}\ \emph {et~al.}(2020)\citenamefont {Chia},
  \citenamefont {Gily\'{e}n}, \citenamefont {Li}, \citenamefont {Lin},
  \citenamefont {Tang},\ and\ \citenamefont {Wang}}]{chia2019sampling}%
  \BibitemOpen
  \bibfield  {author} {\bibinfo {author} {\bibfnamefont {Nai-Hui}\ \bibnamefont
  {Chia}}, \bibinfo {author} {\bibfnamefont {Andr\'{a}s}\ \bibnamefont
  {Gily\'{e}n}}, \bibinfo {author} {\bibfnamefont {Tongyang}\ \bibnamefont
  {Li}}, \bibinfo {author} {\bibfnamefont {Han-Hsuan}\ \bibnamefont {Lin}},
  \bibinfo {author} {\bibfnamefont {Ewin}\ \bibnamefont {Tang}}, \ and\
  \bibinfo {author} {\bibfnamefont {Chunhao}\ \bibnamefont {Wang}},\ }\bibfield
   {title} {\enquote {\bibinfo {title} {{\it Sampling-Based Sublinear Low-Rank
  Matrix Arithmetic Framework for Dequantizing Quantum Machine Learning}},}\
  }in\ \href {\doibase 10.1145/3357713.3384314} {\emph {\bibinfo {booktitle}
  {Proceedings of the 52nd Annual ACM SIGACT Symposium on Theory of
  Computing}}},\ \bibinfo {series and number} {STOC 2020}\ (\bibinfo
  {publisher} {Association for Computing Machinery},\ \bibinfo {address} {New
  York, NY, USA},\ \bibinfo {year} {2020})\ p.\ \bibinfo {pages}
  {387–400}\BibitemShut {NoStop}%
\bibitem [{\citenamefont {Anshu}\ \emph
  {et~al.}(2020{\natexlab{a}})\citenamefont {Anshu}, \citenamefont
  {Arunachalam}, \citenamefont {Kuwahara},\ and\ \citenamefont
  {Soleimanifar}}]{anshu2020sample}%
  \BibitemOpen
  \bibfield  {author} {\bibinfo {author} {\bibfnamefont {Anurag}\ \bibnamefont
  {Anshu}}, \bibinfo {author} {\bibfnamefont {Srinivasan}\ \bibnamefont
  {Arunachalam}}, \bibinfo {author} {\bibfnamefont {Tomotaka}\ \bibnamefont
  {Kuwahara}}, \ and\ \bibinfo {author} {\bibfnamefont {Mehdi}\ \bibnamefont
  {Soleimanifar}},\ }\bibfield  {title} {\enquote {\bibinfo {title} {{\it
  Sample-efficient learning of quantum many-body systems}},}\ }\href@noop {}
  {\bibfield  {journal} {\bibinfo  {journal} {arXiv preprint arXiv:2004.07266}\
  } (\bibinfo {year} {2020}{\natexlab{a}})},\ \Eprint
  {http://arxiv.org/abs/arXiv:2004.07266} {arXiv:2004.07266} \BibitemShut
  {NoStop}%
\bibitem [{\citenamefont {{Brand\~ao}}\ and\ \citenamefont
  {{Svore}}(2017)}]{8104077}%
  \BibitemOpen
  \bibfield  {author} {\bibinfo {author} {\bibfnamefont {F.~G. S.~L.}\
  \bibnamefont {{Brand\~ao}}}\ and\ \bibinfo {author} {\bibfnamefont {K.~M.}\
  \bibnamefont {{Svore}}},\ }\bibfield  {title} {\enquote {\bibinfo {title}
  {{\it Quantum Speed-Ups for Solving Semidefinite Programs}},}\ }in\ \href
  {\doibase 10.1109/FOCS.2017.45} {\emph {\bibinfo {booktitle} {2017 IEEE 58th
  Annual Symposium on Foundations of Computer Science (FOCS)}}}\ (\bibinfo
  {year} {2017})\ pp.\ \bibinfo {pages} {415--426}\BibitemShut {NoStop}%
\bibitem [{\citenamefont {Brand{\~a}o}\ \emph {et~al.}(2019)\citenamefont
  {Brand{\~a}o}, \citenamefont {Kalev}, \citenamefont {Li}, \citenamefont
  {Lin}, \citenamefont {Svore},\ and\ \citenamefont
  {Wu}}]{brando_et_al:LIPIcs:2019:10603}%
  \BibitemOpen
  \bibfield  {author} {\bibinfo {author} {\bibfnamefont {Fernando~GSL}\
  \bibnamefont {Brand{\~a}o}}, \bibinfo {author} {\bibfnamefont {Amir}\
  \bibnamefont {Kalev}}, \bibinfo {author} {\bibfnamefont {Tongyang}\
  \bibnamefont {Li}}, \bibinfo {author} {\bibfnamefont {Cedric Yen-Yu}\
  \bibnamefont {Lin}}, \bibinfo {author} {\bibfnamefont {Krysta~M}\
  \bibnamefont {Svore}}, \ and\ \bibinfo {author} {\bibfnamefont {Xiaodi}\
  \bibnamefont {Wu}},\ }\bibfield  {title} {\enquote {\bibinfo {title} {{\it
  Quantum SDP Solvers: Large Speed-Ups, Optimality, and Applications to Quantum
  Learning}},}\ }in\ \href {\doibase 10.4230/LIPIcs.ICALP.2019.27} {\emph
  {\bibinfo {booktitle} {46th International Colloquium on Automata, Languages,
  and Programming (ICALP 2019)}}}\ (\bibinfo {organization} {Schloss
  Dagstuhl-Leibniz-Zentrum fuer Informatik},\ \bibinfo {year} {2019})\ pp.\
  \bibinfo {pages} {27:1--27:14}\BibitemShut {NoStop}%
\bibitem [{\citenamefont {van Apeldoorn}\ \emph {et~al.}(2020)\citenamefont
  {van Apeldoorn}, \citenamefont {Gily{\'{e}}n}, \citenamefont {Gribling},\
  and\ \citenamefont {de~Wolf}}]{vanApeldoorn2020quantumsdpsolvers}%
  \BibitemOpen
  \bibfield  {author} {\bibinfo {author} {\bibfnamefont {Joran}\ \bibnamefont
  {van Apeldoorn}}, \bibinfo {author} {\bibfnamefont {Andr{\'{a}}s}\
  \bibnamefont {Gily{\'{e}}n}}, \bibinfo {author} {\bibfnamefont {Sander}\
  \bibnamefont {Gribling}}, \ and\ \bibinfo {author} {\bibfnamefont {Ronald}\
  \bibnamefont {de~Wolf}},\ }\bibfield  {title} {\enquote {\bibinfo {title}
  {{\it Quantum {SDP}-{S}olvers: {B}etter upper and lower bounds}},}\ }\href
  {\doibase 10.22331/q-2020-02-14-230} {\bibfield  {journal} {\bibinfo
  {journal} {{Quantum}}\ }\textbf {\bibinfo {volume} {4}},\ \bibinfo {pages}
  {230} (\bibinfo {year} {2020})}\BibitemShut {NoStop}%
\bibitem [{\citenamefont {Eisert}\ \emph {et~al.}(2010)\citenamefont {Eisert},
  \citenamefont {Cramer},\ and\ \citenamefont {Plenio}}]{RevModPhys.82.277}%
  \BibitemOpen
  \bibfield  {author} {\bibinfo {author} {\bibfnamefont {J.}~\bibnamefont
  {Eisert}}, \bibinfo {author} {\bibfnamefont {M.}~\bibnamefont {Cramer}}, \
  and\ \bibinfo {author} {\bibfnamefont {M.~B.}\ \bibnamefont {Plenio}},\
  }\bibfield  {title} {\enquote {\bibinfo {title} {{\it Colloquium: Area laws
  for the entanglement entropy}},}\ }\href {\doibase 10.1103/RevModPhys.82.277}
  {\bibfield  {journal} {\bibinfo  {journal} {Rev. Mod. Phys.}\ }\textbf
  {\bibinfo {volume} {82}},\ \bibinfo {pages} {277--306} (\bibinfo {year}
  {2010})}\BibitemShut {NoStop}%
\bibitem [{\citenamefont {Vidal}\ \emph {et~al.}(2003)\citenamefont {Vidal},
  \citenamefont {Latorre}, \citenamefont {Rico},\ and\ \citenamefont
  {Kitaev}}]{PhysRevLett.90.227902}%
  \BibitemOpen
  \bibfield  {author} {\bibinfo {author} {\bibfnamefont {G.}~\bibnamefont
  {Vidal}}, \bibinfo {author} {\bibfnamefont {J.~I.}\ \bibnamefont {Latorre}},
  \bibinfo {author} {\bibfnamefont {E.}~\bibnamefont {Rico}}, \ and\ \bibinfo
  {author} {\bibfnamefont {A.}~\bibnamefont {Kitaev}},\ }\bibfield  {title}
  {\enquote {\bibinfo {title} {{\it Entanglement in Quantum Critical
  Phenomena}},}\ }\href {\doibase 10.1103/PhysRevLett.90.227902} {\bibfield
  {journal} {\bibinfo  {journal} {Phys. Rev. Lett.}\ }\textbf {\bibinfo
  {volume} {90}},\ \bibinfo {pages} {227902} (\bibinfo {year}
  {2003})}\BibitemShut {NoStop}%
\bibitem [{\citenamefont {Hastings}(2007{\natexlab{a}})}]{Hastings_2007}%
  \BibitemOpen
  \bibfield  {author} {\bibinfo {author} {\bibfnamefont {M~B}\ \bibnamefont
  {Hastings}},\ }\bibfield  {title} {\enquote {\bibinfo {title} {{\it An area
  law for one-dimensional quantum systems}},}\ }\href {\doibase
  10.1088/1742-5468/2007/08/p08024} {\bibfield  {journal} {\bibinfo  {journal}
  {Journal of Statistical Mechanics: Theory and Experiment}\ }\textbf {\bibinfo
  {volume} {2007}},\ \bibinfo {pages} {P08024--P08024} (\bibinfo {year}
  {2007}{\natexlab{a}})}\BibitemShut {NoStop}%
\bibitem [{\citenamefont {Aharonov}\ \emph {et~al.}(2011)\citenamefont
  {Aharonov}, \citenamefont {Arad}, \citenamefont {Vazirani},\ and\
  \citenamefont {Landau}}]{Aharonov_2011}%
  \BibitemOpen
  \bibfield  {author} {\bibinfo {author} {\bibfnamefont {Dorit}\ \bibnamefont
  {Aharonov}}, \bibinfo {author} {\bibfnamefont {Itai}\ \bibnamefont {Arad}},
  \bibinfo {author} {\bibfnamefont {Umesh}\ \bibnamefont {Vazirani}}, \ and\
  \bibinfo {author} {\bibfnamefont {Zeph}\ \bibnamefont {Landau}},\ }\bibfield
  {title} {\enquote {\bibinfo {title} {{\it The detectability lemma and its
  applications to quantum Hamiltonian complexity}},}\ }\href {\doibase
  10.1088/1367-2630/13/11/113043} {\bibfield  {journal} {\bibinfo  {journal}
  {New Journal of Physics}\ }\textbf {\bibinfo {volume} {13}},\ \bibinfo
  {pages} {113043} (\bibinfo {year} {2011})}\BibitemShut {NoStop}%
\bibitem [{\citenamefont {Arad}\ \emph {et~al.}(2012)\citenamefont {Arad},
  \citenamefont {Landau},\ and\ \citenamefont {Vazirani}}]{PhysRevB.85.195145}%
  \BibitemOpen
  \bibfield  {author} {\bibinfo {author} {\bibfnamefont {Itai}\ \bibnamefont
  {Arad}}, \bibinfo {author} {\bibfnamefont {Zeph}\ \bibnamefont {Landau}}, \
  and\ \bibinfo {author} {\bibfnamefont {Umesh}\ \bibnamefont {Vazirani}},\
  }\bibfield  {title} {\enquote {\bibinfo {title} {{\it Improved
  one-dimensional area law for frustration-free systems}},}\ }\href {\doibase
  10.1103/PhysRevB.85.195145} {\bibfield  {journal} {\bibinfo  {journal} {Phys.
  Rev. B}\ }\textbf {\bibinfo {volume} {85}},\ \bibinfo {pages} {195145}
  (\bibinfo {year} {2012})}\BibitemShut {NoStop}%
\bibitem [{\citenamefont {Arad}\ \emph {et~al.}(2013)\citenamefont {Arad},
  \citenamefont {Kitaev}, \citenamefont {Landau},\ and\ \citenamefont
  {Vazirani}}]{arad2013area}%
  \BibitemOpen
  \bibfield  {author} {\bibinfo {author} {\bibfnamefont {Itai}\ \bibnamefont
  {Arad}}, \bibinfo {author} {\bibfnamefont {Alexei}\ \bibnamefont {Kitaev}},
  \bibinfo {author} {\bibfnamefont {Zeph}\ \bibnamefont {Landau}}, \ and\
  \bibinfo {author} {\bibfnamefont {Umesh}\ \bibnamefont {Vazirani}},\
  }\bibfield  {title} {\enquote {\bibinfo {title} {{\it An area law and
  sub-exponential algorithm for 1D systems}},}\ }\href@noop {} {\bibfield
  {journal} {\bibinfo  {journal} {arXiv preprint arXiv:1301.1162}\ } (\bibinfo
  {year} {2013})},\ \Eprint {http://arxiv.org/abs/arXiv:1301.1162}
  {arXiv:1301.1162} \BibitemShut {NoStop}%
\bibitem [{\citenamefont {Brand{\~a}o}\ and\ \citenamefont
  {Horodecki}(2013)}]{brandao2013area}%
  \BibitemOpen
  \bibfield  {author} {\bibinfo {author} {\bibfnamefont {Fernando~GSL}\
  \bibnamefont {Brand{\~a}o}}\ and\ \bibinfo {author} {\bibfnamefont
  {Micha{\l}}\ \bibnamefont {Horodecki}},\ }\bibfield  {title} {\enquote
  {\bibinfo {title} {{\it An area law for entanglement from exponential decay
  of correlations}},}\ }\href {https://doi.org/10.1038/nphys2747} {\bibfield
  {journal} {\bibinfo  {journal} {Nature Physics}\ }\textbf {\bibinfo {volume}
  {9}},\ \bibinfo {pages} {721} (\bibinfo {year} {2013})}\BibitemShut {NoStop}%
\bibitem [{\citenamefont {Abrahamsen}(2019)}]{abrahamsen2019polynomial}%
  \BibitemOpen
  \bibfield  {author} {\bibinfo {author} {\bibfnamefont {Nilin}\ \bibnamefont
  {Abrahamsen}},\ }\bibfield  {title} {\enquote {\bibinfo {title} {{\it A
  polynomial-time algorithm for ground states of spin trees}},}\ }\href@noop {}
  {\bibfield  {journal} {\bibinfo  {journal} {arXiv preprint arXiv:1907.04862}\
  } (\bibinfo {year} {2019})},\ \Eprint {http://arxiv.org/abs/arXiv:1907.04862}
  {arXiv:1907.04862} \BibitemShut {NoStop}%
\bibitem [{\citenamefont {Anshu}\ \emph
  {et~al.}(2020{\natexlab{b}})\citenamefont {Anshu}, \citenamefont {Arad},\
  and\ \citenamefont {Gosset}}]{anshu2019entanglement}%
  \BibitemOpen
  \bibfield  {author} {\bibinfo {author} {\bibfnamefont {Anurag}\ \bibnamefont
  {Anshu}}, \bibinfo {author} {\bibfnamefont {Itai}\ \bibnamefont {Arad}}, \
  and\ \bibinfo {author} {\bibfnamefont {David}\ \bibnamefont {Gosset}},\
  }\bibfield  {title} {\enquote {\bibinfo {title} {{\it Entanglement Subvolume
  Law for 2d Frustration-Free Spin Systems}},}\ }in\ \href {\doibase
  10.1145/3357713.3384292} {\emph {\bibinfo {booktitle} {{\it Proceedings of
  the 52nd Annual ACM SIGACT Symposium on Theory of Computing}, pages =
  {868–874}, numpages = {7}, location = {Chicago, IL, USA}, series = {STOC
  2020}}}}\ (\bibinfo  {publisher} {Association for Computing Machinery},\
  \bibinfo {address} {New York, NY, USA},\ \bibinfo {year} {2020})\BibitemShut
  {NoStop}%
\bibitem [{\citenamefont {Kuwahara}\ and\ \citenamefont
  {Saito}(2020{\natexlab{a}})}]{kuwahara2019area}%
  \BibitemOpen
  \bibfield  {author} {\bibinfo {author} {\bibfnamefont {Tomotaka}\
  \bibnamefont {Kuwahara}}\ and\ \bibinfo {author} {\bibfnamefont {Keiji}\
  \bibnamefont {Saito}},\ }\bibfield  {title} {\enquote {\bibinfo {title} {{\it
  Area law of noncritical ground states in 1D long-range interacting
  systems}},}\ }\href {\doibase 10.1038/s41467-020-18055-x} {\bibfield
  {journal} {\bibinfo  {journal} {Nature Communications}\ }\textbf {\bibinfo
  {volume} {11}},\ \bibinfo {pages} {4478} (\bibinfo {year}
  {2020}{\natexlab{a}})}\BibitemShut {NoStop}%
\bibitem [{\citenamefont {Wolf}\ \emph {et~al.}(2008)\citenamefont {Wolf},
  \citenamefont {Verstraete}, \citenamefont {Hastings},\ and\ \citenamefont
  {Cirac}}]{PhysRevLett.100.070502}%
  \BibitemOpen
  \bibfield  {author} {\bibinfo {author} {\bibfnamefont {Michael~M.}\
  \bibnamefont {Wolf}}, \bibinfo {author} {\bibfnamefont {Frank}\ \bibnamefont
  {Verstraete}}, \bibinfo {author} {\bibfnamefont {Matthew~B.}\ \bibnamefont
  {Hastings}}, \ and\ \bibinfo {author} {\bibfnamefont {J.~Ignacio}\
  \bibnamefont {Cirac}},\ }\bibfield  {title} {\enquote {\bibinfo {title} {{\it
  Area Laws in Quantum Systems: Mutual Information and Correlations}},}\ }\href
  {\doibase 10.1103/PhysRevLett.100.070502} {\bibfield  {journal} {\bibinfo
  {journal} {Phys. Rev. Lett.}\ }\textbf {\bibinfo {volume} {100}},\ \bibinfo
  {pages} {070502} (\bibinfo {year} {2008})}\BibitemShut {NoStop}%
\bibitem [{\citenamefont {Hastings}(2007{\natexlab{b}})}]{PhysRevB.76.201102}%
  \BibitemOpen
  \bibfield  {author} {\bibinfo {author} {\bibfnamefont {M.~B.}\ \bibnamefont
  {Hastings}},\ }\bibfield  {title} {\enquote {\bibinfo {title} {{\it Quantum
  belief propagation: An algorithm for thermal quantum systems}},}\ }\href
  {\doibase 10.1103/PhysRevB.76.201102} {\bibfield  {journal} {\bibinfo
  {journal} {Phys. Rev. B}\ }\textbf {\bibinfo {volume} {76}},\ \bibinfo
  {pages} {201102} (\bibinfo {year} {2007}{\natexlab{b}})}\BibitemShut
  {NoStop}%
\bibitem [{\citenamefont {\ifmmode \check{Z}\else
  \v{Z}\fi{}nidari\ifmmode~\check{c}\else \v{c}\fi{}}\ \emph
  {et~al.}(2008)\citenamefont {\ifmmode \check{Z}\else
  \v{Z}\fi{}nidari\ifmmode~\check{c}\else \v{c}\fi{}}, \citenamefont {Prosen},\
  and\ \citenamefont {Pi\ifmmode~\check{z}\else
  \v{z}\fi{}orn}}]{PhysRevA.78.022103}%
  \BibitemOpen
  \bibfield  {author} {\bibinfo {author} {\bibfnamefont {Marko}\ \bibnamefont
  {\ifmmode \check{Z}\else \v{Z}\fi{}nidari\ifmmode~\check{c}\else
  \v{c}\fi{}}}, \bibinfo {author} {\bibfnamefont {Toma\ifmmode
  \check{z}\else~\v{z}\fi{}}\ \bibnamefont {Prosen}}, \ and\ \bibinfo {author}
  {\bibfnamefont {Iztok}\ \bibnamefont {Pi\ifmmode~\check{z}\else
  \v{z}\fi{}orn}},\ }\bibfield  {title} {\enquote {\bibinfo {title} {{\it
  Complexity of thermal states in quantum spin chains}},}\ }\href {\doibase
  10.1103/PhysRevA.78.022103} {\bibfield  {journal} {\bibinfo  {journal} {Phys.
  Rev. A}\ }\textbf {\bibinfo {volume} {78}},\ \bibinfo {pages} {022103}
  (\bibinfo {year} {2008})}\BibitemShut {NoStop}%
\bibitem [{\citenamefont {Bernigau}\ \emph {et~al.}(2015)\citenamefont
  {Bernigau}, \citenamefont {Kastoryano},\ and\ \citenamefont
  {Eisert}}]{Bernigau_2015}%
  \BibitemOpen
  \bibfield  {author} {\bibinfo {author} {\bibfnamefont {H}~\bibnamefont
  {Bernigau}}, \bibinfo {author} {\bibfnamefont {M~J}\ \bibnamefont
  {Kastoryano}}, \ and\ \bibinfo {author} {\bibfnamefont {J}~\bibnamefont
  {Eisert}},\ }\bibfield  {title} {\enquote {\bibinfo {title} {{\it Mutual
  information area laws for thermal free fermions}},}\ }\href {\doibase
  10.1088/1742-5468/2015/02/p02008} {\bibfield  {journal} {\bibinfo  {journal}
  {Journal of Statistical Mechanics: Theory and Experiment}\ }\textbf {\bibinfo
  {volume} {2015}},\ \bibinfo {pages} {P02008} (\bibinfo {year}
  {2015})}\BibitemShut {NoStop}%
\bibitem [{\citenamefont {Barthel}(2017)}]{barthel2017one}%
  \BibitemOpen
  \bibfield  {author} {\bibinfo {author} {\bibfnamefont {Thomas}\ \bibnamefont
  {Barthel}},\ }\bibfield  {title} {\enquote {\bibinfo {title} {{\it
  One-dimensional quantum systems at finite temperatures can be simulated
  efficiently on classical computers}},}\ }\href@noop {} {\bibfield  {journal}
  {\bibinfo  {journal} {arXiv preprint arXiv:1708.09349}\ } (\bibinfo {year}
  {2017})},\ \Eprint {http://arxiv.org/abs/arXiv:1708.09349} {arXiv:1708.09349}
  \BibitemShut {NoStop}%
\bibitem [{\citenamefont {White}(2009)}]{PhysRevLett.102.190601}%
  \BibitemOpen
  \bibfield  {author} {\bibinfo {author} {\bibfnamefont {Steven~R.}\
  \bibnamefont {White}},\ }\bibfield  {title} {\enquote {\bibinfo {title} {{\it
  Minimally Entangled Typical Quantum States at Finite Temperature}},}\ }\href
  {\doibase 10.1103/PhysRevLett.102.190601} {\bibfield  {journal} {\bibinfo
  {journal} {Phys. Rev. Lett.}\ }\textbf {\bibinfo {volume} {102}},\ \bibinfo
  {pages} {190601} (\bibinfo {year} {2009})}\BibitemShut {NoStop}%
\bibitem [{\citenamefont {Stoudenmire}\ and\ \citenamefont
  {White}(2010)}]{Stoudenmire_2010}%
  \BibitemOpen
  \bibfield  {author} {\bibinfo {author} {\bibfnamefont {E~M}\ \bibnamefont
  {Stoudenmire}}\ and\ \bibinfo {author} {\bibfnamefont {Steven~R}\
  \bibnamefont {White}},\ }\bibfield  {title} {\enquote {\bibinfo {title} {{\it
  Minimally entangled typical thermal state algorithms}},}\ }\href {\doibase
  10.1088/1367-2630/12/5/055026} {\bibfield  {journal} {\bibinfo  {journal}
  {New Journal of Physics}\ }\textbf {\bibinfo {volume} {12}},\ \bibinfo
  {pages} {055026} (\bibinfo {year} {2010})}\BibitemShut {NoStop}%
\bibitem [{\citenamefont {Binder}\ and\ \citenamefont
  {Barthel}(2015)}]{PhysRevB.92.125119}%
  \BibitemOpen
  \bibfield  {author} {\bibinfo {author} {\bibfnamefont {Moritz}\ \bibnamefont
  {Binder}}\ and\ \bibinfo {author} {\bibfnamefont {Thomas}\ \bibnamefont
  {Barthel}},\ }\bibfield  {title} {\enquote {\bibinfo {title} {{\it Minimally
  entangled typical thermal states versus matrix product purifications for the
  simulation of equilibrium states and time evolution}},}\ }\href {\doibase
  10.1103/PhysRevB.92.125119} {\bibfield  {journal} {\bibinfo  {journal} {Phys.
  Rev. B}\ }\textbf {\bibinfo {volume} {92}},\ \bibinfo {pages} {125119}
  (\bibinfo {year} {2015})}\BibitemShut {NoStop}%
\bibitem [{\citenamefont {Rigol}\ \emph {et~al.}(2006)\citenamefont {Rigol},
  \citenamefont {Bryant},\ and\ \citenamefont {Singh}}]{PhysRevLett.97.187202}%
  \BibitemOpen
  \bibfield  {author} {\bibinfo {author} {\bibfnamefont {Marcos}\ \bibnamefont
  {Rigol}}, \bibinfo {author} {\bibfnamefont {Tyler}\ \bibnamefont {Bryant}}, \
  and\ \bibinfo {author} {\bibfnamefont {Rajiv R.~P.}\ \bibnamefont {Singh}},\
  }\bibfield  {title} {\enquote {\bibinfo {title} {{\it Numerical
  Linked-Cluster Approach to Quantum Lattice Models}},}\ }\href {\doibase
  10.1103/PhysRevLett.97.187202} {\bibfield  {journal} {\bibinfo  {journal}
  {Phys. Rev. Lett.}\ }\textbf {\bibinfo {volume} {97}},\ \bibinfo {pages}
  {187202} (\bibinfo {year} {2006})}\BibitemShut {NoStop}%
\bibitem [{\citenamefont {Tang}\ \emph {et~al.}(2013)\citenamefont {Tang},
  \citenamefont {Khatami},\ and\ \citenamefont {Rigol}}]{TANG2013557}%
  \BibitemOpen
  \bibfield  {author} {\bibinfo {author} {\bibfnamefont {Baoming}\ \bibnamefont
  {Tang}}, \bibinfo {author} {\bibfnamefont {Ehsan}\ \bibnamefont {Khatami}}, \
  and\ \bibinfo {author} {\bibfnamefont {Marcos}\ \bibnamefont {Rigol}},\
  }\bibfield  {title} {\enquote {\bibinfo {title} {{\it A short introduction to
  numerical linked-cluster expansions}},}\ }\href {\doibase
  10.1016/j.cpc.2012.10.008} {\bibfield  {journal} {\bibinfo  {journal}
  {Computer Physics Communications}\ }\textbf {\bibinfo {volume} {184}},\
  \bibinfo {pages} {557 -- 564} (\bibinfo {year} {2013})}\BibitemShut {NoStop}%
\bibitem [{\citenamefont {Foulkes}\ \emph {et~al.}(2001)\citenamefont
  {Foulkes}, \citenamefont {Mitas}, \citenamefont {Needs},\ and\ \citenamefont
  {Rajagopal}}]{RevModPhys.73.33}%
  \BibitemOpen
  \bibfield  {author} {\bibinfo {author} {\bibfnamefont {W.~M.~C.}\
  \bibnamefont {Foulkes}}, \bibinfo {author} {\bibfnamefont {L.}~\bibnamefont
  {Mitas}}, \bibinfo {author} {\bibfnamefont {R.~J.}\ \bibnamefont {Needs}}, \
  and\ \bibinfo {author} {\bibfnamefont {G.}~\bibnamefont {Rajagopal}},\
  }\bibfield  {title} {\enquote {\bibinfo {title} {{\it Quantum Monte Carlo
  simulations of solids}},}\ }\href {\doibase 10.1103/RevModPhys.73.33}
  {\bibfield  {journal} {\bibinfo  {journal} {Rev. Mod. Phys.}\ }\textbf
  {\bibinfo {volume} {73}},\ \bibinfo {pages} {33--83} (\bibinfo {year}
  {2001})}\BibitemShut {NoStop}%
\bibitem [{\citenamefont {Sandvik}\ and\ \citenamefont
  {Kurkij\"arvi}(1991)}]{PhysRevB.43.5950}%
  \BibitemOpen
  \bibfield  {author} {\bibinfo {author} {\bibfnamefont {Anders~W.}\
  \bibnamefont {Sandvik}}\ and\ \bibinfo {author} {\bibfnamefont {Juhani}\
  \bibnamefont {Kurkij\"arvi}},\ }\bibfield  {title} {\enquote {\bibinfo
  {title} {{\it Quantum Monte Carlo simulation method for spin systems}},}\
  }\href {\doibase 10.1103/PhysRevB.43.5950} {\bibfield  {journal} {\bibinfo
  {journal} {Phys. Rev. B}\ }\textbf {\bibinfo {volume} {43}},\ \bibinfo
  {pages} {5950--5961} (\bibinfo {year} {1991})}\BibitemShut {NoStop}%
\bibitem [{\citenamefont {Schollw\"ock}(2005)}]{RevModPhys.77.259}%
  \BibitemOpen
  \bibfield  {author} {\bibinfo {author} {\bibfnamefont {U.}~\bibnamefont
  {Schollw\"ock}},\ }\bibfield  {title} {\enquote {\bibinfo {title} {{\it The
  density-matrix renormalization group}},}\ }\href {\doibase
  10.1103/RevModPhys.77.259} {\bibfield  {journal} {\bibinfo  {journal} {Rev.
  Mod. Phys.}\ }\textbf {\bibinfo {volume} {77}},\ \bibinfo {pages} {259--315}
  (\bibinfo {year} {2005})}\BibitemShut {NoStop}%
\bibitem [{\citenamefont {Verstraete}\ \emph {et~al.}(2004)\citenamefont
  {Verstraete}, \citenamefont {Garc\'{\i}a-Ripoll},\ and\ \citenamefont
  {Cirac}}]{PhysRevLett.93.207204}%
  \BibitemOpen
  \bibfield  {author} {\bibinfo {author} {\bibfnamefont {F.}~\bibnamefont
  {Verstraete}}, \bibinfo {author} {\bibfnamefont {J.~J.}\ \bibnamefont
  {Garc\'{\i}a-Ripoll}}, \ and\ \bibinfo {author} {\bibfnamefont {J.~I.}\
  \bibnamefont {Cirac}},\ }\bibfield  {title} {\enquote {\bibinfo {title} {{\it
  Matrix Product Density Operators: Simulation of Finite-Temperature and
  Dissipative Systems}},}\ }\href {\doibase 10.1103/PhysRevLett.93.207204}
  {\bibfield  {journal} {\bibinfo  {journal} {Phys. Rev. Lett.}\ }\textbf
  {\bibinfo {volume} {93}},\ \bibinfo {pages} {207204} (\bibinfo {year}
  {2004})}\BibitemShut {NoStop}%
\bibitem [{\citenamefont {Zwolak}\ and\ \citenamefont
  {Vidal}(2004)}]{PhysRevLett.93.207205}%
  \BibitemOpen
  \bibfield  {author} {\bibinfo {author} {\bibfnamefont {Michael}\ \bibnamefont
  {Zwolak}}\ and\ \bibinfo {author} {\bibfnamefont {Guifr\'e}\ \bibnamefont
  {Vidal}},\ }\bibfield  {title} {\enquote {\bibinfo {title} {{\it Mixed-State
  Dynamics in One-Dimensional Quantum Lattice Systems: A Time-Dependent
  Superoperator Renormalization Algorithm}},}\ }\href {\doibase
  10.1103/PhysRevLett.93.207205} {\bibfield  {journal} {\bibinfo  {journal}
  {Phys. Rev. Lett.}\ }\textbf {\bibinfo {volume} {93}},\ \bibinfo {pages}
  {207205} (\bibinfo {year} {2004})}\BibitemShut {NoStop}%
\bibitem [{\citenamefont {Or\'us}\ and\ \citenamefont
  {Vidal}(2008)}]{PhysRevB.78.155117}%
  \BibitemOpen
  \bibfield  {author} {\bibinfo {author} {\bibfnamefont {R.}~\bibnamefont
  {Or\'us}}\ and\ \bibinfo {author} {\bibfnamefont {G.}~\bibnamefont {Vidal}},\
  }\bibfield  {title} {\enquote {\bibinfo {title} {{\it Infinite time-evolving
  block decimation algorithm beyond unitary evolution}},}\ }\href {\doibase
  10.1103/PhysRevB.78.155117} {\bibfield  {journal} {\bibinfo  {journal} {Phys.
  Rev. B}\ }\textbf {\bibinfo {volume} {78}},\ \bibinfo {pages} {155117}
  (\bibinfo {year} {2008})}\BibitemShut {NoStop}%
\bibitem [{\citenamefont {Li}\ \emph {et~al.}(2011)\citenamefont {Li},
  \citenamefont {Ran}, \citenamefont {Gong}, \citenamefont {Zhao},
  \citenamefont {Xi}, \citenamefont {Ye},\ and\ \citenamefont
  {Su}}]{PhysRevLett.106.127202}%
  \BibitemOpen
  \bibfield  {author} {\bibinfo {author} {\bibfnamefont {Wei}\ \bibnamefont
  {Li}}, \bibinfo {author} {\bibfnamefont {Shi-Ju}\ \bibnamefont {Ran}},
  \bibinfo {author} {\bibfnamefont {Shou-Shu}\ \bibnamefont {Gong}}, \bibinfo
  {author} {\bibfnamefont {Yang}\ \bibnamefont {Zhao}}, \bibinfo {author}
  {\bibfnamefont {Bin}\ \bibnamefont {Xi}}, \bibinfo {author} {\bibfnamefont
  {Fei}\ \bibnamefont {Ye}}, \ and\ \bibinfo {author} {\bibfnamefont {Gang}\
  \bibnamefont {Su}},\ }\bibfield  {title} {\enquote {\bibinfo {title} {{\it
  Linearized Tensor Renormalization Group Algorithm for the Calculation of
  Thermodynamic Properties of Quantum Lattice Models}},}\ }\href {\doibase
  10.1103/PhysRevLett.106.127202} {\bibfield  {journal} {\bibinfo  {journal}
  {Phys. Rev. Lett.}\ }\textbf {\bibinfo {volume} {106}},\ \bibinfo {pages}
  {127202} (\bibinfo {year} {2011})}\BibitemShut {NoStop}%
\bibitem [{\citenamefont {Haegeman}\ \emph {et~al.}(2011)\citenamefont
  {Haegeman}, \citenamefont {Cirac}, \citenamefont {Osborne}, \citenamefont
  {Pi\ifmmode~\check{z}\else \v{z}\fi{}orn}, \citenamefont {Verschelde},\ and\
  \citenamefont {Verstraete}}]{PhysRevLett.107.070601}%
  \BibitemOpen
  \bibfield  {author} {\bibinfo {author} {\bibfnamefont {Jutho}\ \bibnamefont
  {Haegeman}}, \bibinfo {author} {\bibfnamefont {J.~Ignacio}\ \bibnamefont
  {Cirac}}, \bibinfo {author} {\bibfnamefont {Tobias~J.}\ \bibnamefont
  {Osborne}}, \bibinfo {author} {\bibfnamefont {Iztok}\ \bibnamefont
  {Pi\ifmmode~\check{z}\else \v{z}\fi{}orn}}, \bibinfo {author} {\bibfnamefont
  {Henri}\ \bibnamefont {Verschelde}}, \ and\ \bibinfo {author} {\bibfnamefont
  {Frank}\ \bibnamefont {Verstraete}},\ }\bibfield  {title} {\enquote {\bibinfo
  {title} {{\it Time-Dependent Variational Principle for Quantum Lattices}},}\
  }\href {\doibase 10.1103/PhysRevLett.107.070601} {\bibfield  {journal}
  {\bibinfo  {journal} {Phys. Rev. Lett.}\ }\textbf {\bibinfo {volume} {107}},\
  \bibinfo {pages} {070601} (\bibinfo {year} {2011})}\BibitemShut {NoStop}%
\bibitem [{\citenamefont {Czarnik}\ and\ \citenamefont
  {Dziarmaga}(2015)}]{PhysRevB.92.035152}%
  \BibitemOpen
  \bibfield  {author} {\bibinfo {author} {\bibfnamefont {Piotr}\ \bibnamefont
  {Czarnik}}\ and\ \bibinfo {author} {\bibfnamefont {Jacek}\ \bibnamefont
  {Dziarmaga}},\ }\bibfield  {title} {\enquote {\bibinfo {title} {{\it
  Variational approach to projected entangled pair states at finite
  temperature}},}\ }\href {\doibase 10.1103/PhysRevB.92.035152} {\bibfield
  {journal} {\bibinfo  {journal} {Phys. Rev. B}\ }\textbf {\bibinfo {volume}
  {92}},\ \bibinfo {pages} {035152} (\bibinfo {year} {2015})}\BibitemShut
  {NoStop}%
\bibitem [{\citenamefont {Haegeman}\ \emph {et~al.}(2016)\citenamefont
  {Haegeman}, \citenamefont {Lubich}, \citenamefont {Oseledets}, \citenamefont
  {Vandereycken},\ and\ \citenamefont {Verstraete}}]{PhysRevB.94.165116}%
  \BibitemOpen
  \bibfield  {author} {\bibinfo {author} {\bibfnamefont {Jutho}\ \bibnamefont
  {Haegeman}}, \bibinfo {author} {\bibfnamefont {Christian}\ \bibnamefont
  {Lubich}}, \bibinfo {author} {\bibfnamefont {Ivan}\ \bibnamefont
  {Oseledets}}, \bibinfo {author} {\bibfnamefont {Bart}\ \bibnamefont
  {Vandereycken}}, \ and\ \bibinfo {author} {\bibfnamefont {Frank}\
  \bibnamefont {Verstraete}},\ }\bibfield  {title} {\enquote {\bibinfo {title}
  {{\it Unifying time evolution and optimization with matrix product
  states}},}\ }\href {\doibase 10.1103/PhysRevB.94.165116} {\bibfield
  {journal} {\bibinfo  {journal} {Phys. Rev. B}\ }\textbf {\bibinfo {volume}
  {94}},\ \bibinfo {pages} {165116} (\bibinfo {year} {2016})}\BibitemShut
  {NoStop}%
\bibitem [{\citenamefont {Chen}\ \emph {et~al.}(2018)\citenamefont {Chen},
  \citenamefont {Chen}, \citenamefont {Chen}, \citenamefont {Li},\ and\
  \citenamefont {Weichselbaum}}]{PhysRevX.8.031082}%
  \BibitemOpen
  \bibfield  {author} {\bibinfo {author} {\bibfnamefont {Bin-Bin}\ \bibnamefont
  {Chen}}, \bibinfo {author} {\bibfnamefont {Lei}\ \bibnamefont {Chen}},
  \bibinfo {author} {\bibfnamefont {Ziyu}\ \bibnamefont {Chen}}, \bibinfo
  {author} {\bibfnamefont {Wei}\ \bibnamefont {Li}}, \ and\ \bibinfo {author}
  {\bibfnamefont {Andreas}\ \bibnamefont {Weichselbaum}},\ }\bibfield  {title}
  {\enquote {\bibinfo {title} {{\it Exponential Thermal Tensor Network Approach
  for Quantum Lattice Models}},}\ }\href {\doibase 10.1103/PhysRevX.8.031082}
  {\bibfield  {journal} {\bibinfo  {journal} {Phys. Rev. X}\ }\textbf {\bibinfo
  {volume} {8}},\ \bibinfo {pages} {031082} (\bibinfo {year}
  {2018})}\BibitemShut {NoStop}%
\bibitem [{\citenamefont {Poulin}\ and\ \citenamefont
  {Bilgin}(2008)}]{PhysRevA.77.052318}%
  \BibitemOpen
  \bibfield  {author} {\bibinfo {author} {\bibfnamefont {David}\ \bibnamefont
  {Poulin}}\ and\ \bibinfo {author} {\bibfnamefont {Ersen}\ \bibnamefont
  {Bilgin}},\ }\bibfield  {title} {\enquote {\bibinfo {title} {{\it Belief
  propagation algorithm for computing correlation functions in
  finite-temperature quantum many-body systems on loopy graphs}},}\ }\href
  {\doibase 10.1103/PhysRevA.77.052318} {\bibfield  {journal} {\bibinfo
  {journal} {Phys. Rev. A}\ }\textbf {\bibinfo {volume} {77}},\ \bibinfo
  {pages} {052318} (\bibinfo {year} {2008})}\BibitemShut {NoStop}%
\bibitem [{\citenamefont {Bilgin}\ and\ \citenamefont
  {Poulin}(2010)}]{PhysRevB.81.054106}%
  \BibitemOpen
  \bibfield  {author} {\bibinfo {author} {\bibfnamefont {Ersen}\ \bibnamefont
  {Bilgin}}\ and\ \bibinfo {author} {\bibfnamefont {David}\ \bibnamefont
  {Poulin}},\ }\bibfield  {title} {\enquote {\bibinfo {title} {{\it
  Coarse-grained belief propagation for simulation of interacting quantum
  systems at all temperatures}},}\ }\href {\doibase 10.1103/PhysRevB.81.054106}
  {\bibfield  {journal} {\bibinfo  {journal} {Phys. Rev. B}\ }\textbf {\bibinfo
  {volume} {81}},\ \bibinfo {pages} {054106} (\bibinfo {year}
  {2010})}\BibitemShut {NoStop}%
\bibitem [{\citenamefont {Poulin}\ and\ \citenamefont
  {Wocjan}(2009)}]{PhysRevLett.103.220502}%
  \BibitemOpen
  \bibfield  {author} {\bibinfo {author} {\bibfnamefont {David}\ \bibnamefont
  {Poulin}}\ and\ \bibinfo {author} {\bibfnamefont {Pawel}\ \bibnamefont
  {Wocjan}},\ }\bibfield  {title} {\enquote {\bibinfo {title} {{\it Sampling
  from the Thermal Quantum Gibbs State and Evaluating Partition Functions with
  a Quantum Computer}},}\ }\href {\doibase 10.1103/PhysRevLett.103.220502}
  {\bibfield  {journal} {\bibinfo  {journal} {Phys. Rev. Lett.}\ }\textbf
  {\bibinfo {volume} {103}},\ \bibinfo {pages} {220502} (\bibinfo {year}
  {2009})}\BibitemShut {NoStop}%
\bibitem [{\citenamefont {Bilgin}\ and\ \citenamefont
  {Boixo}(2010)}]{PhysRevLett.105.170405}%
  \BibitemOpen
  \bibfield  {author} {\bibinfo {author} {\bibfnamefont {Ersen}\ \bibnamefont
  {Bilgin}}\ and\ \bibinfo {author} {\bibfnamefont {Sergio}\ \bibnamefont
  {Boixo}},\ }\bibfield  {title} {\enquote {\bibinfo {title} {{\it Preparing
  Thermal States of Quantum Systems by Dimension Reduction}},}\ }\href
  {\doibase 10.1103/PhysRevLett.105.170405} {\bibfield  {journal} {\bibinfo
  {journal} {Phys. Rev. Lett.}\ }\textbf {\bibinfo {volume} {105}},\ \bibinfo
  {pages} {170405} (\bibinfo {year} {2010})}\BibitemShut {NoStop}%
\bibitem [{\citenamefont {Temme}\ \emph {et~al.}(2011)\citenamefont {Temme},
  \citenamefont {Osborne}, \citenamefont {Vollbrecht}, \citenamefont {Poulin},\
  and\ \citenamefont {Verstraete}}]{temme2011quantum}%
  \BibitemOpen
  \bibfield  {author} {\bibinfo {author} {\bibfnamefont {Kristan}\ \bibnamefont
  {Temme}}, \bibinfo {author} {\bibfnamefont {Tobias~J}\ \bibnamefont
  {Osborne}}, \bibinfo {author} {\bibfnamefont {Karl~G}\ \bibnamefont
  {Vollbrecht}}, \bibinfo {author} {\bibfnamefont {David}\ \bibnamefont
  {Poulin}}, \ and\ \bibinfo {author} {\bibfnamefont {Frank}\ \bibnamefont
  {Verstraete}},\ }\bibfield  {title} {\enquote {\bibinfo {title} {{\it Quantum
  metropolis sampling}},}\ }\href {https://doi.org/10.1038/nature09770}
  {\bibfield  {journal} {\bibinfo  {journal} {Nature}\ }\textbf {\bibinfo
  {volume} {471}},\ \bibinfo {pages} {87} (\bibinfo {year} {2011})}\BibitemShut
  {NoStop}%
\bibitem [{\citenamefont {Yung}\ and\ \citenamefont
  {Aspuru-Guzik}(2012)}]{Yung754}%
  \BibitemOpen
  \bibfield  {author} {\bibinfo {author} {\bibfnamefont {Man-Hong}\
  \bibnamefont {Yung}}\ and\ \bibinfo {author} {\bibfnamefont {Al{\'a}n}\
  \bibnamefont {Aspuru-Guzik}},\ }\bibfield  {title} {\enquote {\bibinfo
  {title} {{\it A quantum{\textendash}quantum Metropolis algorithm}},}\ }\href
  {\doibase 10.1073/pnas.1111758109} {\bibfield  {journal} {\bibinfo  {journal}
  {Proceedings of the National Academy of Sciences}\ }\textbf {\bibinfo
  {volume} {109}},\ \bibinfo {pages} {754--759} (\bibinfo {year} {2012})},\
  \Eprint {http://arxiv.org/abs/http://www.pnas.org/content/109/3/754.full.pdf}
  {http://www.pnas.org/content/109/3/754.full.pdf} \BibitemShut {NoStop}%
\bibitem [{\citenamefont {Ge}\ \emph {et~al.}(2016)\citenamefont {Ge},
  \citenamefont {Moln\'ar},\ and\ \citenamefont
  {Cirac}}]{PhysRevLett.116.080503}%
  \BibitemOpen
  \bibfield  {author} {\bibinfo {author} {\bibfnamefont {Yimin}\ \bibnamefont
  {Ge}}, \bibinfo {author} {\bibfnamefont {Andr\'as}\ \bibnamefont {Moln\'ar}},
  \ and\ \bibinfo {author} {\bibfnamefont {J.~Ignacio}\ \bibnamefont {Cirac}},\
  }\bibfield  {title} {\enquote {\bibinfo {title} {{\it Rapid Adiabatic
  Preparation of Injective Projected Entangled Pair States and Gibbs
  States}},}\ }\href {\doibase 10.1103/PhysRevLett.116.080503} {\bibfield
  {journal} {\bibinfo  {journal} {Phys. Rev. Lett.}\ }\textbf {\bibinfo
  {volume} {116}},\ \bibinfo {pages} {080503} (\bibinfo {year}
  {2016})}\BibitemShut {NoStop}%
\bibitem [{\citenamefont {Kastoryano}\ and\ \citenamefont
  {Brand{\~a}o}(2016)}]{Kastoryano2016}%
  \BibitemOpen
  \bibfield  {author} {\bibinfo {author} {\bibfnamefont {Michael~J.}\
  \bibnamefont {Kastoryano}}\ and\ \bibinfo {author} {\bibfnamefont {Fernando
  G. S.~L.}\ \bibnamefont {Brand{\~a}o}},\ }\bibfield  {title} {\enquote
  {\bibinfo {title} {{\it Quantum Gibbs Samplers: The Commuting Case}},}\
  }\href {\doibase 10.1007/s00220-016-2641-8} {\bibfield  {journal} {\bibinfo
  {journal} {Communications in Mathematical Physics}\ }\textbf {\bibinfo
  {volume} {344}},\ \bibinfo {pages} {915--957} (\bibinfo {year}
  {2016})}\BibitemShut {NoStop}%
\bibitem [{\citenamefont {Chowdhury}\ and\ \citenamefont
  {Somma}(2017)}]{10.5555/3179483.3179486}%
  \BibitemOpen
  \bibfield  {author} {\bibinfo {author} {\bibfnamefont {Anirban~Narayan}\
  \bibnamefont {Chowdhury}}\ and\ \bibinfo {author} {\bibfnamefont
  {Rolando~D.}\ \bibnamefont {Somma}},\ }\bibfield  {title} {\enquote {\bibinfo
  {title} {{\it Quantum Algorithms for Gibbs Sampling and Hitting-Time
  Estimation}},}\ }\href@noop {} {\bibfield  {journal} {\bibinfo  {journal}
  {Quantum Info. Comput.}\ }\textbf {\bibinfo {volume} {17}},\ \bibinfo {pages}
  {41–64} (\bibinfo {year} {2017})}\BibitemShut {NoStop}%
\bibitem [{\citenamefont {Brand{\~a}o}\ and\ \citenamefont
  {Kastoryano}(2019)}]{Brandao2019}%
  \BibitemOpen
  \bibfield  {author} {\bibinfo {author} {\bibfnamefont {Fernando G. S.~L.}\
  \bibnamefont {Brand{\~a}o}}\ and\ \bibinfo {author} {\bibfnamefont
  {Michael~J.}\ \bibnamefont {Kastoryano}},\ }\bibfield  {title} {\enquote
  {\bibinfo {title} {{\it Finite Correlation Length Implies Efficient
  Preparation of Quantum Thermal States}},}\ }\href {\doibase
  10.1007/s00220-018-3150-8} {\bibfield  {journal} {\bibinfo  {journal}
  {Communications in Mathematical Physics}\ }\textbf {\bibinfo {volume}
  {365}},\ \bibinfo {pages} {1--16} (\bibinfo {year} {2019})}\BibitemShut
  {NoStop}%
\bibitem [{\citenamefont {Gily\'{e}n}\ \emph {et~al.}(2019)\citenamefont
  {Gily\'{e}n}, \citenamefont {Su}, \citenamefont {Low},\ and\ \citenamefont
  {Wiebe}}]{10.1145/3313276.3316366}%
  \BibitemOpen
  \bibfield  {author} {\bibinfo {author} {\bibfnamefont {Andr\'{a}s}\
  \bibnamefont {Gily\'{e}n}}, \bibinfo {author} {\bibfnamefont {Yuan}\
  \bibnamefont {Su}}, \bibinfo {author} {\bibfnamefont {Guang~Hao}\
  \bibnamefont {Low}}, \ and\ \bibinfo {author} {\bibfnamefont {Nathan}\
  \bibnamefont {Wiebe}},\ }\bibfield  {title} {\enquote {\bibinfo {title} {{\it
  Quantum Singular Value Transformation and beyond: Exponential Improvements
  for Quantum Matrix Arithmetics}},}\ }in\ \href {\doibase
  10.1145/3313276.3316366} {\emph {\bibinfo {booktitle} {{\it Proceedings of
  the 51st Annual ACM SIGACT Symposium on Theory of Computing}, pages =
  {193–204}, numpages = {12}, location = {Phoenix, AZ, USA}, series = {STOC
  2019}}}}\ (\bibinfo  {publisher} {Association for Computing Machinery},\
  \bibinfo {address} {New York, NY, USA},\ \bibinfo {year} {2019})\BibitemShut
  {NoStop}%
\bibitem [{\citenamefont {Araki}(1969)}]{Araki1969}%
  \BibitemOpen
  \bibfield  {author} {\bibinfo {author} {\bibfnamefont {Huzihiro}\
  \bibnamefont {Araki}},\ }\bibfield  {title} {\enquote {\bibinfo {title} {{\it
  Gibbs states of a one dimensional quantum lattice}},}\ }\href {\doibase
  10.1007/BF01645134} {\bibfield  {journal} {\bibinfo  {journal}
  {Communications in Mathematical Physics}\ }\textbf {\bibinfo {volume} {14}},\
  \bibinfo {pages} {120--157} (\bibinfo {year} {1969})}\BibitemShut {NoStop}%
\bibitem [{\citenamefont {Gross}(1979)}]{Gross1979}%
  \BibitemOpen
  \bibfield  {author} {\bibinfo {author} {\bibfnamefont {Leonard}\ \bibnamefont
  {Gross}},\ }\bibfield  {title} {\enquote {\bibinfo {title} {{\it Decay of
  correlations in classical lattice models at high temperature}},}\ }\href
  {\doibase 10.1007/BF01562538} {\bibfield  {journal} {\bibinfo  {journal}
  {Communications in Mathematical Physics}\ }\textbf {\bibinfo {volume} {68}},\
  \bibinfo {pages} {9--27} (\bibinfo {year} {1979})}\BibitemShut {NoStop}%
\bibitem [{\citenamefont {Park}\ and\ \citenamefont {Yoo}(1995)}]{Park1995}%
  \BibitemOpen
  \bibfield  {author} {\bibinfo {author} {\bibfnamefont {Yong~Moon}\
  \bibnamefont {Park}}\ and\ \bibinfo {author} {\bibfnamefont {Hyun~Jae}\
  \bibnamefont {Yoo}},\ }\bibfield  {title} {\enquote {\bibinfo {title} {{\it
  Uniqueness and clustering properties of Gibbs states for classical and
  quantum unbounded spin systems}},}\ }\href {\doibase 10.1007/BF02178359}
  {\bibfield  {journal} {\bibinfo  {journal} {Journal of Statistical Physics}\
  }\textbf {\bibinfo {volume} {80}},\ \bibinfo {pages} {223--271} (\bibinfo
  {year} {1995})}\BibitemShut {NoStop}%
\bibitem [{\citenamefont {Ueltschi}(2004)}]{ueltschi2004cluster}%
  \BibitemOpen
  \bibfield  {author} {\bibinfo {author} {\bibfnamefont {Daniel}\ \bibnamefont
  {Ueltschi}},\ }\bibfield  {title} {\enquote {\bibinfo {title} {{\it Cluster
  expansions and correlation functions}},}\ }\href@noop {} {\bibfield
  {journal} {\bibinfo  {journal} {Moscow Mathematical Journal}\ }\textbf
  {\bibinfo {volume} {4}},\ \bibinfo {pages} {511--522} (\bibinfo {year}
  {2004})}\BibitemShut {NoStop}%
\bibitem [{\citenamefont {Kliesch}\ \emph {et~al.}(2014)\citenamefont
  {Kliesch}, \citenamefont {Gogolin}, \citenamefont {Kastoryano}, \citenamefont
  {Riera},\ and\ \citenamefont {Eisert}}]{PhysRevX.4.031019}%
  \BibitemOpen
  \bibfield  {author} {\bibinfo {author} {\bibfnamefont {M.}~\bibnamefont
  {Kliesch}}, \bibinfo {author} {\bibfnamefont {C.}~\bibnamefont {Gogolin}},
  \bibinfo {author} {\bibfnamefont {M.~J.}\ \bibnamefont {Kastoryano}},
  \bibinfo {author} {\bibfnamefont {A.}~\bibnamefont {Riera}}, \ and\ \bibinfo
  {author} {\bibfnamefont {J.}~\bibnamefont {Eisert}},\ }\bibfield  {title}
  {\enquote {\bibinfo {title} {{\it Locality of Temperature}},}\ }\href
  {\doibase 10.1103/PhysRevX.4.031019} {\bibfield  {journal} {\bibinfo
  {journal} {Phys. Rev. X}\ }\textbf {\bibinfo {volume} {4}},\ \bibinfo {pages}
  {031019} (\bibinfo {year} {2014})}\BibitemShut {NoStop}%
\bibitem [{\citenamefont {Fr{\"o}hlich}\ and\ \citenamefont
  {Ueltschi}(2015)}]{frohlich2015some}%
  \BibitemOpen
  \bibfield  {author} {\bibinfo {author} {\bibfnamefont {J{\"u}rg}\
  \bibnamefont {Fr{\"o}hlich}}\ and\ \bibinfo {author} {\bibfnamefont {Daniel}\
  \bibnamefont {Ueltschi}},\ }\bibfield  {title} {\enquote {\bibinfo {title}
  {{\it Some properties of correlations of quantum lattice systems in thermal
  equilibrium}},}\ }\href {\doibase 10.1063/1.4921305} {\bibfield  {journal}
  {\bibinfo  {journal} {Journal of Mathematical Physics}\ }\textbf {\bibinfo
  {volume} {56}},\ \bibinfo {pages} {053302} (\bibinfo {year}
  {2015})}\BibitemShut {NoStop}%
\bibitem [{\citenamefont {Lenci}\ and\ \citenamefont
  {Rey-Bellet}()}]{ref:Marco_LD}%
  \BibitemOpen
  \bibfield  {author} {\bibinfo {author} {\bibfnamefont {Marco}\ \bibnamefont
  {Lenci}}\ and\ \bibinfo {author} {\bibfnamefont {Luc}\ \bibnamefont
  {Rey-Bellet}},\ }\bibfield  {title} {\enquote {\bibinfo {title} {{\it Large
  Deviations in Quantum Lattice Systems: One-Phase Region}},}\ }\href {\doibase
  10.1007/s10955-005-3015-3} {\bibfield  {journal} {\bibinfo  {journal}
  {Journal of Statistical Physics}\ }\textbf {\bibinfo {volume} {119}},\
  \bibinfo {pages} {715--746}}\BibitemShut {NoStop}%
\bibitem [{\citenamefont {Neto{\v{c}}n{\'y}}\ and\ \citenamefont
  {Redig}(2004)}]{Netocny2004}%
  \BibitemOpen
  \bibfield  {author} {\bibinfo {author} {\bibfnamefont {K.}~\bibnamefont
  {Neto{\v{c}}n{\'y}}}\ and\ \bibinfo {author} {\bibfnamefont {F.}~\bibnamefont
  {Redig}},\ }\bibfield  {title} {\enquote {\bibinfo {title} {{\it Large
  Deviations for Quantum Spin Systems}},}\ }\href {\doibase
  10.1007/s10955-004-3452-4} {\bibfield  {journal} {\bibinfo  {journal}
  {Journal of Statistical Physics}\ }\textbf {\bibinfo {volume} {117}},\
  \bibinfo {pages} {521--547} (\bibinfo {year} {2004})}\BibitemShut {NoStop}%
\bibitem [{\citenamefont {Kuwahara}\ and\ \citenamefont
  {Saito}(2020{\natexlab{b}})}]{kuwahara2019gaussian}%
  \BibitemOpen
  \bibfield  {author} {\bibinfo {author} {\bibfnamefont {Tomotaka}\
  \bibnamefont {Kuwahara}}\ and\ \bibinfo {author} {\bibfnamefont {Keiji}\
  \bibnamefont {Saito}},\ }\bibfield  {title} {\enquote {\bibinfo {title} {{\it
  Gaussian concentration bound and Ensemble equivalence in generic quantum
  many-body systems including long-range interactions}},}\ }\href {\doibase
  https://doi.org/10.1016/j.aop.2020.168278} {\bibfield  {journal} {\bibinfo
  {journal} {Annals of Physics}\ }\textbf {\bibinfo {volume} {421}},\ \bibinfo
  {pages} {168278} (\bibinfo {year} {2020}{\natexlab{b}})}\BibitemShut
  {NoStop}%
\bibitem [{\citenamefont {Kato}\ and\ \citenamefont
  {Brand{\~a}o}(2019)}]{kato2016quantum}%
  \BibitemOpen
  \bibfield  {author} {\bibinfo {author} {\bibfnamefont {Kohtaro}\ \bibnamefont
  {Kato}}\ and\ \bibinfo {author} {\bibfnamefont {Fernando G. S.~L.}\
  \bibnamefont {Brand{\~a}o}},\ }\bibfield  {title} {\enquote {\bibinfo {title}
  {{\it Quantum Approximate Markov Chains are Thermal}},}\ }\href {\doibase
  10.1007/s00220-019-03485-6} {\bibfield  {journal} {\bibinfo  {journal}
  {Communications in Mathematical Physics}\ } (\bibinfo {year} {2019}),\
  10.1007/s00220-019-03485-6}\BibitemShut {NoStop}%
\bibitem [{\citenamefont {Kuwahara}\ \emph {et~al.}(2020)\citenamefont
  {Kuwahara}, \citenamefont {Kato},\ and\ \citenamefont
  {Brand\~ao}}]{CMI_clustering}%
  \BibitemOpen
  \bibfield  {author} {\bibinfo {author} {\bibfnamefont {Tomotaka}\
  \bibnamefont {Kuwahara}}, \bibinfo {author} {\bibfnamefont {Kohtaro}\
  \bibnamefont {Kato}}, \ and\ \bibinfo {author} {\bibfnamefont {Fernando G.
  S.~L.}\ \bibnamefont {Brand\~ao}},\ }\bibfield  {title} {\enquote {\bibinfo
  {title} {{\it Clustering of Conditional Mutual Information for Quantum Gibbs
  States above a Threshold Temperature}},}\ }\href {\doibase
  10.1103/PhysRevLett.124.220601} {\bibfield  {journal} {\bibinfo  {journal}
  {Phys. Rev. Lett.}\ }\textbf {\bibinfo {volume} {124}},\ \bibinfo {pages}
  {220601} (\bibinfo {year} {2020})}\BibitemShut {NoStop}%
\bibitem [{\citenamefont {Harrow}\ \emph {et~al.}(2020)\citenamefont {Harrow},
  \citenamefont {Mehraban},\ and\ \citenamefont {Soleimanifar}}]{Mehdi}%
  \BibitemOpen
  \bibfield  {author} {\bibinfo {author} {\bibfnamefont {Aram~W.}\ \bibnamefont
  {Harrow}}, \bibinfo {author} {\bibfnamefont {Saeed}\ \bibnamefont
  {Mehraban}}, \ and\ \bibinfo {author} {\bibfnamefont {Mehdi}\ \bibnamefont
  {Soleimanifar}},\ }\bibfield  {title} {\enquote {\bibinfo {title} {{\it
  Classical Algorithms, Correlation Decay, and Complex Zeros of Partition
  Functions of Quantum Many-Body Systems}},}\ }in\ \href {\doibase
  10.1145/3357713.3384322} {\emph {\bibinfo {booktitle} {{\it Proceedings of
  the 52nd Annual ACM SIGACT Symposium on Theory of Computing}, pages =
  {378–386}, numpages = {9}, series = {STOC 2020}}}}\ (\bibinfo  {publisher}
  {Association for Computing Machinery},\ \bibinfo {address} {New York, NY,
  USA},\ \bibinfo {year} {2020})\BibitemShut {NoStop}%
\bibitem [{\citenamefont {Crosson}\ and\ \citenamefont
  {Slezak}(2020)}]{crosson2020classical}%
  \BibitemOpen
  \bibfield  {author} {\bibinfo {author} {\bibfnamefont {Elizabeth}\
  \bibnamefont {Crosson}}\ and\ \bibinfo {author} {\bibfnamefont {Samuel}\
  \bibnamefont {Slezak}},\ }\bibfield  {title} {\enquote {\bibinfo {title}
  {{\it Classical Simulation of High Temperature Quantum Ising Models}},}\
  }\href@noop {} {\bibfield  {journal} {\bibinfo  {journal} {arXiv preprint
  arXiv:2002.02232}\ } (\bibinfo {year} {2020})},\ \Eprint
  {http://arxiv.org/abs/arXiv:2002.02232} {arXiv:2002.02232} \BibitemShut
  {NoStop}%
\bibitem [{\citenamefont {Mann}\ and\ \citenamefont
  {Helmuth}(2020)}]{mann2020efficient}%
  \BibitemOpen
  \bibfield  {author} {\bibinfo {author} {\bibfnamefont {Ryan~L}\ \bibnamefont
  {Mann}}\ and\ \bibinfo {author} {\bibfnamefont {Tyler}\ \bibnamefont
  {Helmuth}},\ }\bibfield  {title} {\enquote {\bibinfo {title} {{\it Efficient
  Algorithms for Approximating Quantum Partition Functions}},}\ }\href@noop {}
  {\bibfield  {journal} {\bibinfo  {journal} {arXiv preprint arXiv:2004.11568}\
  } (\bibinfo {year} {2020})},\ \Eprint {http://arxiv.org/abs/2004.11568}
  {2004.11568} \BibitemShut {NoStop}%
\bibitem [{\citenamefont {Barahona}(1982)}]{Barahona_1982}%
  \BibitemOpen
  \bibfield  {author} {\bibinfo {author} {\bibfnamefont {F}~\bibnamefont
  {Barahona}},\ }\bibfield  {title} {\enquote {\bibinfo {title} {{\it On the
  computational complexity of Ising spin glass models}},}\ }\href {\doibase
  10.1088/0305-4470/15/10/028} {\bibfield  {journal} {\bibinfo  {journal}
  {Journal of Physics A: Mathematical and General}\ }\textbf {\bibinfo {volume}
  {15}},\ \bibinfo {pages} {3241--3253} (\bibinfo {year} {1982})}\BibitemShut
  {NoStop}%
\bibitem [{\citenamefont {Goldberg}\ and\ \citenamefont
  {Jerrum}(2015)}]{Goldberg2015}%
  \BibitemOpen
  \bibfield  {author} {\bibinfo {author} {\bibfnamefont {Leslie~Ann}\
  \bibnamefont {Goldberg}}\ and\ \bibinfo {author} {\bibfnamefont {Mark}\
  \bibnamefont {Jerrum}},\ }\bibfield  {title} {\enquote {\bibinfo {title} {A
  complexity classification of spin systems with an external field},}\ }\href
  {\doibase 10.1073/pnas.1505664112} {\bibfield  {journal} {\bibinfo  {journal}
  {Proceedings of the National Academy of Sciences of the United States of
  America}\ }\textbf {\bibinfo {volume} {112}},\ \bibinfo {pages}
  {13161--13166} (\bibinfo {year} {2015})}\BibitemShut {NoStop}%
\bibitem [{Note1()}]{Note1}%
  \BibitemOpen
  \bibinfo {note} {More in detail, we mean the following statement. Let $Z$ be
  a partition function as $Z:={\protect \rm tr}(e^{-\beta H})$ with $H$ the
  system Hamiltonian. Then, there exists a constant $\zeta >0$ such that
  approximating $\protect \qopname \relax o{log}(Z)$ up to an error $\zeta n$
  ($n$: system size) is the NP-hard problem~\cite {Goldberg2015}.}\BibitemShut
  {Stop}%
\bibitem [{\citenamefont {Jerrum}\ and\ \citenamefont
  {Sinclair}(1993)}]{doi:10.1137/0222066}%
  \BibitemOpen
  \bibfield  {author} {\bibinfo {author} {\bibfnamefont {Mark}\ \bibnamefont
  {Jerrum}}\ and\ \bibinfo {author} {\bibfnamefont {Alistair}\ \bibnamefont
  {Sinclair}},\ }\bibfield  {title} {\enquote {\bibinfo {title} {{\it
  Polynomial-Time Approximation Algorithms for the Ising Model}},}\ }\href
  {\doibase 10.1137/0222066} {\bibfield  {journal} {\bibinfo  {journal} {SIAM
  Journal on Computing}\ }\textbf {\bibinfo {volume} {22}},\ \bibinfo {pages}
  {1087--1116} (\bibinfo {year} {1993})}\BibitemShut {NoStop}%
\bibitem [{\citenamefont {Bravyi}\ and\ \citenamefont
  {Gosset}(2017)}]{PhysRevLett.119.100503}%
  \BibitemOpen
  \bibfield  {author} {\bibinfo {author} {\bibfnamefont {Sergey}\ \bibnamefont
  {Bravyi}}\ and\ \bibinfo {author} {\bibfnamefont {David}\ \bibnamefont
  {Gosset}},\ }\bibfield  {title} {\enquote {\bibinfo {title} {{\it
  Polynomial-Time Classical Simulation of Quantum Ferromagnets}},}\ }\href
  {\doibase 10.1103/PhysRevLett.119.100503} {\bibfield  {journal} {\bibinfo
  {journal} {Phys. Rev. Lett.}\ }\textbf {\bibinfo {volume} {119}},\ \bibinfo
  {pages} {100503} (\bibinfo {year} {2017})}\BibitemShut {NoStop}%
\bibitem [{\citenamefont {Lamm}\ and\ \citenamefont
  {Lawrence}(2018)}]{PhysRevLett.121.170501}%
  \BibitemOpen
  \bibfield  {author} {\bibinfo {author} {\bibfnamefont {Henry}\ \bibnamefont
  {Lamm}}\ and\ \bibinfo {author} {\bibfnamefont {Scott}\ \bibnamefont
  {Lawrence}},\ }\bibfield  {title} {\enquote {\bibinfo {title} {{\it
  Simulation of Nonequilibrium Dynamics on a Quantum Computer}},}\ }\href
  {\doibase 10.1103/PhysRevLett.121.170501} {\bibfield  {journal} {\bibinfo
  {journal} {Phys. Rev. Lett.}\ }\textbf {\bibinfo {volume} {121}},\ \bibinfo
  {pages} {170501} (\bibinfo {year} {2018})}\BibitemShut {NoStop}%
\bibitem [{\citenamefont {Beach}\ \emph {et~al.}(2019)\citenamefont {Beach},
  \citenamefont {Melko}, \citenamefont {Grover},\ and\ \citenamefont
  {Hsieh}}]{PhysRevB.100.094434}%
  \BibitemOpen
  \bibfield  {author} {\bibinfo {author} {\bibfnamefont {Matthew J.~S.}\
  \bibnamefont {Beach}}, \bibinfo {author} {\bibfnamefont {Roger~G.}\
  \bibnamefont {Melko}}, \bibinfo {author} {\bibfnamefont {Tarun}\ \bibnamefont
  {Grover}}, \ and\ \bibinfo {author} {\bibfnamefont {Timothy~H.}\ \bibnamefont
  {Hsieh}},\ }\bibfield  {title} {\enquote {\bibinfo {title} {{\it Making
  trotters sprint: A variational imaginary time ansatz for quantum many-body
  systems}},}\ }\href {\doibase 10.1103/PhysRevB.100.094434} {\bibfield
  {journal} {\bibinfo  {journal} {Phys. Rev. B}\ }\textbf {\bibinfo {volume}
  {100}},\ \bibinfo {pages} {094434} (\bibinfo {year} {2019})}\BibitemShut
  {NoStop}%
\bibitem [{\citenamefont {Yuan}\ \emph {et~al.}(2019)\citenamefont {Yuan},
  \citenamefont {Endo}, \citenamefont {Zhao}, \citenamefont {Li},\ and\
  \citenamefont {Benjamin}}]{Yuan2019theoryofvariational}%
  \BibitemOpen
  \bibfield  {author} {\bibinfo {author} {\bibfnamefont {Xiao}\ \bibnamefont
  {Yuan}}, \bibinfo {author} {\bibfnamefont {Suguru}\ \bibnamefont {Endo}},
  \bibinfo {author} {\bibfnamefont {Qi}~\bibnamefont {Zhao}}, \bibinfo {author}
  {\bibfnamefont {Ying}\ \bibnamefont {Li}}, \ and\ \bibinfo {author}
  {\bibfnamefont {Simon~C.}\ \bibnamefont {Benjamin}},\ }\bibfield  {title}
  {\enquote {\bibinfo {title} {{\it Theory of variational quantum
  simulation}},}\ }\href {\doibase 10.22331/q-2019-10-07-191} {\bibfield
  {journal} {\bibinfo  {journal} {{Quantum}}\ }\textbf {\bibinfo {volume}
  {3}},\ \bibinfo {pages} {191} (\bibinfo {year} {2019})}\BibitemShut {NoStop}%
\bibitem [{\citenamefont {McArdle}\ \emph {et~al.}(2019)\citenamefont
  {McArdle}, \citenamefont {Jones}, \citenamefont {Endo}, \citenamefont {Li},
  \citenamefont {Benjamin},\ and\ \citenamefont {Yuan}}]{McArdle2019}%
  \BibitemOpen
  \bibfield  {author} {\bibinfo {author} {\bibfnamefont {Sam}\ \bibnamefont
  {McArdle}}, \bibinfo {author} {\bibfnamefont {Tyson}\ \bibnamefont {Jones}},
  \bibinfo {author} {\bibfnamefont {Suguru}\ \bibnamefont {Endo}}, \bibinfo
  {author} {\bibfnamefont {Ying}\ \bibnamefont {Li}}, \bibinfo {author}
  {\bibfnamefont {Simon~C.}\ \bibnamefont {Benjamin}}, \ and\ \bibinfo {author}
  {\bibfnamefont {Xiao}\ \bibnamefont {Yuan}},\ }\bibfield  {title} {\enquote
  {\bibinfo {title} {{\it Variational ansatz-based quantum simulation of
  imaginary time evolution}},}\ }\href {\doibase 10.1038/s41534-019-0187-2}
  {\bibfield  {journal} {\bibinfo  {journal} {npj Quantum Information}\
  }\textbf {\bibinfo {volume} {5}},\ \bibinfo {pages} {75} (\bibinfo {year}
  {2019})}\BibitemShut {NoStop}%
\bibitem [{\citenamefont {Yeter-Aydeniz}\ \emph {et~al.}(2020)\citenamefont
  {Yeter-Aydeniz}, \citenamefont {Pooser},\ and\ \citenamefont
  {Siopsis}}]{yeter2019practical}%
  \BibitemOpen
  \bibfield  {author} {\bibinfo {author} {\bibfnamefont {K{\"u}bra}\
  \bibnamefont {Yeter-Aydeniz}}, \bibinfo {author} {\bibfnamefont {Raphael~C.}\
  \bibnamefont {Pooser}}, \ and\ \bibinfo {author} {\bibfnamefont {George}\
  \bibnamefont {Siopsis}},\ }\bibfield  {title} {\enquote {\bibinfo {title}
  {{\it Practical quantum computation of chemical and nuclear energy levels
  using quantum imaginary time evolution and Lanczos algorithms}},}\ }\href
  {\doibase 10.1038/s41534-020-00290-1} {\bibfield  {journal} {\bibinfo
  {journal} {npj Quantum Information}\ }\textbf {\bibinfo {volume} {6}},\
  \bibinfo {pages} {63} (\bibinfo {year} {2020})}\BibitemShut {NoStop}%
\bibitem [{\citenamefont {Love}(2020)}]{love2020cooling}%
  \BibitemOpen
  \bibfield  {author} {\bibinfo {author} {\bibfnamefont {Peter~J}\ \bibnamefont
  {Love}},\ }\bibfield  {title} {\enquote {\bibinfo {title} {{\it Cooling with
  imaginary time}},}\ }\href {\doibase 10.1038/s41567-019-0709-z} {\bibfield
  {journal} {\bibinfo  {journal} {Nature Physics}\ }\textbf {\bibinfo {volume}
  {16}},\ \bibinfo {pages} {130--131} (\bibinfo {year} {2020})}\BibitemShut
  {NoStop}%
\bibitem [{\citenamefont {Nishi}\ \emph {et~al.}(2020)\citenamefont {Nishi},
  \citenamefont {Kosugi},\ and\ \citenamefont
  {Matsushita}}]{nishi2020implementation}%
  \BibitemOpen
  \bibfield  {author} {\bibinfo {author} {\bibfnamefont {Hirofumi}\
  \bibnamefont {Nishi}}, \bibinfo {author} {\bibfnamefont {Taichi}\
  \bibnamefont {Kosugi}}, \ and\ \bibinfo {author} {\bibfnamefont {Yu-ichiro}\
  \bibnamefont {Matsushita}},\ }\bibfield  {title} {\enquote {\bibinfo {title}
  {{\it Implementation of quantum imaginary-time evolution method on NISQ
  devices: Nonlocal approximation}},}\ }\href@noop {} {\bibfield  {journal}
  {\bibinfo  {journal} {arXiv preprint arXiv:2005.12715}\ } (\bibinfo {year}
  {2020})},\ \Eprint {http://arxiv.org/abs/arXiv:2005.12715} {arXiv:2005.12715}
  \BibitemShut {NoStop}%
\bibitem [{Note2()}]{Note2}%
  \BibitemOpen
  \bibinfo {note} {Above this threshold temperature, one can ensure that there
  are no quantum/classical phase transitions.}\BibitemShut {Stop}%
\bibitem [{\citenamefont {Gottesman}\ and\ \citenamefont
  {Hastings}(2010)}]{Gottesman_2010}%
  \BibitemOpen
  \bibfield  {author} {\bibinfo {author} {\bibfnamefont {Daniel}\ \bibnamefont
  {Gottesman}}\ and\ \bibinfo {author} {\bibfnamefont {M~B}\ \bibnamefont
  {Hastings}},\ }\bibfield  {title} {\enquote {\bibinfo {title} {{\it
  Entanglement versus gap for one-dimensional spin systems}},}\ }\href
  {\doibase 10.1088/1367-2630/12/2/025002} {\bibfield  {journal} {\bibinfo
  {journal} {New Journal of Physics}\ }\textbf {\bibinfo {volume} {12}},\
  \bibinfo {pages} {025002} (\bibinfo {year} {2010})}\BibitemShut {NoStop}%
\bibitem [{\citenamefont {{Aharonov}}\ \emph {et~al.}(2014)\citenamefont
  {{Aharonov}}, \citenamefont {{Harrow}}, \citenamefont {{Landau}},
  \citenamefont {{Nagaj}}, \citenamefont {{Szegedy}},\ and\ \citenamefont
  {{Vazirani}}}]{6979009}%
  \BibitemOpen
  \bibfield  {author} {\bibinfo {author} {\bibfnamefont {D.}~\bibnamefont
  {{Aharonov}}}, \bibinfo {author} {\bibfnamefont {A.~W.}\ \bibnamefont
  {{Harrow}}}, \bibinfo {author} {\bibfnamefont {Z.}~\bibnamefont {{Landau}}},
  \bibinfo {author} {\bibfnamefont {D.}~\bibnamefont {{Nagaj}}}, \bibinfo
  {author} {\bibfnamefont {M.}~\bibnamefont {{Szegedy}}}, \ and\ \bibinfo
  {author} {\bibfnamefont {U.}~\bibnamefont {{Vazirani}}},\ }\bibfield  {title}
  {\enquote {\bibinfo {title} {{\it Local Tests of Global Entanglement and a
  Counterexample to the Generalized Area Law}},}\ }in\ \href {\doibase
  10.1109/FOCS.2014.34} {\emph {\bibinfo {booktitle} {2014 IEEE 55th Annual
  Symposium on Foundations of Computer Science}}}\ (\bibinfo {year} {2014})\
  pp.\ \bibinfo {pages} {246--255}\BibitemShut {NoStop}%
\bibitem [{\citenamefont {Bravyi}(2007)}]{PhysRevA.76.052319}%
  \BibitemOpen
  \bibfield  {author} {\bibinfo {author} {\bibfnamefont {Sergey}\ \bibnamefont
  {Bravyi}},\ }\bibfield  {title} {\enquote {\bibinfo {title} {{\it Upper
  bounds on entangling rates of bipartite Hamiltonians}},}\ }\href {\doibase
  10.1103/PhysRevA.76.052319} {\bibfield  {journal} {\bibinfo  {journal} {Phys.
  Rev. A}\ }\textbf {\bibinfo {volume} {76}},\ \bibinfo {pages} {052319}
  (\bibinfo {year} {2007})}\BibitemShut {NoStop}%
\bibitem [{\citenamefont {Eisert}\ and\ \citenamefont
  {Osborne}(2006)}]{PhysRevLett.97.150404}%
  \BibitemOpen
  \bibfield  {author} {\bibinfo {author} {\bibfnamefont {Jens}\ \bibnamefont
  {Eisert}}\ and\ \bibinfo {author} {\bibfnamefont {Tobias~J.}\ \bibnamefont
  {Osborne}},\ }\bibfield  {title} {\enquote {\bibinfo {title} {{\it General
  Entanglement Scaling Laws from Time Evolution}},}\ }\href {\doibase
  10.1103/PhysRevLett.97.150404} {\bibfield  {journal} {\bibinfo  {journal}
  {Phys. Rev. Lett.}\ }\textbf {\bibinfo {volume} {97}},\ \bibinfo {pages}
  {150404} (\bibinfo {year} {2006})}\BibitemShut {NoStop}%
\bibitem [{\citenamefont {Van~Acoleyen}\ \emph {et~al.}(2013)\citenamefont
  {Van~Acoleyen}, \citenamefont {Mari\"en},\ and\ \citenamefont
  {Verstraete}}]{PhysRevLett.111.170501}%
  \BibitemOpen
  \bibfield  {author} {\bibinfo {author} {\bibfnamefont {Karel}\ \bibnamefont
  {Van~Acoleyen}}, \bibinfo {author} {\bibfnamefont {Micha\"el}\ \bibnamefont
  {Mari\"en}}, \ and\ \bibinfo {author} {\bibfnamefont {Frank}\ \bibnamefont
  {Verstraete}},\ }\bibfield  {title} {\enquote {\bibinfo {title} {{\it
  Entanglement Rates and Area Laws}},}\ }\href {\doibase
  10.1103/PhysRevLett.111.170501} {\bibfield  {journal} {\bibinfo  {journal}
  {Phys. Rev. Lett.}\ }\textbf {\bibinfo {volume} {111}},\ \bibinfo {pages}
  {170501} (\bibinfo {year} {2013})}\BibitemShut {NoStop}%
\bibitem [{\citenamefont {Mari{\"e}n}\ \emph {et~al.}(2016)\citenamefont
  {Mari{\"e}n}, \citenamefont {Audenaert}, \citenamefont {Van~Acoleyen},\ and\
  \citenamefont {Verstraete}}]{Marien2016}%
  \BibitemOpen
  \bibfield  {author} {\bibinfo {author} {\bibfnamefont {Micha{\"e}l}\
  \bibnamefont {Mari{\"e}n}}, \bibinfo {author} {\bibfnamefont {Koenraad
  M.~R.}\ \bibnamefont {Audenaert}}, \bibinfo {author} {\bibfnamefont {Karel}\
  \bibnamefont {Van~Acoleyen}}, \ and\ \bibinfo {author} {\bibfnamefont
  {Frank}\ \bibnamefont {Verstraete}},\ }\bibfield  {title} {\enquote {\bibinfo
  {title} {{\it Entanglement Rates and the Stability of the Area Law for the
  Entanglement Entropy}},}\ }\href {\doibase 10.1007/s00220-016-2709-5}
  {\bibfield  {journal} {\bibinfo  {journal} {Communications in Mathematical
  Physics}\ }\textbf {\bibinfo {volume} {346}},\ \bibinfo {pages} {35--73}
  (\bibinfo {year} {2016})}\BibitemShut {NoStop}%
\bibitem [{\citenamefont {Sachdeva}\ and\ \citenamefont
  {Vishnoi}(2014)}]{TCS-065}%
  \BibitemOpen
  \bibfield  {author} {\bibinfo {author} {\bibfnamefont {Sushant}\ \bibnamefont
  {Sachdeva}}\ and\ \bibinfo {author} {\bibfnamefont {Nisheeth~K.}\
  \bibnamefont {Vishnoi}},\ }\bibfield  {title} {\enquote {\bibinfo {title}
  {{\it Faster Algorithms via Approximation Theory}},}\ }\href {\doibase
  10.1561/0400000065} {\bibfield  {journal} {\bibinfo  {journal} {Foundations
  and Trends® in Theoretical Computer Science}\ }\textbf {\bibinfo {volume}
  {9}},\ \bibinfo {pages} {125--210} (\bibinfo {year} {2014})}\BibitemShut
  {NoStop}%
\bibitem [{\citenamefont {Hastings}(2006)}]{PhysRevB.73.085115}%
  \BibitemOpen
  \bibfield  {author} {\bibinfo {author} {\bibfnamefont {M.~B.}\ \bibnamefont
  {Hastings}},\ }\bibfield  {title} {\enquote {\bibinfo {title} {{\it Solving
  gapped Hamiltonians locally}},}\ }\href {\doibase 10.1103/PhysRevB.73.085115}
  {\bibfield  {journal} {\bibinfo  {journal} {Phys. Rev. B}\ }\textbf {\bibinfo
  {volume} {73}},\ \bibinfo {pages} {085115} (\bibinfo {year}
  {2006})}\BibitemShut {NoStop}%
\bibitem [{\citenamefont {Molnar}\ \emph {et~al.}(2015)\citenamefont {Molnar},
  \citenamefont {Schuch}, \citenamefont {Verstraete},\ and\ \citenamefont
  {Cirac}}]{PhysRevB.91.045138}%
  \BibitemOpen
  \bibfield  {author} {\bibinfo {author} {\bibfnamefont {Andras}\ \bibnamefont
  {Molnar}}, \bibinfo {author} {\bibfnamefont {Norbert}\ \bibnamefont
  {Schuch}}, \bibinfo {author} {\bibfnamefont {Frank}\ \bibnamefont
  {Verstraete}}, \ and\ \bibinfo {author} {\bibfnamefont {J.~Ignacio}\
  \bibnamefont {Cirac}},\ }\bibfield  {title} {\enquote {\bibinfo {title} {{\it
  Approximating Gibbs states of local Hamiltonians efficiently with projected
  entangled pair states}},}\ }\href {\doibase 10.1103/PhysRevB.91.045138}
  {\bibfield  {journal} {\bibinfo  {journal} {Phys. Rev. B}\ }\textbf {\bibinfo
  {volume} {91}},\ \bibinfo {pages} {045138} (\bibinfo {year}
  {2015})}\BibitemShut {NoStop}%
\bibitem [{\citenamefont {Osborne}(2012)}]{0034-4885-75-2-022001}%
  \BibitemOpen
  \bibfield  {author} {\bibinfo {author} {\bibfnamefont {Tobias~J}\
  \bibnamefont {Osborne}},\ }\bibfield  {title} {\enquote {\bibinfo {title}
  {{\it Hamiltonian complexity}},}\ }\href
  {http://stacks.iop.org/0034-4885/75/i=2/a=022001} {\bibfield  {journal}
  {\bibinfo  {journal} {Reports on Progress in Physics}\ }\textbf {\bibinfo
  {volume} {75}},\ \bibinfo {pages} {022001} (\bibinfo {year}
  {2012})}\BibitemShut {NoStop}%
\bibitem [{\citenamefont {Gharibian}\ \emph {et~al.}(2015)\citenamefont
  {Gharibian}, \citenamefont {Huang}, \citenamefont {Landau},\ and\
  \citenamefont {Shin}}]{gharibian2015quantum}%
  \BibitemOpen
  \bibfield  {author} {\bibinfo {author} {\bibfnamefont {Sevag}\ \bibnamefont
  {Gharibian}}, \bibinfo {author} {\bibfnamefont {Yichen}\ \bibnamefont
  {Huang}}, \bibinfo {author} {\bibfnamefont {Zeph}\ \bibnamefont {Landau}}, \
  and\ \bibinfo {author} {\bibfnamefont {Seung~Woo}\ \bibnamefont {Shin}},\
  }\bibfield  {title} {\enquote {\bibinfo {title} {Quantum hamiltonian
  complexity},}\ }\href {\doibase 10.1561/0400000066} {\bibfield  {journal}
  {\bibinfo  {journal} {Foundations and Trends in Theoretical Computer
  Science}\ }\textbf {\bibinfo {volume} {10}},\ \bibinfo {pages} {159–282}
  (\bibinfo {year} {2015})}\BibitemShut {NoStop}%
\bibitem [{\citenamefont {Landau}\ \emph {et~al.}(2015)\citenamefont {Landau},
  \citenamefont {Vazirani},\ and\ \citenamefont
  {Vidick}}]{landau2015polynomial}%
  \BibitemOpen
  \bibfield  {author} {\bibinfo {author} {\bibfnamefont {Zeph}\ \bibnamefont
  {Landau}}, \bibinfo {author} {\bibfnamefont {Umesh}\ \bibnamefont
  {Vazirani}}, \ and\ \bibinfo {author} {\bibfnamefont {Thomas}\ \bibnamefont
  {Vidick}},\ }\bibfield  {title} {\enquote {\bibinfo {title} {{\it A
  polynomial time algorithm for the ground state of one-dimensional gapped
  local Hamiltonians}},}\ }\href {https://doi.org/10.1038/nphys3345} {\bibfield
   {journal} {\bibinfo  {journal} {Nature Physics}\ }\textbf {\bibinfo {volume}
  {11}},\ \bibinfo {pages} {566} (\bibinfo {year} {2015})}\BibitemShut
  {NoStop}%
\bibitem [{\citenamefont {Arad}\ \emph {et~al.}(2017)\citenamefont {Arad},
  \citenamefont {Landau}, \citenamefont {Vazirani},\ and\ \citenamefont
  {Vidick}}]{Arad2017}%
  \BibitemOpen
  \bibfield  {author} {\bibinfo {author} {\bibfnamefont {Itai}\ \bibnamefont
  {Arad}}, \bibinfo {author} {\bibfnamefont {Zeph}\ \bibnamefont {Landau}},
  \bibinfo {author} {\bibfnamefont {Umesh}\ \bibnamefont {Vazirani}}, \ and\
  \bibinfo {author} {\bibfnamefont {Thomas}\ \bibnamefont {Vidick}},\
  }\bibfield  {title} {\enquote {\bibinfo {title} {{\it Rigorous RG Algorithms
  and Area Laws for Low Energy Eigenstates in 1D}},}\ }\href {\doibase
  10.1007/s00220-017-2973-z} {\bibfield  {journal} {\bibinfo  {journal}
  {Communications in Mathematical Physics}\ }\textbf {\bibinfo {volume}
  {356}},\ \bibinfo {pages} {65--105} (\bibinfo {year} {2017})}\BibitemShut
  {NoStop}%
\bibitem [{\citenamefont {las Cuevas}\ \emph {et~al.}(2013)\citenamefont {las
  Cuevas}, \citenamefont {Schuch}, \citenamefont {P{\'{e}}rez-Garc{\'{\i}}a},\
  and\ \citenamefont {Cirac}}]{de2013purifications}%
  \BibitemOpen
  \bibfield  {author} {\bibinfo {author} {\bibfnamefont {Gemma~De}\
  \bibnamefont {las Cuevas}}, \bibinfo {author} {\bibfnamefont {Norbert}\
  \bibnamefont {Schuch}}, \bibinfo {author} {\bibfnamefont {David}\
  \bibnamefont {P{\'{e}}rez-Garc{\'{\i}}a}}, \ and\ \bibinfo {author}
  {\bibfnamefont {J~Ignacio}\ \bibnamefont {Cirac}},\ }\bibfield  {title}
  {\enquote {\bibinfo {title} {{\it Purifications of multipartite states:
  limitations and constructive methods}},}\ }\href {\doibase
  10.1088/1367-2630/15/12/123021} {\bibfield  {journal} {\bibinfo  {journal}
  {New Journal of Physics}\ }\textbf {\bibinfo {volume} {15}},\ \bibinfo
  {pages} {123021} (\bibinfo {year} {2013})}\BibitemShut {NoStop}%
\bibitem [{\citenamefont {De~las Cuevas}\ \emph {et~al.}(2016)\citenamefont
  {De~las Cuevas}, \citenamefont {Cubitt}, \citenamefont {Cirac}, \citenamefont
  {Wolf},\ and\ \citenamefont {Pérez-García}}]{de2016fundamental}%
  \BibitemOpen
  \bibfield  {author} {\bibinfo {author} {\bibfnamefont {G.}~\bibnamefont
  {De~las Cuevas}}, \bibinfo {author} {\bibfnamefont {T.~S.}\ \bibnamefont
  {Cubitt}}, \bibinfo {author} {\bibfnamefont {J.~I.}\ \bibnamefont {Cirac}},
  \bibinfo {author} {\bibfnamefont {M.~M.}\ \bibnamefont {Wolf}}, \ and\
  \bibinfo {author} {\bibfnamefont {D.}~\bibnamefont {Pérez-García}},\
  }\bibfield  {title} {\enquote {\bibinfo {title} {{\it Fundamental limitations
  in the purifications of tensor networks}},}\ }\href {\doibase
  10.1063/1.4954983} {\bibfield  {journal} {\bibinfo  {journal} {Journal of
  Mathematical Physics}\ }\textbf {\bibinfo {volume} {57}},\ \bibinfo {pages}
  {071902} (\bibinfo {year} {2016})}\BibitemShut {NoStop}%
\bibitem [{\citenamefont {Sch\"on}\ \emph {et~al.}(2005)\citenamefont
  {Sch\"on}, \citenamefont {Solano}, \citenamefont {Verstraete}, \citenamefont
  {Cirac},\ and\ \citenamefont {Wolf}}]{PhysRevLett.95.110503}%
  \BibitemOpen
  \bibfield  {author} {\bibinfo {author} {\bibfnamefont {C.}~\bibnamefont
  {Sch\"on}}, \bibinfo {author} {\bibfnamefont {E.}~\bibnamefont {Solano}},
  \bibinfo {author} {\bibfnamefont {F.}~\bibnamefont {Verstraete}}, \bibinfo
  {author} {\bibfnamefont {J.~I.}\ \bibnamefont {Cirac}}, \ and\ \bibinfo
  {author} {\bibfnamefont {M.~M.}\ \bibnamefont {Wolf}},\ }\bibfield  {title}
  {\enquote {\bibinfo {title} {{\it Sequential Generation of Entangled
  Multiqubit States}},}\ }\href {\doibase 10.1103/PhysRevLett.95.110503}
  {\bibfield  {journal} {\bibinfo  {journal} {Phys. Rev. Lett.}\ }\textbf
  {\bibinfo {volume} {95}},\ \bibinfo {pages} {110503} (\bibinfo {year}
  {2005})}\BibitemShut {NoStop}%
\bibitem [{\citenamefont {Liu}\ \emph {et~al.}(2019)\citenamefont {Liu},
  \citenamefont {Zhang}, \citenamefont {Wan},\ and\ \citenamefont
  {Wang}}]{PhysRevResearch.1.023025}%
  \BibitemOpen
  \bibfield  {author} {\bibinfo {author} {\bibfnamefont {Jin-Guo}\ \bibnamefont
  {Liu}}, \bibinfo {author} {\bibfnamefont {Yi-Hong}\ \bibnamefont {Zhang}},
  \bibinfo {author} {\bibfnamefont {Yuan}\ \bibnamefont {Wan}}, \ and\ \bibinfo
  {author} {\bibfnamefont {Lei}\ \bibnamefont {Wang}},\ }\bibfield  {title}
  {\enquote {\bibinfo {title} {{\it Variational quantum eigensolver with fewer
  qubits}},}\ }\href {\doibase 10.1103/PhysRevResearch.1.023025} {\bibfield
  {journal} {\bibinfo  {journal} {Phys. Rev. Research}\ }\textbf {\bibinfo
  {volume} {1}},\ \bibinfo {pages} {023025} (\bibinfo {year}
  {2019})}\BibitemShut {NoStop}%
\bibitem [{\citenamefont {Ran}(2020)}]{PhysRevA.101.032310}%
  \BibitemOpen
  \bibfield  {author} {\bibinfo {author} {\bibfnamefont {Shi-Ju}\ \bibnamefont
  {Ran}},\ }\bibfield  {title} {\enquote {\bibinfo {title} {{\it Encoding of
  matrix product states into quantum circuits of one- and two-qubit gates}},}\
  }\href {\doibase 10.1103/PhysRevA.101.032310} {\bibfield  {journal} {\bibinfo
   {journal} {Phys. Rev. A}\ }\textbf {\bibinfo {volume} {101}},\ \bibinfo
  {pages} {032310} (\bibinfo {year} {2020})}\BibitemShut {NoStop}%
\bibitem [{Note3()}]{Note3}%
  \BibitemOpen
  \bibinfo {note} {We start from the basic formula for the Chebyshev
  polynomials: $xT_r(x)= [T_{r+1}(x)+T_{r-1}(x)]/2$. Let $Y$ be a random
  variable taking values $1$ or $-1$ with the probability $1/2$. We then obtain
  $xT_r(x) =\protect \mathbb {E}_{Y_1} [T_{r+Y_1}(x)]=
  [T_{r+1}(x)+T_{r-1}(x)]/2$. In the same way, we can obtain $x^2T_r(x)
  =\protect \mathbb {E}_{Y_1} [xT_{r+Y_1}(x)]=\protect \mathbb {E}_{Y_1,Y_2}
  [T_{r+Y_1+Y_2}(x)]$. By repeating the process, we obtain $x^jT_r(x) =\protect
  \mathbb {E}_{Y_1,\protect \ldots Y_j} [T_{r+D_j}(x)]$ with
  $D_j=Y_1+Y_2+\protect \cdots +Y_j$. The probability distribution of $D_j$
  obeys the binomial distribution as $2^{-j} \protect \binom {j}{(j+D_j)/2}$,
  which gives Eq.~\protect \textup {\hbox {\mathsurround \z@ \protect
  \normalfont (\ignorespaces \ref {second_expansion_Cheby}\unskip \@@italiccorr
  )}}.}\BibitemShut {Stop}%
\bibitem [{\citenamefont {Fannes}(1973)}]{fannes1973}%
  \BibitemOpen
  \bibfield  {author} {\bibinfo {author} {\bibfnamefont {M.}~\bibnamefont
  {Fannes}},\ }\bibfield  {title} {\enquote {\bibinfo {title} {{\it A
  continuity property of the entropy density for spin lattice systems}},}\
  }\href {https://projecteuclid.org:443/euclid.cmp/1103859037} {\bibfield
  {journal} {\bibinfo  {journal} {Comm. Math. Phys.}\ }\textbf {\bibinfo
  {volume} {31}},\ \bibinfo {pages} {291--294} (\bibinfo {year}
  {1973})}\BibitemShut {NoStop}%
\bibitem [{\citenamefont {Hastings}(2007{\natexlab{c}})}]{PhysRevB.76.035114}%
  \BibitemOpen
  \bibfield  {author} {\bibinfo {author} {\bibfnamefont {M.~B.}\ \bibnamefont
  {Hastings}},\ }\bibfield  {title} {\enquote {\bibinfo {title} {{\it Entropy
  and entanglement in quantum ground states}},}\ }\href {\doibase
  10.1103/PhysRevB.76.035114} {\bibfield  {journal} {\bibinfo  {journal} {Phys.
  Rev. B}\ }\textbf {\bibinfo {volume} {76}},\ \bibinfo {pages} {035114}
  (\bibinfo {year} {2007}{\natexlab{c}})}\BibitemShut {NoStop}%
\bibitem [{\citenamefont {Masanes}(2009)}]{PhysRevA.80.052104}%
  \BibitemOpen
  \bibfield  {author} {\bibinfo {author} {\bibfnamefont {Llu\'{\i}s}\
  \bibnamefont {Masanes}},\ }\bibfield  {title} {\enquote {\bibinfo {title}
  {{\it Area law for the entropy of low-energy states}},}\ }\href {\doibase
  10.1103/PhysRevA.80.052104} {\bibfield  {journal} {\bibinfo  {journal} {Phys.
  Rev. A}\ }\textbf {\bibinfo {volume} {80}},\ \bibinfo {pages} {052104}
  (\bibinfo {year} {2009})}\BibitemShut {NoStop}%
\bibitem [{\citenamefont {Kuwahara}\ and\ \citenamefont
  {Saito}(2020{\natexlab{c}})}]{PhysRevLett.124.200604}%
  \BibitemOpen
  \bibfield  {author} {\bibinfo {author} {\bibfnamefont {Tomotaka}\
  \bibnamefont {Kuwahara}}\ and\ \bibinfo {author} {\bibfnamefont {Keiji}\
  \bibnamefont {Saito}},\ }\bibfield  {title} {\enquote {\bibinfo {title} {{\it
  Eigenstate Thermalization from the Clustering Property of Correlation}},}\
  }\href {\doibase 10.1103/PhysRevLett.124.200604} {\bibfield  {journal}
  {\bibinfo  {journal} {Phys. Rev. Lett.}\ }\textbf {\bibinfo {volume} {124}},\
  \bibinfo {pages} {200604} (\bibinfo {year} {2020}{\natexlab{c}})}\BibitemShut
  {NoStop}%
\bibitem [{\citenamefont {{Haah}}\ \emph {et~al.}(2018)\citenamefont {{Haah}},
  \citenamefont {{Hastings}}, \citenamefont {{Kothari}},\ and\ \citenamefont
  {{Low}}}]{HaahHKL18}%
  \BibitemOpen
  \bibfield  {author} {\bibinfo {author} {\bibfnamefont {J.}~\bibnamefont
  {{Haah}}}, \bibinfo {author} {\bibfnamefont {M.}~\bibnamefont {{Hastings}}},
  \bibinfo {author} {\bibfnamefont {R.}~\bibnamefont {{Kothari}}}, \ and\
  \bibinfo {author} {\bibfnamefont {G.~H.}\ \bibnamefont {{Low}}},\ }\bibfield
  {title} {\enquote {\bibinfo {title} {{\it Quantum Algorithm for Simulating
  Real Time Evolution of Lattice Hamiltonians}},}\ }in\ \href {\doibase
  10.1109/FOCS.2018.00041} {\emph {\bibinfo {booktitle} {2018 IEEE 59th Annual
  Symposium on Foundations of Computer Science (FOCS)}}}\ (\bibinfo {year}
  {2018})\ pp.\ \bibinfo {pages} {350--360}\BibitemShut {NoStop}%
\bibitem [{\citenamefont {Vidal}(2004)}]{PhysRevLett.93.040502}%
  \BibitemOpen
  \bibfield  {author} {\bibinfo {author} {\bibfnamefont {Guifr\'e}\
  \bibnamefont {Vidal}},\ }\bibfield  {title} {\enquote {\bibinfo {title} {{\it
  Efficient Simulation of One-Dimensional Quantum Many-Body Systems}},}\ }\href
  {\doibase 10.1103/PhysRevLett.93.040502} {\bibfield  {journal} {\bibinfo
  {journal} {Phys. Rev. Lett.}\ }\textbf {\bibinfo {volume} {93}},\ \bibinfo
  {pages} {040502} (\bibinfo {year} {2004})}\BibitemShut {NoStop}%
\bibitem [{\citenamefont {Schollw{\"o}ck}(2011)}]{SCHOLLWOCK201196}%
  \BibitemOpen
  \bibfield  {author} {\bibinfo {author} {\bibfnamefont {Ulrich}\ \bibnamefont
  {Schollw{\"o}ck}},\ }\bibfield  {title} {\enquote {\bibinfo {title} {{\it The
  density-matrix renormalization group in the age of matrix product states}},}\
  }\href {\doibase 10.1016/j.aop.2010.09.012} {\bibfield  {journal} {\bibinfo
  {journal} {Annals of Physics}\ }\textbf {\bibinfo {volume} {326}},\ \bibinfo
  {pages} {96 -- 192} (\bibinfo {year} {2011})},\ \bibinfo {note} {january 2011
  Special Issue}\BibitemShut {NoStop}%
\bibitem [{\citenamefont {Verstraete}\ and\ \citenamefont
  {Cirac}(2006)}]{VC06mps}%
  \BibitemOpen
  \bibfield  {author} {\bibinfo {author} {\bibfnamefont {F.}~\bibnamefont
  {Verstraete}}\ and\ \bibinfo {author} {\bibfnamefont {J.~I.}\ \bibnamefont
  {Cirac}},\ }\bibfield  {title} {\enquote {\bibinfo {title} {{\it Matrix
  product states represent ground states faithfully}},}\ }\href {\doibase
  10.1103/PhysRevB.73.094423} {\bibfield  {journal} {\bibinfo  {journal} {Phys.
  Rev. B}\ }\textbf {\bibinfo {volume} {73}},\ \bibinfo {pages} {094423}
  (\bibinfo {year} {2006})}\BibitemShut {NoStop}%
\bibitem [{\citenamefont {Terhal}\ \emph {et~al.}(2002)\citenamefont {Terhal},
  \citenamefont {Horodecki}, \citenamefont {Leung},\ and\ \citenamefont
  {DiVincenzo}}]{TerhalHLD02}%
  \BibitemOpen
  \bibfield  {author} {\bibinfo {author} {\bibfnamefont {Barbara~M.}\
  \bibnamefont {Terhal}}, \bibinfo {author} {\bibfnamefont {Michal}\
  \bibnamefont {Horodecki}}, \bibinfo {author} {\bibfnamefont {Debbie~W.}\
  \bibnamefont {Leung}}, \ and\ \bibinfo {author} {\bibfnamefont {David~P.}\
  \bibnamefont {DiVincenzo}},\ }\bibfield  {title} {\enquote {\bibinfo {title}
  {{\it The entanglement of purification}},}\ }\href {\doibase
  10.1063/1.1498001} {\bibfield  {journal} {\bibinfo  {journal} {Journal of
  Mathematical Physics}\ }\textbf {\bibinfo {volume} {43}},\ \bibinfo {pages}
  {4286--4298} (\bibinfo {year} {2002})},\ \Eprint
  {http://arxiv.org/abs/https://doi.org/10.1063/1.1498001}
  {https://doi.org/10.1063/1.1498001} \BibitemShut {NoStop}%
\bibitem [{\citenamefont {Guth~Jarkovsk\'y}\ \emph {et~al.}(2020)\citenamefont
  {Guth~Jarkovsk\'y}, \citenamefont {Moln\'ar}, \citenamefont {Schuch},\ and\
  \citenamefont {Cirac}}]{jarkovsky2020efficient}%
  \BibitemOpen
  \bibfield  {author} {\bibinfo {author} {\bibfnamefont {Ji\ifmmode
  \check{r}\else~\v{r}\fi{}\'{\i}}\ \bibnamefont {Guth~Jarkovsk\'y}}, \bibinfo
  {author} {\bibfnamefont {Andr\'as}\ \bibnamefont {Moln\'ar}}, \bibinfo
  {author} {\bibfnamefont {Norbert}\ \bibnamefont {Schuch}}, \ and\ \bibinfo
  {author} {\bibfnamefont {J.~Ignacio}\ \bibnamefont {Cirac}},\ }\bibfield
  {title} {\enquote {\bibinfo {title} {{\it Efficient Description of Many-Body
  Systems with Matrix Product Density Operators}},}\ }\href {\doibase
  10.1103/PRXQuantum.1.010304} {\bibfield  {journal} {\bibinfo  {journal} {PRX
  Quantum}\ }\textbf {\bibinfo {volume} {1}},\ \bibinfo {pages} {010304}
  (\bibinfo {year} {2020})}\BibitemShut {NoStop}%
\bibitem [{\citenamefont {Bagchi}\ and\ \citenamefont
  {Pati}(2015)}]{PhysRevA.91.042323}%
  \BibitemOpen
  \bibfield  {author} {\bibinfo {author} {\bibfnamefont {Shrobona}\
  \bibnamefont {Bagchi}}\ and\ \bibinfo {author} {\bibfnamefont {Arun~Kumar}\
  \bibnamefont {Pati}},\ }\bibfield  {title} {\enquote {\bibinfo {title} {{\it
  Monogamy, polygamy, and other properties of entanglement of purification}},}\
  }\href {\doibase 10.1103/PhysRevA.91.042323} {\bibfield  {journal} {\bibinfo
  {journal} {Phys. Rev. A}\ }\textbf {\bibinfo {volume} {91}},\ \bibinfo
  {pages} {042323} (\bibinfo {year} {2015})}\BibitemShut {NoStop}%
\bibitem [{\citenamefont {Berta}\ \emph {et~al.}(2018)\citenamefont {Berta},
  \citenamefont {Brand\~ao}, \citenamefont {Haegeman}, \citenamefont {Scholz},\
  and\ \citenamefont {Verstraete}}]{PhysRevB.98.235154}%
  \BibitemOpen
  \bibfield  {author} {\bibinfo {author} {\bibfnamefont {Mario}\ \bibnamefont
  {Berta}}, \bibinfo {author} {\bibfnamefont {Fernando G. S.~L.}\ \bibnamefont
  {Brand\~ao}}, \bibinfo {author} {\bibfnamefont {Jutho}\ \bibnamefont
  {Haegeman}}, \bibinfo {author} {\bibfnamefont {Volkher~B.}\ \bibnamefont
  {Scholz}}, \ and\ \bibinfo {author} {\bibfnamefont {Frank}\ \bibnamefont
  {Verstraete}},\ }\bibfield  {title} {\enquote {\bibinfo {title} {{\it Thermal
  states as convex combinations of matrix product states}},}\ }\href {\doibase
  10.1103/PhysRevB.98.235154} {\bibfield  {journal} {\bibinfo  {journal} {Phys.
  Rev. B}\ }\textbf {\bibinfo {volume} {98}},\ \bibinfo {pages} {235154}
  (\bibinfo {year} {2018})}\BibitemShut {NoStop}%
\bibitem [{\citenamefont {Leviatan}\ \emph {et~al.}(2017)\citenamefont
  {Leviatan}, \citenamefont {Pollmann}, \citenamefont {Bardarson},
  \citenamefont {Huse},\ and\ \citenamefont {Altman}}]{leviatan2017quantum}%
  \BibitemOpen
  \bibfield  {author} {\bibinfo {author} {\bibfnamefont {Eyal}\ \bibnamefont
  {Leviatan}}, \bibinfo {author} {\bibfnamefont {Frank}\ \bibnamefont
  {Pollmann}}, \bibinfo {author} {\bibfnamefont {Jens~H.}\ \bibnamefont
  {Bardarson}}, \bibinfo {author} {\bibfnamefont {David~A.}\ \bibnamefont
  {Huse}}, \ and\ \bibinfo {author} {\bibfnamefont {Ehud}\ \bibnamefont
  {Altman}},\ }\href@noop {} {\enquote {\bibinfo {title} {{\it Quantum
  thermalization dynamics with Matrix-Product States}},}\ } (\bibinfo {year}
  {2017}),\ \Eprint {http://arxiv.org/abs/1702.08894} {arXiv:1702.08894
  [cond-mat.stat-mech]} \BibitemShut {NoStop}%
\bibitem [{Note4()}]{Note4}%
  \BibitemOpen
  \bibinfo {note} {At the inequality (9) in Ref.~\cite {TerhalHLD02}, the upper
  bound of $E_{f,\alpha }(\sigma ) \le E_{p,\alpha }(\sigma )$ is obtained for
  the case of $\alpha =1$. However, the proof therein can be easily extended to
  generic $\alpha $.}\BibitemShut {Stop}%
\bibitem [{\citenamefont {Osborne}(2006)}]{PhysRevLett.97.157202}%
  \BibitemOpen
  \bibfield  {author} {\bibinfo {author} {\bibfnamefont {Tobias~J.}\
  \bibnamefont {Osborne}},\ }\bibfield  {title} {\enquote {\bibinfo {title}
  {{\it Efficient Approximation of the Dynamics of One-Dimensional Quantum Spin
  Systems}},}\ }\href {\doibase 10.1103/PhysRevLett.97.157202} {\bibfield
  {journal} {\bibinfo  {journal} {Phys. Rev. Lett.}\ }\textbf {\bibinfo
  {volume} {97}},\ \bibinfo {pages} {157202} (\bibinfo {year}
  {2006})}\BibitemShut {NoStop}%
\bibitem [{\citenamefont {Hastings}(2008)}]{PhysRevB.77.144302}%
  \BibitemOpen
  \bibfield  {author} {\bibinfo {author} {\bibfnamefont {M.~B.}\ \bibnamefont
  {Hastings}},\ }\bibfield  {title} {\enquote {\bibinfo {title} {{\it
  Observations outside the light cone: Algorithms for nonequilibrium and
  thermal states}},}\ }\href {\doibase 10.1103/PhysRevB.77.144302} {\bibfield
  {journal} {\bibinfo  {journal} {Phys. Rev. B}\ }\textbf {\bibinfo {volume}
  {77}},\ \bibinfo {pages} {144302} (\bibinfo {year} {2008})}\BibitemShut
  {NoStop}%
\bibitem [{\citenamefont {Childs}\ \emph {et~al.}(2021)\citenamefont {Childs},
  \citenamefont {Su}, \citenamefont {Tran}, \citenamefont {Wiebe},\ and\
  \citenamefont {Zhu}}]{childs2019theory}%
  \BibitemOpen
  \bibfield  {author} {\bibinfo {author} {\bibfnamefont {Andrew~M.}\
  \bibnamefont {Childs}}, \bibinfo {author} {\bibfnamefont {Yuan}\ \bibnamefont
  {Su}}, \bibinfo {author} {\bibfnamefont {Minh~C.}\ \bibnamefont {Tran}},
  \bibinfo {author} {\bibfnamefont {Nathan}\ \bibnamefont {Wiebe}}, \ and\
  \bibinfo {author} {\bibfnamefont {Shuchen}\ \bibnamefont {Zhu}},\ }\bibfield
  {title} {\enquote {\bibinfo {title} {{\it Theory of Trotter Error with
  Commutator Scaling}},}\ }\href {\doibase 10.1103/PhysRevX.11.011020}
  {\bibfield  {journal} {\bibinfo  {journal} {Phys. Rev. X}\ }\textbf {\bibinfo
  {volume} {11}},\ \bibinfo {pages} {011020} (\bibinfo {year}
  {2021})}\BibitemShut {NoStop}%
\bibitem [{\citenamefont {Sutter}(2018)}]{sutter2018approximate}%
  \BibitemOpen
  \bibfield  {author} {\bibinfo {author} {\bibfnamefont {David}\ \bibnamefont
  {Sutter}},\ }\bibfield  {title} {\enquote {\bibinfo {title} {Approximate
  quantum markov chains},}\ }\href {\doibase 10.1007/978-3-319-78732-9_5}
  {\bibfield  {journal} {\bibinfo  {journal} {SpringerBriefs in Mathematical
  Physics}\ ,\ \bibinfo {pages} {75–100}} (\bibinfo {year}
  {2018})}\BibitemShut {NoStop}%
\bibitem [{\citenamefont {Eckart}\ and\ \citenamefont
  {Young}(1936)}]{Eckart1936}%
  \BibitemOpen
  \bibfield  {author} {\bibinfo {author} {\bibfnamefont {Carl}\ \bibnamefont
  {Eckart}}\ and\ \bibinfo {author} {\bibfnamefont {Gale}\ \bibnamefont
  {Young}},\ }\bibfield  {title} {\enquote {\bibinfo {title} {{\it The
  approximation of one matrix by another of lower rank}},}\ }\href {\doibase
  10.1007/BF02288367} {\bibfield  {journal} {\bibinfo  {journal}
  {Psychometrika}\ }\textbf {\bibinfo {volume} {1}},\ \bibinfo {pages}
  {211--218} (\bibinfo {year} {1936})}\BibitemShut {NoStop}%
\bibitem [{\citenamefont {Kuwahara}\ \emph {et~al.}(2016)\citenamefont
  {Kuwahara}, \citenamefont {Mori},\ and\ \citenamefont
  {Saito}}]{kuwahara2015floquet}%
  \BibitemOpen
  \bibfield  {author} {\bibinfo {author} {\bibfnamefont {Tomotaka}\
  \bibnamefont {Kuwahara}}, \bibinfo {author} {\bibfnamefont {Takashi}\
  \bibnamefont {Mori}}, \ and\ \bibinfo {author} {\bibfnamefont {Keiji}\
  \bibnamefont {Saito}},\ }\bibfield  {title} {\enquote {\bibinfo {title} {{\it
  Floquet-Magnus theory and generic transient dynamics in periodically driven
  many-body quantum systems}},}\ }\href {\doibase 10.1016/j.aop.2016.01.012}
  {\bibfield  {journal} {\bibinfo  {journal} {Annals of Physics}\ }\textbf
  {\bibinfo {volume} {367}},\ \bibinfo {pages} {96 -- 124} (\bibinfo {year}
  {2016})}\BibitemShut {NoStop}%
\bibitem [{\citenamefont {Lieb}\ and\ \citenamefont
  {Robinson}(1972)}]{ref:LR-bound72}%
  \BibitemOpen
  \bibfield  {author} {\bibinfo {author} {\bibfnamefont {ElliottH.}\
  \bibnamefont {Lieb}}\ and\ \bibinfo {author} {\bibfnamefont {DerekW.}\
  \bibnamefont {Robinson}},\ }\bibfield  {title} {\enquote {\bibinfo {title}
  {{\it The finite group velocity of quantum spin systems}},}\ }\href {\doibase
  10.1007/BF01645779} {\bibfield  {journal} {\bibinfo  {journal}
  {Communications in Mathematical Physics}\ }\textbf {\bibinfo {volume} {28}},\
  \bibinfo {pages} {251--257} (\bibinfo {year} {1972})}\BibitemShut {NoStop}%
\bibitem [{\citenamefont {Bravyi}\ \emph {et~al.}(2006)\citenamefont {Bravyi},
  \citenamefont {Hastings},\ and\ \citenamefont
  {Verstraete}}]{PhysRevLett.97.050401}%
  \BibitemOpen
  \bibfield  {author} {\bibinfo {author} {\bibfnamefont {S.}~\bibnamefont
  {Bravyi}}, \bibinfo {author} {\bibfnamefont {M.~B.}\ \bibnamefont
  {Hastings}}, \ and\ \bibinfo {author} {\bibfnamefont {F.}~\bibnamefont
  {Verstraete}},\ }\bibfield  {title} {\enquote {\bibinfo {title} {{\it
  Lieb-Robinson Bounds and the Generation of Correlations and Topological
  Quantum Order}},}\ }\href {\doibase 10.1103/PhysRevLett.97.050401} {\bibfield
   {journal} {\bibinfo  {journal} {Phys. Rev. Lett.}\ }\textbf {\bibinfo
  {volume} {97}},\ \bibinfo {pages} {050401} (\bibinfo {year}
  {2006})}\BibitemShut {NoStop}%
\bibitem [{\citenamefont {Alicki}\ and\ \citenamefont
  {Fannes}(2004)}]{Alicki_2004}%
  \BibitemOpen
  \bibfield  {author} {\bibinfo {author} {\bibfnamefont {R}~\bibnamefont
  {Alicki}}\ and\ \bibinfo {author} {\bibfnamefont {M}~\bibnamefont {Fannes}},\
  }\bibfield  {title} {\enquote {\bibinfo {title} {{\it Continuity of quantum
  conditional information}},}\ }\href {\doibase 10.1088/0305-4470/37/5/l01}
  {\bibfield  {journal} {\bibinfo  {journal} {Journal of Physics A:
  Mathematical and General}\ }\textbf {\bibinfo {volume} {37}},\ \bibinfo
  {pages} {L55--L57} (\bibinfo {year} {2004})}\BibitemShut {NoStop}%
\end{thebibliography}%

\appendix

\newpage

\section{More detailed setup}  \label{sec:More detailed setup}

We here recall the setup.
We consider a quantum spin system with $n$ spins, where each of the spin sits on a vertex of the $\sd$-dimensional graph (or $\sd$-dimensional lattice) with $\Lambda$ the total spin set, namely $|\Lambda|=n$.
We assume that a finite dimensional Hilbert space ($\spin$-dimension) is assigned to each of the spins. 
For a partial set $X\subseteq \Lambda$, we denote the cardinality, that is, the number of vertices contained in $X$, by $|X|$ (e.g. $X=\{i_1,i_2,\ldots, i_{|X|}\}$).
We also denote the complementary subset of $X$ by $X^\co := \Lambda\setminus X$.
We denote the Hilbert space of a subset $X \subseteq \Lambda$ and its dimension by $\mathcal{H}_X$ and $\mathcal{D}_X$, respectively.

For arbitrary subsets $X, Y \subseteq \Lambda$, we define $\dist_{X,Y}$ as the shortest path length on the graph that connects $X$ and $Y$; that is, if $X\cap Y \neq \emptyset$, $\dist_{X,Y}=0$. 
When $X$ is composed of only one element (i.e., $X=\{i\}$), we denote $\dist_{\{i\},Y}$ by $\dist_{i,Y}$ for the simplicity.
We also define $\diam(X)$ as follows: 
\begin{align}
\diam(X):  =1+ \max_{i,j\in X} (\dist_{i,j}).
\end{align}

\subsection{One-dimensional $k$-local Hamiltonian} 
Let us now define one-dimensional systems, where the Hamiltonian $H$ is given by the general $k$-local operator:
\begin{align}
H = \sum_{X\subset \Lambda, \diam(X) \le k} h_X,\quad 
\max_{i\in \Lambda} \sum_{X:X\ni i}\|h_X\| \le g,  \label{sup_def:Hamiltonian}
\end{align}
where $h_X$ are the interaction terms acting on the subset $X$. 
Here, $\sum_{X:X\ni i}$ means the summation which picks up all the subsets $X\subset \Lambda$ such that  $X\ni i$.
In the main text, we have considered the Hamiltonian in the form of Eq.~\eqref{def:Hamiltonian}.
By choosing $k=2$ and $g=1$, the Hamiltonian~\eqref{sup_def:Hamiltonian} reduces to the form of~\eqref{def:Hamiltonian}.


We here define $\Lambda_{\le i}$ ($\Lambda_{>i}$) for an arbitrary $i\in \Lambda$ as the subset $\{j\}_{j\le i}$ ($\{j\}_{j> i}$). 
We denote $v_i$ by the interaction between $\Lambda_{\le i}$ and $\Lambda_{>i}$: 
\begin{align}
\label{sup_def_v_i_1D}
v_i= \sum_{X:X\cap \Lambda_{\le i} \neq \emptyset,X\cap\Lambda_{>i} \neq \emptyset } h_X .
\end{align}
For the Hamiltonian~\eqref{def:Hamiltonian} in the main text, $v_i$ is simply given by $h_{i,i+1}$.
We  then define the Schmidt rank ${\rm SR} (v_i,i)$ as $\locd$:
  \begin{align}
{\rm SR} (v_i,i) \le \locd , \label{sup_locd_1D}
\end{align}
where $\locd$ is at most of $\spin^{\orderof{k}}$.

\subsection{High-dimensional $k$-local Hamiltonian}

In considering $\sd$-dimensional systems, we also consider the $k$-local operator: 
\begin{align}
H = \sum_{\substack{X\subset \Lambda, |X| \le k \\  \diam(X) \le k}} h_X,\quad 
\max_{i\in \Lambda} \sum_{X:X\ni i}\|h_X\| \le g \label{sup_def:Hamiltonian_2D}.
\end{align} 
We slice the total system $\Lambda$ into $l_\Lambda$ pieces:
\begin{align}
&\Lambda = \Lambda_1 \sqcup \Lambda_2\sqcup \cdots \sqcup \Lambda_{l_\Lambda}, \notag  \\
&|\Lambda_j | \le | \partial \Lambda | = \orderof{n^{\sd -1 /\sd}} ,
\label{sup_decomp_2D}
\end{align}
where $l_\Lambda$ is the system length, namely $l_\Lambda = \orderof{n^{1/\sd}}$, and we define $| \partial \Lambda |$ as an integer which gives the upper bounds for $|\Lambda_j|$.

Similar to the one-dimensional case, we define $\Lambda_{\le i}$ ($\Lambda_{>i}$) for an arbitrary $i\in \Lambda$ as the subset $\bigsqcup_{j\le i} \Lambda_j$ ($\bigsqcup_{j>i} \Lambda_j$). 
We then define the Schmidt rank ${\rm SR} (O,i)$ in the same way as Eq.~\eqref{Scgmidt_rank_def}.
We also define $v_i$ as the interaction between $\Lambda_{\le i}$ and $\Lambda_{>i}$:
\begin{align}
v_i= \sum_{X:X\cap \Lambda_{\le i} \neq \emptyset,X\cap\Lambda_{>i} \neq \emptyset } h_X . \label{sup_def_v_i_high_D}
\end{align} 
Here, each of the $\{v_i\}_{i=1}^{l_\Lambda}$ consists of at most of $\orderof{|\partial \Lambda|}$ local interaction terms $h_X$. 
We define $\locd$ as the upper bound for the  Schmidt ranks of $\{v_i\}$:
  \begin{align}
{\rm SR} (v_i,i) \le \locd = \spin^{\orderof{k}} |\partial \Lambda|.
\label{sup_def:D_loc}
\end{align}

\section{Basic analytical tools} \label{appdx_Basic analytical tools}

\subsection{Generalized H\"older inequality for Schatten norm} 
For a general Schatten $p$ norm, we can prove the following generalized H\"older inequality (see Prop. 2.5 in Ref.~\cite{sutter2018approximate}):
\begin{align}
\left \| \prod_{j=1}^s O_j \right \|_p \le \prod_{j=1}^s \|O_j \|_{p_j} , \label{sup_generalized_Holder_ineq}
\end{align}
where $\sum_{j=1}^s 1/p_j =1/p$.
From the inequality, we can immediately obtain 
\begin{align}
\left \| O_1 O_2 \right \|_p \le \|O_1\|_p \|O_2\|,
 \label{sup_generalized_Holder_ineq_operator_norm}
\end{align}
where we set $p_1=p$ and $p_2=\infty$ in \eqref{sup_generalized_Holder_ineq}.

\subsection{The Eckart-Young theorem}

We here show the Eckart-Young theorem~\cite{Eckart1936} without the proof:
\begin{lemma}[\bf The Eckart-Young theorem] \label{lemma:The Eckart-Young theorem_0}
Let us consider a normalized state $\ket{\psi}$ and give its Schmidt decomposition as
 \begin{align}
\ket{\psi} = \sum_{m=1}^{D_\psi} \mu_m  \ket{\psi_{1,m}} \otimes \ket{\psi_{2,m}},
\end{align}
where $\mu_1\ge\mu_2 \ge \mu_3 \cdots \ge  \mu_{D_\psi}$, and $\{\ket{\psi_{1,m}}\}_{m=1}^{D_\psi}$ and $\{\ket{\psi_{2,m}}\}_{m=1}^{D_\psi}$ are orthonormal states, respectively.
We then consider another quantum state $\ket{\hat{\psi}}$ with its Schmidt rank $D$ and define the overlap with the state $\ket{\psi}$ as 
$
\| \ket{\psi} -\ket{\hat{\psi}}\|.
$
Then, for the Schmidt rank truncation as 
\begin{align}
\ket{\psi_D} = \sum_{m\le D} \mu_m  \ket{\psi_{1,m}} \otimes \ket{\psi_{2,m}},
\end{align}
the Eckart-Young theorem gives the following inequality:
 \begin{align}
\|\ket{\psi}-\ket{\psi_D}  \|^2 = \sum_{m>D} \mu_m^2 \le \| \ket{\psi} -\ket{\hat{\psi}}\|^2 , \label{sup_thm:The Eckart-Young theorem}
\end{align}
where $\ket{\hat{\psi}}$ can be unnormalized. 
\end{lemma}
\noindent
We note that the Eckart-Young theorem can be also applied to operator by regarding it as the vector with $\mathcal{D}_\Lambda^2$ elements. 
For an operator $O$, we can obtain the Schmidt decomposition as 
 \begin{align}
O = \sum_{m=1}^{D_O} \mu_m  O_{1,m} \otimes O_{2,m}, \label{sup_Schmidt_decomp_O}
\end{align}
where $\{O_{1,m}\}$ and $\{O_{2,m}\}$ are operator bases with the property of $\| O_{1,m}\|_2 = 1$ and $\tr(O_{1,m}O_{1,m'})=$ for $m\neq m'$.
For an arbitrary operator $\hat{O}$ with its Schmidt rank $D$, we obtain 
 \begin{align}
\|O-O_D  \|_2^2 =\sum_{m>D} \mu_m^2 \le \| O-\hat{O}\|_2^2 , \label{sup_thm:The Eckart-Young theorem_op}
\end{align}
where we defined $O_D:=\sum_{m\le D} \mu_m  O_{1,m} \otimes O_{2,m}$.
We note that in applying the operator the Eckart-Young theorem is only applied to the Schatten $2$-norm.
As far as we know, the Eckart-Young theorem cannot be extended to general Schatten $p$-norm.
%

\subsection{Approximation of square operators} \label{sup_Sec:Approximation of square operators}

In the analyses, we often use the following lemma, which connects the closeness between two operators to that between square of the two operators:
\begin{lemma}\label{lem:Approximation of square operators_0}
Let $O$ and $\tilde{O}$ be operators which are close to each other in the following sense:
\begin{align}
\|O - \tilde{O} \|_{2p} \le  \delta \|O\|_{2p} \quad (\delta \le 1) .\label{condition_Approximation of square operators_0}
\end{align}
Then, the square of the operator $O$, which is $O^\dagger O$, is close to $\tilde{O}^\dagger \tilde{O}$ as follows: 
\begin{align}
\|O^\dagger O- \tilde{O}^\dagger \tilde{O}\|_{p} \le 3 \delta \|O^\dagger O\|_{p} .\label{ineq_Approximation of square operators_0}
\end{align}
\end{lemma}
\noindent
The proof is straightforward by extending the result in Ref.~\cite{PhysRevB.91.045138}, where the positivity of $O$ has been assumed.
We show the proof in the following.

\subsubsection{Proof of Lemma~\ref{lem:Approximation of square operators_0}} \label{sup_proof_lem:Approximation of square operators}

Following Ref.~\cite{PhysRevB.91.045138}, we start from 
\begin{align}
\|O^\dagger O- \tilde{O}^\dagger \tilde{O}\|_{p} 
=& \|O^\dagger (O- \tilde{O}) - (\tilde{O}^\dagger - O^\dagger) \tilde{O}\|_{p}   \notag\\
\le& \|O^\dagger (O- \tilde{O}) \|_p + \|(\tilde{O}^\dagger - O^\dagger) \tilde{O}\|_{p} , \notag
\end{align}
where the inequality is derived from the triangle inequality.
By using the H\"older inequality~\eqref{sup_generalized_Holder_ineq} with $p_1=p_2=2p$, we obtain 
\begin{align}
\|O^\dagger (O- \tilde{O}) \|_p \le \|O^\dagger\|_{2p} \| O- \tilde{O} \|_{2p} \le \delta \|O\|_{2p}^2, \notag
\end{align}
where we use the inequality~\eqref{condition_Approximation of square operators_0} and $\|O^\dagger\|_{2p}=  \|O\|_{2p}$.
In the same way, we obtain
\begin{align}
\|(\tilde{O}^\dagger - O^\dagger) \tilde{O}\|_{p} \le \delta  \|O\|_{2p}\|\tilde{O}\|_{2p}\le  \delta  \|O\|^2_{2p}(1+ \delta ),\notag
\end{align}
where the last inequality is derived from $\|\tilde{O}\|_{2p} =\|\tilde{O} - O + O\|_{2p} \le \|\tilde{O} - O\|_{2p} +\| O\|_{2p}  \le \|O\|_{2p}(1+ \delta )$. 
The definition of the Schatten norm~\eqref{def:Schatten p norm} implies 
\begin{align}
\|O\|^2_{2p} := \left[\tr (O^\dagger O)^{p} \right]^{1/{p}}  &=\left\{\tr [ (O^\dagger O) (O^\dagger O)^\dagger]^{p/2} \right\}^{1/p}  \notag \\
&=\|O^\dagger O\|_{p},\notag
\end{align}
where we use hermiticity of $O^\dagger O$.
By combining all the above inequalities, we arrive at the inequality of
\begin{align}
\|O^\dagger O- \tilde{O}^\dagger \tilde{O}\|_{p} \le \delta  (2+ \delta )  \|O^\dagger O\|_{p} \le 3\delta \|O^\dagger O\|_{p},\notag
\end{align}
where we use the condition $\delta\le 1$ in the last inequality.
This completes the proof of the inequality~\eqref{ineq_Approximation of square operators_0}. $\square$

\subsection{Approximation of $q$th power of operators}

The statement in Lemma~\ref{lem:Approximation of square operators_0} is extended to arbitrary powers:
\begin{lemma}\label{lem:Approximation of qth power of operators_0}
Let $O$ and $\tilde{O}$ be operators which satisfy the inequality
\begin{align}
\|O - \tilde{O} \|_{2qp} \le  \delta \|O\|_{2qp} \quad (\delta \le 1) .\label{sup_condition_Approximation of square operators_2}
\end{align}
Then, the $p$th power of the operator $O^\dagger O$ is close to $(\tilde{O}^\dagger \tilde{O})^p$ as follows:
\begin{align}
\|(O^\dagger O)^q- (\tilde{O}^\dagger \tilde{O})^q \|_p \le 3\delta q e^{3\delta q} \| (O^\dagger O)^q\|_{p} .\label{ineq_Approximation of product operators_0}
\end{align}
\end{lemma}
\noindent
The proof is a simple generalization of Proposition~1 in Ref.~\cite{PhysRevB.91.045138} to arbitrary Schatten-$p$ norms.

\subsubsection{Proof of Lemma~\ref{lem:Approximation of qth power of operators_0}} \label{sup_proof_lem:Approximation of qth power of operators}

Following Ref.~\cite{PhysRevB.91.045138}, we start from the equation as follows:
\begin{align}
&(O^\dagger O)^q- (\tilde{O}^\dagger \tilde{O})^q  \notag \\
&= \sum_{s=1}^q (O^\dagger O)^{q-s} \left (O^\dagger O  - \tilde{O}^\dagger \tilde{O} \right) (\tilde{O}^\dagger \tilde{O})^{s-1}.
\end{align}
We can easily check that the above equation holds for arbitrary $q$.
By using the triangle inequality for the Schatten norm, we have 
\begin{align}
&\| (O^\dagger O)^q - (\tilde{O}^\dagger \tilde{O})^q  \|_p \notag \\
&\le  \sum_{s=1}^q \left\| (O^\dagger O)^{q-s} \left (O^\dagger O  - \tilde{O}^\dagger \tilde{O} \right) (\tilde{O}^\dagger \tilde{O})^{s-1} \right\|_p.
\label{sup_ineq:(O^dagger O)^p- (tilde{O}^dagger...}
\end{align}
Then, our task is to estimate the upper bound of the norm of $(O^\dagger O)^{q-s} \left (O^\dagger O  - \tilde{O}^\dagger \tilde{O} \right) (\tilde{O}^\dagger \tilde{O})^{s-1}$.
From the generalized H\"older inequality~\eqref{sup_generalized_Holder_ineq}, we obtain 
\begin{align}
&\left\| (O^\dagger O)^{q-s} \left (O^\dagger O  - \tilde{O}^\dagger \tilde{O} \right) (\tilde{O}^\dagger \tilde{O})^{s-1} \right\|_p  \notag \\
&\le \| (O^\dagger O)^{q-s}\|_{\frac{pq}{q-s}} \|O^\dagger O  - \tilde{O}^\dagger \tilde{O}\|_{pq} \| (\tilde{O}^\dagger \tilde{O})^{s-1}\|_{\frac{pq}{s-1}} \notag \\
&\le  \| O^\dagger O \|_{pq}^{q-s} \cdot   3\delta \| O^\dagger O \|_{pq} \cdot   \| \tilde{O}^\dagger \tilde{O}\|_{pq}^{s-1} ,
\label{sup_Ineq: left| (O^dagger O)^p-s ...}
\end{align}
where the equations $\| (O^\dagger O)^{q-s}\|_{\frac{pq}{q-s}}=\| O^\dagger O \|_{pq}^{q-s}$ and $\| (\tilde{O}^\dagger \tilde{O})^{s-1}\|_{\frac{pq}{s-1}}=\| \tilde{O}^\dagger \tilde{O}\|_{pq}^{s-1} $ are straightforwardly derived from the definition~\eqref{def:Schatten p norm}, and we use 
the inequality~\eqref{ineq_Approximation of square operators_0} for $\|O^\dagger O  - \tilde{O}^\dagger \tilde{O}\|_{pq}$. 
Furthermore, by using 
$
\| \tilde{O}^\dagger \tilde{O}\|_{pq} =\|\tilde{O}^\dagger \tilde{O} - O^\dagger O + O^\dagger O\|_{pq} 
\le \|\tilde{O}^\dagger \tilde{O} - O^\dagger O\|_{pq}    + \| O^\dagger O\|_{pq}  \le (3\delta +1)\| O^\dagger O\|_{pq}  ,
$
the inequality~\eqref{sup_Ineq: left| (O^dagger O)^p-s ...} reduces to
\begin{align}
&\left\| (O^\dagger O)^{q-s} \left (O^\dagger O  - \tilde{O}^\dagger \tilde{O} \right) (\tilde{O}^\dagger \tilde{O})^{s-1} \right\|_p\notag \\
& \le   3\delta  (3\delta +1)^q \| O^\dagger O\|_{pq}^q  \le 3\delta  e^{3\delta q}\| (O^\dagger O)^q\|_{p}  .
\label{sup_Ineq: left| (O^dagger O)^p-s ...2}
\end{align}
By applying the inequality~\eqref{sup_Ineq: left| (O^dagger O)^p-s ...2} to \eqref{sup_ineq:(O^dagger O)^p- (tilde{O}^dagger...}, we obtain the main inequality~\eqref{ineq_Approximation of product operators_0}.
This completes the proof of the inequality~\eqref{ineq_Approximation of product operators_0}. $\square$

\subsection{Upper bound on the norm for multi-commutators} \label{sup_Sec:Upper bound of the commutator norm}

For the norm of multi-commutators, we can prove the following lemma (see Lemma~3 in Ref.~\cite{kuwahara2015floquet}):
\begin{lemma} \label{sup_norm_multi_commutator1}
Let $\{A_s\}_{s=1}^M$ be $k_s$-local operators such that 
\begin{align}
A_s = \sum_{|X|\le k_s} a_{s,X},\quad \max_{i\in \Lambda}\sum_{X:X\ni i} \|a_{s,X}\| \le g_s.
\end{align}
Then, for an arbitrary operator $O_X$ supported on a subset $X$, the norm of the multi-commutator is bounded from above by
\begin{align}
\| \ad_{A_M} \ad_{A_{M-1}} \cdots \ad_{A_1}(O_X ) \| \le \prod_{m=1}^M (2  g_{m} K_m) \|O_X\|, \label{sup_fundamental_ineq}
\end{align}
where $K_m := |X|+ \sum_{s\le m-1} k_s$.
\end{lemma}

For $g_1=g_2=\cdots = g_M=g$ and $k_1=k_2=\cdots=k_M=k$, we have 
\begin{align}
&\| \ad_{A_M} \ad_{A_{M-1}} \cdots \ad_{A_1}(O_X ) \|  \notag \\
&\le (2g k)^M \frac{|X|}{k}\br{\frac{|X|}{k}+1} \cdots \br{\frac{|X|}{k}+M-1} \|O_X\|. \label{sup_fundamental_ineq_0}
\end{align}

\section{Full proof of Proposition~\ref{main_thm_area_law}} \label{sec:proof_Main proposition and the proof}

In this section, we show the proof outline of Proposition~\ref{main_thm_area_law} in Sec.~\ref{sec:proof_Main statement and the proof} 
which plays key roles in the proofs of the main results (Theorems~\ref{main_thm_area_law2}, \ref{main_thm_MPO_approximation} and \ref{thm:Renyi entanglement of purification}). 
We prove it based on several essential Lemmas~\ref{sup_lemma_rho_belief}, \ref{sup_lem:low_deg_poly}, \ref{sup_third_lemma_approx_est} and \ref{sup_fourth_lemma_schmidt}. 
Throughout the proof, while considering the Schmidt rank for a target decomposition $\Lambda=L\cup R$, we denote ${\rm SR} (O,i_0)$ by ${\rm SR} (O)$ for simplicity.

\subsection{Proof strategy}

\begin{figure*}
\centering
{
\includegraphics[clip, scale=0.5]{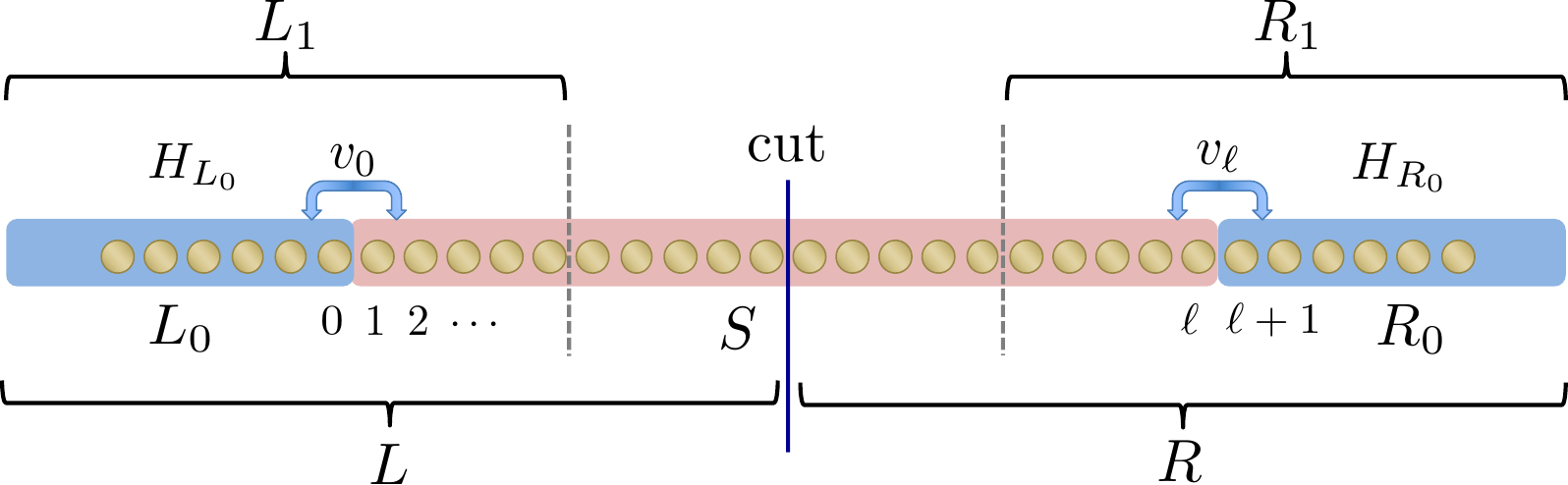}
}
\caption{The decomposition of the system considered in the proof.
}
\label{sup_fig:system}
\end{figure*}

We here relabel each of the sites such that $L=\{i\}_{i\le \ell/2}$ and $R=\{i\}_{i\ge \ell/2+1}$, where the length $\ell$ is a multiple of $4$ to be chosen later.
We can arbitrarily extend the system size $\Lambda\to \Lambda \sqcup \delta \Lambda$ without changing the Hamiltonian. 
We only have to add zero operators: 
\begin{align}
H_{\Lambda \sqcup \delta \Lambda}' = H_{\Lambda} + \hat{0}_{\delta \Lambda} , \label{extending_Hamiltonian_system_size}
\end{align}
where $\hat{0}_{\delta \Lambda}$ is the zero operator acting on $\delta \Lambda$. 
Note that $\hat{0}_{\delta \Lambda}$ still satisfies the form of~\eqref{sup_def:Hamiltonian}. 

We then decompose the total system into three pieces $L_0$, $S$ and $R_0$ (see Fig.~\ref{sup_fig:system}), where $L_0=\{i\}_{i \le 0}$, $S=\{i\}_{1\le i \le \ell}$ and $R_0=\{i\}_{i\ge \ell+1}$.
Accordingly, we also decompose the Hamiltonian as follows:
\begin{align}
&H= H_S + H_{L_0} +H_{R_0} + v_0+v_\ell ,  \quad H_S :=\sum_{X\subset S} h_X ,  \notag\\
&\quad H_{L_0} :=\sum_{X\subset L_0} h_X\quad  H_{R_0} :=\sum_{X\subset R_0} h_X,
\end{align}
where $v_0$ and $v_\ell$ have been defined by Eq.~\eqref{sup_def_v_i_1D}. 
We note that $H_S$, $H_{L_0}$ and $H_{R_0}$ commute with each other. 
By shifting the energy origin appropriately, we set 
\begin{align}
&H_S \succeq 0, \label{sup_assump_H_S}
\end{align}
where $\succeq$ means that $H_S$ is positive semidefinite.
We will divide $\beta$ into $2\M$ pieces ($\M\in \mathbb{N}$) and introduce 
\begin{align}
\rho_0 := e^{-\beta_0 H}, \quad \beta_0:=\beta/(2\M). \notag 
\end{align}

The first step of the proof is the approximation of $\rho_0$, which is in the following form: 
\begin{align}
&\tilde{\rho}_0:=\tilde{\Phi}_0^\dagger e^{-\beta_0 (H_{L_0} + H_{R_0})} \func_m(\beta_0 H_S) \tilde{\Phi}_0, \notag  \\
&\tilde{\Phi}_0:= \Phi_{L_1} \otimes\Phi_{R_1},
\end{align}
where $\Phi_{L_1}$ and $\Phi_{R_1}$ are operators supported on $L_1$ and $R_1$, respectively (i.e., $L_1=\{i\}_{i\le \ell/4}$ and $R_1=\{i\}_{i\ge 3\ell/4+1}$), and the degree $m$ polynomial $\func_m(x)$ approximates the exponential function $e^{-\beta_0 x}$. For every $\delta\le 1/(3\M)$, we will estimate the length $\ell$ and the degree $m$ such that: 
\begin{align}
\label{sup_eq:rho0approximation}
\| \rho_0- \tilde{\rho}_0 \|_{2\M p} \le \delta \|\rho_0\|_{2\M p}.
\end{align}
Then, by applying inequality~\eqref{ineq_Approximation of product operators_0}, we have
\begin{align}
\| e^{-\beta H}- \tilde{\rho}_0^{2\M} \|_{p} \le 3 \delta qe^{3\delta \M}  \|\rho_0^{2\M}\|_p \le \epsilon\|e^{-\beta H}\|_p
\label{sup_e^-beta H-tilde_rho_0^2M|_p}
\end{align}
with $\epsilon=3e\M \delta$,
where we use $\rho_0^{2\M}= e^{-\beta H}$ and $3 \delta qe^{3\delta \M} \le 3 e q  \delta $ from $\delta \le 1/(3q)$.
Therefore, by choosing $\hat{\rho}=\tilde{\rho}_0^{2\M}/\tr(e^{-\beta H})$, we can achieve the bound~\eqref{ineq_hat_rho_p}. 
Note that the condition $\epsilon \le e$ in Proposition~\ref{main_thm_area_law} is due to the equations $\epsilon=3e\M \delta$ and $\delta \le 1/(3\M)$.

The second step is to estimate the upper bound of the Schmidt rank of $\tilde{\rho}_0^{2\M}$, which is given by
\begin{align}
\label{sup_eq:finalapproximation}
\left[  \tilde{\Phi}_0^\dagger e^{-\beta_0 (H_{L_0} + H_{R_0})} \func_m(\beta_0H_S)\tilde{\Phi}_0 \right]^{2\M} . 
\end{align}
Then, the sufficient Schmidt rank to achieve the inequality~\eqref{sup_e^-beta H-tilde_rho_0^2M|_p} is given by a function of $q$ (see Ineq.~\eqref{sup_Schmidt_rank_of_tilde_rho} below for more details). 
By choosing $q$ so that the Schmidt rank is minimum, we will show that the Schmidt rank is upper bounded by \eqref{ineq_on_schmidt_rank_thm}.
We thus prove Proposition~\ref{main_thm_area_law}.
In the following, we are going to show the details of the above arguments.

\subsection{Approximation of $\rho_0$}
In the following, we define a parameter $\nu$ as follows:
\begin{align}
\nu= \max[ \beta_0,\log(6/\delta) ]. \label{sup_def:nu_para}
\end{align}
In addition, we choose $\M$ such that 
 \begin{align}
\M^2 \ge \beta  . \label{sup_basic_additional_assumptions}
\end{align}
Let $H_0:= H_S+H_{L_0}+H_{R_0}$. 
We first relate the two operators $\rho_0=e^{-\beta_0 H}$ and $e^{-\beta_0 H_0}$. 
We can formally write the following:
\begin{align}
\rho_0 =  \Phi_0 e^{- \beta_0 H_0} \Phi_0^\dagger,
\end{align}
where $\Phi_0$ is usually highly non-local operator. 
The first lemma ensures that the $\Phi$ is approximated by an operator supported on $L_1 \sqcup R_1$:
\begin{lemma} \label{sup_lemma_rho_belief}
There exists an operator $\tilde{\Phi}_0=\Phi_{L_1} \otimes\Phi_{R_1}$ such that for 
\begin{align}
\rho'_0 =\tilde{\Phi}_0 e^{- \beta_0 H_0}\tilde{\Phi}_0^\dagger ,
\label{sup_def_rho_0'}
\end{align}
we have
\begin{align}
\| \rho'_0 - \rho_0\|_{p_0} \le  3 e^{-c_1\ell/\beta_0 + c_2\beta_0} \|\rho_0\|_{p_0} \label{sup_Ineq:lemma_rho_belief}
\end{align}
for arbitrary $p_0\in\mathbb{N}$,
where we assume $-c_1\ell/\beta_0 + c_2\beta_0 \le 0$, and $c_1$ and $c_2$ are $\orderof{1}$ constants.
\end{lemma}
The proof of this lemma is based on the belief propagation~\cite{PhysRevB.76.201102,kato2016quantum} and the Lieb-Robinson bound~\cite{ref:LR-bound72,PhysRevLett.97.050401}.

\subsubsection{Proof of Lemma~\ref{sup_lemma_rho_belief}}

For the proof, we start from the belief propagation~\cite{PhysRevB.76.201102}, which gives
\begin{align}
\rho_0= e^{- \beta_0 H}= \Phi_0 e^{- \beta_0 H_0} \Phi_0^\dagger,
\label{sup_BP_rho_0}
\end{align}
where the operator $\Phi_0$ is defined as 
\begin{align}
&\Phi_0:= \mathcal{T} e^{\int_0^{1} \phi (\tau) d\tau} , \notag \\
&\phi (\tau):= \frac{-\beta_0v_0-\beta_0 v_\ell}{2}  \notag \\
&\quad \quad+ i\beta_0 \int_{-\infty}^\infty g(t)[  v_0 (t,H_\tau)+v_\ell (t,H_\tau)]dt, 
\label{sup_Def:Phi_0_phi}
\end{align}
where $H_\tau = H_0 + \tau (v_0+v_\ell)$, $\mathcal{T}$ denotes the ordering operator, $v_0 (t,H_\tau)= e^{itH_\tau}v_0e^{-itH_\tau}$, $v_\ell (t,H_\tau)= e^{itH_\tau}v_\ell e^{-itH_\tau}$, and $g(t)$ is defined as
\begin{align}
 g(t)
 := {\rm sign}(t)  \frac{e^{-2\pi |t|/\beta_0}}{1-e^{-2\pi |t|/\beta_0}}.
 \label{sup_def_g_t_exp_decay}
\end{align}
Note that the function $g(t)$ decays exponentially with $t$ and hence the operator $\phi (\tau)$ is quasi-local due to the Lieb-Robinson bound~\cite{ref:LR-bound72,PhysRevLett.97.050401}.
We aim to obtain the approximation $\Phi_0 \approx \Phi_{L_1} \otimes \Phi_{R_1}=:\tilde{\Phi}_0$, and consider the norm difference of 
\begin{align}
\epsilon_\Phi:=\left\|\tilde{\Phi}_0 e^{- \beta H_0}\tilde{\Phi}_0^\dagger -e^{- \beta_0 H} \right\|_{p_0} 
\end{align}
for arbitrary $p_0\in \mathbb{N}$.

In order to quantitatively evaluate the quasi-locality of $\phi (\tau)$, we first define $v_0 (t,H_\tau,L_1)$ as an approximation of $v_0(t,H_\tau)$ in the region $L_1$:
\begin{align}
v_0 (t,H_\tau,L_1) := \frac{1}{\mathcal{D}_{\Lambda\setminus L_1}} \tr_{\Lambda\setminus L_1}[v_0(t,H_\tau)] \otimes \hat{1}_{\Lambda\setminus L_1}. \notag 
\end{align}
We define $v_\ell (t,H_\tau,R_1)$ in the same way.
By utilizing the Lieb-Robinson bound~\cite{PhysRevLett.97.050401}, we obtain the approximation error of 
\begin{align}
&\| v_0 (t,H_\tau) - v_0 (t,H_\tau,L_1) \| \le c |t| e^{-c'(\ell/4-vt)} ,  \label{sup_v_0_tau_L_1}
\end{align}
where $c$, $c'$ and $v$ are constants of $\orderof{1}$ and we obtain the same upper bound for $\| v_\ell (t,H_\tau) - v_\ell (t,H_\tau,R_1) \|$.
By using the notations of $v_0 (t,H_\tau,L_1)$ and $v_\ell (t,H_\tau,R_1)$, we define $\tilde{\phi}_{L_1}(\tau) $ and $\tilde{\phi}_{R_1}(\tau)$ as follows:
\begin{align}
&\tilde{\phi}_{L_1}(\tau):= \frac{-\beta_0}{2} v_0 +  i \beta_0 \int_{-\infty}^\infty g(t)v_0 (t,H_\tau,L_1)dt, \notag \\
&\tilde{\phi}_{R_1}(\tau):= \frac{-\beta_0}{2}v_\ell +  i \beta_0 \int_{-\infty}^\infty g(t)v_\ell (t,H_\tau,R_1)dt.  \notag 
\end{align}
We notice that $\tilde{\phi}_{L_1}(\tau)$ and $\tilde{\phi}_{R_1}(\tau)$ are supported on the subsets $L_1$ and $R_1$, respectively.
We then approximate $\phi(\tau)$ by $\tilde{\phi} (\tau)= \tilde{\phi}_{L_1}(\tau) +\tilde{\phi}_{R_1}(\tau)$ with an error of
\begin{align}
\| \phi(\tau)-\tilde{\phi} (\tau)\| \le c_0 \beta_0^2 e^{-c_1\ell/\beta_0} \quad (0\le \tau \le \beta_0)
\label{sup_approx_error_phi_tilde_phi}
\end{align}
with $c_0$ and $c_1$ constants of $\orderof{1}$,
where the inequality is derived from the approximation error in \eqref{sup_v_0_tau_L_1} and the exponential decay of $g(t)$ as in Eq.~\eqref{sup_def_g_t_exp_decay}.

From the approximation of $\phi(\tau)$ by $\tilde{\phi} (\tau)$, we define $\tilde{\Phi}_0$ as 
\begin{align}
\tilde{\Phi}_0:= \mathcal{T} e^{-\int_0^1 \tilde{\phi} (\tau) d\tau}=\Phi_{L_1} \otimes \Phi_{R_1} ,
\label{sup_def:phi_L1_Phi_R1}
\end{align} 
where we define $\Phi_{L_1}:=\mathcal{T} e^{-\int_0^1 \tilde{\phi}_{L_1} (\tau) d\tau}$ and $\Phi_{R_1}:= \mathcal{T} e^{-\int_0^1 \tilde{\phi}_{R_1} (\tau) d\tau}$.
By using the inequality~\eqref{sup_approx_error_phi_tilde_phi}, we can obtain the approximation error of $\Phi_0$ by 
\begin{align}
\| 1-\tilde{\Phi}_0 \Phi_0^{-1}\| 
&\le c_0 \beta_0^2e^{-c_1\ell/\beta_0} e^{2\int_0^1 \| \phi (\tau)\| d\tau}  \notag \\
&\le c_0 \beta_0^2 e^{-c_1\ell/\beta_0 + 2 c_1' \beta_0}, \label{sup_approx_belief_propagation}
\end{align} 
with $c_1'$ an $\orderof{1}$ constant, 
where the upper bound $\| \phi (\tau)\| \le c_1' \beta_0$ can be derived by following Ref.~\cite{kato2016quantum} (see Eq.~(42) therein).
By letting $O_0:=\tilde{\Phi}_0 \Phi_0^{-1}$, we have, using the triangle inequality
\begin{align}
&\left\|\Phi_0 e^{- \beta_0 H_0} \Phi_0^\dagger  -\tilde{\Phi}_0 e^{- \beta_0 H_0} \tilde{\Phi}_0^\dagger    \right\|_{p_0} 
=\|\rho_0 -O_0\rho_0 O_0^{\dagger} \|_{p_0} \notag \\
&\le\| (1 -O_0)\rho_0 O_0^{\dagger}\|_{p_0}  + 
\| O_0\rho_0 (1-O_0^{\dagger})\|_{p_0}   \notag \\
&\quad + \| (1 -O_0)\rho_0 (1-O_0^{\dagger}) \|_{p_0} ,
\label{sup_Phi_0_Phi_0_tilde_ineq}
\end{align}
From the upper bound~\eqref{sup_approx_belief_propagation}, the norm of $1- O_0$ satisfies the following inequality:
\begin{align}
\| 1- O_0 \|  = \| 1-\tilde{\Phi}_0 \Phi_0^{-1}\| 
&\le   e^{-c_1\ell/\beta_0 + c_2 \beta_0}, \notag 
\end{align} 
where we choose $c_2=\orderof{1}$ such that $c_0 \beta_0^2 e^{2 c_1' \beta_0} \le e^{c_2 \beta_0}$. 
Then, the condition $-c_1\ell/\beta_0 + c_2\beta_0 \le 0$ in the lemma implies $\| 1- O_0 \| \le 1$.
Therefore, by applying the H\"older inequality~\eqref{sup_generalized_Holder_ineq_operator_norm} to each of the terms in \eqref{sup_Phi_0_Phi_0_tilde_ineq}, we obtain 
\begin{align}
&\left\|\Phi_0 e^{- \beta_0 H_0} \Phi_0^\dagger  -\tilde{\Phi}_0 e^{- \beta_0 H_0} \tilde{\Phi}_0^\dagger    \right\|_{p_0} \notag \\
&\le\|\rho_0\|_{p_0} \left(\| 1- O_0 \|^2 + 2\| 1- O_0 \|    \right)   \notag \\
&\le 3  e^{-c_1\ell/\beta_0 + c_2 \beta_0}  \|\rho_0\|_{p_0} ,
\label{sup_last_ineq_Lemma_brief_prop}
\end{align}
where  in the second inequality, we get $\| 1- O_0 \|^2 \le \| 1- O_0 \| $ due to $\| 1- O_0 \| \le 1$.
This completes the proof. $\square$

 {~}

\hrulefill{\bf [ End of Proof of Lemma~\ref{sup_lemma_rho_belief}] }

{~}

The lemma implies that as the length $\ell$ becomes large, the approximation error decays exponentially with $e^{-\orderof{\ell/\beta_0}}$.
Thus, in order to achieve the inequality
\begin{align}
\| \rho'_0 - \rho_0\|_{2\M p} \le \frac{\delta}{2}\|\rho_0\|_{2\M p}.
\label{sup_first_approx}
\end{align}
we need to choose $\ell$ as 
\begin{align}
\ell\ge \frac{c_2}{c_1} \beta_0^2 + \frac{\beta_0}{c_1} \log(6/\delta) .
\label{sup_ell_c_2_c_1_beta_0}
\end{align}
By using the parameter $\nu$ in Eq.~\eqref{sup_def:nu_para}, we can write
\begin{align}
\label{sup_eq:ellalternatechoice}
\ell= \tilde{c}_1 \beta_0 \nu =\tilde{c}_1 \nu \beta/(2q)  ,
\end{align}
where $ \tilde{c}_1$ is a constant of $\orderof{1}$.

Second, we approximate $e^{- \beta_0 H_0}$ by an operator with small Schmidt rank. 
For this purpose, we use the fact that $H_S$, $H_{L_0}$ and $H_{R_0}$ commute with each other, and write $e^{- \beta_0 H_0}=e^{- \beta_0 (H_{L_0}+H_{R_0})}e^{-\beta_0 H_S}$. Then, we approximate $e^{-\beta_0 H_S}$ by a low degree polynomial of $H_S$. 
The most straightforward approximation is given by the truncation of the Taylor expansion, which gives a good approximation of $e^{-\beta_0 H_S}$ by taking the polynomial degree as large as $\|\beta_0 H_S\| + \log(1/\delta_0)$ with $\delta_0$ the precision error. 
Unfortunately, we cannot get any improvement of the thermal area law if we utilize the Taylor expansion.

One of the key aspects of our proof is the use of the following Lemma from Ref.~\cite[Theorem 4.1]{TCS-065}, which allows us to achieve the improved thermal area law:
\begin{lemma}
\label{sup_lem:low_deg_poly}
Let $\delta_0\in (0,1)$. For any $m$ satisfying 
 \begin{align}
 \label{sup_lower_m_lem:low_deg_poly}
m > c_f \sqrt{\max[\beta_0\|H_S\|, \log(1/\delta_0)] \log(1/\delta_0)} ,
\end{align}
(with $c_f=\orderof{1}$) there exists a polynomial $\func_m(x)$ with degree $m$ that satisfies
 \begin{align}
| \func_m(x) - e^{-x} | \le \delta_0  \for x\in [0,\beta_0\|H_S\|]. \label{sup_ineq_delta_poly_approx}
\end{align} 
When $\beta_0\|H_S\| \gg \log(1/\delta_0)$, the above estimation gives a significantly better polynomial degree than that from the Taylor expansion.
\end{lemma}
\noindent
We recall that this polynomial approximation is obtained from the Chebyshev polynomial expansion~\eqref{random_walk_exponential} in Sec.~\ref{sec:Physical intuition from the random walk behavior}, which is characterized by the random walk behavior (see Fig.~\ref{fig:Random_walk}).

By using the polynomial $\func_m(x)$ defined above, we approximate the operator $\rho'_0$ in Eq.~\eqref{sup_def_rho_0'} as 
\begin{align}
\tilde{\rho}_0 :=\tilde{\Phi}_0 e^{- \beta_0 (H_{L_0}+H_{R_0})}\func_m(\beta_0 H_S)\tilde{\Phi}_0^\dagger .
\label{sup_def_tilde_rho_0_0}
\end{align}
Because of \eqref{sup_assump_H_S}, the spectrum of $\beta_0 H_S$ is included in the span of $[0,\beta_0\|H_S\|]$, and hence the inequality~\eqref{sup_ineq_delta_poly_approx} gives
 \begin{align}
 \label{app_better_poly_H_s}
\|\func_m(\beta_0 H_S) - e^{-\beta_0 H_S} \| \le \delta_0 .
\end{align} 
We note that the current approximation~\eqref{app_better_poly_H_s} is obtained in terms of the operator norm (i.e., Schatten $\infty$-norm) instead of the generic Schatten $p$-norm.
The next problem is to estimate the approximation error $\|\rho_0'-\tilde{\rho}_0\|_{p_0}$ for arbitrary Schatten $p_0$-norm.  
We prove the following lemma:
\begin{lemma} \label{sup_third_lemma_approx_est}
Let $p_0\in\mathbb{N}$ and $\delta_0\in (0,1)$. Under the choice of $\Phi_{L_1} \otimes\Phi_{R_1}$ in Lemma \ref{sup_lemma_rho_belief}, $\ell$ in Eq.~\eqref{sup_ell_c_2_c_1_beta_0} and $m, \func_m(x)$ in Lemma \ref{sup_lem:low_deg_poly}, we have
 \begin{align}
\|\rho_0'-\tilde{\rho}_0\|_{p_0} \le \mathcal{D}_S^{1/p_0}\delta_0 e^{c_3 \beta_0} \left \| \rho_0\right\|_{p_0} ,
\label{sup_main_ineq:third_lemma_approx_est}
\end{align}
where $c_3$ is an $\orderof{1}$ constant.
\end{lemma}

\subsubsection{Proof of Lemma~\ref{sup_third_lemma_approx_est}}
From the definitions~\eqref{sup_def_rho_0'} and \eqref{sup_def_tilde_rho_0_0} of $\rho_0'$ and $\tilde{\rho}_0$, respectively, we start from the inequality 
 \begin{align}
&\|\rho_0'-\tilde{\rho}_0\|_{p_0}   \notag \\
&=\left\| \tilde{\Phi}_0^\dagger e^{- \beta_0 (H_{L_0}+H_{R_0})} \left( \func_m(H_S) -e^{-\beta_0 H_S}\right) \tilde{\Phi}_0 \right\|_{p_0} \notag \\
&\le \left \| e^{- \beta_0 (H_{L_0}+H_{R_0})} \left( \func_m(H_S) -e^{-\beta_0 H_S}\right)\right\|_{p_0} \cdot \left\| \tilde{\Phi}_0\right\|^2  , 
\label{sup_rho_0'_minus_tilde_rho_0}
\end{align}
where we used H\"older's inequality~\eqref{sup_generalized_Holder_ineq_operator_norm}.
From the definition~\eqref{sup_def:phi_L1_Phi_R1} of $ \Phi_{L_1} \otimes\Phi_{R_1}$, we obtain 
  \begin{align}
  \label{sup_norm_tilde_phi}
 \left\|\tilde{\Phi}_0 \right\|^2  = \left\| \Phi_{L_1} \otimes\Phi_{R_1}\right\|^2  \le e^{2c' \beta_0}.
\end{align}
We next consider 
 \begin{align}
 &\left \| e^{- \beta_0 (H_{L_0}+H_{R_0})} \left( \func_m(H_S) -e^{-\beta_0 H_S}\right)\right\|_{p_0}^{p_0} \notag \\
 =& \sum_{s=1}^{\mathcal{D}_{L_0R_0}}
 \sum_{s'=1}^{\mathcal{D}_{S}} e^{- p_0\beta_0 \tilde{E}_s}  \left| \func_m( \varepsilon_{s'}) -e^{-\beta_0 \varepsilon_{s'}}\right|^{p_0},
 \label{sup_p_norm_p_power_third_lemma}
\end{align}
where $\{\tilde{E}_s\}_{s=1}^{\mathcal{D}_{L_0R_0}}$ and $\{\varepsilon_{s'}\}_{s'=1}^{\mathcal{D}_{S}}$ are eigenvalues of $H_{L_0}+H_{R_0}$ and $H_S$, respectively.
Note that the Hamiltonians $H_{L_0}$, $H_{R_0}$ and $H_S$ commute with each other and are diagonalizable simultaneously.  
From the assumption~\eqref{sup_assump_H_S}, we have $\varepsilon_{1}=0$, and $\varepsilon_{\mathcal{D}_{S}}\le \|H_S\|$.

From the inequality~\eqref{sup_ineq_delta_poly_approx} which is 
 \begin{align}
| \func_m(x) - e^{-x} | \le \delta_0  \for x\in [0,\beta_0\|H_S\|], 
\end{align} 
we have 
 \begin{align}
\left|\func_m( \varepsilon_{s'}) -e^{-\beta_0 \varepsilon_{s'}}\right|^{p_0} \le \delta_0^{p_0}.
\end{align}
By applying the above inequality to \eqref{sup_p_norm_p_power_third_lemma}, we obtain 
 \begin{align}
& \left \| e^{- \beta_0 (H_{L_0}+H_{R_0})} \left( \func_m(H_S) -e^{-\beta_0 H_S}\right)\right\|_{p_0}^{p_0}   \notag \\
 \le&  \sum_{s=1}^{\mathcal{D}_{L_0R_0}} \sum_{s'=1}^{\mathcal{D}_{S}} e^{- p_0\beta_0 \tilde{E}_s}\delta_0^{p_0} \notag \\
\le & \frac{\mathcal{D}_S\delta_0^{p_0}}{ \sum_{s'=1}^{\mathcal{D}_{S}}e^{-p_0\beta_0 \varepsilon_{s'}}} \sum_{s=1}^{\mathcal{D}_{L_0R_0}} \sum_{s'=1}^{\mathcal{D}_{S}} e^{- p_0\beta_0 \tilde{E}_s}e^{-p_0\beta_0 \varepsilon_{s'}} \notag \\
\le& \mathcal{D}_S\delta_0^{p_0}  \left \| e^{- \beta_0 (H_{L_0}+H_{R_0}+H_S)} \right\|_{p_0}^{p_0} ,
\label{sup_third_lemma_p_norm_bound_1}
\end{align}
where we use $\sum_{s'=1}^{\mathcal{D}_{S}}e^{-p_0\beta_0 \varepsilon_{s'}} \ge e^{-p_0\beta_0 \varepsilon_{1}}=1$.

We next consider the upper bound of $\| e^{- \beta_0 H_0} \|_{p_0}$ in terms of $\| e^{- \beta_0 H} \|_{p_0}$. 
Recall that $H_0=H_{L_0}+H_{R_0}+H_S$ and hence 
$
 e^{- \beta_0 H_0}=e^{-\beta_0 (H-v_0-v_\ell)}
$.
By using the Golden-Thompson inequality, we have
\begin{align}
\tr( e^{- p_0 \beta_0 H_0} )  
&\le \tr \br{e^{-p_0 \beta_0 H}\cdot e^{-p_0 \beta_0(v_0+v_\ell)}} \notag \\
&\le e^{p_0 \beta_0 \| v_0+v_\ell\|} \tr \br{e^{-p_0 \beta_0 H}} \notag \\
&\le e^{2 gk p_0 \beta_0}  \|e^{-\beta_0 H}\|_{p_0}^{p_0} ,
 \label{sup_third_lemma_p_norm_bound_2}
\end{align}
where we use $\|v_0+v_\ell\| \le 2gk$ from the condition in Eq.~\eqref{sup_def:Hamiltonian}.
Note that $\tr( e^{- p_0 \beta_0 H_0} )  = \|e^{-\beta_0 H_0}\|_{p_0}^{p_0}$.
By combining the inequalities~\eqref{sup_third_lemma_p_norm_bound_1} and \eqref{sup_third_lemma_p_norm_bound_2}, we arrive at the inequality 
 \begin{align}
& \left \| e^{- \beta_0 (H_{L_0}+H_{R_0})} \left( \func_m(H_S) -e^{-\beta_0 H_S}\right)\right\|_{p_0}^{p_0}  \notag \\
&\le \mathcal{D}_S\delta_0^{p_0} e^{2gkp_0 \beta_0} \left \| e^{- \beta_0 H} \right\|_{p_0}^{p_0} .  
\label{sup_rho_0'_minus_tilde_rho_0_second_term}
\end{align}
By applying the inequalities~\eqref{sup_norm_tilde_phi} and \eqref{sup_rho_0'_minus_tilde_rho_0_second_term} to \eqref{sup_rho_0'_minus_tilde_rho_0}, 
we obtain the main inequality~\eqref{sup_main_ineq:third_lemma_approx_est} with $c_3=2c' +2gk$. 
This completes the proof. $\square$

{~}

\hrulefill{\bf [ End of Proof of Lemma~\ref{sup_third_lemma_approx_est}] }

{~}

Let us substitute $p_0=2\M p$ in Lemma \ref{sup_third_lemma_approx_est} and choose $\delta_0$ that satisfies 
 \begin{align}
 \mathcal{D}_S^{1/(2\M p)}\delta_0 e^{c_3 \beta_0} = \frac{\delta}{2}. \notag 
\end{align}
This ensures that $\|\rho_0'-\tilde{\rho}_0\|_{2\M p} \le (\delta/2)\left\| \rho_0\right\|_{2\M p}$ and we conclude 
 \begin{align}
\|\rho_0-\tilde{\rho}_0\|_{2\M p} &\le \|\rho_0-\rho_0'\|_{2\M p}  + \|\rho_0'-\tilde{\rho}_0\|_{2\M p}  \le  \delta \|\rho_0\|_{2\M p}, \notag 
\end{align}
where we use the inequality~\eqref{sup_first_approx}. 
Therefore, the choice of $\tilde{\rho}_0$ as in Eq.~\eqref{sup_def_tilde_rho_0_0} achieves the inequality~\eqref{sup_eq:rho0approximation}.

Let us simplify the expression for all the parameters appearing so far. 
We consider 
 \begin{align}
&\delta_0 = \frac{\delta}{2} e^{-c_3 \beta_0} \mathcal{D}_S^{-1/(2\M p)}=\frac{\delta}{2} e^{-c_3 \beta_0-c_{\spin} \ell/(2\M p)}   \notag \\
&\to \log(1/\delta_0) =  \log(2/\delta) +  c_3 \beta_0 + \frac{c_{\spin} \tilde{c}_1 \nu \beta_0}{2\M p},
\end{align}
where we define $\mathcal{D}_S= \spin^{\ell} =: e^{c_{\spin} \ell}$ and use the expression of $\ell$ in Eq.~\eqref{sup_eq:ellalternatechoice}. From the assumption~\eqref{sup_basic_additional_assumptions}, 
we have $\beta_0=\frac{\beta}{2\M} \le \M/2$. This yields
\begin{align}
\log(2/\delta) +  c_3 \beta_0 \leq \log(1/\delta_0) 
&\leq \log(2/\delta) +  c_3 \beta_0 + \frac{c_{\spin} \tilde{c}_1 \nu}{4p} \notag \\
& \leq \log(2/\delta) +  c_3 \beta_0 + \frac{c_{\spin} \tilde{c}_1 \nu}{4}, \notag 
\end{align}
where we use $p\ge1$.
Using the definition~\eqref{sup_def:nu_para} of $\nu$, we can thus write
 \begin{align}
\log(1/\delta_0) =  \tilde{c}_2\nu 
\end{align}
for some constant $\tilde{c}_2$. From the choice of $\ell$ in \eqref{sup_ell_c_2_c_1_beta_0}, we have 
 \begin{align}
\max[\beta_0\|H_S\|, \log(1/\delta_0)]  =  \orderof{\beta_0 \ell},
\end{align}
where we use $\|H_S\| \le g \ell$ from the condition in~\eqref{sup_def:Hamiltonian}.
Hence, we obtain the following simpler form of $m$:
 \begin{align}
 \label{sup_eq:finalchoiceofm}
m&=\left \lceil c_f \sqrt{\max[\beta_0\|H_S\|, \log(1/\delta_0)] \log(1/\delta_0)} \right \rceil   \notag \\
&= \tilde{c}_2'  \sqrt{\nu\beta_0\ell} .
\end{align}

\subsection{Schmidt rank analysis} \label{app:Schmidt rank analysis}

\begin{figure}
\centering
{
\includegraphics[clip, scale=0.35]{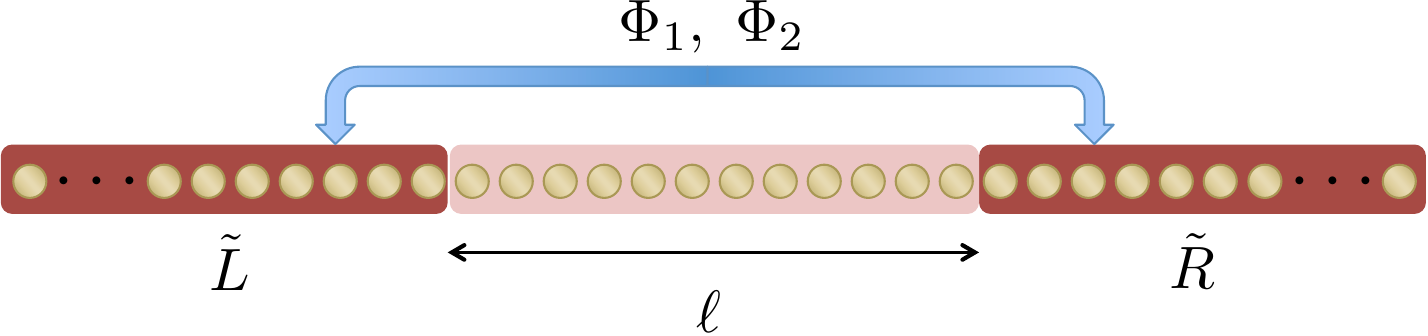}
}
\caption{The decomposition of the system in the Schmidt rank analysis}
\label{sup_fig:Schmidt}
\end{figure}

The remaining task is to estimate the Schmidt rank of the operator $\tilde{\rho}_{0}^{2q}$ which is given by \eqref{sup_eq:finalapproximation}.
For this purpose, we consider the following more general problem for the simplicity of notation. We also utilize the lemma in the subsequent sections.
Let us define a decomposition of the total system into $\tilde{L}$, $S$ and $\tilde{R}$ (see Fig.~\ref{sup_fig:Schmidt}).
We then aim to estimate the Schmidt rank of an operator of the form
 \begin{align}
\hat{\gunc}_{m,M}= [\Phi_1 \gunc_m(H_S) \Phi_2 ]^M,
\label{sup_Schmidt_edge_perturb}
\end{align}
where $\gunc_m(x)$ is an arbitrary degree $m$ polynomial, the operators $\Phi_1$ and $\Phi_2$ are supported on $\tilde{L}$ and $\tilde{R}$ respectively and $H_S$ is a local Hamiltonian on the subset $S$ ($|S|=\ell$).
The Schmidt rank estimation for an arbitrary polynomial of $H$ has been given in  Ref.~\cite{arad2013area}.
However, in the present case, the additional operators $\Phi_1$ and $\Phi_2$ prohibit the direct application of that results to \eqref{sup_Schmidt_edge_perturb}.
In the following lemma, we can obtain the modified version of the Schmidt rank estimation in Ref.~\cite{arad2013area}. 
For the generalization to high-dimensional systems in Sec.~\ref{sup_sec_Improved thermal area law in high dimensions} we consider the high-dimensional Hamiltonian~\eqref{sup_def:Hamiltonian_2D}.
\begin{lemma} \label{sup_fourth_lemma_schmidt}
For an arbitrary operator in the form of \eqref{sup_Schmidt_edge_perturb}, the Schmidt rank across the bi-partition of the system to the left and right at the point $i\in S$ is upper bounded by 
\begin{align}
{\rm SR} (\hat{\gunc}_{m,M},i) \le \min_{\tilde{\ell}: \tilde{\ell} \le \ell} \left[ \spin^{\tilde{\ell} |\partial \Lambda|}\br{10mM\locd}^{2M +2\tilde{\ell} + \frac{2kmM}{\tilde{\ell}}}\right] ,
\end{align}
where $\partial \Lambda$ and $\locd$ are defined in \eqref{sup_decomp_2D} and \eqref{sup_def:D_loc}, respectively.
If we consider one-dimensional Hamiltonian with two-body interactions ($k=2$), we have $|\partial \Lambda|=1$ and $\locd\le \spin$.
\end{lemma}

We can further extend Lemma~\ref{sup_fourth_lemma_schmidt} to the following operator:
 \begin{align}
\hat{\gunc}^{(p)}_{m,M}= [\Phi_1 \gunc_m(H_{S_1}) \gunc_m(H_{S_2}) \cdots  \gunc_m(H_{S_p}) \Phi_2 ]^M,
\notag
\end{align}
where $S_j \subseteq S$ ($j=1,2,\ldots,p$) with $|S|=\ell$. 
We then obtain the following corollary: 
\begin{corol} \label{sup_fourth_lemma_schmidt_corol}
For an arbitrary operator in the form of \eqref{sup_Schmidt_edge_perturb}, the Schmidt rank across the bi-partition of the system to the left and right at the point $i\in S$ is upper-bounded by 
\begin{align}
\label{sup_main_ineq_fourth_lemma_schmidt_corol}
&{\rm  SR} (\hat{\gunc}^{(p)}_{m,M},i)  \notag \\
&\le \min_{\tilde{\ell}: \tilde{\ell} \le \ell} \left[ \spin^{\tilde{\ell} |\partial \Lambda|}\br{10mM\locd}^{2pM +2p\tilde{\ell} + \frac{2pkmM}{\tilde{\ell}}}\right] ,
\end{align}
where $\partial \Lambda$ and $\locd$ are defined in \eqref{sup_decomp_2D} and \eqref{sup_def:D_loc}, respectively.
\end{corol}

\textit{Proof of Corollary~\ref{sup_fourth_lemma_schmidt_corol}.}
The proof is the same as that of Lemma~\ref{sup_fourth_lemma_schmidt}.
The difference is that the inequality~\eqref{sup_eq:polyinterSR} is replaced by
\begin{align}
&{\rm SR}(\hat{\gunc}^{(p)}_{m,M},i)  \notag \\
&\leq 
\min_{\tilde{\ell}:\tilde{\ell} \le \ell}
\left[ \spin^{\tilde{\ell} |\partial \Lambda|} \prod_{j=1}^p {mM \choose \tilde{\ell}}^2 \max_{s\in [l_0]}{\rm SR_s}\left(\hat{\gunc}_{m,M}^{(j),\leq \frac{mM}{l_0},s}\right) \right ],   
\notag
\end{align}
where $\hat{\gunc}_{m,M}^{(j),\leq \frac{mM}{l_0},s}$ for the Hamiltonian $H_{S_j}$ is defined in the same way as $\hat{\gunc}_{m,M}^{\leq \frac{mM}{l_0},s}$ in \eqref{sup_eq:polyinterSR} for the Hamiltonian $H_S$.
We then obtain the same inequality as \eqref{sup_upp_ineq_for the Schmidt_main}, and prove the inequality~\eqref{sup_main_ineq_fourth_lemma_schmidt_corol}. 
This completes the proof. $\square$

\begin{figure}[bb]
\centering
{
\includegraphics[clip, scale=0.3]{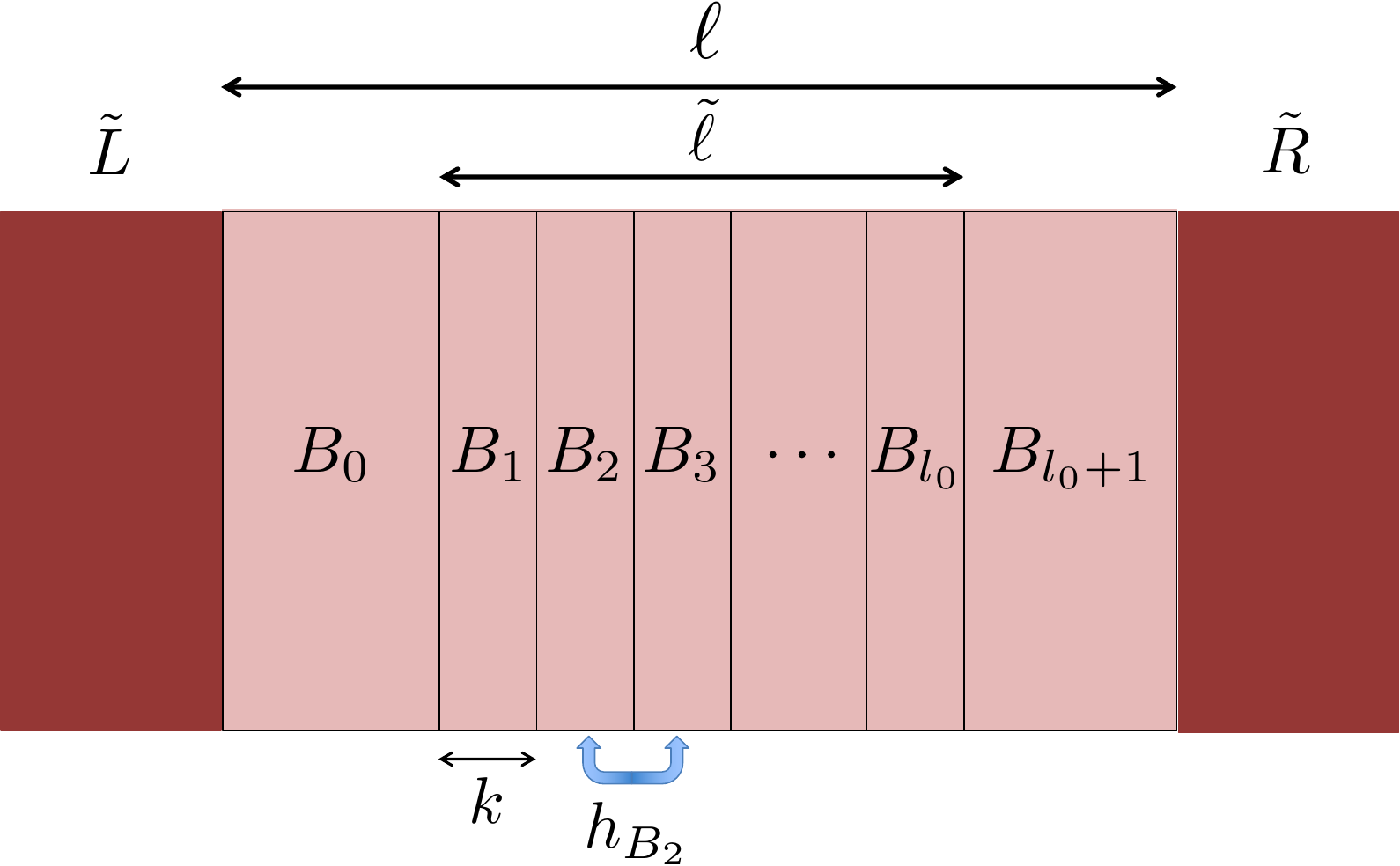}
}
\caption{The decomposition of the subset $S$ into blocks.}
\label{sup_fig:Schmidt_block}
\end{figure}

\subsubsection{Proof of Lemma~\ref{sup_fourth_lemma_schmidt}} 
We apply an analysis similar to that in Ref.~\cite{anshu2019entanglement}, which modified the proof in \cite{arad2013area} for the Schmidt rank estimation.

First, we decompose $S$ into $(l_0+2)$ blocks $\{B_s\}_{s=0}^{l_0+1}$ with $|B_s|= k$ ($s=1,2,\ldots, l_0$) and $l_0 = \tilde{\ell}/k$ (see Fig.~\ref{sup_fig:Schmidt_block}).
Here, $\tilde{\ell}$ is a control parameter such that $\tilde{\ell} \le \ell$.  
We then decompose the Hamiltonian $H_S$ as 
\begin{align}
H_S =h_{B_0} +h_{B_{l_0+1}}  + \sum_{s=1}^{l_0}h_{B_s}, 
\label{sup_hamiltonian_decomp}
\end{align}
where $h_{B_s}$ is comprised of the internal interactions in $B_s$ and block-block interactions between $B_s$ and $B_{s+1}$.
Note that the interaction length is at most $k$, and hence only adjacent blocks can interact with each other.
Also, from the inequality\eqref{sup_def:D_loc}, the Schmidt rank of $h_{B_s}$ is upper-bounded by $\locd =\spin^{\orderof{k}} |\partial \Lambda|$.

\begin{widetext}

We expand $\gunc_m(H_S)=\sum_{j=0}^m a_j (H_S)^j$ by using the decomposition~\eqref{sup_hamiltonian_decomp}. 
Using the polynomial interpolation argument in \cite{arad2013area}, it holds that (see \cite[Lemma 5.2,5.3]{anshu2019entanglement})
\begin{align}
\label{sup_eq:polyinterSR}
{\rm SR}(\hat{\gunc}_{m,M},i) \leq \min_{\tilde{\ell}:\tilde{\ell} \le \ell}\left[ {mM \choose \tilde{\ell}}^2\spin^{\tilde{\ell} |\partial \Lambda|}\max_{s\in [l_0]}{\rm SR_s}\left(\hat{\gunc}_{m,M}^{\leq \frac{mM}{l_0},s}\right) \right],   
\quad l_0 = \tilde{\ell}/k
\end{align}
where ${\rm SR_s}(\cdots)$ is the Schmidt rank across the bi-partition between $B_s$ and $B_{s+1}$. Also, the operator $\hat{\gunc}_{m,M}^{\leq \frac{mM}{l_0},s}$ is derived from $\hat{\gunc}_{m,M}$ by considering only those terms in which $h_{B_s}$ occurs at most $(mM/l_0)$ times. Let us $H_S= P + h_{B_s} + Q$, where $P$ is to the ``left'' of $h_{B_s}$ and $Q$ is to the ``right'' of $h_{B_s}$ and expand the powers $H_S$. 
From $[P,Q]=0$, any particular power $(H_S)^{T}$ is a linear combination of the following terms:
$$
\br{P^{p_{1}}Q^{q_{1}}} h_{B_s} \br{P^{p_{2}}Q^{q_{2}}} h_{B_s} \ldots \br{P^{p_{T'-1}}Q^{q_{T'-1}}}h_{B_s}\br{P^{p_{T'}}Q^{q_{T'}}}
$$
with $\sum_{i=1}^{T'} \br{p_i+q_i}\leq T$ and $T'\leq T$. This allows us to expand $\hat{\gunc}_{m,M}$ as a linear combination of the following terms:
\begin{align}
\label{sup_eq:lcmonomials}
    &\Phi_1\br{P^{p_{1,1}}Q^{q_{1,1}}} h_{B_s} \br{P^{p_{1,2}}Q^{q_{1,2}}} h_{B_s} \ldots \br{P^{p_{1,T_1-1}}Q^{q_{1,T_1-1}}} h_{B_s} \br{P^{p_{1,T_1}}Q^{q_{1,T_1}}}\notag\\
    &\Phi_2\Phi_1\br{P^{p_{2,1}}Q^{q_{2,1}}} h_{B_s} \br{P^{p_{2,2}}Q^{q_{2,2}}} h_{B_s} \ldots \br{P^{p_{2,T_2-1}}Q^{q_{2,T_2-1}}} h_{B_s}\br{P^{p_{2,T_2}}Q^{q_{2,T_2}}}\notag\\
    & \ldots \notag\\ 
    &\Phi_2\Phi_1 \br{P^{p_{M,1}}Q^{q_{M,1}}} h_{B_s}\br{P^{p_{M,2}}Q^{q_{M,2}}} h_{B_s} \ldots \br{P^{p_{M,T_M-1}}Q^{q_{M,T_M-1}}}h_{B_s} \br{P^{p_{M,T_M}}Q^{q_{M,T_M}}} \Phi_2.
\end{align}
\end{widetext}
Above, the positive integers $T_i$ and the powers $p_{i,k}, q_{i,k}\geq 0$ are such that
\begin{align}
    \sum_{i=1}^M\sum_{k=1}^{T_i} (p_{i,k} + q_{i,k}) \leq mM,
    \label{sup_eq:sumofpowers}
\end{align}
since the total degree is $mM$. But, recall that we are interested in $\hat{\gunc}_{m,M}^{\leq \frac{mM}{l_0}, s}$ where $h_{B_s}$ occur at most $(mM/l_0)$ times, which enforces the following constraint
\begin{align*}
    & \sum_{i=1}^M (T_i-1) \leq \frac{mM}{l_0} \implies \sum_{i=1}^M T_i \leq \frac{mM}{l_0}+M. 
\end{align*}
The number of the combinations of positive integers $\{T_1, T_2, \ldots T_M\}$ satisfying $\sum_{i=1}^M T_i = r$ is smaller than $r$-multicombination from a set of $M$ elements, and hence is upper-bounded by
\begin{align*}
\multiset{M}{r} = {M+r-1 \choose r} \leq 2^{M+r-1}.
\end{align*}
Then, the combinations of positive integers $\{T_1, T_2, \ldots T_M\}$ satisfying $\sum_{i=1}^M T_i \le \frac{mM}{l_0}+M$ is smaller than
\begin{align*}
\sum_{r=0}^{\frac{mM}{l_0}+M}2^{M+r-1} \le 2^{2M+\frac{mM}{l_0}}.
\end{align*}

When a tuple $\{T_1, T_2, \ldots T_M\}$ is given, the number of the non-zero integers in $\{p_{i,k}, q_{i,k}\}_{i\in [M], k\in [T_i]}$ which appears in Eq.~\eqref{sup_eq:lcmonomials} is equal to 
\begin{align*}
\sum_{i=1}^M 2T_i \leq 2\left(\frac{mM}{l_0}+M\right) .
\end{align*} 
Therefore, for a fixed $\{T_1, T_2, \ldots T_M\}$, the number of the combinations of positive integers $\{p_{i,k}, q_{i,k}\}_{i\in [M], k\in [T_i]}$ satisfying Eq.~\eqref{sup_eq:sumofpowers}  is upper-bounded by $(mM)$-multicombination from a set of $(\sum_{i=1}^M 2T_i)$ elements:
\begin{align*}
&\multiset{\sum_{i=1}^M 2T_i }{mM} \leq \multiset{\frac{2mM}{l_0} +2M }{mM}  \notag \\
&= \binom{\frac{2mM}{l_0} +2M+mM-1}{mM} \le \br{5mM}^{\frac{2mM}{l_0}+2M}.
\end{align*}
For each of the non-zero integers in $\{p_{i,k}, q_{i,k}\}_{i\in [M], k\in [T_i]}$, the Schmidt rank of the expression in Eq. \eqref{sup_eq:lcmonomials}, across the cut between $B_s$ and $B_{s+1}$, is at most $\locd^{\frac{mM}{l_0}}$. 
It is because only $h_{B_s}$ increases the Schmidt rank across the cut between $B_s$ and $B_{s+1}$ and the number of $h_{B_s}$ appearing in Eq. \eqref{sup_eq:lcmonomials} is smaller than $(mM/l_0)$ from the definition of $\hat{\gunc}_{m,M}^{\leq \frac{mM}{l_0},s}$.
Therefore, we finally arrive at the inequality of  
\begin{align}
{\rm SR_s}\left(\hat{\gunc}_{m,M}^{\leq \frac{mM}{l_0},s}\right)
&\leq 2^{2M+\frac{mM}{l_0}}\br{5mM}^{\frac{2mM}{l_0}+2M} \locd^{\frac{mM}{l_0}} \notag \\
&\leq \br{10mM\locd}^{2M + \frac{2mM}{l_0}} \notag \\
&= \br{10mM\locd}^{2M + \frac{2kmM}{\tilde{\ell}}}  \notag ,
\end{align}
where we use $l_0 = \tilde{\ell}/k$ in the last equation.
By applying the above inequality to~\eqref{sup_eq:polyinterSR}, we obtain
\begin{align}
\label{sup_upp_ineq_for the Schmidt_main}
 &{mM \choose \tilde{\ell}}^2 \spin^{\tilde{\ell}| \partial \Lambda|}\max_{s\in [l_0]}{\rm SR_s}\left(\hat{\gunc}_{m,M}^{\leq \frac{mM}{l_0},s}\right)  \notag \\
 & \leq {mM \choose \tilde{\ell}}^2 \spin^{\tilde{\ell} |\partial \Lambda|}\br{10mM\locd}^{2M + \frac{2kmM}{\tilde{\ell}}} \notag \\
 &\leq \spin^{\tilde{\ell} |\partial \Lambda|}\br{10mM\locd}^{2M +2\tilde{\ell} + \frac{2kmM}{\tilde{\ell}}} ,
\end{align}
where we use ${mM \choose \ell}^2\le (mM)^{2\tilde{\ell}}$.
This completes the proof.
$\square$

{~}

\hrulefill{\bf [ End of Proof of Lemma~\ref{sup_fourth_lemma_schmidt}] }

{~}

In considering one-dimensional systems in Lemma~\ref{sup_fourth_lemma_schmidt} with $\tilde{\ell}=\ell$, we have
\begin{align}
{\rm SR} (\hat{\gunc}_{m,M}) 
&\le \spin^{\ell}\br{10mM \spin^k}^{2M +2\ell + \frac{2kmM}{\ell}} \notag \\
& \le \br{10mM \spin^k}^{2M +3\ell + \frac{2kmM}{\ell}} ,
\end{align}
because of $|\partial \Lambda|=1$ and $\locd \le \spin^k$ [see also the inequality~\eqref{sup_locd_1D}].
By applying the above inequality to 
$\tilde{\rho}_{0}$ in Eq.~\eqref{sup_def_tilde_rho_0_0} with $M=2q$ [Eq.~\eqref{sup_eq:finalapproximation}], $m=\tilde{c}_2'  \sqrt{\nu\beta_0\ell}$ [Eq. \ref{sup_eq:finalchoiceofm}] and $\ell=\tilde{c}_1 \nu \beta_0 =\tilde{c}_1 \nu \beta/(2q)$ [Eq.~\eqref{sup_eq:ellalternatechoice}], 
we obtain 
\begin{align}
{\rm SR} (\tilde{\rho}_{0}^{2\M}) &\le 
\br{20mq \spin^k}^{4q +3\tilde{c}_1 \nu \beta/(2q) + 4\tilde{c}_2'kq\sqrt{\nu\beta_0/\ell} } \notag \\
&=\br{20mq \spin^k}^{\left(4 + 4k \tilde{c}_2'/\sqrt{\tilde{c}_1}\right)  q +(3/2)\tilde{c}_1 \beta \nu /q} . 
\label{sup_Schmidt_rank_of_tilde_rho}
\end{align}
Now, we specify the choice of $q$ by solving for
\begin{align}
q^2 = \beta \nu = \beta \max \left(\log(2/\delta), \frac{\beta}{2q} \right),
\end{align}
where we use the definition of $\nu$ in Eq.~\eqref{sup_def:nu_para}.
This gives the result of
\begin{align}
q \propto \max \left(\beta^{2/3}, [\beta\log(2/\delta)]^{1/2} \right),
\end{align}
where we choose $q$ appropriately so that the condition~\eqref{sup_basic_additional_assumptions} may be satisfied (i.e., $\beta \le q^2$). 
From $\delta = \epsilon/(3eq) = \orderof{\epsilon/\beta}$, by applying the notation of $q^\ast_\epsilon$ in Eq.~\eqref{def_q_ast} to \eqref{sup_Schmidt_rank_of_tilde_rho}, we finally obtain 
\begin{align}
{\rm SR} (\tilde{\rho}_0^{2q}) &\le e^{q^\ast_\epsilon \log (q^\ast_\epsilon)}.
\end{align}
This completes the proof of Proposition~\ref{main_thm_area_law}. $\square$

%


\section{Proof of Theorem~\ref{main_thm_area_law2} in high dimensional cases} \label{sup_sec_Improved thermal area law in high dimensions}
We here prove the improved thermal area law for high-dimensional Hamiltonians~\eqref{sup_def:Hamiltonian_2D}.  

\subsection{Restatement}
For the convenience of the reader, we restate the statement in the form of the following theorem:
{~}\\
\begin{theorem} 
Let us consider $\sd$-dimensional lattice and a vertical cut of the total system: $\Lambda=L \sqcup R$ with $L=\Lambda_1\sqcup\Lambda_2\sqcup \cdots \sqcup \Lambda_i$ and $L=\Lambda_{i+1}\sqcup\Lambda_{i+2}\sqcup \cdots \sqcup \Lambda_{l_\Lambda}$, where we use the notation in Eq.~\eqref{sup_decomp_2D}.
Then, we obtain the improved area law for the mutual information as follows:
\begin{align}
\label{sup_ineq_high_dim_improved_thermal}
I(L:R)_{\rho_\beta} \le C |\partial \Lambda|  \beta^{2/3} \log^{2/3} (\beta |\partial \Lambda|),
\end{align} 
where $C$ is a constant which depends on $k$, $g$, $\spin$ and $\sd$.  
\end{theorem}

\noindent
{\bf Remark.} 
The above upper bound is qualitatively better than the established thermal area law of $I(L:R)_{\rho_\beta} \lesssim   \beta  |\partial \Lambda|$ for $\beta \gtrsim \log^2(|\partial \Lambda|)$.
For the simplicity, we here consider a vertical cut of the total system, but the generalization to rectangular cut is straightforward. 

We notice that the logarithmic correction originates from the super-exponential dependence of $m$ in Lemma~\ref{sup_fourth_lemma_schmidt}.
If we can improve the $m$-independence in Lemma~\ref{sup_fourth_lemma_schmidt} 
\begin{align}
\br{10mM\locd}^{2M +2\tilde{\ell} + \frac{2kmM}{\tilde{\ell}}} \to ({\rm const.} )^{2M +2\tilde{\ell} + \frac{2kmM}{\tilde{\ell}}}, \notag 
\end{align}
we can prove the improved area law in the form of $I(L:R)_{\rho_\beta} \le C |\partial \Lambda|  \beta^{2/3}$.

\subsection{High-level overview} \label{sup_sec_High-level overview}

We, in the following, restrict ourselves to the inverse temperature such that
\begin{align}
\beta \ge \log^2(|\partial \Lambda|),
\end{align}
since the regime of $\beta < \log^2(|\partial \Lambda|)$ in \eqref{sup_ineq_high_dim_improved_thermal} has been already covered by the previous thermal area law~\cite{PhysRevLett.100.070502}. 

The proof strategy is very close to that in one-dimensional case.
We here relabel each of the sites such that $L=\{\Lambda_i\}_{i\le \ell/2}$ and $R=\{\Lambda_i\}_{i\ge \ell/2+1}$ (see Eq.~\eqref{sup_decomp_2D} for the definition of $\Lambda_i$), where the length $\ell$ is an integer which is multiple of $4$ to be chosen later. 
We then decompose the total system into three pieces $L_0$, $S$ and $R_0$ (see Fig.~\ref{sup_fig:system}), where $L_0=\{\Lambda_i\}_{i \le 0}$, $S=\{\Lambda_i\}_{1\le i \le \ell}$ and $R_0=\{\Lambda_i\}_{i\ge \ell+1}$.
Accordingly, we also decompose the Hamiltonian as follows:
\begin{align}
&H= H_S + H_{L_0} +H_{R_0} + v_0+v_\ell ,   \quad H_S :=\sum_{X: X\subset S} h_X , \notag \\
&H_{L_0} :=\sum_{X: X\subset L_0} h_X, \quad H_{R_0} :=\sum_{X: X\subset R_0} h_X ,
\end{align}
where $v_i$ is defined in Eq.~\eqref{sup_def_v_i_high_D}.
We note that $H_S$, $H_{L_0}$ and $H_{R_0}$ commute with each other. 
As in the one dimensional case, by shifting the energy origin appropriately, we set 
\begin{align}
&H_S \succeq 0, \label{sup_assump_H_S_high}
\end{align}
where $\succeq$ means that $H_S$ is positive semidefinite,
We divide $\beta$ into $2\M$ pieces ($\M\in \mathbb{N}$) and introduce
\begin{align}
\rho_{0} := e^{-\beta_0 H}, \quad \beta_0:=\beta/(2\M) .
\end{align}

The first difference from the one dimensional case is that we cannot derive Lemma~\ref{sup_lemma_rho_belief} as in the case of 1D, since we cannot utilize the belief propagation technique \cite{PhysRevB.76.201102} in high-dimensional systems. In high-dimensional cases, the operator $\phi(\tau)$ in Eq.~\eqref{sup_Def:Phi_0_phi} has the norm of $\orderof{\beta_0 |\partial \Lambda|}$, while in one dimensional case, it has the norm of $\orderof{\beta_0}$. This fact reduces the approximation error in \eqref{sup_Ineq:lemma_rho_belief} to $e^{-\orderof{\ell/\beta_0} + \orderof{\beta_0 |\partial \Lambda|}} \|\rho_0\|_{p_0}$ in high-dimensional systems.
Hence, we need to choose $\ell = \orderof{\beta_0^2 |\partial \Lambda|}$ to ensure a good approximation error, but this is too large to be utilized in the derivation of the improved thermal area law. 

In order to overcome this difficulty, we choose $q=\orderof{\beta}$ such that 
\begin{align}
\beta_0:=\beta/(2\M) \le \frac{1}{32gk}.
\end{align}
As shown in Lemma~\ref{sup_lemma_rho_belief_high} below, this condition allows us to construct the operator $\tilde{\rho}_{0}$ as in \eqref{sup_eq:rho0approximation} (i.e., $\|\rho_{0}- \tilde{\rho}_{0}\|_{2\M p}\le \delta \|\rho_{0}\|_{2\M p} $) in the following form:
\begin{align}
\label{sup_eq:first_approximation_high}
&\tilde{\rho}_{0}:=\tilde{\Phi}_0 e^{-\beta_0 (H_{L_0} + H_{R_0})} \func_m(\beta_0 H_S), \notag  \\
&\tilde{\Phi}_0:= \Phi_{L_1} \otimes\Phi_{R_1},
\end{align}
where $\Phi_{L_1}$ and $\Phi_{R_1}$ are operators supported on $L_1$ and $R_1$, respectively (i.e., $L_1=\{\Lambda_i\}_{i\le \ell/4}$ and $L_1=\{\Lambda_i\}_{i\ge 3\ell/4+1}$), and the degree $m$ polynomial $\func_m(x)$ approximates the exponential function $e^{-\beta_0 x}$. 
As in the inequality~\eqref{sup_e^-beta H-tilde_rho_0^2M|_p}, this operator gives the approximation 
\begin{align}
\label{sup_approximation_error_M_p}
\| e^{-\beta H}- \tilde{\rho}_0^{2\M} \|_{p} \le \epsilon\|e^{-\beta H}\|_p \quad {\rm with} \quad \epsilon:=3e\M \delta.
\end{align}

The mutual information is roughly determined by the upper bound of the Schmidt rank of $\tilde{\rho}_0^{2\M}$  which is given by
\begin{align}
\label{sup_eq:finalapproximation_high}
\left[ \tilde{\Phi}_0 e^{-\beta_0 (H_{L_0} + H_{R_0})} \func_m(\beta_0H_S) \right]^{2\M} . 
\end{align}
The Schmidt rank of the above operator is $(m\M)^{\orderof{\M} + \orderof{\ell |\partial \Lambda|} + \orderof{m \M/\ell}}$ from Lemma~\ref{sup_fourth_lemma_schmidt}.
In the one-dimensional case, for $\M=\orderof{\beta}$, this estimation gives the Schmidt rank of $e^{\orderof{\beta}}$ and spoils the improved thermal area law. 
However, in high-dimensional systems, the contribution of $e^{\orderof{\beta}}$ is much smaller than $e^{\beta^{2/3}|\partial \Lambda| }$ as long as $\beta\le |\partial \Lambda|^3$. 
Therefore, it is still possible to derive an improved area law from the approximation by \eqref{sup_eq:finalapproximation_high}. 
This point is the second difference between 1D case and high-dimensional cases.

In the proof of the area law, we roughly choose (see Appendix~\ref{completing_high_D_improved_area_law} in more details)
\begin{align}
m \approx |\partial \Lambda| \sqrt{\ell}, \quad \M\approx\beta , \quad  \ell \approx \beta^{2/3}, 
\end{align}
which gives the Schmidt rank of the operator~\eqref{sup_eq:finalapproximation_high} as 
$(m\M)^{\orderof{\M} + \orderof{\ell|\partial \Lambda|} + \orderof{m \M/\ell}}\approx \exp[\beta^{2/3} |\partial \Lambda|  \log(\beta|\partial \Lambda|)]$.
We thus obtain the inequality~\eqref{sup_ineq_high_dim_improved_thermal}. 

In the following, we show how the basic lemmas in one-dimensional case are extended to the high-dimensional cases.

\subsection{Approximation of $\rho_0$ of \eqref{sup_eq:first_approximation_high}}

We relate the two operators $\rho_0=e^{-\beta_0 H}$ and $e^{-\beta_0 H_0}$. 
We can formally write 
\begin{align}
\rho_0 = \Phi_0 e^{-\beta_0 H_0}   ,
\end{align}
where $\Phi_0=e^{-\beta_0 H} e^{\beta_0 H_0} $ is usually highly non-local operator. 
The lemma below ensures that the $\Phi_0$ is approximated by an operator supported on $L_1 \sqcup R_1$:
\begin{lemma} \label{sup_lemma_rho_belief_high}
Let $\tilde{H}$ and $\tilde{H}_0$ be Hamiltonians as follows:
\begin{align}
&\tilde{H}= H_{L_1} + H_{R_1} , \notag \\
& \tilde{H}_0= H_{L_1} + H_{R_1} - v_0 - v_\ell.
\end{align}
We then define 
\begin{align}
&\tilde{\Phi}_0:= \tilde{\Phi}_{L_1} \otimes \tilde{\Phi}_{R_1} = e^{-\beta_0\tilde{H}} e^{\beta_0\tilde{H}_0},\notag \\ 
&\tilde{\Phi}_{L_1} :=e^{-\beta_0 H_{L_1}} e^{\beta_0 (H_{L_1} - v_0)},\notag \\ 
&\tilde{\Phi}_{R_1} :=e^{-\beta_0 H_{R_1}} e^{\beta_0 (H_{R_1} - v_\ell)}. 
\label{sup_def_tilde_phi_high}
\end{align}
Then, for $\beta _0\le 1/(32gk)$ and $\ell \ge 2k \log(|\partial \Lambda|)$, the approximated operator
\begin{align}
\rho'_{0} =\tilde{\Phi}_0 \rho_0 =\tilde{\Phi}_0e^{-\beta_0 H_0} ,
\label{sup_def_rho_0'_high}
\end{align}
satisfies 
\begin{align}
\| \rho'_{0} - \rho_{0}\|_{p_0} \le 2|\partial \Lambda|  e^{-\ell /(2k) }  \|\rho_0\|_{p_0} \label{sup_Ineq:lemma_rho_belief_high}
\end{align}
for arbitrary $p_0\in\mathbb{N}$.
\end{lemma}
As has been mentioned in the previous subsection, this decomposition has an advantage over the belief propagation method used in Lemma~\ref{sup_lemma_rho_belief}. By using the decomposition~\eqref{sup_def_rho_0'_high}, we can achieve the upper bound of $|\partial \Lambda|  e^{-\orderof{\ell }}$ instead of $e^{-\orderof{\ell} + \orderof{|\partial \Lambda|}}$.


\subsubsection{Proof of Lemma~\ref{sup_lemma_rho_belief_high}}


We first define $V$ as $V:=H- H_0=\tilde{H} - \tilde{H}_0$, namely
\begin{align}
V=  v_0 + v_\ell.
\end{align}
We also define $V':= H_0-\tilde{H}_0$.

We here aim to prove 
\begin{align}
\|  e^{-\beta_0 \tilde{H}} e^{\beta_0 \tilde{H}_0} e^{-\beta_0 H_0} e^{\beta_0 H} -1  \| \le 2|\partial \Lambda|  e^{-\ell /(2k) }.
\label{sup_main_ineq_proof_of_lemma_high_d}
\end{align}
When we obtain the above upper bound, we arrive at the main inequality as follows:
\begin{align}
&\|\tilde{\Phi}_0 \rho'_{0} - \rho_0  \| \notag \\
&=\| e^{-\beta_0 \tilde{H}} e^{\beta_0 \tilde{H}_0}e^{-\beta_0 H_0}   -e^{-\beta_0 H} \|_p  \notag \\
&= \| e^{-\beta_0 \tilde{H}} e^{\beta_0 \tilde{H}_0} e^{-\beta_0 H_0} e^{\beta_0 H} e^{-\beta_0 H}   -e^{-\beta_0 H} \|_p \notag \\
&\le \| e^{-\beta_0 \tilde{H}} e^{\beta_0 \tilde{H}_0} e^{-\beta_0 H_0} e^{\beta_0 H} -1 \| \cdot \| e^{-\beta_0 H} \|_p \notag \\
&\le 2|\partial \Lambda|  e^{-\ell /(2k) } \| e^{-\beta_0 H} \|_p,
\end{align}
where the first inequality comes from H\"older's inequality~\eqref{sup_generalized_Holder_ineq_operator_norm}.

\begin{widetext}
In order to derive the inequality~\eqref{sup_main_ineq_proof_of_lemma_high_d}, we define $\mathcal{G}(\tau)$ as 
\begin{align}
\mathcal{G}(\tau):=e^{-\tau \tilde{H}} e^{\tau \tilde{H}_0}  e^{-\tau H_0} e^{\tau H} .   \label{sup_Def:mathcal_G_tau}
\end{align}
We then obtain 
\begin{align}
\frac{d}{d\tau}\mathcal{G}(\tau)&= e^{-\tau \tilde{H}} (\tilde{H}_0- \tilde{H}) e^{\tau \tilde{H}}\mathcal{G}(\tau) + 
e^{-\tau \tilde{H}} e^{\tau \tilde{H}_0}  e^{-\tau H_0} (H- H_0)e^{\tau H_0} e^{-\tau \tilde{H}_0}  e^{\tau \tilde{H}} \mathcal{G}(\tau)  \notag \\
&= - e^{-\tau \tilde{H}} \left( V -e^{\tau \tilde{H}_0}  e^{-\tau H_0}Ve^{\tau H_0} e^{-\tau \tilde{H}_0}  \right) e^{\tau \tilde{H}}\mathcal{G}(\tau),
\end{align}
where we use the definition of $H- H_0=\tilde{H} - \tilde{H}_0=V$.
The solution of the above differential equation is given by
\begin{align}
\mathcal{G}(\beta_0)= \mathcal{T} \exp \left( - \int_0^{\beta_0} e^{-\tau \tilde{H}} \left( V -e^{\tau \tilde{H}_0}  e^{-\tau H_0}Ve^{\tau H_0} e^{-\tau \tilde{H}_0}  \right) e^{\tau \tilde{H}} d\tau \right) ,
\end{align}
where $\mathcal{T}$ is the ordering operator.
From the above equation, we obtain the upper bound of $\| e^{-\beta_0 \tilde{H}} e^{\beta_0 \tilde{H}_0} e^{-\beta_0 H_0} e^{\beta_0 H} -1 \| =\| \mathcal{G}(\beta_0) -1 \| $ as 
\begin{align}
\| \mathcal{G}(\beta_0) - 1 \| \le \exp \left( \int_0^{\beta_0} \left\| e^{-\tau \tilde{H}} \left( V -e^{\tau \tilde{H}_0}  e^{-\tau H_0}Ve^{\tau H_0} e^{-\tau \tilde{H}_0}  \right) e^{\tau \tilde{H}} \right\| d\tau \right) -1 .
\label{sup_mathcal_G_beta_-1}
\end{align}
We can prove the following upper bound (see below for the proof): 
 \begin{align}
\left \|  e^{-\tau \tilde{H}} \left( V -e^{\tau \tilde{H}_0}  e^{-\tau H_0}Ve^{\tau H_0} e^{-\tau \tilde{H}_0}  \right) e^{\tau \tilde{H}} \right\| 
\le  \frac{g k  |\partial \Lambda|}{2} 8^{-s_\ell } ,\quad s_\ell:= \frac{\ell}{4k}-2
\label{sup_imaginary_time_evo_norm_V_0}
\end{align}
For $\ell \ge 2k \log(|\partial \Lambda|)$, we have 
 \begin{align}
\frac{g k \beta |\partial \Lambda|}{2} 8^{-s_\ell } \le  |\partial \Lambda| 8^{-\ell/(4k)} \le  |\partial \Lambda| e^{-\ell/(2k)} \le 1 ,
\end{align}
where we use $\beta_0 \le 1/(32gk)$ and $e^2<8$.
We thus use the inequality~\eqref{sup_imaginary_time_evo_norm_V_0} to reduce the inequality~\eqref{sup_mathcal_G_beta_-1} to
\begin{align}
\| \mathcal{G}(\beta_0) - 1 \| \le  \exp\left ( \frac{g k \beta_0 |\partial \Lambda|}{2} 8^{-s_\ell } \right) -1 \le  g k \beta_0  |\partial \Lambda|  8^{-s_\ell } \le 2|\partial \Lambda|  8^{-\ell /(4k) } \le 
2|\partial \Lambda|  e^{-\ell /(2k) } ,
\end{align}
where we use $e^x -1 \le 2x$ for $0\le x\le 1$ and $gk\beta_0 \le 1/32$.
This completes the proof. $\square$

\subsubsection{Proof of the inequality~\eqref{sup_imaginary_time_evo_norm_V_0}}
We start from the following equation:
\begin{align}
V -e^{\tau \tilde{H}_0}  e^{-\tau H_0}Ve^{\tau H_0} e^{-\tau \tilde{H}_0}  
  =&- \int_0^\tau  \frac{d}{dx} \left(e^{x \tilde{H}_0}  e^{-x H_0}Ve^{x H_0} e^{-x \tilde{H}_0}  \right) dx \notag \\
=&- \int_0^\tau \left(e^{x \tilde{H}_0}  e^{-x H_0} [ e^{x H_0} \tilde{H}_0e^{-x H_0} - H_0,   V  ] e^{x H_0} e^{-x \tilde{H}_0}  \right) dx \notag \\ 
=&\int_0^\tau  \left(e^{x \tilde{H}_0}  e^{-x H_0} [ e^{x H_0} V' e^{-x H_0} ,   V  ] e^{x H_0} e^{-x \tilde{H}_0}  \right) dx ,
\end{align}
where we use the definition $ V':=  H_0-\tilde{H}_0$. 
This yields
 \begin{align}
\left \|  e^{-\tau \tilde{H}} \left( V -e^{\tau \tilde{H}_0}  e^{-\tau H_0}Ve^{\tau H_0} e^{-\tau \tilde{H}_0}  \right) e^{\tau \tilde{H}} \right\| 
\le &\int_0^\tau  \left\| e^{-\tau \tilde{H}}  e^{x \tilde{H}_0}  e^{-x H_0} [ e^{x H_0} V' e^{-x H_0} ,   V  ] e^{x H_0} e^{-x \tilde{H}_0} e^{\tau \tilde{H}}\right\| dx\notag \\
=&\int_0^\tau  \left\| e^{-\tau \tilde{H}}  e^{x \tilde{H}_0} [  V' ,   e^{-x H_0} V e^{x H_0}  ]  e^{-x \tilde{H}_0} e^{\tau \tilde{H}}\right\| dx . 
\label{sup_third_inequality_proof_1}
\end{align}
The commutator $ [  V' ,   e^{-x H_0} V e^{x H_0}  ] $ is decomposed by the Baker-Campbell-Hausdorff expansion:
\begin{align}
 [  V' ,   e^{-x H_0} V e^{x H_0}  ] &= \sum_{s=0}^\infty \frac{(-x)^s}{s!} \ad_{V'} \ad_{H_0}^s (V) .
\label{sup_third_inequality_proof_2}
\end{align}
Because the supports of $V$ and $V'$ are separated at least by a distance of $\ell/4 -2k$, we have 
\begin{align}
\ad_{V'} \ad_{H_0}^s (V) =0  \for s \le \frac{\ell/4 -2k}{k} =: s_\ell  .
\end{align}

Furthermore, we have 
 \begin{align}
e^{-\tau \tilde{H}}  e^{x \tilde{H}_0} [  V' ,   e^{-x H_0} V e^{x H_0}  ]  e^{-x \tilde{H}_0} e^{\tau \tilde{H}} 
=\sum_{m_2=0}^\infty \frac{(-\tau)^{m_2}}{m_2!} \sum_{m_1=0}^\infty \frac{x^{m_1}}{m_1!} \sum_{s= s_\ell+1}^\infty \frac{(-x)^s}{s!} \ad_{\tilde{H}}^{m_2} \ad_{\tilde{H}_0}^{m_1} \ad_{V'} \ad_{H_0}^s (V) .
\label{sup_inequality_imaginary_time_evo_ution_1}
\end{align}
From Lemma~\ref{sup_norm_multi_commutator1} or inequality~\eqref{sup_fundamental_ineq_0},  the norm of the multi-commutator is upper-bounded by
\begin{align}
\| \ad_{\tilde{H}}^{m_2} \ad_{\tilde{H}_0}^{m_1} \ad_{V'} \ad_{H_0}^s (h_X)  \| \le (2gk)^{m_1+m_2+s+1} (m_1+m_2+s+1)! \|h_X\| ,
\end{align}
where $h_X$ is supported on $X$ such that $|X|\le k$. 
Then, because of the definition of $v_i$ in Eq.~\eqref{sup_def_v_i_high_D}, we have 
\begin{align}
\| \ad_{\tilde{H}}^{m_2} \ad_{\tilde{H}_0}^{m_1} \ad_{V'} \ad_{H_0}^s (v_i)  \| 
&\le (2gk)^{m_1+m_2+s+1} (m_1+m_2+s+1)!  \sum_{X:X\cap \Lambda_{\le i} \neq \emptyset,X\cap\Lambda_{>i} \neq \emptyset } \|h_X\|  \notag \\
&\le (2gk)^{m_1+m_2+s+1} (m_1+m_2+s+1)! \sum_{j\in \Lambda_i \sqcup \Lambda_{i-1}\sqcup \cdots  \Lambda_{i-k+1} } \sum_{X: X \ni j} \|h_X\| \notag \\
&\le gk |\partial \Lambda| (2gk)^{m_1+m_2+s+1} (m_1+m_2+s+1)! ,
\label{sup_upper_bound_multi_commu_v_1_i}
\end{align}
where we use $|\Lambda_i| + |\Lambda_{i-1}|+\cdots + |\Lambda_{i-k+1}| \le  k |\partial \Lambda|$.
Because of $V=  v_0 + v_\ell$, we obtain 
\begin{align}
\| \ad_{\tilde{H}}^{m_2} \ad_{\tilde{H}_0}^{m_1} \ad_{V'} \ad_{H_0}^s (V)  \| 
\le |\partial \Lambda| (2gk)^{m_1+m_2+s+2} (m_1+m_2+s+1)! .
\end{align}
By using the inequality 
\begin{align}
\frac{ (m_1+m_2+s+1)! }{m_1! m_2! s!} = \frac{(m_1+1)!}{m_1! 1!} \frac{(m_1+m_2+1)!}{(m_1+ 1)!m_2!}  \frac{(m_1+m_2 + s+1)!}{(m_1+m_2+ 1)!m_2!} \le 8^{m_1+1} 4^{m_2} 2^s,
\end{align}
we have 
 \begin{align}
&\sum_{m_2=0}^\infty \frac{\tau^{m_2}}{m_2!} \sum_{m_1=0}^\infty \frac{x^{m_1}}{m_1!} \sum_{s= s_\ell+1}^\infty \frac{x^s}{s!} \| \ad_{\tilde{H}}^{m_2} \ad_{\tilde{H}_0}^{m_1} \ad_{V'} \ad_{H_0}^s (V) \|   \notag \\
\le &\sum_{m_2=0}^\infty \tau^{m_2}\sum_{m_1=0}^\infty x^{m_1}\sum_{s= s_\ell+1}^\infty x^s |\partial \Lambda| (2gk)^{m_1+m_2+s+2} \frac{(m_1+m_2+s+1)! }{m_1!m_2!s!}  \notag \\
\le &\frac{32g^2k^2  |\partial \Lambda|}{1-16gk x}\frac{1}{1-8gk\tau} \frac{(4gkx)^{s_\ell+1}}{1-4gkx}.
\end{align}
From $\max(\tau, x) \le \beta_0 \le 1/(32gk)$, the above inequality reduces to
 \begin{align}
\sum_{m_2=0}^\infty \frac{\tau^{m_2}}{m_2!} \sum_{m_1=0}^\infty \frac{x^{m_1}}{m_1!} \sum_{s= s_\ell+1}^\infty \frac{x^s}{s!} \| \ad_{\tilde{H}}^{m_2} \ad_{\tilde{H}_0}^{m_1} \ad_{V'} \ad_{H_0}^s (V) \|  \le &13 g^2 k^2  |\partial \Lambda|  8^{-s_\ell } .
\label{sup_inequality_imaginary_time_evo_ution_2}
\end{align}
By combining the inequalities~\eqref{sup_inequality_imaginary_time_evo_ution_1} and \eqref{sup_inequality_imaginary_time_evo_ution_2}, we obtain
 \begin{align}
\| e^{-\tau \tilde{H}}  e^{x \tilde{H}_0} [  V' ,   e^{-x H_0} V e^{x H_0}  ]  e^{-x \tilde{H}_0} e^{\tau \tilde{H}} \|  \le13 g^2 k^2  |\partial \Lambda|  8^{-s_\ell }.
\end{align}
From the inequality~\eqref{sup_third_inequality_proof_1}, we thus prove the inequality of 
 \begin{align}
\left \|  e^{-\tau \tilde{H}} \left( V -e^{\tau \tilde{H}_0}  e^{-\tau H_0}Ve^{\tau H_0} e^{-\tau \tilde{H}_0}  \right) e^{\tau \tilde{H}} \right\| 
\le  13 g^2 k^2 \tau |\partial \Lambda|  8^{-s_\ell }\le \frac{g k  |\partial \Lambda|}{2} 8^{-s_\ell } ,
\end{align}
where we use $\tau \le \beta_0 \le 1/(32gk)$ in the second inequality.
This completes the proof. $\square$

 {~}

\hrulefill{\bf [ End of Proof of Lemma~\ref{sup_lemma_rho_belief_high}] }

{~}
\end{widetext}

The lemma implies that as the length $\ell$ becomes large, the approximation error decays exponentially with $e^{-\orderof{\ell}}$.
Thus, in order to achieve the inequality
\begin{align}
\| \rho'_0 - \rho_0\|_{2\M p} \le \frac{\delta}{2}\|\rho_0\|_{2\M p},
\label{sup_first_approx_high}
\end{align}
we need to choose $\ell$ as 
\begin{align}
\ell\ge 2k \log (4|\partial \Lambda| / \delta).
\label{sup_ell_c_2_c_1_beta_0_high}
\end{align}

We approximate $e^{- \beta_0 H_0}$ by an operator with small Schmidt rank. 
For this purpose, we use the fact that $H_S$, $H_{L_0}$ and $H_{R_0}$ commute with each other, and write $e^{- \beta_0 H_0}=e^{- \beta_0 (H_{L_0}+H_{R_0})}e^{-\beta_0 H_S}$. Then, we approximate $e^{-\beta_0 H_S}$ by using the polynomial of $H_S$ in Lemma~\ref{sup_lem:low_deg_poly}.
By using the polynomial $\func_m(x)$ defined there, we approximate the operator $\rho'_0$ in Eq.~\eqref{sup_def_rho_0'_high} by
\begin{align}
\tilde{\rho}_0 :=\tilde{\Phi}_0 e^{- \beta_0 (H_{L_0}+H_{R_0})}\func_m(\beta_0 H_S) .
\label{sup_def_tilde_rho_0}
\end{align}
Because of \eqref{sup_assump_H_S}, the spectrum of $\beta_0 H_S$ is included in the span of $[0,\beta_0\|H_S\|]$, and hence the inequality~\eqref{sup_ineq_delta_poly_approx} gives
 \begin{align}
 \label{sup_func_m(beta_0 H_S) - e^-beta_0 H_S}
\|\func_m(\beta_0 H_S) - e^{-\beta_0 H_S} \| \le \delta_0 .
\end{align} 
by choosing $m$ appropriately following Lemma~\ref{sup_lem:low_deg_poly}.
The next problem is to estimate the approximation error $\|\rho_0'-\tilde{\rho}_0\|_{p_0}$ for arbitrary Schatten $p_0$-norm.  
We prove the following lemma, which is similar to Lemma~\ref{sup_third_lemma_approx_est}:
\begin{lemma} \label{sup_third_lemma_approx_est_high}
Let $p_0\in\mathbb{N}$ and $\delta_0\in (0,1)$. Under the choice of $\Phi_{L_1} \otimes\Phi_{R_1}$ in Lemma \ref{sup_lemma_rho_belief}, $\ell$ in Eq.~\eqref{sup_ell_c_2_c_1_beta_0_high} and $m, \func_m(x)$ in Lemma \ref{sup_lem:low_deg_poly}, we have
 \begin{align}
\|\rho_0'-\tilde{\rho}_0\|_{p_0} \le \mathcal{D}_S^{1/p_0}\delta_0 e^{|\partial \Lambda|/7} \left \| \rho_0\right\|_{p_0} 
\label{sup_main_ineq:third_lemma_approx_est_high}
\end{align}
for $\beta_0 \le 1/(32gk)$.
\end{lemma}

\subsubsection{Proof of Lemma~\ref{sup_third_lemma_approx_est_high}}
From the definitions~\eqref{sup_def_rho_0'_high} and \eqref{sup_def_tilde_rho_0} of $\rho_0'$ and $\tilde{\rho}_0$, respectively, we start from the inequality of 
 \begin{align}
&\|\rho_0'-\tilde{\rho}_0\|_{p_0}   \notag \\
&=\left\| \tilde{\Phi}_0  e^{- \beta_0 (H_{L_0}+H_{R_0})} \left( \func_m(H_S) -e^{-\beta_0 H_S}\right)\right\|_{p_0} \notag \\
&\le \left \| e^{- \beta_0 (H_{L_0}+H_{R_0})} \left( \func_m(H_S) -e^{-\beta_0 H_S}\right)\right\|_{p_0} \cdot \| \tilde{\Phi}_0\| , 
\label{sup_rho_0'_minus_tilde_rho_0_high}
\end{align}
where we used H\"older's inequality~\eqref{sup_generalized_Holder_ineq_operator_norm} in the second line.
From the definition~\eqref{sup_def_tilde_phi_high} of $\tilde{\Phi}_0$, we obtain the following upper bound (see below for the proof):
  \begin{align}
  \label{sup_norm_tilde_phi_high}
\| \tilde{\Phi}_0\|\le e^{|\partial \Lambda| /15}.
\end{align}

Next, we obtain the same inequality as \eqref{sup_third_lemma_p_norm_bound_1}, which gives the upper bound of
 \begin{align}
 &\left \| e^{- \beta_0 (H_{L_0}+H_{R_0})} \left( \func_m(H_S) -e^{-\beta_0 H_S}\right)\right\|_{p_0}^{p_0}  \notag \\
 &\le   \mathcal{D}_S\delta_0^{p_0}  \left \| e^{- \beta_0 (H_{L_0}+H_{R_0}+H_S)} \right\|_{p_0}^{p_0} .
\label{sup_third_lemma_p_norm_bound_1_high}
\end{align}
In order to estimate the upper bound of $\| e^{- \beta_0 (H_{L_0}+H_{R_0}+H_S)} \|_{p_0}$ in terms of $\| e^{- \beta_0 H} \|_{p_0}$, 
we use the Golden-Thompson inequality to derive
\begin{align}
\tr( e^{- p_0 \beta_0 H_0} )  
&\le \tr\br{e^{-p_0 \beta_0(v_0+v_\ell)} \cdot e^{-p_0 \beta_0 H}} \notag \\
&\le e^{2 p_0 \beta_0 gk |\partial \Lambda| }  \|e^{-\beta_0 H}\|_{p_0}^{p_0}  \notag \\
&\le e^{p_0 |\partial \Lambda| /16}  \|e^{-\beta_0 H}\|_{p_0}^{p_0} 
,
 \label{sup_third_lemma_p_norm_bound_2_high}
\end{align}
where we use $\tr( e^{- p_0 \beta_0 H_0} )  = \|e^{-\beta_0 H_0}\|_{p_0}^{p_0}$, $\beta_0 \le 1/(32 gk)$, and derive the upper bound of $\| v_i\|$ from the definition~\eqref{sup_def_v_i_high_D}
\begin{align}
\| v_i \| &\le  \sum_{X:X\cap \Lambda_{\le i} \neq \emptyset,X\cap\Lambda_{>i} \neq \emptyset } \|h_X\|    \notag \\
&\le \sum_{j\in \Lambda_i \sqcup \Lambda_{i-1}\sqcup \cdots  \Lambda_{i-k+1} } \sum_{X: X \ni j} \|h_X\|  \le gk |\partial \Lambda| . \notag
\end{align}

By combining the inequalities~\eqref{sup_third_lemma_p_norm_bound_1_high} and \eqref{sup_third_lemma_p_norm_bound_2_high}, we arrive at the inequality of 
 \begin{align}
& \left \| e^{- \beta_0 (H_{L_0}+H_{R_0})} \left( \func_m(H_S) -e^{-\beta_0 H_S}\right)\right\|_{p_0}^{p_0}  \notag \\
&\le \mathcal{D}_S\delta_0^{p_0}e^{p_0 |\partial \Lambda| /16} \left \| e^{- \beta_0 H} \right\|_{p_0}^{p_0} .  
\label{sup_rho_0'_minus_tilde_rho_0_second_term_high}
\end{align}
By applying the inequalities~\eqref{sup_norm_tilde_phi_high} and \eqref{sup_rho_0'_minus_tilde_rho_0_second_term_high} to \eqref{sup_rho_0'_minus_tilde_rho_0_high}, 
we obtain the main inequality~\eqref{sup_main_ineq:third_lemma_approx_est}. 
This completes the proof. $\square$

\subsubsection{Proof of the inequality~\eqref{sup_norm_tilde_phi_high}}

By using Eq.~\eqref{sup_def_tilde_phi_high}, we have
\begin{align}
\| \tilde{\Phi}_0 \| \le \left \| e^{-\beta_0 H_{L_1}} e^{\beta_0 (H_{L_1} - v_0)}\right \|\cdot \left\|e^{-\beta_0 H_{R_1}} e^{\beta_0 (H_{R_1} - v_\ell)}\right\|. \notag 
\end{align}
We here consider 
\begin{align}
&e^{-\beta_0 H_{L_1}} e^{\beta_0 (H_{L_1} - v_0)} \notag \\
&= \mathcal{T} \exp \left ( -\int_0^{\beta_0} e^{-xH_{L_1}} v_0 e^{xH_{L_1}} dx \right), \notag 
\end{align}
which gives rise to the inequality of
\begin{align}
&\left\|  e^{-\beta_0 H_{L_1}} e^{\beta_0 (H_{L_1} - v_0)} \right\|  \notag \\
&\le \exp \left ( \int_0^{\beta_0} \| e^{-xH_{L_1}} v_0 e^{xH_{L_1}} \| dx \right).
\label{sup_exp_H_L_1H_L1_v0}
\end{align}
We thus aim to derive the upper bound of $\| e^{-xH_{L_1}} v_0 e^{xH_{L_1}} \| $. 

By using the Baker-Campbell-Hausdorff expansion, we have 
\begin{align}
\| e^{-xH_{L_1}} v_0 e^{xH_{L_1}} \|  \le \sum_{m=0}^\infty \frac{x^m}{m!} \left\| \ad_{H_{L_1}}^m(v_0)\right\| .
\end{align}
By using Lemma~\ref{sup_norm_multi_commutator1} or the inequality~\eqref{sup_fundamental_ineq_0},  the norm of $\left\| \ad_{H_{L_1}}^m(v_0)\right\|$ is upper-bounded as follows:
\begin{align}
\left\| \ad_{H_{L_1}}^m(v_0)\right\| \le gk |\partial \Lambda| (2gk)^{m}m!,
\end{align}
where we use an analysis similar to \eqref{sup_upper_bound_multi_commu_v_1_i}. 
Hence, we calculate the upper bound of $\| e^{-xH_{L_1}} v_0 e^{xH_{L_1}} \|$ as 
\begin{align}
\| e^{-xH_{L_1}} v_0 e^{xH_{L_1}} \|  &\le \sum_{m=0}^\infty \frac{x^m}{m!} \cdot  gk |\partial \Lambda|  (2gk)^{m}m!  \notag \\
&= \frac{gk |\partial \Lambda| }{1-2gk x} \le \frac{16gk}{15} |\partial \Lambda| ,
\end{align}
where we use $x\le \beta_0 \le 1/(32gk)$. 
By applying this inequality to \eqref{sup_exp_H_L_1H_L1_v0}, we have 
\begin{align}
\left\|  e^{-\beta_0 H_{L_1}} e^{\beta_0 (H_{L_1} - v_0)} \right\| \le e^{\frac{16gk\beta_0}{15} |\partial \Lambda| } \le e^{\frac{1}{30} |\partial \Lambda| }.
\end{align}
We obtain the same inequality for $\left\|e^{-\beta_0 H_{R_1}} e^{\beta_0 (H_{R_1} - v_\ell)}\right\|$.
This completes the proof.

{~}

\hrulefill{\bf [ End of Proof of Lemma~\ref{sup_third_lemma_approx_est_high}]}

{~}

Let us substitute $p_0=2\M p$ in Lemma \ref{sup_third_lemma_approx_est_high} and choose $\delta_0$ such that satisfies 
 \begin{align}
 \mathcal{D}_S^{1/(2\M p)}\delta_0e^{|\partial \Lambda|/7} \le \frac{\delta}{2}.
 \label{sup_ineq_for_delta_0}
\end{align}
This ensures that $\|\rho_0'-\tilde{\rho}_0\|_{2\M p} \le (\delta/2)\left\| \rho_0\right\|_{2\M p}$ and we conclude 
 \begin{align}
\|\rho_0-\tilde{\rho}_0\|_{2\M p} &\le \|\rho_0-\rho_0'\|_{2\M p}  + \|\rho_0'-\tilde{\rho}_0\|_{2\M p}   \notag \\
&\le  \delta \|\rho_0\|_{2\M p},
\end{align}
where we use the inequality~\eqref{sup_first_approx_high}.

Let us simplify the expression for all the parameters appearing so far. 
We first consider 
 \begin{align}
\mathcal{D}_S = \spin^{|S|} \le e^{\ell |\partial \Lambda|  \log(\spin)},
\end{align}
and hence from \eqref{sup_ineq_for_delta_0} with $q=\beta/(2\beta_0)$ and $p\ge1$, we can choose $\delta_0$ as 
 \begin{align}
 \label{sup_def_nu'_eq}
\log(1/\delta_0) &=  \log(2/\delta) +   |\partial \Lambda| \left(\frac{1}{7} +  \beta_0 \log(\spin) \frac{\ell}{\beta}\right) \notag \\
&=
 |\partial \Lambda| \left( \frac{1}{7} +  \beta_0 \log(\spin) \frac{\ell}{\beta} + \frac{\log(2/\delta)}{ |\partial \Lambda| }      \right)   \notag \\
& = \nu' |\partial \Lambda| 
\end{align}
with
 \begin{align}
\nu':=\frac{1}{7} +  \beta_0 \log(\spin) \frac{\ell}{\beta} + \frac{\log(2/\delta)}{ |\partial \Lambda| } \le {\const.}\times \ell .
\end{align}

Also, the norm of the Hamiltonian $\beta_0 H_S$ is bounded from above by 
 \begin{align}
\beta_0\|H_S\| \le \beta_0 g |S| \le \beta_0 g \ell | \partial \Lambda | ,
\end{align}
where we used the definition~\eqref{sup_decomp_2D} of $| \partial \Lambda |$. 
Because of the above upper bound, we have 
 \begin{align}
 \label{sup_upper_bound/beta_0H_S_1_delta_0}
\max[\beta_0\|H_S\|, \log(1/\delta_0)]  &\le   | \partial \Lambda |  \max[ \beta_0 g \ell,\nu' ] \notag \\
&=\orderof{\ell | \partial \Lambda |}.
\end{align}
Hence, from the inequality~\eqref{sup_lower_m_lem:low_deg_poly} in Lemma~\ref{sup_lem:low_deg_poly},  
we obtain the following form of $m$ to achieve the inequality~\eqref{sup_func_m(beta_0 H_S) - e^-beta_0 H_S}:
 \begin{align}
 \label{sup_eq:finalchoiceofm_high}
m&=\left \lceil c_f \sqrt{\max[\beta_0\|H_S\|, \log(1/\delta_0)] \log(1/\delta_0)} \right \rceil   \notag \\
&= \tilde{c} | \partial \Lambda |  \sqrt{\nu' \ell} ,
\end{align}
where $\tilde{c}$ is a constant of $\orderof{1}$.

%
%

Finally, we apply Lemma~\ref{sup_fourth_lemma_schmidt} to $\tilde{\rho}_0^{2\M}$.
We have $\locd\le \spin^k |\partial \Lambda| $, and hence
\begin{align}
\label{sup_upper_bound_Schmidt_rank}
{\rm SR} (\hat{\gunc}_{m,M}) \le \min_{\tilde{\ell}:\tilde{\ell} \le \ell}
\left[\spin^{\tilde{\ell} |\partial \Lambda|}\br{10mM \spin^k |\partial \Lambda| }^{2M +2\tilde{\ell} + \frac{2kmM}{\tilde{\ell}}}
\right]. 
\end{align}
Under the choice of
\begin{align}
\label{choice_parameters_high_D}
&\hat{\gunc}_{m,M} =\tilde{\rho}_0^{2\M} ,\quad M=2\M=(\beta/\beta_0),\quad m= \tilde{c} | \partial \Lambda |  \sqrt{\nu' \ell}, \notag \\
&\ell \ge  2k \log (4|\partial \Lambda| / \delta) \ge \tilde{c}' \log (|\partial \Lambda| / \delta),
\end{align}
\begin{widetext}
we reduce the upper bound of \eqref{sup_upper_bound_Schmidt_rank} to
\begin{align}
\label{sup_high_D_Schmidt}
{\rm SR} (\tilde{\rho}_0^{2q}) &\le \min_{\tilde{\ell}:\tilde{\ell} \le \ell}
\left[  
\spin^{\tilde{\ell} |\partial \Lambda|} 
\br{\frac{10\tilde{c}\beta \spin^k |\partial \Lambda|^2\sqrt{\nu'\ell}}{ \beta_0} }^{2\beta/\beta_0 +2 \tilde{\ell} + 2\tilde{c}k (\beta/\beta_0)  |\partial \Lambda| \tilde{\ell}^{-1}\sqrt{\nu'\ell} }\right] .
\end{align}
\end{widetext}

\subsection{Choice of polynomial degree $m$ and region length $\ell$} \label{completing_high_D_improved_area_law}

We here consider how to choose the parameters $m$ and $\ell$. 
We assume $|R| \ge |L|$ ($\ge |\partial \Lambda|$) and choose $\delta$ as $\delta = 1/|L|^2$, 
and the condition for $\ell$ in \eqref{choice_parameters_high_D} reads 
\begin{align}
\label{sup_assumption_ell___1st}
\ell \ge 2k \log (4|\partial \Lambda| / \delta)   \ge  \tilde{c}_1 \log (|L|) , 
\end{align}
where $\tilde{c}_1$ is a constant which only depends on $g$, $k$ and $\sd$.
Then, under the condition of $\beta \ge \log^2 (|\partial \Lambda|) \propto \log^2 (|L|)$, we can choose $\ell$ such that 
\begin{align}
\ell \le  \tilde{c}_1' \beta . \label{sup_assumption_ell_2nd}
\end{align}
We then obtain the upper bound of $\nu'$ in \eqref{sup_def_nu'_eq} as 
 \begin{align}
\nu'=\frac{1}{7} +   \tilde{c}'_1   \beta_0 \log(\spin) + \frac{\log(2|L|^2)}{ |\partial \Lambda| }  \le \tilde{c}_2,
\end{align}
where $\tilde{c}_2$ is a constant which only depends on $g$, $k$, $\sd$ and $\spin$.

We here denote
 \begin{align}
\frac{10\tilde{c}\beta \spin^k |\partial \Lambda|^2\sqrt{\nu'\ell}}{ \beta_0}  \le e^{\tilde{c}_3 \log (\beta |\partial \Lambda|)}
\end{align}
with $\tilde{c}_3$ an $\orderof{1}$ constant.
\begin{widetext}
Then, the upper bound~\eqref{sup_high_D_Schmidt} is simplified as 
\begin{align}
\label{sup_high_D_Schmidt_specific} 
{\rm SR} (\tilde{\rho}_0^{2q}) 
&\le \min_{\tilde{\ell}:\tilde{\ell} \le \ell}
\left[  
e^{2\tilde{c}_3 (1/\beta_0 +\tilde{c}'_1)  \beta \log (\beta |\partial \Lambda|)}
\cdot
e^{\tilde{\ell} \log(\spin) |\partial \Lambda| + 2\tilde{c}\sqrt{\tilde{c}_2}k (\beta/\beta_0) \tilde{c}_3 \log (\beta |\partial \Lambda|) \tilde{\ell}^{-1} \ell^{1/2}   |\partial \Lambda| } \right] . \notag \\
&= e^{\tilde{c}_4 \beta \log (\beta |\partial \Lambda|)} \min_{\tilde{\ell}:\tilde{\ell} \le \ell}
\left[  e^{\tilde{\ell} \log(\spin) |\partial \Lambda| +\tilde{c}_5\beta \log (\beta |\partial \Lambda|) \tilde{\ell}^{-1} \ell^{1/2}   |\partial \Lambda| } \right] ,
\end{align}
where we define $\tilde{c}_4:=2\tilde{c}_3 (1/\beta_0 +\tilde{c}'_1)$ and $\tilde{c}_5:=2\tilde{c}\sqrt{\tilde{c}_2} \tilde{c}_3k/ \beta_0$.
\end{widetext}

In the above upper bound, we would like to choose 
\begin{align}
\tilde{\ell} =  \left \lceil \left( \frac{\tilde{c}_5}{\log(\spin) } \beta \log (\beta |\partial \Lambda|) \right)^{1/2} \ell^{1/4}\right \rceil .
\end{align}
In order that the choice above is consistent with $\tilde{\ell}\le \ell$, the length $\ell$ should satisfy
\begin{align}
\label{sup_condition_for_ell_should_satisfy}
\ell \ge  \left( \frac{\tilde{c}_5}{\log(\spin) } \beta \log (\beta |\partial \Lambda|) \right)^{2/3}.
\end{align}
We note that this choice of $\ell$ exists under the constraints of \eqref{sup_assumption_ell___1st} and \eqref{sup_assumption_ell_2nd} because of $\beta \ge \log^2(|\partial \Lambda|)$.
By applying the above choice of $\tilde{\ell}$ with \eqref{sup_condition_for_ell_should_satisfy} to the upper bound~\eqref{sup_high_D_Schmidt_specific},   
we finally arrive at the inequality
\begin{align}
\label{sup_final_upp_thermal_area_high_D}
&{\rm SR} (\tilde{\rho}_0^{2q})  \notag \\
&\le \exp\left[ \tilde{c}_4 \beta \log (\beta |\partial \Lambda|) + \tilde{c}_6|\partial \Lambda|  \beta^{2/3} \log^{2/3} (\beta |\partial \Lambda|) \right].
\end{align}
The inequality $\beta \log (\beta |\partial \Lambda|) \ge |\partial \Lambda|  \beta^{2/3} \log^{2/3} (\beta |\partial \Lambda|)$ holds for $\beta \gtrsim  |\partial \Lambda|^3$.
However, when $\beta=\orderof{|\partial \Lambda|^3}$, the upper bound gives $e^{|\partial \Lambda|^{3}}$ and is worse than the trivial upper bound $e^{\orderof{n}}$ because of $|\partial \Lambda|= \orderof{n^{\frac{\sd-1}{\sd}}}$. 
We thus conclude that the second term in \eqref{sup_final_upp_thermal_area_high_D} is more dominant than the first term.

We have chosen $\delta=1/|L|^2$ and hence the inequality~\eqref{sup_approximation_error_M_p} for $p=1$ ensures 
\begin{align}
\| e^{-\beta H}- \tilde{\rho}_0^{2\M} \|_{1} \le  \frac{3e \beta}{2\beta_0|L|^2},
\end{align} 
where we set $\|e^{-\beta H}\|_1=1$.
Then, by using the Alicki-Fannes inequality~\cite{fannes1973, Alicki_2004}, the main inequality~\eqref{sup_ineq_high_dim_improved_thermal} is obtained:
\begin{align}
I(L:R)_{\rho_\beta} &\le I(L:R)_{\tilde{\rho}_0^{2\M}} +  \orderof{\beta/|L|}  \notag \\
&\le 2 \log[ {\rm SR} (\tilde{\rho}_0^{q}) ] + \orderof{\beta/|L|}  \notag\\
&\le C |\partial \Lambda|  \beta^{2/3} \log^{2/3} (\beta |\partial \Lambda|),
\end{align} 
where the inequality $I(L:R)_{\tilde{\rho}_0^{2\M}}  \le 2 \log[ {\rm SR} (\tilde{\rho}_0^{q}) ]$ is derived from the purification of $\tilde{\rho}_0^{2q}$ as 
\begin{align}
\ket{\psi} = (\tilde{\rho}_0^{q}\otimes \hat{1}) \sum_{j=1}^{\mathcal{D}_\Lambda} \ket{j}_{\Lambda}\otimes \ket{j}_{\Lambda'},
\end{align}
where $\{\ket{j}\}_{j=1}^{\mathcal{D}_\Lambda}$ is an arbitrary orthonormal basis  (see also Sec.~\ref{proof_thm:Renyi entanglement of purification}).
The mutual information $I(L:R)_{\tilde{\rho}_0^{2\M}}$ is smaller than two times of the entanglement entropy for $\ket{\psi}$~(see the inequality~\eqref{mutual_info_renyi_purifi} in the main text), which is trivially smaller than $2\log[ {\rm SR} (\tilde{\rho}_0^{q}) ]$.
This completes the proof. $\square$

%

\section{Proofs of Proposition~\ref{Prop:small_beta} and Lemma~\ref{lem:Schmidt_rank_M_beta_q}} \label{sec:app_quasi_poly_alg}

\subsection{Proposition~\ref{Prop:small_beta} for general $k$-local Hamiltonian~\eqref{sup_def:Hamiltonian}}\label{sec:app_quasi_poly_alg_prop}
We here prove the following statement about high temperatures which plays a crucial role in obtaining the quasi-linear time algorithm.

\begin{prop} \label{Prop:small_beta_general}
For $\beta \le 1/(8gk)$, we can construct a matrix product representation $M_\beta$ of $\rho_\beta$ up to an error 
\begin{align}
\| M_\beta - e^{-\beta H} \|_p \le  \epsilon \|e^{-\beta H}\|_p 
\end{align}
for an arbitrary positive $p$, where $M_\beta$ has the bond dimension of $e^{\tOrder \left(\sqrt{\log(n/\epsilon)}\right) }$.
The sufficient computational time for this construction is given by
\begin{align}
n e^{\tOrder \left(\sqrt{\log(n/\epsilon)}\right) }.
\end{align}
We notice that the computational cost does not depend on $p$.
\end{prop}

We here consider general $k$-local Hamiltonians. 
In the main text, the Hamiltonian~\eqref{def:Hamiltonian} is considered. 
By choosing 
\begin{align}
k=2, \quad g=1 
\end{align}
in Proposition~\ref{Prop:small_beta_general}, we can obtain Proposition~\ref{Prop:small_beta}. 
Here, the equation $g=1$ is derived from the condition $\max_{i\in[n]} (\|h_{i-1,i}\|+\|h_{i,i+1}\|) \le g=1$ in Eq.~\eqref{def:Hamiltonian}.

\begin{figure*}
\centering
{
\includegraphics[clip, scale=0.5]{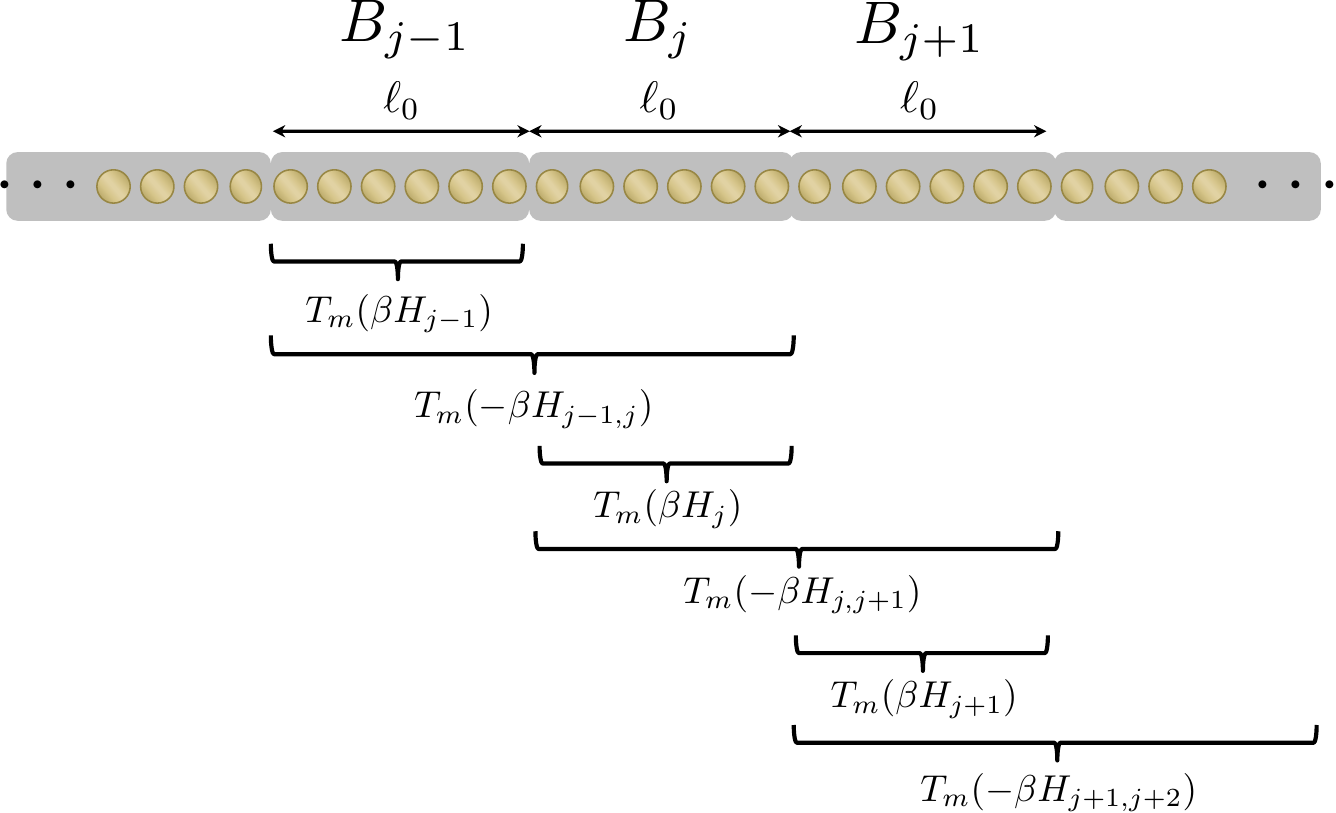}
}
\caption{Basic strategy for the approximation of $e^{-\beta H}$ ($\beta \le 1/(8gk)$). We decomposition the total systems into blocks. 
We then take the two-step approximation: 
i) $e^{-\beta H} \approx \tilde{\Phi}_{1:n_0}$ and ii) $\tilde{\Phi}_{1:n_0} \approx \tilde{\Phi}_{1:n_0}^{(m)}$, which yield $e^{-\beta H} \approx  \tilde{\Phi}_{1:n_0}^{(m)}$.
Here, the approximated quantum Gibbs state $\tilde{\Phi}_{1:n_0}^{(m)}$ is constructed from the polynomials $T_m (\beta H_{j-1})$ and $T_m (-\beta H_{j-1,j})$ as in Eq.~\eqref{sup_def_tilde_phi_j_H_j_(m)}
}
\label{sup_fig:decomp_block}
\end{figure*}

\subsection{Proof strategy}
We aim to give an explicit algorithm to obtain the MPO approximation of $e^{-\beta H}$ for $\beta \le 1/(8gk)$. 
We decompose the total system into small blocks $\{B_s\}_{s=1}^{n_0}$ with length $\ell_0$ (i.e., $|B_s|=\ell_0$), which gives $\|\Lambda\|=n_0\ell_0$ (see Fig.~\ref{sup_fig:decomp_block}).
In fact, we may not be able to find an integer $n_0$ satisfying $n=n_0\ell_0$, but we can arbitrary extend the system size $\Lambda\to \Lambda \sqcup \delta \Lambda$ without changing the Hamiltonian. We only have to add the zero operators and the form of \eqref{sup_def:Hamiltonian} is still retained as follows: 
\begin{align}
H = \sum_{X\subset\Lambda \sqcup \delta \Lambda,  |X|\le k} h_X, 
\quad \sup_{i\in \Lambda \sqcup \delta \Lambda} \sum_{X:X\ni i}\|h_X\| \le g , \notag
\end{align}
where $h_X=\hat{0}_X$ if $X\cap \delta \Lambda \neq \emptyset$. 

We then define $H_{1:j}$ as 
\begin{align}
H_{1:j} = \sum_{X\subset B_{\le j} }h_X  ,\quad B_{\le j}:= B_1 \sqcup  B_2 \sqcup \cdots \sqcup B_j .
\end{align}
By using this notation, we define operators $\Phi_j $ and $\Phi_{1:j}$ as follows:
\begin{align}
\Phi_j := e^{\beta H_{1:j-1}} e^{-\beta H_{1:j}} ,\quad \Phi_{1:j} =\Phi_1\Phi_2 \cdots \Phi_j.
\end{align}
where we define $H_{1:0}=\hat{0}$.
Note that each of $\{\Phi_j\}_{j=1}^{n_0}$ may be highly non-local. 
By using $\{\Phi_j\}_{j=1}^{n_0}$, we have 
\begin{align}
e^{-\beta H} =\Phi_{1:n_0}.
\label{sup_Phi_1_n_0_def}
\end{align}

We, in the following, derive efficient approximations for $\{\Phi_j\}_{j=1}^{n_0}$.
For the purpose, we define $\tilde{\Phi}_j$ and $\tilde{\Phi}_{1:j}$ as follows:
\begin{align}
&\tilde{\Phi}_j := e^{\beta H_{j-1}} e^{-\beta H_{j-1,j}} , \quad \tilde{\Phi}_{1:j} =\tilde{\Phi}_1\tilde{\Phi}_2 \cdots\tilde{\Phi}_j, \notag \\
&H_j := \sum_{X:X\subset B_j} h_X +  \sum_{X: X\cap B_j \neq \emptyset,X\cap B_{j+1} \neq \emptyset } h_X , \notag \\
&H_{j,j+1} :=H_j + H_{j+1}.
\label{sup_def_tilde_phi_j_H_j}
\end{align} 
Here, $H_j$ is comprised of the internal interaction in the block $B_j$ and the block-block interactions between $B_j$ and $B_{j+1}$.
We first approximate $e^{-\beta H}$ by $\tilde{\Phi}_{1:n_0}$.
Then, we approximate $\tilde{\Phi}_{1:n_0}$ by using polynomial approximations as
\begin{align}
&\tilde{\Phi}_{j}^{(m)} := T_m (\beta H_{j-1}) T_m (-\beta H_{j-1,j}) , \notag \\
&\tilde{\Phi}^{(m)}_{1:j} =\tilde{\Phi}_1^{(m)}\tilde{\Phi}_2^{(m)} \cdots\tilde{\Phi}_j^{(m)},
\label{sup_def_tilde_phi_j_H_j_(m)}
\end{align}
where $T_m (x) =\sum_{s=0}^m x^m/m!$ is the truncated Taylor expansion.
In the following, we estimate the parameters $\ell_0$ and $m$ to achieve the precision of 
\begin{align}
\left\| \tilde{\Phi}_{1:n_0}^{(m)} e^{\beta H} -1 \right\| \le \epsilon.
\label{sup_operator_norm_quasi_linear}
\end{align}
This above upper bound yields, for arbitrary Schatten $p$-norm ,
\begin{align}
\left\|\tilde{\Phi}_{1:n_0}^{(m)}  - e^{-\beta H} \right\|_p 
&\le \left\|  \left(\tilde{\Phi}_{1:n_0}^{(m)}  e^{\beta H} -1 \right )e^{-\beta H} \right \|_p  \notag \\
&\le \left\|\tilde{\Phi}_{1:n_0}^{(m)}  e^{\beta H} -1  \right\| \cdot \| e^{-\beta H}  \|_p \notag \\
&\le \epsilon \| e^{-\beta H}  \|_p ,
\label{sup_approx_error_tilde_Phi_m}
\end{align}
where we use the  H\"older inequality~\eqref{sup_generalized_Holder_ineq_operator_norm} in the second step.

In Appendix~\ref{sec:proof of choice_quasi_linear}, we prove that the inequality~\eqref{sup_operator_norm_quasi_linear} is achieved by choosing $\ell_0$ and $m$ as 
\begin{align}
\ell_0 = c_0 k\log(6n/\epsilon) \quad {\rm and} \quad m= c_1 \log(6n/\epsilon),
\label{sup_choice_of_ell_0_m}
\end{align}
where $c_0$ and $c_1$ is a constant of $\orderof{1}$.
Under the choice above, we estimate the Schmidt rank $\tilde{\Phi}_{1:n_0}^{(m)}$ across an arbitrary cut.
\begin{widetext}
Here,  $\tilde{\Phi}_{1:n_0}^{(m)}$ is given by
\begin{align}
\label{sup_explicit_form_tilde_Phi_1_n0_m}
\tilde{\Phi}_{1:n_0}^{(m)}= T_m (-\beta H_1)T_m (\beta H_1) T_m (-\beta H_{1,2}) T_m (\beta H_2) T_m (-\beta H_{2,3}) \cdots  T_m (\beta H_{n_0-1}) T_m (-\beta H_{n_0-1,n_0}) .
\end{align}
Let us consider a cut between $\Lambda_{\le i}$ and $\Lambda_{>i}$ for a fixed $i\in \Lambda$. 
Then, at most five polynomials contribute to the Schmidt rank of ${\rm SR}(\tilde{\Phi}_{1:n_0}^{(m)},i)$ (see Fig.~\ref{sup_fig:decomp_block}), where we denote them as 
$T_m (-\beta H_{j-1,j})$, $T_m (\beta H_j)$, $T_m (-\beta H_{j,j+1})$,  $T_m (\beta H_{j+1})$ and $T_m (-\beta H_{j+1,j+2})$ ($j\in [n_0]$). 
We thus obtain 
\begin{align}
\label{sup_upper_bound_of_SR_Phi_m_i}
\log \left[{\rm SR}(\tilde{\Phi}_{1:n_0}^{(m)},i) \right] \le& \log \left[{\rm SR}(T_m (-\beta H_{j-1,j}),i) \right] + \log \left[{\rm SR}(T_m (\beta H_j),i) \right] + \log \left[{\rm SR}(T_m (-\beta H_{j,j+1}),i) \right]\notag \\
&+  \log \left[{\rm SR}(T_m (\beta H_{j+1}),i) \right] + \log \left[{\rm SR}(T_m (-\beta H_{j+1,j+2}),i) \right] .
\end{align}
\end{widetext}
By using Lemma~\ref{sup_fourth_lemma_schmidt} with $\Phi_1=\Phi_2=1$ and $M=1$, we obtain from Eq.~\eqref{sup_choice_of_ell_0_m}
\begin{align}
\log \left[{\rm SR}(T_m (\beta H_j),i) \right] &\le C \max(m/\ell_0, \sqrt{m}) \log (\spin m)  \notag \\
&= \tOrder \left(\sqrt{\log(n/\epsilon)}\right) ,
\end{align}
where $C$ is a constant of $\orderof{1}$ which depends on $k$. 
Therefore, for an arbitrary cut,  $\log \left[{\rm SR}(\tilde{\Phi}_{1:n_0}^{(m)},i) \right] $ is bounded from above by $\tOrder \left(\sqrt{\log(n/\epsilon)}\right) $.
This ensures that the operator $\tilde{\Phi}_{1:n_0}^{(m)}$ is expressed by a matrix product operator with bond dimension of $e^{\tOrder \left(\sqrt{\log(n/\epsilon)}\right)}$.  
Because the operator $\tilde{\Phi}_{1:n_0}^{(m)}$ satisfies the approximation error of \eqref{sup_approx_error_tilde_Phi_m}, we prove the first part of the statement in Proposition~\ref{Prop:small_beta_general}.

In order to prove the second part of the statement, we consider the computational cost to construct the MPO of $\tilde{\Phi}_{1:n_0}^{(m)}$. 
We first note that each of the polynomials $T_m (\beta H_j)$ and $T_m (\beta H_{j,j+1})$ is described by a local MPO with bond dimension $D=e^{\tOrder \left(\sqrt{\log(n/\epsilon)}\right)}$.
In the computations of $H_j^q$ and $H_{j,j+1}^q$ ($q\le m$), we can utilize the compression of the MPO which is based on the singular value decomposition. Next, recall that we can express arbitrary local Hamiltonians by the MPO with a constant bond dimension~\cite{SCHOLLWOCK201196}.
Using this, we recursively construct the power of the Hamiltonian $H_j^q$ from $H_j^{q-1}$.
At each stage of this recursion, we ensure that the bond dimension is smaller than $D$, by compressing the MPO using the singular value decomposition. By representing the MPO in the canonical form~\cite{SCHOLLWOCK201196}, this can be performed efficiently with a computational cost of $\poly(D)$ (since the Schmidt coefficient beyond the rank $D$ is exactly equal to zero, and the error in this compression is equal to zero).
These procedures allow us to construct the local MPO of  $T_m (\beta H_j)$ and $T_m (\beta H_{j,j+1})$ with a runtime of $\poly(D)=e^{\tOrder \left(\sqrt{\log(n/\epsilon)}\right)}$. 

The remaining task is to connect all the local MPOs of $T_m (\beta H_j)$ and $T_m (\beta H_{j,j+1})$ to construct the operator $\tilde{\Phi}_{1:n_0}^{(m)}$ in  \eqref{sup_explicit_form_tilde_Phi_1_n0_m}.
From the inequality~\eqref{sup_upper_bound_of_SR_Phi_m_i}, the bond dimension is at most $D^5$, and hence the iterative multiplications of the functions $T_m (\beta H_j)$ and $T_m (\beta H_{j,j+1})$ requires $\poly(D)$ computational time, which results in the total computational time of $n\times \poly(D) =n e^{\tOrder \left(\sqrt{\log(n/\epsilon)}\right)}$.
This completes the proof of Proposition~\ref{Prop:small_beta_general}. $\square$

%

\subsection{Proof of the choice~\eqref{sup_choice_of_ell_0_m}}
\label{sec:proof of choice_quasi_linear}

We prove that the choice of \eqref{sup_choice_of_ell_0_m} achieves the approximation error~\eqref{sup_operator_norm_quasi_linear}.
In order to estimate the LHS in~\eqref{sup_operator_norm_quasi_linear}, we recursively estimate
\begin{align}
\epsilon_j:= \| \tilde{\Phi}_{1:j}^{(m)} \Phi_{1:j}^{-1} -1 \| ,
\end{align}
where we set $\tilde{\Phi}_{1:0}=\Phi_{1:0}=1$. 
Because of $\Phi_{1:n_0}=e^{-\beta H}$ as in \eqref{sup_Phi_1_n_0_def}, we have $\epsilon_{n_0}=\| \tilde{\Phi}_{1:n_0}^{(m)} e^{\beta H} -1\|$.
By using $\epsilon_j$, we can calculate the upper bound of $\epsilon_{j+1}$.
From $\Phi_{1:j}= e^{-\beta H_{1:j}} $, we have
\begin{align}
\tilde{\Phi}_{1:{j+1}}^{(m)} \Phi_{1:{j+1}}^{-1}
=&\tilde{\Phi}_{1:j}^{(m)}  \Phi_{1:j}^{-1} \left( e^{-\beta H_{1:j}}  \tilde{\Phi}_{j+1}^{(m)} \Phi_{j+1}^{-1}  e^{\beta H_{1:j}}\right)\notag \\
=&\tilde{\Phi}_{1:j}^{(m)}  \Phi_{1:j}^{-1} \tilde{\Psi}_j, 
\end{align}
where $\tilde{\Psi}_j:=e^{-\beta H_{1:j}}  \tilde{\Phi}_{j+1}^{(m)} \Phi_{j+1}^{-1}  e^{\beta H_{1:j}}$. 
We then obtain 
\begin{align}
&\tilde{\Phi}_{1:{j+1}}^{(m)} \Phi_{1:{j+1}}^{-1}-1 \notag \\
&=(\tilde{\Phi}_{1:j}^{(m)}  \Phi_{1:j}^{-1} -1) (\tilde{\Psi}_j-1)  + (\tilde{\Psi}_j-1 )+( \tilde{\Phi}_{1:j}^{(m)}  \Phi_{1:j}^{-1} -1), \notag 
\end{align}
and hence 
\begin{align}
\epsilon_{j+1} \le  \epsilon_j \delta_j+ \epsilon_j +\delta_j,
\end{align}
where $\delta_j:= \|\tilde{\Psi}_j-1\|$.
When we obtain $\delta_j \le \bar{\delta}$, we have $\epsilon_{j+1} \le  (1+\bar{\delta}) \epsilon_j + \bar{\delta}$, which yields $\epsilon_{n_0} \le  (1+\bar{\delta})^{n_0}-1$.
We here use $\epsilon_0=0$.
For $\bar{\delta} \le 1/n_0$, we have 
\begin{align}
\epsilon_{n_0} \le 2 n_0 \bar{\delta}. \label{sup_upper_bound_epsilon_n_0}
\end{align}
Therefore, the problem reduces to the estimation of $\delta_j$. 

The operator $\Psi_j$ includes the imaginary time evolution by $e^{-\beta H_{1:j}}$, but the high-temperature assumption of $\beta<1/(8gk)$ allows us to prove $\delta_j \ll 1$.
In order to calculate the upper bound of $\|\tilde{\Psi}_j -1\|$, we define
\begin{align}
&\Psi_j^{(m)}:= e^{-\beta H_{1:j}}  \tilde{\Phi}_{j+1}^{(m)} \tilde{\Phi}_{j+1}^{-1}  e^{\beta H_{1:j}}, \notag \\
&\Psi_j :=  e^{-\beta H_{1:j}}  \tilde{\Phi}_{j+1} \Phi_{j+1}^{-1}    e^{\beta H_{1:j}},  \notag \\
&\delta_{j,1} = \|\Psi_j^{(m)}-1\| ,\quad \delta_{j,2} = \|\Psi_j -1\|.
\end{align}
The above definition implies $\tilde{\Psi}_j -1 = \Psi_j^{(m)} \Psi_j -1 $, and hence 
\begin{align}
&\|\tilde{\Psi}_j -1\|  \notag \\
&\le \|(\Psi_j^{(m)} -1) (\Psi_j -1) +  (\Psi_j -1) + (\Psi_j^{(m)} -1) \|  \notag \\
&\le \delta_{j,1} \delta_{j,2} + \delta_{j,2} +\delta_{j,1}.
\label{sup_upper_delta_j_ineq}
\end{align}

Indeed, we prove the following lemmas:
\begin{lemma}\label{sup_lemma1:tilde_Phi_approx_Phi}
Under the assumption of $\beta \le 1/(8gk)$, we obtain the upper bound of
\begin{align}
\delta_{j,1} \le  (4/3)^{(2\ell_0/k) +1}\|\tilde{\Phi}_{j+1}^{(m)} \tilde{\Phi}_{j+1}^{-1}-1\|  .
\label{sup_upp_bound_delta_j_1}
\end{align}
Here, $m$ is a control parameter and can be chosen appropriately. 
\end{lemma}
\begin{lemma}\label{sup_lemma2:tilde_Phi_approx_Phi}
Under the assumption of $\beta \le 1/(8gk)$, we obtain the upper bound of
\begin{align}
\delta_{j,2} \le10g\ell_0 2^{-\ell_0/k}e^{10g \beta \ell_0/3} \le 10g\ell_0 2^{-\ell_0/(3k)},
\end{align}
where the second inequality is derived from $e^{10gk \beta/3} \le e^{5/12} < 2^{2/3}$.
\end{lemma}

Based on the above lemma, we choose the block size $\ell_0$ as 
\begin{align}
\ell_0 = c_0 k\log(1/\tilde{\epsilon}), \label{sup_choice_ell_0_1}
\end{align}
where $c_0$ is a constant such that $\delta_{j,2} \le10g\ell_0 2^{-\ell_0/(3k)} \le \tilde{\epsilon}$ and we fix $\tilde{\epsilon}$ $(<1)$ afterwards.
Also, in order to upper-bound $\delta_{j,1}$ in \eqref{sup_upp_bound_delta_j_1}, we need to estimate the norm of 
\begin{align}
&\tilde{\Phi}_{j+1}^{(m)} \tilde{\Phi}_{j+1}^{-1} -1 \notag \\
&= T_m (\beta H_{j-1}) T_m (-\beta H_{j-1,j})  e^{\beta H_{j-1,j}} e^{-\beta H_{j-1}} -1. \notag 
\end{align}
\begin{widetext}
We then obtain 
\begin{align}
\left \| \tilde{\Phi}_{j+1}^{(m)} \tilde{\Phi}_{j+1}^{-1} -1 \right\|
&\le 
\left\|  T_m (\beta H_{j-1}) \left[ T_m (-\beta H_{j-1,j})  e^{\beta H_{j-1,j}}-1\right]  e^{-\beta H_{j-1}} + 
 T_m (\beta H_{j-1}) e^{-\beta H_{j-1}} -1 \right \| \notag \\
&\le  \|  T_m (\beta H_{j-1})\| \cdot \|  e^{-\beta H_{j-1}} \| \cdot  \|  T_m (-\beta H_{j-1,j})  e^{\beta H_{j-1,j}}-1\| + 
\| T_m (\beta H_{j-1}) e^{-\beta H_{j-1}} -1 \| . \notag 
\end{align}
\end{widetext}
Because of $\|H_{j-1}\| \le g\ell_0$ and $\|H_{j-1,j}\|\le 2g\ell_0$, we have $\|  T_m (\beta H_{j-1})\| \le e^{\orderof{\beta g \ell_0}} = e^{\orderof{\ell_0/k}}$
and $\|  e^{-\beta H_{j-1}} \| \le e^{\orderof{\ell_0/k}}$.
In order to achieve $\|\tilde{\Phi}_{j+1}^{(m)} \tilde{\Phi}_{j+1}^{-1}-1\|  \le \tilde{\epsilon} (4/3)^{-(2\ell_0/k) -1}$ (or $\delta_{j,1}\le \tilde{\epsilon} $), we need to choose $m$ such that 
\begin{align}
 \|  T_m (-\beta H_{j-1,j})  e^{\beta H_{j-1,j}}-1\| \le \tilde{\epsilon} e^{- \orderof{\ell_0/k}} .
\end{align}
From $\|\beta H_{j-1,j}\| \lesssim \beta g \ell_0 =\orderof{\ell_0/k}$ for $\beta \le 1/(8gk)$, the above inequality is satisfied by choosing
$m= \orderof{\ell_0/k} + \orderof{\log(1/\tilde{\epsilon})}$.
The choice of Eq.~\eqref{sup_choice_ell_0_1} implies 
\begin{align}
m= c_1 \log(1/\tilde{\epsilon}),
\end{align}
where $c_1$ is a constant of $\orderof{1}$.

Under the above choices of $\ell_0$ and $m$, we obtain $\delta_{j,1}\le \tilde{\epsilon} $ and $\delta_{j,2}\le \tilde{\epsilon} $, and hence, from the inequality~\eqref{sup_upper_delta_j_ineq}, we have
\begin{align}
\|\tilde{\Psi}_j -1\| \le 3\tilde{\epsilon}.
\end{align}
We thus obtain $\bar{\delta}=3 \tilde{\epsilon}$, which reduces the inequality~\eqref{sup_upper_bound_epsilon_n_0} to
\begin{align}
\epsilon_{n_0} \le 6 \tilde{\epsilon}  n_0 \le 6n\tilde{\epsilon} . 
\end{align}
By choosing $\tilde{\epsilon}=\epsilon/(6n)$, we can obtain the desired precision~\eqref{sup_operator_norm_quasi_linear} between $\tilde{\Phi}_{1:n_0}^{(m)}$ and $e^{-\beta H}$.
This completes the proof.  $\square$

\subsection{Proof of Lemma~\ref{sup_lemma1:tilde_Phi_approx_Phi}}

We here consider an arbitrary operator $O_S$ supported on $S$, and derive the upper bound of
\begin{align}
e^{-\beta H_{1:j}} O_S e^{\beta H_{1:j}} = \sum_{m=0}^\infty \frac{(-\beta)^m}{m!} \ad^m_{H_{1:j}} (O_S).
\end{align}
By using Lemma~\ref{sup_norm_multi_commutator1} or the inequality~\eqref{sup_fundamental_ineq_0}, we can derive
\begin{align}
\|\ad^m_{H_{1:j}} (O_S) \| \le  (2gk)^m \|O_S\| \prod_{s=1}^m  [ |S|/k + (s-1)], \label{sup_Lemma_3_KMS2016}
\end{align}
where we use the condition that $H_{1:j}$ and $H_j$ are $k$-local operators as in Eq.~\eqref{sup_def:Hamiltonian}.
We then obtain 
\begin{align}
&\| e^{-\beta H_{1:j}} O_S e^{\beta H_{1:j}}\|  \notag \\
&\le  \|O_S\|  \sum_{m=0}^\infty \frac{(2gk\beta)^m}{m!} \prod_{s=1}^m [ |S|/k + s-1] \notag \\
&= \|O_S\|  (1-2gk\beta)^{-|S|/k},
 \label{sup_imaginary_time_evoluation_O_S}
\end{align}
where we use the equation of $(1-x)^{-y}=\sum_{m=0}^\infty x^m/m! \prod_{s=1}^m (y+ s-1)$.

We then choose $O_S$ as $\tilde{\Phi}_{j+1}^{(m)} \tilde{\Phi}_{j+1}^{-1}-1$, which yields
\begin{align}
e^{-\beta H_{1:j}} O_S e^{\beta H_{1:j}} = \Psi_j^{(m)}-1.
\end{align}
From the definitions~\eqref{sup_def_tilde_phi_j_H_j} and \eqref{sup_def_tilde_phi_j_H_j_(m)}, we have
\begin{align}
\tilde{\Phi}_{j+1}^{(m)} \tilde{\Phi}_{j+1}^{-1} = T_m (\beta H_j) T_m (-\beta H_{j,j+1}) 
e^{\beta H_{j,j+1}} e^{\beta H_{j}} , \notag
\end{align}
and hence the support of this operator satisfies 
\begin{align}
\left| {\rm Supp} \left( \tilde{\Phi}_{j+1}^{(m)} \tilde{\Phi}_{j+1}^{-1}\right) \right| \le 2 \ell_0 +k .
\end{align}
Therefore, by using the inequality~\eqref{sup_imaginary_time_evoluation_O_S} with $|S|=2\ell_0+k$, we have
\begin{align}
\| \Psi_j^{(m)}-1\| \le (4/3)^{(2\ell_0/k) +1}\|\tilde{\Phi}_{j+1}^{(m)} \tilde{\Phi}_{j+1}^{-1}-1\| ,
\end{align}
where we use $1-2gk\beta \ge 3/4$ because of $\beta \le 1/(8gk)$.
This completes the proof of Lemma~\ref{sup_lemma1:tilde_Phi_approx_Phi}. $\square$

\subsection{Proof of Lemma~\ref{sup_lemma2:tilde_Phi_approx_Phi}}

We here estimate the norm of 
\begin{align}
&\Psi_j-1  \notag \\
&= e^{-\beta H_{1:j}}  \tilde{\Phi}_{j+1} \Phi_{j+1}^{-1}  e^{\beta H_{1:j}}-1 \notag \\
&=\left(e^{-\beta H_{1:j}}  \tilde{\Phi}_{j+1} e^{\beta H_{1:j}} \right) \left( e^{-\beta H_{1:j}}\Phi_{j+1}^{-1}  e^{\beta H_{1:j}}\right)-1. \notag 
\end{align}
For the estimation, we are going to simplify the operators $e^{-\beta H_{1:j}}  \tilde{\Phi}_{j+1} e^{\beta H_{1:j}}$ and $e^{-\beta H_{1:j}}\Phi_{j+1}^{-1}  e^{\beta H_{1:j}}$.

We first consider $e^{-\beta H_{1:j}}  \tilde{\Phi}_{j+1} e^{\beta H_{1:j}}$, and start from the equation of
\begin{align}
e^{\beta H_{1:j+1}} = \br{ \mathcal{T} e^{-\int_0^\beta  e^{\tau H_{1:j}} H_{j+1}  e^{-\tau H_{1:j}} d\tau} }e^{\beta H_{1:j}},
\end{align}
where $\mathcal{T}$ is the ordering operator.
Then, from $\Phi_{j+1}^{-1}=e^{\beta H_{1:j+1}}  e^{-\beta H_{1:j}}$, the above equation reduces $ e^{-\beta H_{1:j}} \Phi_{j+1}^{-1}  e^{\beta H_{1:j}} $ to the following form:
\begin{align}
 &e^{-\beta H_{1:j}} \Phi_{j+1}^{-1}  e^{\beta H_{1:j}} =\mathcal{T} e^{-\int_0^\beta  H_{j+1}^{(\tau)}d\tau} ,  \notag\\
 &H_{j+1}^{(\tau)}:=e^{-(\beta-\tau) H_{1:j}} H_{j+1} e^{(\beta-\tau) H_{1:j}}.
 \label{sup_Phi_j+1_-1_express}
\end{align}
In a similar way, we can represent $e^{-\beta H_{j:j+1}}$ as 
\begin{align}
e^{-\beta H_{j:j+1}} = \br{ \mathcal{T} e^{-\int_0^\beta  e^{-\tau H_j} H_{j+1} e^{\tau H_j} d\tau}}e^{-\beta H_j}  ,
\end{align}
and hence we have from $\tilde{\Phi}_{j+1} := e^{\beta H_j} e^{-\beta H_{j,j+1}}$
\begin{align}
&e^{-\beta H_{1:j}}  \tilde{\Phi}_{j+1} e^{\beta H_{1:j}}  \notag \\
&= e^{-\beta H_{1:j}} \left(  \mathcal{T} e^{-\int_0^\beta e^{(\beta-\tau) H_j} H_{j+1} e^{-(\beta-\tau) H_j} d\tau}\right) e^{\beta H_{1:j}}\notag \\
&= \mathcal{T} e^{-\int_0^\beta e^{-\beta H_{1:j}} e^{(\beta-\tau) H_j} H_{j+1} e^{-(\beta-\tau) H_j} e^{\beta H_{1:j}}d\tau}  \notag\\
&=\mathcal{T} e^{-\int_0^\beta  \tilde{H}_{j+1}^{(\tau)} d\tau}, 
 \label{sup_tilde_Phi_j+1_express}
\end{align}
where we define $\tilde{H}_{j+1}^{(\tau)}$ as 
\begin{align}
\tilde{H}_{j+1}^{(\tau)}:=e^{-\beta H_{1:j}} e^{(\beta-\tau) H_j} H_{j+1} e^{-(\beta-\tau) H_j} e^{\beta H_{1:j}}.
 \label{sup_Def:tilde_H_j+1_tau}
\end{align}

We now prove the following claim:
\begin{claim} \label{sup_claim:inverse_operator}
Let $\{A_j\}_{j=1}^N$ and $\{B_j\}_{j=1}^N$ be arbitrary operators.
We also define $\Phi_{A,j} :=e^{A_1} e^{A_2} \cdots e^{A_j}$ and $\Phi_{B,j}:=e^{B_j} \cdots e^{B_2}e^{B_1}$.
We then obtain the following upper bound as 
\begin{align}
\| \Phi_{A,N} \Phi_{B,N}  -1\|  \le \bar{\Phi} \sum_{s=1}^N \| e^{A_s} e^{B_s} -1\|  ,
\label{sup_main_ineq:inverse_operator}
\end{align}
where $\bar{\Phi}:= \exp [ \sum_{s=1}^N (\|A_s\|+ \|B_s\|)]$.
\end{claim}

\textit{Proof of Claim~\ref{sup_claim:inverse_operator}.}
By using the triangle inequality, we first obtain 
\begin{align}
&\| \Phi_{A,N} \Phi_{B,N}  -1\|  \notag \\
&\le \| \Phi_{A,N-1} \Phi_{B,N-1} -1  \notag \\
&\quad \ \  + \Phi_{A,N-1}  (e^{A_N} e^{B_N} -1)\Phi_{B,N-1}\|  \notag \\
&\le \| \Phi_{A,N-1} \Phi_{B,N-1} -1\| + \bar{\Phi}\|e^{A_N} e^{B_N} -1\|,
\end{align}
where we use $\| \Phi_{A,N-1}\| \cdot \|\Phi_{B,N-1}\|\le \bar{\Phi}$.
By iteratively applying the above inequality to $\| \Phi_{A,s} \Phi_{B,s} -1\| $, we arrive at the main inequality~\eqref{sup_main_ineq:inverse_operator}. $\square$

{~}\\

\noindent 
By using the Trotter decomposition in the expressions of \eqref{sup_Phi_j+1_-1_express} and \eqref{sup_tilde_Phi_j+1_express}, we can assign as $\Phi_{A,N} \to e^{-\beta H_{1:j}}  \tilde{\Phi}_{j+1} e^{\beta H_{1:j}} $ and 
$\Phi_{B,N} \to e^{-\beta H_{1:j}} \Phi_{j+1}^{-1}  e^{\beta H_{1:j}}$ in the limit of $N\to \infty$. 
Then, from Lemma~\ref{sup_claim:inverse_operator}, we obtain 
\begin{align}
\| \Psi_j -1\| \le  \bar{\Phi}_\beta \int_0^\beta \| H_{j+1}^{(\tau)} - \tilde{H}_{j+1}^{(\beta-\tau)}\| d\tau , \label{sup_upper_bound_Psi_j}
\end{align}
where we define $\bar{\Phi}_\beta$ as 
\begin{align}
\bar{\Phi}_\beta:=\exp\left(\int_0^\beta  \| H_{j+1}^{(\tau)}\| +  \|\tilde{H}_{j+1}^{(\tau)}\|  d\tau \right). \label{sup_bar_Phi_beta}
\end{align}
To complete the proof, we need to show the following claim:  

\begin{claim} \label{sup_claim:upper_bound_of_two_quantities_Phi}
Under the assumption of $\beta < 1/(8gk)$, the following upper bounds hold: 
\begin{align}
\| H_{j+1}^{(\tau)} - \tilde{H}_{j+1}^{(\beta-\tau)}\| \le 10 g \ell_0 2^{-\ell_0/k} \label{sup_ineq1:upper_bound_of_two_quantities_Phi}
\end{align}
and 
\begin{align}
\bar{\Phi}_\beta\le e^{10g \beta \ell_0/3} . \label{sup_ineq2:upper_bound_of_two_quantities_Phi}
\end{align}
\end{claim}

{~} \\
\noindent
By applying the above claim to \eqref{sup_upper_bound_Psi_j}, we prove Lemma~\ref{sup_lemma2:tilde_Phi_approx_Phi}. $\square$

\begin{widetext}
\subsubsection{Proof of Claim~\ref{sup_claim:upper_bound_of_two_quantities_Phi}.}
We first estimate the norm of $H_{j+1}^{(\tau)} - \tilde{H}_{j+1}^{(\tau)}$. 
For this purpose, we first note that the $H_{j-1}$ is supported on the subset $B_{j-1} \sqcup \{j\ell_0 +1 ,j\ell_0 +2 , \cdots, j\ell_0 +k-1 \}$, namely 
 \begin{align}
{\rm Supp} (H_{j-1}) \subset B_j \sqcup \{j\ell_0 +1 ,j\ell_0 +2 , \cdots, j\ell_0 +k \}, \notag 
\end{align}
where ${\rm Supp} (\cdots)$ denotes the support of the operator.
On the other hand, because $H_j$ includes at most $k$-body interactions, the support of $\ad_{H_j}^{q} (H_{j+1})$ is given by
 \begin{align}
{\rm Supp} [\ad_{H_j}^{q} (H_{j+1})] \subset  \{(j+1)\ell_0 - qk  ,(j+1)\ell_0 - qk +1 , \cdots,(j+1)\ell_0 \}  \sqcup  B_{j+1} \sqcup B_{j+2}.
\end{align}
Therefore, we have 
\begin{align}
\left[ H_{j-1} ,  \ad_{H_j}^{m} (H_{j+1}) \right]  =0 \quad {\rm if } \quad k+ m k \le \ell_0 .
\end{align}
This implies 
\begin{align}
\ad_{H_j}^{m} (H_{j+1})  = \ad_{H_{1:j}}^{m} (H_{j+1})  \quad {\rm for} \quad m  \le \ell_0/k-1 .
\end{align}

Hence, from the definition~\eqref{sup_Def:tilde_H_j+1_tau} of $\tilde{H}_{j+1}^{(\beta-\tau)}$, we have 
\begin{align}
\tilde{H}_{j+1}^{(\beta-\tau)} &=\sum_{m=0}^\infty \sum_{m_1+m_2=m}\frac{(-\beta)^{m_1}}{m_1!}\frac{(\beta-\tau)^{m_2}}{m_2!} \ad_{H_{1:j}}^{m_1} \ad_{H_j}^{m_2} (H_{j+1}) \notag \\
&=\sum_{m\le \ell_0/k-1} \frac{(-\tau)^m}{m!} \ad_{H_{1:j}}^{m} (H_{j+1}) +  \sum_{m> \ell_0/k-1}^{\infty}\sum_{m_1+m_2=m}\frac{(-\beta)^{m_1}}{m_1!}\frac{(\beta-\tau)^{m_2}}{m_2!}  \ad_{H_{1:j}}^{m_1} \ad_{H_j}^{m_2} (H_{j+1}) .
\end{align}
Therefore, we have the upper bound of
\begin{align}
\| H_{j+1}^{(\tau)} - \tilde{H}_{j+1}^{(\beta-\tau)}\|  \le \sum_{m> \ell_0/k-1} \left(\frac{\tau^m}{m!} \ad_{H_{1:j}}^m (H_{j+1}) + \sum_{m_1+m_2=m}\frac{\beta^{m_1}}{m_1!}\frac{|\beta-\tau|^{m_2}}{m_2!}  \ad_{H_{1:j}}^{m_1} \ad_{H_j}^{m_2} (H_{j+1}) \right).
\label{sup_| H_j+1^(tau) - tilde{H}_j+1^(beta-tau)|_ineq_1}
\end{align}
The remaining task is to estimate the summations.
By applying the inequality~\eqref{sup_Lemma_3_KMS2016} with $O_S=h_X$, we obtain 
\begin{align}
\ad_{H_{1:j}}^{m_1} \ad_{H_j}^{m_2} (h_X) \le (2gk)^{m_1+m_2} (m_1+m_2)! \|h_X\|,
\end{align}
where $h_X$ is an interaction operator in $H_{j+1}$.
From this inequality with $m_2=0$ and the definition of $H_{j+1}$, we have 
\begin{align}
\sum_{m> \ell_0/k-1} \frac{\tau^m}{m!}\| \ad_{H_{1:j}}^m (h_X)\|\le 
\|h_X\| \sum_{m> \ell_0/k-1} (2g k \tau)^{m}\le \|h_X\| \frac{(2g k \beta)^{\ell_0/k-1}}{1-2g k \beta},
\label{sup_| H_j+1^(tau) - tilde{H}_j+1^(beta-tau)|_ineq_2}
\end{align}
where we use $\tau \le \beta$. 
From the definition of $H_{j+1}$ in Eq.~\eqref{sup_def_tilde_phi_j_H_j}, we have 
\begin{align}
\sum_{m> \ell_0/k-1} \frac{\tau^m}{m!}\| \ad_{H_{1:j}}^m (H_{j+1})\|\le 
\frac{(2g k \beta)^{\ell_0/k-1}}{1-2g k \beta} \sum_{X:X\cap B_j \neq \emptyset}  \|h_X\| 
\le \frac{g \ell_0(2g k \beta)^{\ell_0/k-1}}{1-2g k \beta} ,
\end{align}
where we use the $\sum_{X:X\cap B_j \neq \emptyset}  \|h_X\| \le \sum_{i\in B_j} \sum_{X:X\ni i}  \|h_X\| \le g |B_j|$ with the condition in Eq.~\eqref{sup_def:Hamiltonian}. 

In a similar way, we calculate 
\begin{align}
&\sum_{m> \ell_0/k-1} \sum_{m_1+m_2=m}\frac{\beta^{m_1}}{m_1!}\frac{|\beta-\tau|^{m_2}}{m_2!}  \|\ad_{H_{1:j}}^{m_1} \ad_{H_j}^{m_2} (H_{j+1}) \| \notag \\
\le &g\ell_0 \sum_{m> \ell_0/k-1}  (2g k \beta)^{m}  \sum_{m_1+m_2=m}\frac{(m_1+m_2)!}{m_1!m_2!}=\frac{g\ell_0  (4gk\beta)^{\ell_0/k-1}}{1-4gk\beta},
\label{sup_| H_j+1^(tau) - tilde{H}_j+1^(beta-tau)|_ineq_3}
\end{align}
where we use $\sum_{m_1+m_2=m}\frac{(m_1+m_2)!}{m_1!m_2!}=2^m$.
By applying the inequalities~\eqref{sup_| H_j+1^(tau) - tilde{H}_j+1^(beta-tau)|_ineq_2} and \eqref{sup_| H_j+1^(tau) - tilde{H}_j+1^(beta-tau)|_ineq_3} to \eqref{sup_| H_j+1^(tau) - tilde{H}_j+1^(beta-tau)|_ineq_1}, we obtain 
\begin{align}
\| H_{j+1}^{(\tau)} - \tilde{H}_{j+1}^{(\beta-\tau)}\|  &\le \frac{g \ell_0(2g k \beta)^{\ell_0/k-1}}{1-2g k \beta}  + \frac{g\ell_0  (4gk\beta)^{\ell_0/k-1}}{1-4gk\beta} .
\end{align}
Therefore, by using the assumption $\beta \le 1/(8gk)$, we prove the inequality~\eqref{sup_ineq1:upper_bound_of_two_quantities_Phi}.

The above analyses can also be utilized to estimate the norms of $\| H_{j+1}^{(\tau)}\|$ and $\|\tilde{H}_{j+1}^{(\tau)}\|$.
From the inequality~\eqref{sup_| H_j+1^(tau) - tilde{H}_j+1^(beta-tau)|_ineq_2}, we first obtain 
\begin{align}
\| H_{j+1}^{(\tau)}\| \le \sum_{m=0}^{\infty} \frac{\tau^m}{m!}\| \ad_{H_{1:j}}^m (H_{j+1})\|\le \frac{g\ell_0}{1-2gk\beta}.
\end{align}
From the inequality~\eqref{sup_| H_j+1^(tau) - tilde{H}_j+1^(beta-tau)|_ineq_3}, we can also derive
\begin{align}
\|\tilde{H}_{j+1}^{(\tau)}\| 
\le \sum_{m=0}^\infty \sum_{m_1+m_2=m}\frac{\beta^{m_1}}{m_1!}\frac{|\beta-\tau|^{m_2}}{m_2!}  \|\ad_{H_{1:j}}^{m_1} \ad_{H_j}^{m_2} (H_{j+1}) \| 
\le \frac{g\ell_0}{1-4gk\beta}.
\end{align}
By applying the above two inequalities to Eq.~\eqref{sup_bar_Phi_beta} under the assumption $\beta \le 1/(8gk)$, we prove the inequality~\eqref{sup_ineq2:upper_bound_of_two_quantities_Phi}.
This completes the proof of Claim~\ref{sup_claim:upper_bound_of_two_quantities_Phi}. $\square$

\subsection{Proof of Lemma~\ref{lem:Schmidt_rank_M_beta_q}} \label{sec:app_quasi_poly_alg_lemma}

We here prove Lemma~\ref{lem:Schmidt_rank_M_beta_q} in the main text, which gives the upper bound of Schmidt rank of $M_\beta^q$ with 
$M_\beta$ equal to $\tilde{\Phi}_{1:n_0}^{(m)}$ in Eq.~\eqref{sup_explicit_form_tilde_Phi_1_n0_m}: 
\begin{align}
M_\beta= T_m (-\beta H_1)T_m (\beta H_1) T_m (-\beta H_{1,2}) T_m (\beta H_2) T_m (-\beta H_{2,3}) \cdots  T_m (\beta H_{n_0-1}) T_m (-\beta H_{n_0-1,n_0}) ,
\end{align}
where $m$ and $\ell_0$ are chosen as in Eq.~\eqref{sup_choice_of_ell_0_m}.
Our purpose is to prove that, for arbitrary $q\in \mathbb{N}$, the Schmidt rank of the $q$-th power of $M_\beta$ is upper-bounded by 
\begin{align}
\label{sup_upp_SR_M_beta_q}
\log[ {\rm SR} (M_\beta^q)] \le C \max (q, \sqrt{mq}) \log(mq).
\end{align}

As shown in the inequality~\eqref{sup_upper_bound_of_SR_Phi_m_i}, for an arbitrary cut, at most five polynomials contribute to the Schmidt rank.
We denote them as $T_m (-\beta H_{j-1,j})$, $T_m (\beta H_j)$, $T_m (-\beta H_{j,j+1})$,  $T_m (\beta H_{j+1})$ and $T_m (-\beta H_{j+1,j+2})$ ($j\in [n_0]$). 
We then denote $M_\beta$ by
\begin{align}
M_\beta= \Phi_1 T_m (-\beta H_{j-1,j}) T_m (\beta H_j)T_m (-\beta H_{j,j+1})T_m (\beta H_{j+1})T_m (-\beta H_{j+1,j+2}) \Phi_2,
\end{align}
where 
\begin{align}
&\Phi_1 =  T_m (-\beta H_1)T_m (\beta H_1) T_m (-\beta H_{1,2})  \cdots T_m (-\beta H_{j-2,j-1}) T_m (\beta H_{j-1})  ,\notag \\
&\Phi_2 =T_m (\beta H_{j+2}) T_m (-\beta H_{j+2,j+3})  \cdots T_m (\beta H_{n_0-1}) T_m (-\beta H_{n_0-1,n_0}) . 
\end{align}
Note that the Hamiltonians $H_j$ and $H_{j,j+1}$ are defined on the subsets $B_j$ and $B_j\sqcup B_{j+1}$, respectively (see Fig.~\ref{sup_fig:decomp_block}).
We then apply Corollary~\ref{sup_fourth_lemma_schmidt_corol} to $M_\beta^q$ with $p=5$ and $\ell=2\ell_0$. 
The inequality~\eqref{sup_main_ineq_fourth_lemma_schmidt_corol} gives 
\begin{align}
{\rm  SR} (M_\beta^q) \le  \min_{\tilde{\ell}: \tilde{\ell} \le 2\ell_0} \left[\spin^{\tilde{\ell}}\br{10mq \spin^k}^{10q +10\tilde{\ell} + \frac{10kmq}{\tilde{\ell}}}\right] 
\le \min_{\tilde{\ell}: \tilde{\ell} \le 2\ell_0} \left[\br{10mq\spin^{2k}}^{10q +10\tilde{\ell} + \frac{10kmq}{\tilde{\ell}}}\right]. 
\end{align}
\end{widetext}

We now choose $\tilde{\ell}$ as 
 \begin{align}
 \label{sup_choice_tilde_ell_Lemma6}
\tilde{\ell} = \sqrt{kmq} = \sqrt{\frac{c_1}{c_0} \ell_0 q},
\end{align}
where the second equation comes from the choice of~\eqref{sup_choice_of_ell_0_m}. 
Because of the constraint $\tilde{\ell}\le 2\ell_0$, the exponent $q$ should satisfy 
 \begin{align}
q\le \frac{4c_0}{c_1} \ell_0 = \frac{4c_0^2 k}{c_1} \log(6n/\epsilon).
\end{align}
Under this condition, we can choose $\tilde{\ell}$ as in Eq.~\eqref{sup_choice_tilde_ell_Lemma6} and hence we obtain 
\begin{align}
\label{sup_upp_1_Lemma6}
\log[{\rm  SR} (M_\beta^q)] 
&\le C' \log (mq) [q + \sqrt{mq}]  \notag \\
&\le C \sqrt{mq} \log (mq) , 
\end{align}
with $C'$ and $C$ constants of $\orderof{1}$,
where we use $q \lesssim m$ because of $q\le \frac{2c_0}{c_1} \ell_0$ and $\ell_0 \propto m$ from Eq.~\eqref{sup_choice_of_ell_0_m}. 

On the other hand, for $q>\frac{4c_0}{c_1} \ell_0$, we cannot choose $\ell_0$ as in \eqref{sup_choice_tilde_ell_Lemma6}. 
We here choose $\tilde{\ell}=2\ell_0$, and obtain 
\begin{align}
10q +10\tilde{\ell} + \frac{10kmq}{\tilde{\ell}} &= 10q +20\ell_0 + \frac{5kmq}{\ell_0} \notag \\
&\le \left(10 + \frac{5c_1}{c_0} + \frac{5c_1}{c_0} \right) q,
\end{align}
where we use $\ell_0< c_1q/(4c_0)$ and $m/\ell_0=c_1/(c_0k)$ from Eq.~\eqref{sup_choice_of_ell_0_m}.
We thus obtain 
\begin{align}
\label{sup_upp_2_Lemma6}
\log[{\rm  SR} (M_\beta^q)] \le C q \log (mq) . 
\end{align}
By combining the inequalities~\eqref{sup_upp_1_Lemma6} and \eqref{sup_upp_2_Lemma6}, we obtain the main inequality~\eqref{sup_upp_SR_M_beta_q}. 
This completes the proof. $\square$

%


%

\end{document}